# THESE

Présentée pour obtenir le grade de

**DOCTEUR DE L'ECOLE POLYTECHNIQUE**

**Spécialité : Physique du Solide**

Par

**Emmanuel LHUILLIER**

# TRANSPORT ELECTRONIQUE DANS LES SUPER RESEAUX : applications aux détecteurs infrarouges à grandes longueur d'onde

**Thèse soutenue le 18 octobre 2010 devant le jury composé de :**

| | | |
|---|---|---|
| Pr Vincent BERGER | PARIS DIDEROT | Président du jury |
| Dr Robson FERREIRA | LPA | Rapporteur |
| Dr François JULIEN | IEF | Rapporteur |
| Pr Jérôme FAIST | ETH | Examinateur |
| Dr Alexandru NEDELCU | ALCATEL THALES III-V Lab | Examinateur |
| Pr Emmanuel ROSENCHER | ONERA | Directeur de thèse |
| Dr Isabelle RIBET-MOHAMED | ONERA | Encadrant de thèse |

# RESUME

L'imagerie infrarouge bas flux requiert des détecteurs grandes longueurs d'onde de hautes performances. Les détecteurs à puits quantiques (QWIP), de par la maturité de GaAs, la facilité à ajuster la longueur d'onde détectée sur une très large gamme et la possibilité de réaliser de larges matrices uniformes constituent d'excellents candidats pour ces applications. Afin de confirmer leur intérêt nous avons procédé à la caractérisation électro-optique fine d'un composant QWIP détectant à 15µm. Les performances mesurées ont été utilisées pour simuler celles d'une caméra basée sur ce détecteur et dédiée à un scénario faible flux et ont permis de valider la capacité de la filière QWIP à répondre à de telles missions infrarouges. Ces simulations ont aussi mis en évidence le rôle extrêmement préjudiciable joué par le courant d'obscurité. Nous avons alors mis au point une simulation basée sur un code de diffusion entre états localisés qui nous a permis de mieux appréhender le transport dans ces structures. Un important travail de développement de l'outil de simulation a été nécessaire. Ce code a révélé le rôle déterminant du profil de dopage sur le niveau de courant d'obscurité. Nous avons ainsi pu réaliser de nouvelles structures aux profils de dopage optimisés et dont le niveau de courant d'obscurité est abaissé de 50%. Nous avons par ailleurs pu apporter une interprétation quantique à la forme des courbes *I(V)* observée. Mais notre code de simulation s'avère plus généralement un outil puissant de simulation du transport dans les hétérostructures. L'influence des défauts de croissance (défauts d'interface et désordre) a pu être quantifiée et nous avons pu apporter les premières prédictions de performances de QCD THz. Enfin l'influence des effets non locaux sur le transport a été étudiée. L'observation de dents de scie sur les courbes *I(V)* de QWIP a pu être modélisée et son influence sur la détectivité évaluée.

# MOTS CLEFS

Infrarouge, super réseaux, puits quantiques, transport électronique, QWIP, QCD, diffusion.

---

# ABSTRACT

The low flux infrared imaging needs performant high wavelength detectors. Quantum Well Infrared Photodetectors (QWIP), thanks to the maturity of GaAs, the possibility to adjust the detected wavelength on a large range and to realize large uniform matrix are good candidate for such applications. In order to validate this interest, we have performed an electro-optic characterization of a 15µm sample. These measurements have been used to simulate the performance of a camera based on this QWIP and used in a low infrared photons flux scenario. We predict that this QWIP would succeed. Nevertheless these simulations also underline the detrimental role of the dark current. Thus we have developed a simulation tool based on a hoping approach between localized states, which provide us a better understanding of the transport in these heterostructures. The code has in particular underlines the role plays by the electron –ionized impurities interaction, which make the dark current very sensitive to the doping profile. Using this tool we have designed new structures, with optimized doping profile, in which the scattering rate has been decreased by a factor two. Moreover we have identified a quantum origin to the plateau shape of the I(V) curve. This code is more generally a useful simulation tool for the transport in hétérostructures. The influence of growth defects (non ideal interface and disorder) has been quantized and we have performed the first evaluation of The $R_0A$ in a THz QCD. Finally non local transport effects have been investigated. Saw teeth observation on the I(V) curves have been modeled and their influence on the detectivty estimated.

# KEY WORDS

Infrared, Superlattice, QWIP, QCD, doping, hopping, scattering, electronic transport,

# REMERCIEMENTS

*Au meilleur d'entre nous,*

Etonnant paradoxes que ces remerciements, quelques pages qui viennent donc conclure ces trois années et dix-huit jours de travail. Dernières écrites et pourtant placées au début. Si peu pertinentes et pourtant une bonne partie des lecteurs n'ira pas au-delà. Ce devrait être un plaisir à écrire, celui de la dernière brique apportée à l'édifice, mais pourtant leur écriture peut devenir étonnement « angoissante ». J'imagine déjà qu'à la lecture de ce dernier mot, certains peuvent avoir des sueurs froides à l'idée de ce qui va suivre, mais je vais tâcher de rester sage.

Il est souvent de coutume de commencer par remercier les membres du jury mais je vais déroger une première fois aux traditions pour commencer par remercier ceux qui m'ont amené ici. Nous disposons finalement assez rarement d'une telle tribune pour ne pas manquer l'occasion quand celle-ci se présente à nous. Aussi loin que je ne me souvienne je n'ai pas eu de rêve d'enfant consistant à vouloir devenir pompier ou astronaute. Mon premier désir de ce type était de devenir trader. J'ai néanmoins dû renoncer à ce rêve le jour ou j'ai compris que j'avais fort peu de compétence pour demander de l'argent à mon prochain. Néanmoins à la vue de certains événements récents je me dis que ce n'était pas un si mauvais choix. J'ai donc alors dû trouver une vocation palliative, or mes compétences littéraires ayant également montré leurs limites c'est donc vers les sciences que je me suis tourné. Et c'est finalement en cours de physique et chimie que j'ai réussi à trouver mon bonheur. Je dois l'avouer Mme Bonzon, Mr Morand, Mr et Mme Mercier et Mr Héritier m'ont procuré un certain nombre d'orgasmes intellectuels Je dois certainement associer à mes remerciements Jérôme Lesueur sans qui j'aurais sans doute passé ma thèse au fond d'un tunnel à la frontière Franco-Suisse.

Je me retrouve ainsi à 23 ans, encore plein de belles illusions, à la recherche d'une thèse. C'et lors d'une visite au LPA que je trouve que le principe de mettre en résonance des niveaux dans une hétérostructure polarisée est une idée géniale qui pourrait bien m'occuper pour les trois prochaines années. C'est donc comme cela que, dans l'énorme recueil de sujet possible du master CFP, je me suis retrouvé à prendre contact avec Vincent Berger. Et comme de plus il sait très bien vendre son sujet voilà comment je me suis retrouvé embrigadé. Ce que j'ignorais c'est que le monde de la détection infrarouge soit si petit et je me suis rapidement retrouvé sur un autre projet commun à l'Onera, Paris 7 et le III-V lab.

Trois ans plus tard, je peux dire que ce sujet m'a beaucoup plu. De Gennes l'aurait certainement qualifié de fruit un peu mur, mais il constitue en même temps un excellent complément de formation. Ce qui de plus a rendu cette thèse agréable est la totale liberté dont j'ai pu profiter au cours de ces années. Merci pour m'avoir encourager à enseigner, m'avoir laissé suivre autant de formation si variée et pour m'avoir laissé une grande liberté dans les orientations du sujet.

Cet historique étant à présent fini, je peux enfin passer aux remerciements.

*Aux membres du jury :*

Je tiens tout d'abord à remercier Robson Ferreira et François Julien pour avoir accepter d'être les rapporteurs de ma thèse. J'ai réellement apprécié le temps que vous avec pris pour me recevoir et discuter avant ma soutenance de cette thèse.

Merci également à Jérôme Faist d'avoir accepter d'être dans mon jury. Mais mes remerciements à votre égard portent également sur la collaboration que nous avons eue pendant cette thèse.

Bruno Desruelle a été mon correspondant DGA, nous avons eu l'occasion de nous rencontrer a plusieurs reprises (doctoriale, soutenance de thèse ou lors des journées DGA) à chaque fois il a eu la sympathie de venir échanger quelques mots sur l'avancée de ma thèse ou de me prodiguer quelques précieux conseils, je l'en remercie. Je lui souhaite bonne chance dans sa nouvelle carrière.

*A mes encadrants :*

Tout d'abord Emmanuel qui a été mon directeur de thèse pendant ces trois années, j'ai beaucoup apprécié ton dynamisme, même si il m'a demandé ensuite de passer beaucoup de temps à faire le tri dans ton abondance d'idée. Je te remercie pour m'avoir fait profiter de ton carnet d'adresse et pour les conseils que tu as pu me prodiguer. La prochaine étape de ma carrière à Chicago te doit sans aucun doute beaucoup…mille fois merci.

Isabelle a été mon encadrante à l'Onera, il nous aura fallu un peu de temps avant de trouver notre rythme de croisière mais je crois finalement que ce fut une collaboration efficace. Je te remercie également de la liberté que tu m'as donnée dans les orientations de cette thèse. Chers membres du jury c'est à elle que vous devez d'avoir lu un manuscrit acceptable.

Vincent, merci à toi de m'avoir orienté vers cette thèse, et d'y avoir impulser les bonnes idées nécessaires au démarrage de cette thèse. Je garderais un très bon souvenir de notre road-trip américain. Je n'en dirais pas plus et terminerais mes remerciements à ton égard par un simple « Bravo et merci ».

Enfin Alexandru. Je crois que j'ai vraiment eu une bonne idée de rendre ma carte d'étudiant à l'accueil de Thales en ta présence la première fois que nous nous sommes rencontré. Nous avons ainsi découvert que nous étions deux PC avec dix promos d'écart, je crois que cela a définitivement influencé notre relation. Mais je veux surtout te remercier pour m'avoir fait partager ta connaissance du détecteur QWIP, tes relectures non-complaisantes et pour avoir permis une collaboration efficace entre nos deux équipes.

*A tous ceux avec qui j'ai pu travailler :*

Ce travail a majoritairement été réalisé au sein de l'équipe CIO du DOTA. Je tiens tout d'abord à remercier Michel, le géo trouve tout du labo, la partie expérimentale de mon travail lui doit beaucoup. Ses conseils pour réaliser les mesures et sa disponibilité m'ont permis d'avancer rapidement. Ce labo ne serait pas le même sans Michel ! Je remercie également Marc pour ses conseils de programmation qui m'ont ainsi permis de passer plus de temps à faire de la physique que des constructeurs par copie. Merci à Sylvain pour m'avoir appris à me servir du FTIR. Je n'oublie pas non plus Eric, l'homme par lequel une idée de manip devient réalité. J'ai beaucoup apprécié ses réalisations toujours expresses de mes pièces mécaniques. J'inclus également dans cette liste Brigitte, Valérie et Christelle qui m'on rendu ces trois années beaucoup plus simples.

Ses trois années à l'ONERA n'auraient sans aucun douten pas été les mêmes sans mes deux collègues de bureau qu'ont été Bruno et Gregory. Nos conversations improbables de toute nature ont instauré une bonne humeur. Merci à Bruno pour son million de petits utilitaires (jabref, wmftoeps…) et autres conseils informatiques. Mes collègues de bureau suivants n'ont pas su apprécier à leur juste valeur les grognements (LAAAAPIN, Heeeather…) et autres délires du genre (C&C) aussi bien que vous ! Le travail nous a quelque peu éloigné, mais j'avoue toujours apprécier d'objectivement discuter en votre présence.

Je voudrais également remercier Riad, étoile montante, pour nos échanges dans le RER et m'avoir fait partager ta motivation intellectuelle. J'avoue avoir été impressionné par ta capacité à faire avancer les choses.

Je n'ai plus qu'a souhaiter bonne continuation à tous les jeunes thésards qui continuent leur thèse à l'Onera : (Salut) Fred et (Salut) Patrick, Charly (PC Style)…

Je remercie également Laetitia Doyennette a qui je dois la technos des mes derniers échantillons. J'ai également apprécié les discussions que nous avons pu échanger notamment lors de nos transhumances entre Palaiseau et Paris 7. Tes conseils sur le monitorat et sur les concours m'ont été très utiles.

Bien que mes passages à Paris 7 furent assez clairsemés au cours de ces trois années les permanents (Guisepe, Sara, Ivan.) et les thésards (Erwan, Xavier, Alberto…) m'ont toujours réservé un accueil très chaleureux. Je pense plus spécifiquement à Alexandre qui, à la fin de cette année, sera le dernier étudiant représentant la communauté QWIP à P7, garde la niak ! Une pensée particulière va à Angela pour ses conseils physiques et pour m'avoir offert ses codes k·p. J'ai grandement apprécié ta disponibilité pour répondre à mes questions.

*A tous ceux avec qui j'ai pu collaborer :*

Bien que notre collaboration ait commencé dès le début de ma thèse, c'est lors des conférences QSIP que j'ai vraiment pu faire connaissance avec l'équipe détection de Thales (Philippe, Eric, Mathieu, Virginie, Thomas, Arnaud….). Cette thèse vous doit sans aucun doute beaucoup. Merci pour les connaissances que vous avez pu m'apporter sur les III-V et plus généralement sur les détecteurs infrarouges. Je dois bien sur dire quelques mots particuliers sur mes compagnons de thèse que sont Amandine et Vincent. Merci à vous deux pour m'avoir appris à mieux travailler avec Paris 7 et Thales. J'espère seulement que nos échanges vous auront autant apportés qu'ils m'en ont apporté…Il ne me reste plus qu'à vous souhaiter bonne route.

J’ai rencontré Gilles Patriarche grâce au projet d’Erwan sur la rugosité, et c’est donc finalement moi qui aurait eu les retombés de cette collaboration. Merci Gilles d’avoir fait pour moi ses observations STEM si vite. Ce fut un échange d’autant plus enrichissant qu’il n’était justement pas avec quelqu’un de la communauté détecteur. Et surtout merci, car ces mesures m’ont permis d’écrire deux papiers.

J’ai eu l’occasion à deux reprises de rencontrer Nicolas Péré-Laperne, une première alors qu’il était encore en thèse au LPA et une seconde alors qu’il revenait de sa truculente escapade à Northwestern. Ton passage à l’ONERA aura constitué pour moi un havre de QWIP dans ce monde de MCT.

Je voudrais de nouveau remercier Jerome Faist et Fabrizio Castellano pour leur collaboration et leur aide dans la compréhension des phénomènes de transport. Leur culture complémentaire nous a permis d’apporter un point de vue alternatif. Et la visite que j’ai pu faire de leur groupe à Zurich, avec Emmanuel, restera comme un moment intellectuellement bouillonnant de cette thèse.

*Aux autres :*

Il y a également ceux que je ne souhaite pas remercier, au premier rang desquels la RATP et ses (non) conducteur de la ligne B. En plus de déjà y passer deux heures par jour, je ne les remercie pas pour ses 300 et quelques heures supplémentaires que leurs grèves m’auront permis de chaudement placer en compagnie d’improbables compagnons d’infortune. Je leur dessers une mention spéciale pour leur grève reconductible en période de soutenance.

Enfin merci à mes parents pour m’avoir offert le cerveau adéquat à la réalisation de cette thèse. Pater mille fois merci pour tes corrections.

Et pour finir, Sandrine, merci pour ses trois années (enfin rassure toi, ce n’est qu’un début puisque tu as signé pour beaucoup plus) durant lesquels tu as su me supporter, particulièrement ces soirées post journées ratées. Nous nous étions fixé trois objectifs pour cette année et je crois que c’est vraiment en bonne voie. Mais cette belle aventure ne fait que commencer…

*A ceux qui auront apprécié ma folie*

# TABLE DES MATIERES

# NOTATIONS

## Constantes physiques

| Symbole | Valeur | Description |
|---|---|---|
| e | $1.6\times10^{-19}$ C | Charge du proton |
| c | 299 792 458 $m\cdot s^{-1}$ | Vitesse de la lumière |
| h | $6.62\times10^{-34}$ J·s | Constante de Planck |
| ħ | $1.05\times10^{-34}$ J·s | Constante de Planck réduite |
| $k_b$ | $1.38\times10^{-23}$ $J\cdot K^{-1}$ | Constante de Boltzmann |
| $m_0$ | $9.1\times10^{-31}$ kg | Masse de l'électron dans le vide |
| $\varepsilon_0$ | $8.85\times10^{-12}$ $m^{-3}\cdot kg\cdot s^{4}\cdot A^{2}$ | Permittivité du vide |
| eV | $1.6\times10^{-19}$ J | Electron volt |
| meV | $1.6\times10^{-22}$ J | milli-electronvolt |

## Notations propres aux hétérostructures

| Symbole | Description |
|---|---|
| $L_b$ | Dimension d'une barrière |
| $L_w$ | Dimension d'un puits |
| $L_d$ ou d | Dimension d'une période |
| $E_1$ | Energie du niveau fondamental |
| $E_2$ | Energie du premier niveau excité |
| T | Température |
| $E_f$ | Energie de Fermi |
| $n_{2D}$ | Dopage surfacique du puits |
| $n_{3D}$ | Dopage volumique du puits |
| $n_{3D}^{c}$ | Dopage volumique des contacts |
| F | Champ électrique sur la structure |

# ABBREVIATIONS

Voici un certain nombre d'acronymes et abréviations que j'utiliserai par la suite :

THz : térahertz

QWIP : Quantum Well Infrared Photodetector ou détecteur à Multi Puits Quantiques (MPQ).

MBE : Molecular Beam Epitaxy ou épitaxie par jet moléculaire.

MOCVD : MetalOrganic Chemichal Vapor Deposition.

MOVPE : MetalOrganic Vapour Phase Epitaxy

QCD : Quantum Cascade Detector

QCL : Quantum Cascade Laser

MCT : Mercury Cadmium Telluride

FTIR : Fourrier Transform infrared spectrometer.

SWIR : Short Wavelength InfraRed (1µm → 2.5µm).

MWIR : Middle Wavelength InfraRed (3µm → 5µm).

LWIR : Long Wavelength InfraRed (8µm → 12µm).

VLWIR : Very Long Wavelength InfraRed (>12µm).

RDN : Résistance différentielle négative.

# INTRODUCTION

Quand, dans les années 70, Esaki a réalisé la première hétérostructure de semi-conducteurs, il a réussi à faire passer le monde des basses dimensionnalités d'hypothèse de théoricien à l'une des physiques les plus riches de la fin du XX$^{éme}$ siècles. Il a ainsi pavé la route à la fabrication contrôlée de nanostructures, réalisant ainsi le fantasme de Feynman[1].

Une quinzaine d'années plus tard, ces hétérostructures ont commencé à être utilisées pour l'émission et la détection de rayonnement infrarouge. Le puits quantique, brique de base de ces dispositifs, permet alors un contrôle des propriétés optiques via des propriétés géométriques, ce qui constitue une véritable révolution.

Sautons encore une vingtaine d'année : la technologie des détecteurs QWIP est devenue mature et est commercialisée couramment. Elle permet d'atteindre d'excellentes performances en LWIR et en régime thermique. Les détecteurs à puits quantiques (QWIP) profitent de la maturité de GaAs, ce qui permet la réalisation de larges plans focaux et une facilité à ajuster la longueur d'onde détectée. Les QWIP sont de plus l'une des rares filières qui permette d'adresser les très grandes longueurs d'onde. Le VLWIR dédié au bas flux est un domaine spectral extrêmement exigeant pour les détecteurs en raison des faibles flux de photons qui lui sont associés. Il en résulte un régime de fonctionnement très spécifique (tunnel) qui nécessite une étude à part entière qui sera développée au cours de ce manuscrit.

L'objectif de cette thèse était de comprendre le transport dans les super-réseaux faiblement couplés en régime tunnel afin d'optimiser les performances des détecteurs reposant sur cette structure, au premier lieu desquels les QWIP. Pour cela je suis parti de la caractérisation d'un détecteur pour identifier les points sur lesquels devait se focaliser le travail d'optimisation. Un important travail de modélisation, du transport tunnel dans les QWIP hautes longueurs d'onde, a alors pu commencer. Ce travail a permis d'améliorer notre compréhension des phénomènes de transport en jeu et il m'a amené au développement d'un outil de simulation du transport quantique. Cet outil a ensuite permis de développer des structures optimisées répondant aux besoins identifiés lors de la caractérisation préliminaire.

Dès le départ cette thèse a été conçue comme une collaboration tripartite. L'ONERA a apporté sa compétence pour l'étude et la caractérisation des détecteurs infrarouges. Alcatel-Thales III-V Lab a travaillé à la réalisation et sur la physique des composants et enfin le laboratoire Matériaux et phénomènes quantiques (MPQ) de l'université Paris 7 a amené son expertise dans la modélisation de ces structures.

# ORGANISATION DU MANUSCRIT

Le mémoire se décompose en six grandes parties. Le premier chapitre commence par des notions de base à propos des hétérostructures et de l'infrarouge. J'y présente les applications et les filières de détection qui sont propres aux grandes longueurs d'onde. La fin de ce chapitre se focalise sur le fonctionnement de la filière QWIP qui est l'objet de cette thèse. Un état de l'art des performances des QWIP hautes longueurs d'onde est également présenté.

La seconde partie s'intéresse à la caractérisation d'un QWIP à grande longueur d'onde (15µm) dans des conditions d'utilisation aussi proches que possible de celles définies par les applications. J'y présente en particulier le banc de caractérisation aux faibles flux que j'ai pu réaliser. Enfin nous utilisons les résultats de la caractérisation pour estimer, à partir d'un code de simulation, les performances d'une caméra reprenant comme détecteur le composant QWIP 15µm. Ce code nous a également permis d'identifier que le courant d'obscurité reste encore le point faible des QWIP utilisés dans des applications à faibles flux.

Fort du constat fait au chapitre précèdent, il fallait alors modéliser le transport tunnel dans ces structures. Dans ce troisième chapitre je présente les différentes approches du transport que nous avons pu envisager pour le traitement du transport à basse température. J'insiste particulièrement sur la physique contenue dans chacun des modèles. A la fin de ce chapitre j'essaie de dresser une vision cohérente de l'utilisation de ces différents modèles.

Le quatrième chapitre se focalise sur le modèle de transport que nous avons finalement retenu. C'est une approche de transport par hopping (saut entre états localisés). Je détaille dans ce chapitre les procédures d'évaluation des fonctions d'onde et des taux de diffusion entre états. J'apporte également une interprétation quantique à la forme des courbes *I(V)* mesurées au chapitre deux. L'un des enjeux de ce chapitre est d'identifier le(s) mécanisme(s) dominant(s) à l'origine du transport et sur lequel (lesquels) il faut jouer pour réduire le courant d'obscurité.

Le chapitre cinq présente les principaux résultats issus de l'exploitation de l'outil de simulation présenté au chapitre quatre. Il se subdivise en trois sous-parties. La première s'intéresse à l'influence du profil de dopage sur le niveau de courant d'obscurité. L'outil de simulation m'a permis de dessiner des structures optimisées que nous avons pu faire réaliser par le III-V Lab et dont je présente ici les performances. La suite de ce chapitre montre comment cet outil de simulation peut être utilisé dans des structures plus complexes que le QWIP idéal. J'y étudie en particulier les effets des défauts d'interfaces et de l'introduction de désordre Coulombien sur les propriétés optiques et de transport des structures QWIP hautes longueurs d'onde. Enfin la fin de ce chapitre est dédiée à la prédiction des performances des QCD THz.

Dans les cinq premiers chapitres de cette thèse je me suis focalisé sur la compréhension du transport local au coeur de la structure. Dans le dernier chapitre j'inclus les effets liés aux contacts et à la distribution inhomogène du champ électrique. Ceci doit nous permettre de rendre compte de l'observation de dents de scie sur les courbes *I(V)* de certains QWIP. J'étudie aussi leur influence sur la détectivité du composant.

# 1. HETEROSTRUCTURES ET DETECTION INFRAROUGE

## Sommaire

Ce chapitre introductif va se décomposer de la façon suivante, je commencerai par quelques rappels sur le rayonnement infrarouge. En parallèle je m'intéresserai au développement des nano-structures de basse dimensionnalité et je montrerai comment ces deux branches de la physique ont fini par se croiser pour conduire à des systèmes complexes de détection infrarouge. Je dresserai ensuite un état de l'art des différentes filières de détection quantique infrarouge en me focalisant particulièrement sur la problématique des hautes longueur d'onde, de 12µm jusqu'au THz.

# 1.1. L'Infrarouge

## 1.1.1. Historique

Au cours du XIX$^{\text{éme}}$ siècle, chaque pays de la vieille Europe, alors en proie à l'euphorie de la révolution industrielle et dans un contexte de nationalisme exacerbé, s'enorgueillissait de la découverte d'une partie du spectre électromagnétique : En 1800 l'anglais (d'origine allemande) Herschel fit la découverte de l'infrarouge. Cette découverte est suivie de près par celle de l'allemand Ritter, des ultraviolets en 1801. En 1864 ce fut au tour de l'Ecosse d'être à l'honneur grâce à la théorie de l'électromagnétisme de Maxwell[2]. L'allemand Hertz valida cette théorie en réalisant un dispositif d'émission d'ondes radio (1886). Le siècle finit en fanfare avec la découverte par Röntgen (1895) des rayons X, puis des rayons gamma par le français Villard (1900).

La détection infrarouge n'en est alors qu'à ses balbutiements, puisque l'instrument de Herschel est constitué d'un monochromateur dont le détecteur est un simple thermomètre. C'est sous la pression des astronomes que la détection infrarouge va petit à petit tendre vers des instruments performants spécifiques à l'infrarouge. Jusqu'au milieu du XX$^{\text{eme}}$ siècle, la détection est de type thermique et repose sur des effets thermoélectriques (effet Seebeck), pyroélectriques ou thermomécaniques. La détection infrarouge connaît un véritable pas en avant avec l'apparition des premiers photo-détecteurs au début du XX$^{\text{éme}}$. Après la seconde guerre mondiale et dans un contexte de guerre froide, les militaires voient dans la détection infrarouge la possibilité de détecter « l'invisible ». Les armées deviennent alors les premiers utilisateurs de ce type de détection et sont à la source des besoins les plus poussés.

## 1.1.2. Physique de l'infrarouge

### *1.1.2.1. De la catastrophe ultraviolette à la théorie des quanta*

Au début du XX$^{\text{ème}}$ siècle Rayleigh et Jeans ont démontré que les lois de la physique classique conduisaient à une énergie infinie pour le spectre du corps noir. Ce phénomène, qui porte le nom de catastrophe ultraviolette a conduit, avec l'effet photo électrique, au développement de la mécanique quantique. Planck développa alors la notion de quanta d'énergie, le photon était né ! La quantification de l'énergie des modes du champ électromagnétique conduit Planck à la loi[3] qui porte aujourd'hui son nom et qui donne la luminance spectrique énergétique rayonnée par un corps noir, par unité de surface, d'angle solide et par unité de longueur d'onde :

$$\left[\frac{\partial L(\lambda,T)}{\partial\lambda}\right]_{CN}^{T} = \frac{2hc^2\lambda^{-5}}{\exp^{\frac{hc}{\lambda k_b T}}-1} \qquad (1\text{-}1)$$

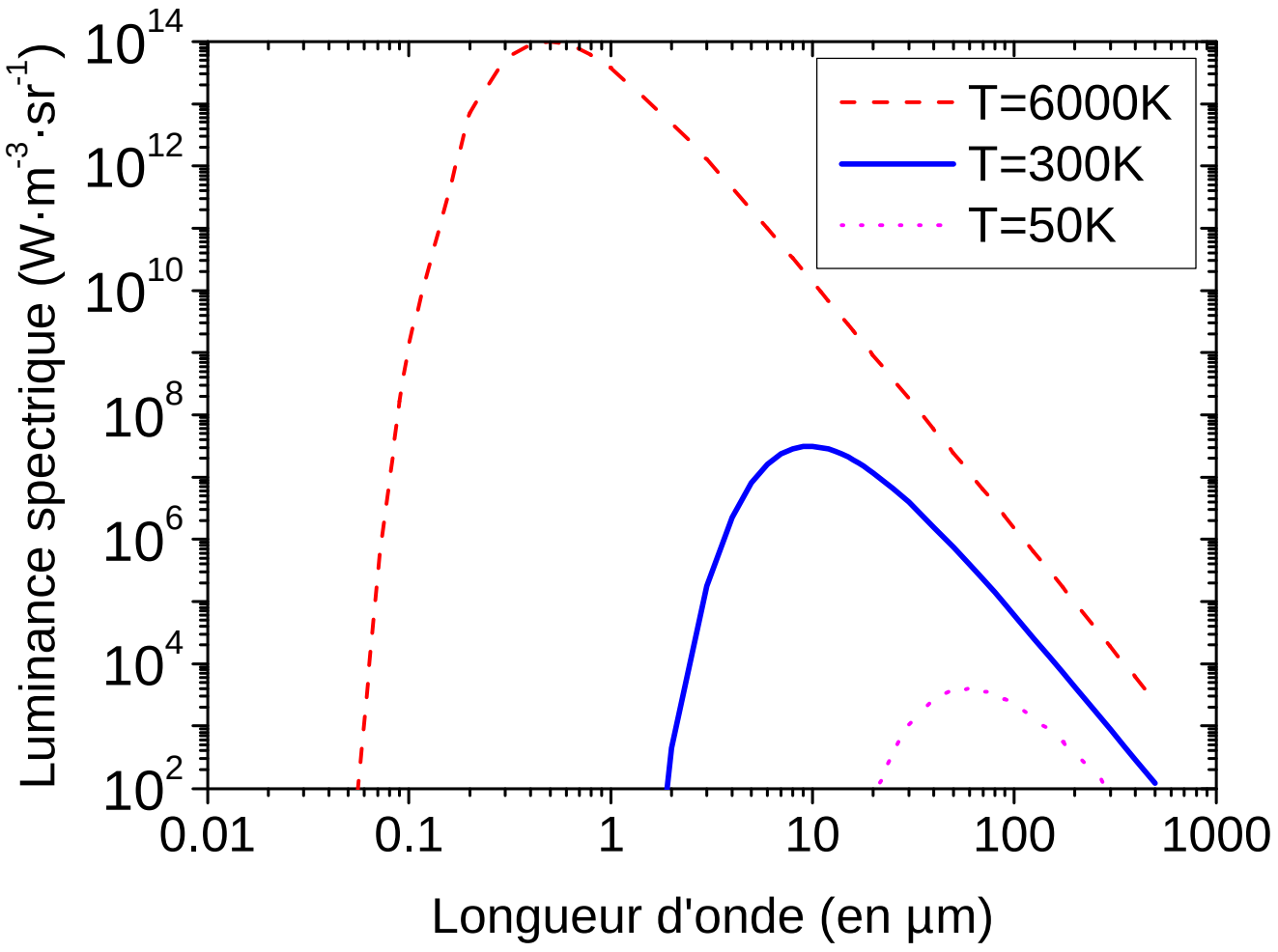

*FIG. 1-1 Loi de Planck pour trois températures de corps noir.*

Dans l’expression (1-1), T est la température du corps noir considéré et λ la longueur d’onde. La figure FIG. 1-1 montre la loi du rayonnement du corps noir pour trois températures typiques. 6000K correspond à la température équivalente du soleil, la nature a donc adaptée la longueur d’onde perceptible par l’œil humain (0.4µm→0.8µm) au maximum du rayonnement de notre principale source d’éclairement. 300K correspond à la température de notre environnement (0.5µm), l’imagerie thermique classique a donc tendance à se focaliser sur le maximum de rayonnement situé autour de 10µm. Enfin si la température de la source diminue, le rayonnement est décalé vers les grandes longueurs d’onde (au delà de 12µm), c’est le cas d’une source à 50K. Notons que pour deux températures de corps noir différentes, les courbes de rayonnement ne se croisent pas, c’est un avantage en spectrométrie car un spectre permet de remonter de façon univoque à la température de la source. Mais c’est une grande difficulté pour l’imagerie d’objet froid, dans la mesure ou d’importants problèmes de flux parasites sont à prévoir.

### *1.1.2.2. Loi de Wien et loi de Stefan*

En intégrant et en dérivant la loi de Planck, nous obtenons respectivement la loi de Stefan et la loi du déplacement de Wien. La loi de Stefan donne l’excitance totale d’un corps noir pour une température donnée :

$$[Me]_{CN}^{T} = \int_{demi-espace} d\Omega \int_0^{\infty} \frac{\partial L}{\partial \lambda}(\lambda, T) d\lambda = \sigma T^4 \qquad (1\text{-}2)$$

Où $\sigma$ est la constante de Stefan et vaut $\sigma = \frac{2\pi^5 {k_b}^4}{15c^2h^3} = 5.67.10^{-8} Wm^{-2}K^{-4}$. La loi du déplacement de Wien donne la longueur d'onde du maximum d'émission d'un corps noir pour une température donnée $\lambda_{\max} = \frac{2898}{T}$ µm (avec T la température en Kelvin).

### 1.1.3. Intérêt des grandes longueurs d'onde

#### *1.1.3.1. Considérations générales*

L'intérêt pour l'infrarouge résulte du fait que chaque corps est une source de rayonnement et cela par opposition au visible où la majorité des corps ne font que réfléchir le rayonnement d'une source primaire. Il est alors possible d'entrevoir des applications inimaginables en rayonnement visible, telles que : « voir la nuit », repérer des personnes malades au sein d'une foule, observer des étoiles froides…

En dessous de 3µm, le spectre électromagnétique d'un corps résulte majoritairement de la réflexion du rayonnement de corps plus chauds. Ce n'est qu'au delà de 3µm que son émission propre devient prépondérante. Il en résulte que plus un corps est froid, plus la longueur d'onde d'émission à laquelle son rayonnement propre prévaut est importante. Il se dessine alors un intérêt pour la détection haute longueur d'onde lors de la détection d'objet froid.

A ces considérations propres à la loi de Planck vient s'ajouter la notion de transmission atmosphérique. En effet les vibrations des molécules de l'air tombent dans cette gamme de fréquence, ce qui a deux conséquences. Tout d'abord faire de la spectroscopie infrarouge se révèle très intéressant, dans la mesure où il va être possible de sonder la composition atmosphérique. En contrepartie il ne sera pas possible de faire des activités d'imagerie à toutes les longueurs d'onde puisque l'atmosphère, via principalement l'eau et le dioxyde de carbone, absorbe. Il est usuel de définir des plages/bandes de transparence de l'atmosphère dont les deux plus utilisées pour l'imagerie sont la bande II (3-5µm) et la bande III (8-12µm), voir la FIG. 1-2. En dehors de ces plages, les activités d'imagerie infrarouge devront être soit très courtes portées, soit embarquées sur satellite ou porteur.

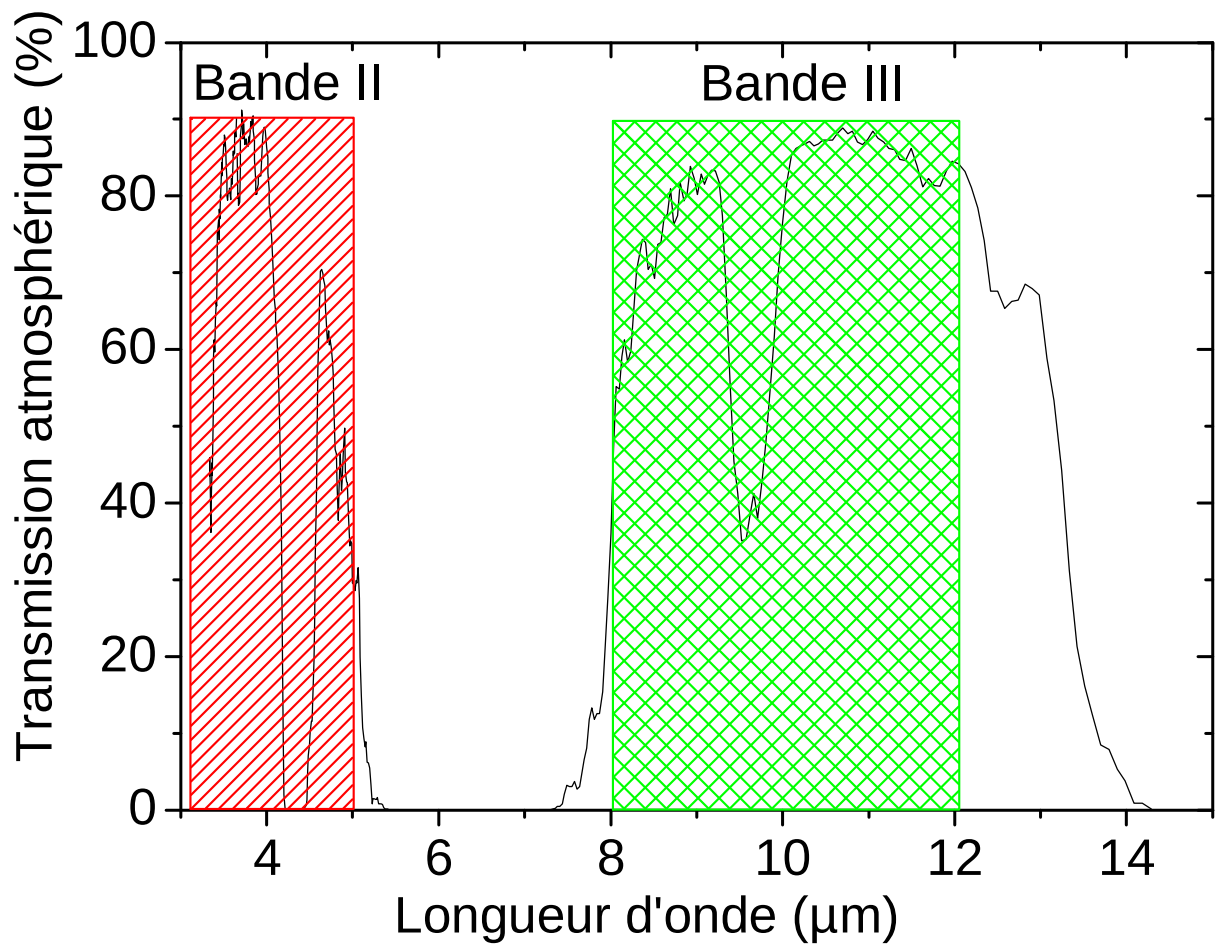

*FIG. 1-2 Transmission atmosphérique en fonction de la longueur d'onde pour un satellite regardant la terre, condition US standard, d'après code ONERA Matisse, fourni par P. Simoneau.*

#### 1.1.3.2. Motivation du travail en cours

Trois grands types d'applications sont actuellement envisagées à l'ONERA sur les détecteurs aux hautes longueurs d'onde : l'astronomie, les applications de défense et l'imagerie en soufflerie.

L'ONERA en tant que spécialiste français des souffleries a été amené à s'investir dans la construction de la soufflerie cryogénique européenne[4] (ETW). L'imagerie thermique de l'écoulement froid nécessite de l'imagerie infrarouge haute longueur d'onde. C'est une application relativement marginale mais qui à l'avantage d'être purement académique et donc non contraint par des règles de confidentialité. C'est donc cette application que nous étudierons plus avant au chapitre deux.

L'infrarouge est aussi une gamme de détection très commode en astronomie, notamment pour la détection d'exoplanète. Citons la mission DARWIN de l'ESA[5] dont l'objectif est l'observation de super-terre. Pour une telle mission l'observation dans le visible est difficile à envisager car le flux de photons provenant de l'étoile autour de laquelle gravite la planète est un milliard de fois plus élevé que celui de la planète elle-même. Tandis que dans l'infrarouge ce ratio des flux de photons n'est plus que de dix mille à cent mille. La mission DARWIN envisage donc l'utilisation d'une constellation de six télescopes pour l'observation de telles planètes, voir FIG. 1-3. Les applications spatiales, embarquées sur satellite, cherchent généralement de grands formats matriciels pour observer de large zone et des températures de fonctionnement élevées pour augmenter la durée de vie de la machine à froid et diminuer leur volume ainsi que leur consommation électrique. DARWIN est certainement l'une des missions infrarouge les plus exigeantes quant aux performances du détecteur[6] ($J_{dark}$=4.5×10$^{-13}$A.cm$^{-2}$, η≈50% et $T_{det}$ entre 15K et 40K).

Mais loin de ne servir qu'à des applications d'astronomie, les détecteurs infrarouges grandes longueurs d'onde sont essentiellement dédiés à des applications militaires et cela devrait le rester pour les années à venir. En effet le livre blanc[7] qui fixe les orientations stratégiques et militaires de la France pour la prochaine décennie a défini la détection de missiles balistiques comme l'une des priorités françaises en terme de défense. Or là ou il est acceptable d'avoir sur un champ de bataille, voir FIG. 1-3, un dispositif anti-missile (type Patriot) de courte portée avec une fiabilité de l'ordre de quelques dizaines de pourcent[8], il n'est évidement pas acceptable de disposer d'un bouclier anti-missile balistique, sous entendu nucléaire, dont le taux d'échec puisse être même de l'ordre de quelques pourcents. La détection de ces missiles par des détecteurs infrarouges hautes performances est donc une priorité.

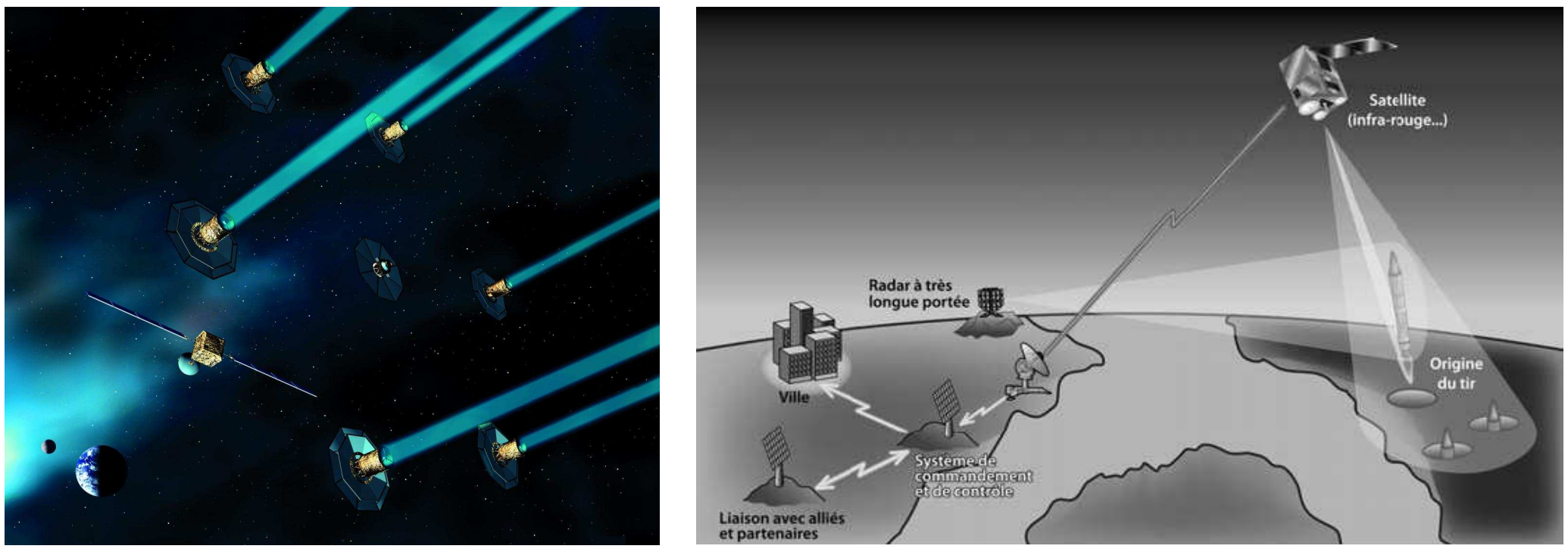

*FIG. 1-3 A gauche : constellation de satellites pour la mission DARWIN. A droite : Schéma de la détection antimissile, d'après ref 7.*

## 1.2. Systèmes de basse dimensionnalité

L'invention au début des années 1970 des hétérostructures[9] à base de GaAs, matériau roi de l'optoélectronique grâce à son gap direct, fut non seulement une prouesse technologique, mais a également permis l'exploration expérimentale du monde des basses dimensionnalités, jusque là réservé au théoricien. La FIG. 1-4 montre une interface de GaAs/AlGaAs dopée où se forme un gaz bidimensionnel d'électrons.

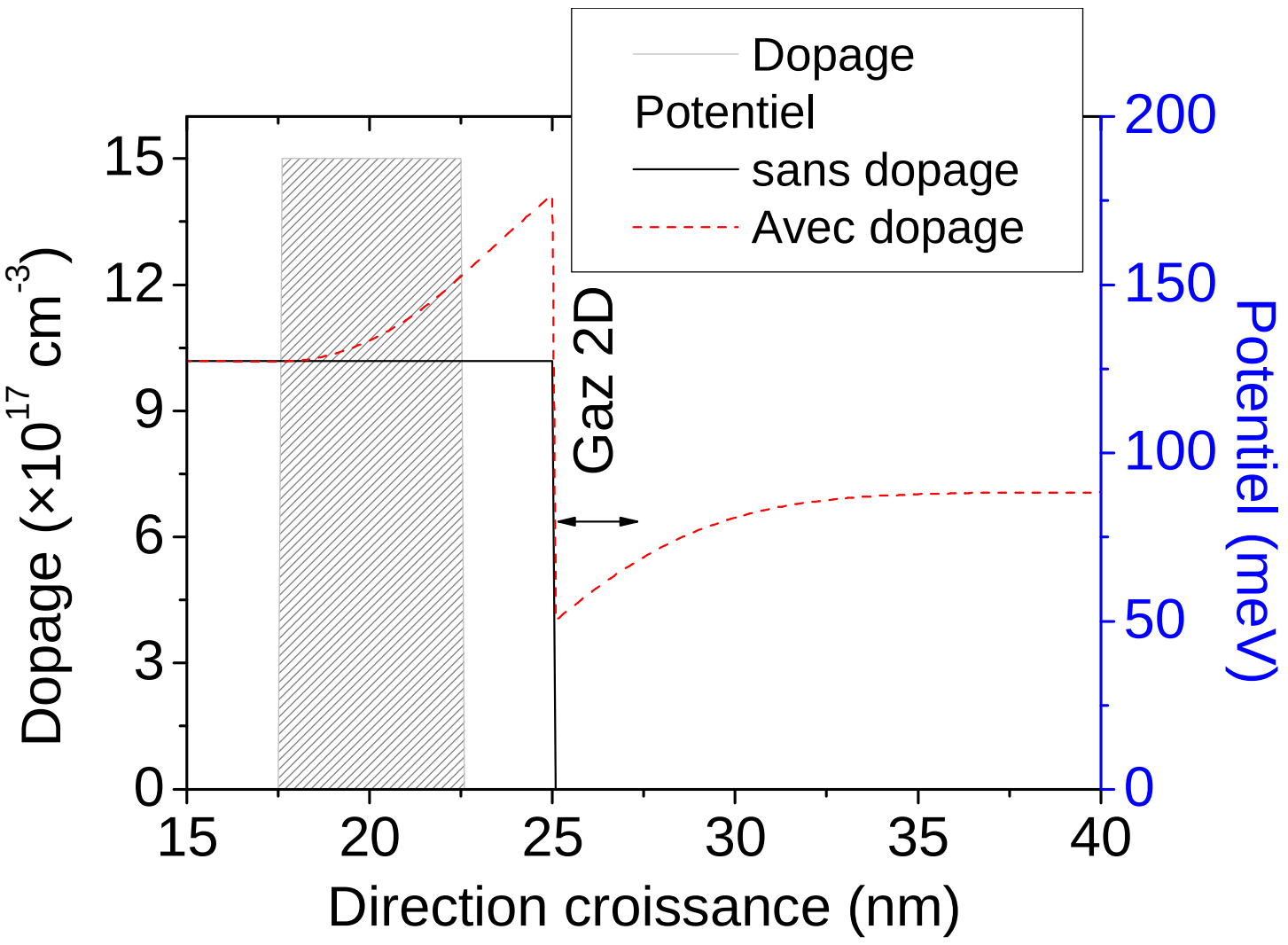

*FIG. 1-4 : Profil de potentiel d'une interface de $Al_{0.15}Ga_{0.85}As$/GaAs en présence d'un fort dopage dans la barrière. Présence d'un gaz bidimensionnel d'électrons à l'interface entre GaAs et AlGaAs. De tels systèmes sont à présent disponibles de façon commerciale au prix de quelques centaines d'euros pour un wafer de trois pouces.*

Les hétéro-jonctions ont pavé la voie pour toute une série de structures plus complexes allant du simple puits quantique aux structures les plus originales comme les super réseaux[9] et les lasers à cascade quantique[10] en passant par les doubles barrières tunnel (résonantes).

Depuis, d'autres matériaux ont permis de s'intéresser aux systèmes de basse dimensionnalité, il est possible de citer :

- A deux dimensions : Les supraconducteurs hautes $T_c$[11] où le transport se fait essentiellement via les plans d'oxyde de cuivre. Plus récemment, en 2004, le graphène[12] a démontré l'existence de cristaux 2D et permet l'étude de la mécanique quantique relativiste grâce à sa relation de dispersion linéaire.

- A une dimension. La découverte en 1991 des nanotubes de carbone[13] a développé l'intérêt pour les structures 1D. Ce matériau a l'avantage de conserver des propriétés intéressantes, tant mécaniques qu'électroniques, à température ambiante. De nombreuses applications sont d'ores et déjà à l'étude (utilisation de nanotubes comme canon à électrons, transistor à nanotubes[14], renforcement de matériaux composites, réalisation de bolomètre). Certains supra conducteurs organiques comme les sels de Bechgaard[15] offrent également la possibilité de tester le monde 1D. Enfin, plus récemment, l'amélioration des techniques d'épitaxie a permis la croissance de nanofils de semiconducteur. Dans ces structures 1D le comportement des électrons ne suit plus celui du liquide de Fermi mais plutôt celui d'un liquide de Lünthinger.

- A dimension nulle. Les boîtes quantiques ont connu un développement récent (deuxième moitié des années 1990) et sont elles aussi promises à un bel avenir. D'autant que la possibilité de les faire croître par des voies chimiques[16], moins lourdes que celles habituellement utilisées (MBE ou MOVPE) pourrait encore renforcer l'intérêt pour celles-ci.

La majorité des systèmes cités précédemment ont été utilisés afin de détecter le rayonnement infrarouge avec un succès plus ou moins important, dans la suite nous allons nous intéresser aux filières qui ont réussi à s'imposer.

## 1.3. Détection infrarouge par des hétero-structures

Bien que tout système dont l'énergie typique soit de l'ordre de grandeur de la dizaine à quelques centaines de meV soit potentiellement un détecteur infrarouge, seul un nombre limité de détecteurs ont réussi à s'imposer. Nous allons dans ce paragraphe dresser une liste des filières en présence.

La détection infrarouge oppose deux grandes filières : la filière thermique et la filière quantique. La première repose sur la détection du flux d'énergie incident. Ce qui a l'avantage d'être très simple d'utilisation (pas de refroidissement et donc bas coût) mais lente du point de vue du temps de réponse. Les bolomètres[17] semblent être la voie la plus développée à ce jour. A l'opposé, les filières quantiques proposent des temps de réponses plus courts mais nécessitent un refroidissement. Les détecteurs quantiques reposent pour la plupart sur des structures plus ou moins complexes, allant de la simple jonction PN aux hétéro-structures les plus complexes comme les détecteurs à cascade quantique. Ces détecteurs reposent tous sur le même principe, à savoir l'absorption d'un photon qui fait passer un électron d'un état isolant vers un état conducteur. Je me limiterai volontairement dans la suite aux filières qui sont dédiées aux grandes longueurs d'onde ($\lambda$>12µm). Quatre technologies s'affrontent sur ce créneau : le MCT (ou diode à HgCdTe), les détecteurs à puits quantiques ou QWIP, les super réseaux de type II et les détecteurs à bande d'impuretés bloquantes (BIB).

### 1.3.1. Détecteur quantique

#### *1.3.1.1. Filières en présence*

Nous allons rapidement passer en revue le principe de détection des quatre filières de détecteurs quantiques citées préalablement. Je ne vais pas ici rentrer dans le détail de fonctionnement de ces filières. Le but est de donner le principe de fonctionnement de chacune d'elle et de donner leurs avantages et inconvénients. Le lecteur intéressé pourra se reporter à la récente revue de Rogalski et al[18].

Les détecteurs MCT, à base d'HgCdTe, sont basés sur une simple transition bande de valence-bande de conduction, voir FIG. 1-5 (a). Ils sont le plus souvent insérés dans une diode de type p-i-n. Le gap, et donc la longueur d'onde de détection, sont ajustés via le contrôle du taux de mercure. C'est la filière la plus mature en dessous de 12µm et elle vient de fêter en 2009 ses cinquante ans.

Les détecteurs QWIP sont de type unipolaire (le plus souvent n) et reposent sur des transitions entre niveaux discrets de la bande de conduction, voir FIG. 1-5 (b). L'électron dérive ensuite jusqu'à l'électrode, au sein du continuum. Cette filière profite de la maturité technologique de GaAs.

Pour le super réseau de type II la détection se fait entre mini bandes : l'une d'entre elles est contenue dans la bande de valence (et sert au transport des trous) tandis que l'autre est située dans la bande de conduction (et permet le transport des électrons), voir FIG. 1-5 (c). Ce confinement des porteurs dans deux matériaux différent a pour avantage d'augmenter le temps de vie de la paire électron-trou. Cette filière a, sur le papier, une très forte potentialité (fonctionnement photovoltaïque, fort rendement quantique, accordabilité en longueur d'onde…) et est actuellement source de nombreuses recherches de par le monde. Néanmoins, la maturité technologique de cette filière est encore très loin d'atteindre celle des QWIP ou du MCT. Elle se heurtait encore récemment à des problèmes de courant de fuite sur les flancs des mesas[19].

Il existe une gamme de détecteur haute longueur d'onde, allant jusqu'au THz, dont le principe est basé sur l'utilisation de niveaux d'impuretés dans une matrice de semiconducteur[20,21], voir FIG. 1-5 (d), on parle de détecteur à bande d'impuretés bloquantes (BIB). Le détecteur repose sur une jonction n-i. La partie n est celle qui est optiquement active tandis que la partie i est utilisée afin de réduire le courant d'obscurité. Le principe de fonctionnement est le suivant[22] : un électron se trouvant sur un niveau donneur de la zone n est promu dans la bande de conduction du semi-conducteur hôte. Si la structure est polarisée, un courant photonique traverse la structure. Le courant d'obscurité est lui très faible en raison de l'absence de niveau donneur dans la zone i. L'intérêt de ces structures est qu'elles peuvent être fortement dopées, afin d'obtenir une forte absorption, tout en conservant un courant d'obscurité faible. Un tel détecteur doit fonctionner à froid, tout d'abord pour une raison optique afin que les électrons des impuretés ne soient pas ionisés. Et deuxièmement pour des raisons de transport, il ne faut pas promouvoir les électrons par absorption de phonon. Il en résulte une température d'utilisation inférieure à 10K. L'absorption est large bande et l'ajustement de la longueur d'onde de coupure se fait via le choix du matériau et du niveau de dopage, ce qui est loin d'être aussi aisé qu'avec les précédentes filières. Les matériaux les plus courants sont : le silicium dopé arsenic, le silicium dopé antimoine, le silicium dopé gallium (jonction p-i dans ce cas) et le germanium dopé gallium. La maturité des matériaux IV permet de faire des matrices de grande taille. Du point de vue de la fabrication c'est une filière essentiellement confinée aux Etats-Unis et qui sert à des applications de défense et d'astronomie[22].

Le tableau tab. 1-1 compare les propriétés matricielles de ces différentes filières.

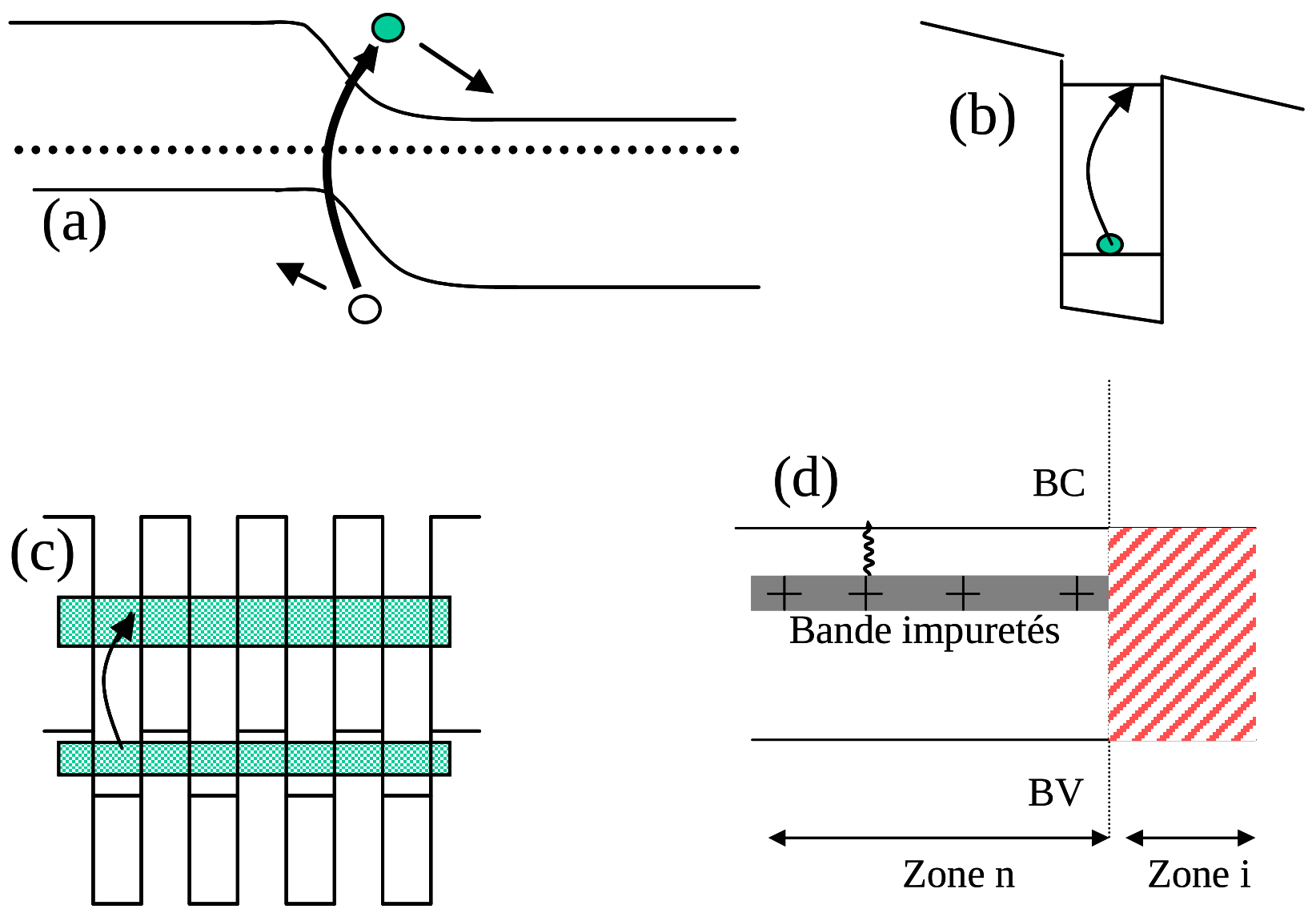

*FIG. 1-5 (a) Détection dans un MCT : transition entre bande de valence et bande de conduction. (b) Détection dans un QWIP : transition entre sous-bandes de la bande de conduction, puis dérive dans le continuum. (c) Détection dans un super réseau de type II : transition entre mini-bandes. (d) Détection dans un détecteur à bande d'impuretés bloquantes : transition entre niveau d'impuretés et bande de conduction.*

### 1.3.2. Aspects technologiques

Maintenant que la physique de chacune de ces filières a été présentée, intéressons-nous à quelques propriétés plus globales des détecteurs et de leurs caméras.

#### *1.3.2.1. Bande large vs bande étroite*

Les détecteurs basés sur des transitions inter-bandes absorbent une large plage de longueurs d'onde (toutes les énergies au-dessus du gap), tandis que les détecteurs inter-sousbandes sont fortement résonants et n'absorbent qu'une bande étroite. Pour des applications faible flux il est tentant d'utiliser des détecteurs bande large afin de ne se priver d'aucun photon. Toutefois dans la perspective d'observation d'objet dont la signature (émissivité) est variable avec la longueur d'onde, l'utilisation de détecteur bande large peut aboutir à un faible contraste avec le fond de la scène, ce genre de problème peut être résolu via le recours à des détecteurs bande étroite.

#### 1.3.2.2. Détecteur multi-longueurs d'onde

Il sortirait largement du cadre de cette thèse d'aborder l'intérêt des détecteurs multi longueurs d'onde, voire hyperspectraux. Je me contenterai donc de signaler que les trois grandes filières (MCT, QWIP et Super-réseaux de type II) ont démontré la faisabilité technologique de détecteurs sur au moins deux plages de longueur d'onde (MWIR et LWIR par exemple). Des QWIP détectant quatre longueurs d'onde ont également été réalisés[23]. Signalons également la réalisation de détecteurs mégapixel bicolores[24]. Il est clair que pour des raisons technologiques la multiplication des longueurs d'onde s'accompagne généralement de la perte de la simultanéité spatiale (un type de pixel par longueur d'onde) ou temporelle (les pixels détectent les différentes longueurs d'onde de façon séquentiel).

#### 1.3.2.3. Format matriciel et taille de pixels

La tendance générale est d'aller vers une augmentation du nombre de pixels qui s'accompagne d'une diminution de la taille des mono-éléments, le but étant bien sûr d'obtenir un champ toujours plus large et une résolution toujours meilleure. Des matrices 2000×2000 (pas de 18µm) ont été réalisées aussi bien en filière MCT que QWIP. La possibilité d'étendre la taille de matrice jusqu'à 4000×4000 est actuellement envisagée pour les QWIP[18]. En attendant certains utilisateurs rapportent l'utilisation de « matrices de matrices » (2×2 matrice de 2000×2000). Les applications hautes longueurs d'onde ne sont toutefois pas forcément de bons candidats pour les plus petits pixels dans la mesure où la taille du pixel doit être adaptée à « celle du photon ». Actuellement l'augmentation de la taille des matrices est limitée par des problèmes d'hybridation de la couche active avec le circuit de lecture. Il existe un problème de planéité des matrices de grande taille auquel s'ajoute le fait que grande matrice rime avec petit pixel ce qui augmente encore la difficulté.

| | MCT | QWIP | SL II | BIB |
|---|---|---|---|---|
| Type de transition | Interbande | Inter-sous bande | Inter-minibande | Niveau d'impuretés vers bande de conduction |
| Plage absorption | Bande large | Bande étroite | Bande large | Bande large |
| Taille matrice maximale | 2000×2000 | 2000×2000 | 1000×1000 | 2000×2000 |
| Type de porteur | Electron-trou | Electron | Electron-trou | Electron |
| Homogénéité | Faible | Excellente | Faible | |
| Fabriquant | Rockwell, BAE systems, Raytheon, Sofradir | Thales, IR nova, QWIPTECH | AIM Raytheon | |

*tab. 1-1 Récapitulatif des performances des quatre principales filières hautes longueurs d'onde.*

## 1.4. Les détecteurs QWIP

Dans cette thèse nous nous sommes principalement focalisés sur la filière QWIP et les détecteurs qui en découlent directement : les QCD (détecteurs à cascade quantique). Nous allons donc, dans cette fin de chapitre, nous intéresser plus avant à leur fonctionnement.

Les QWIP, pour Quantum Well Infrared Photodetector, permettent de détecter des longueurs d'onde entre 3µm et le THz (>30µm). Ils se présententent sous la forme d'un empilement de plusieurs dizaines de périodes de GaAs/AlGaAs, prises en sandwich entre deux contacts dopés en GaAs. Cet empilement périodique de deux matériaux est appelé super-réseau, le nom venant du fait que la structure reproduit à une échelle supérieure la périodicité du réseau atomique. La position relative des bandes de valence et conduction fait que l'on parle de super-réseau de type I. Le confinement des électrons et des trous se fait alors dans le même matériau.

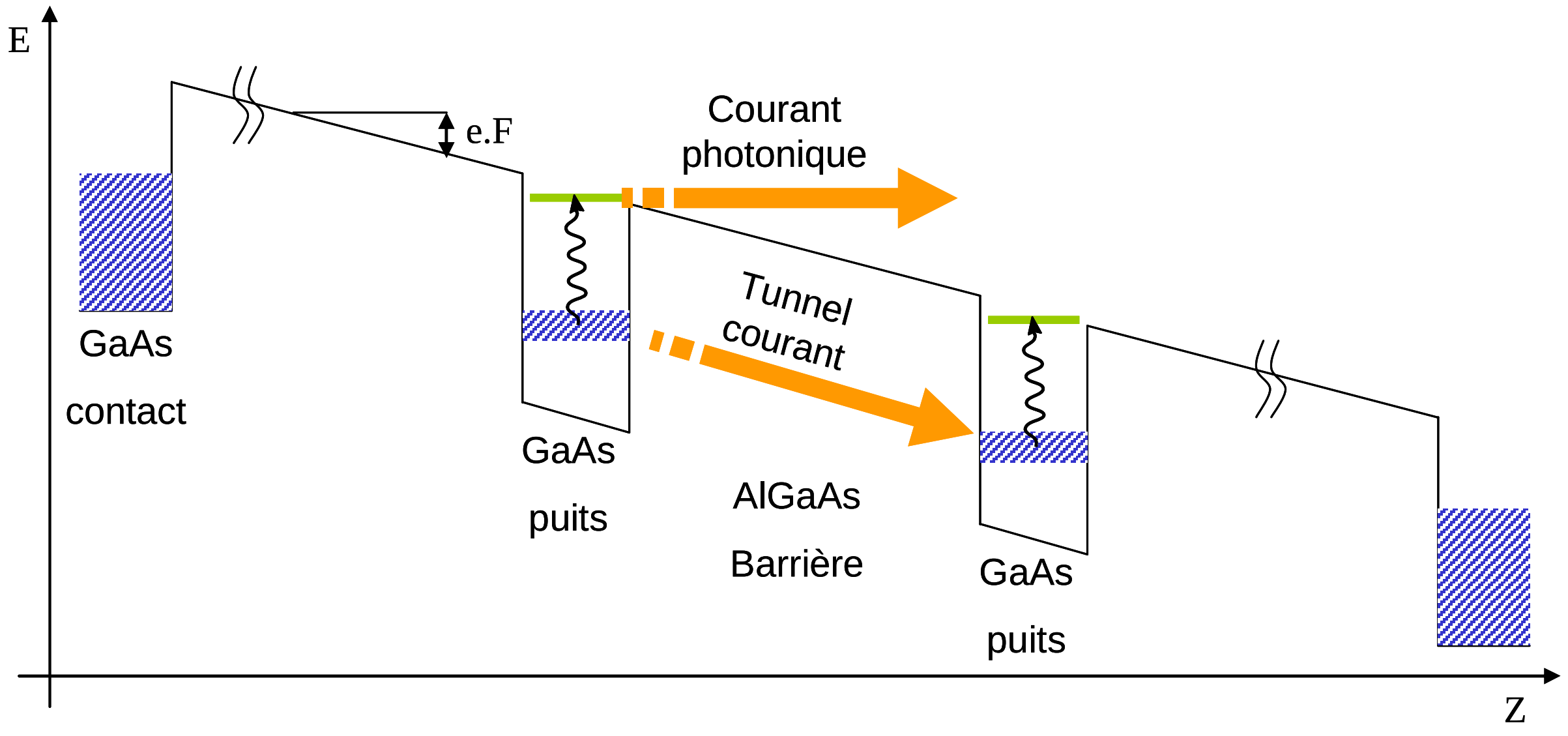

*FIG. 1-6 Schéma de la structure d'un QWIP*

La croissance des couches est obtenue par MBE ou par MOVPE, l'empilement peut atteindre quelques micromètres. Les matériaux utilisés sont GaAs et $Al_xGa_{1-x}As$, dans lequel la concentration en aluminium contrôle la hauteur des barrières. Le pourcentage d'aluminium est compris entre 1 et 40% ; au-delà le gap devient indirect (gap $\Gamma \rightarrow X$). En bande II le recours à un ajout d'indium est souvent nécessaire[25]. Les couches de GaAs forment des puits tandis que les barrières sont constituées de $Al_xGa_{1-x}As$, la figure FIG. 1-6 schématise la structure d'un QWIP. La taille des puits, autour de 100 Å, contrôle principalement le nombre de niveaux présents et leur écart en énergie. Généralement les QWIP sont conçus pour qu'il y ait deux niveaux par puits. Signalons que les QWIP destinés à l'absorption multi-photons[26] peuvent en contenir un plus grand nombre. La taille des barrières agit directement sur le couplage optique et électronique de la structure. Les barrières sont généralement épaisses, entre 100 et 1000Å. Enfin la structure est le plus souvent dopée n, au niveau des contacts et des puits. Le transport dans les QWIP est unipolaire, seuls les porteurs majoritaires y contribuent. Le silicium est l'impureté la plus couramment utilisée, car elle présente l'avantage de pouvoir doper n (en substitution du gallium) ou doper p (en substitution de l'arsenic). D'autres dopants, comme le Béryllium, sont également utilisés. Les concentrations typiques en dopant sont de $10^{17}/10^{18}cm^{-3}$ pour les contacts et de $10^{10}$ à $10^{12}cm^{-2}$ pour les puits. Le dopage dans la structure est un compromis entre l'absorption, qui croît avec le nombre de porteurs, et le courant d'obscurité qui croît également avec le dopage.

### 1.4.1. Principe de détection

#### *1.4.1.1. Courant photonique*

Le QWIP est un photoconducteur, il nécessite donc d'être polarisé pour fonctionner. Le signal se présente sous forme d'un courant à deux composantes, l'une dite photonique, l'autre d'obscurité (ou dark en anglais). Le rapport entre le courant photonique et le flux incident définit la réponse du détecteur, en ampère par watt. A faible nombre de puits, cette dernière est indépendante du nombre de puits[29], puis décroît en 1/N quand le nombre de puits augmente.

Le principe de détection est le suivant : la structure étant dopée n, il existe des électrons au sein de la bande de conduction. Il se forme ainsi un niveau de Fermi au dessus du niveau fondamental du puits. Ces électrons peuvent absorber un photon dont l'énergie correspond à la transition entre le niveau fondamental et le premier niveau excité. Une fois sur le niveau excité l'électron passe dans le continuum au sein duquel il dérive sous champ, formant un courant. Cette vision idyllique se complique quand plusieurs puits composent la structure, car l'électron peut être re-capturé au sein des puits suivants.

Dans un QWIP classique, dédié à la détection d'une longueur d'onde donnée, le courant photonique est lié à la nature de la transition entre le niveau fondamental et le niveau excité. La transition peut être de type lié-lié, les deux niveaux sont situés au sein du puits, dans ce cas le couplage entre les niveaux électroniques est bon (recouvrement entre niveaux purement 2D), mais l'extraction vers le continuum est médiocre. Si la transition est liée libre, la force d'oscillateur est réduite par rapport au cas précédent (recouvrement d'un niveau 2D avec un niveau 3D), mais l'extraction s'en trouve améliorée. Le meilleur compromis est donc la transition lié-quasi-lié, dans laquelle le niveau excité est résonant avec la barrière.

#### *1.4.1.2. Courant d'obscurité*

Le courant d'obscurité est le courant qui existe en l'absence d'éclairement, le détecteur émet alors autant de photons qu'il en reçoit. Il peut se décomposer en deux composantes : la première dite thermoionique dans laquelle les électrons suivent le même chemin que dans le cas du courant photonique. La promotion de l'électron vers le niveau excité est cette fois-ci assurée par les phonons. Cette composante est dominante à haute température, et c'est dans ce régime que fonctionnent actuellement les caméras commerciales en bande II et III. La seconde composante réside dans le couplage tunnel entre puits. La compréhension de ce courant fait l'objet d'une grande partie de cette thèse. Elle a la particularité d'être indépendante de la température, c'est donc elle qui subsiste à très basse température.

Un courant d'obscurité trop important remplirait les capacités de lecture, ce qui réduit le temps d'intégration, tout en ajoutant un bruit parasite de génération recombinaison. Il est donc nécessaire pour obtenir un détecteur performant de réduire autant que possible le niveau du courant d'obscurité. Cette affirmation est d'autant plus vraie que l'on s'intéresse à des applications faibles flux. Le courant d'obscurité est souvent le point limitant des QWIP notamment pour certaines applications hautes longueurs d'onde.

### 1.4.2. Electromagnétisme dans les QWIP

#### *1.4.2.1. Règle de sélection*

Il existe dans les QWIP une règle de sélection qui n'autorise pas l'absorption du champ électromagnétique en incidence normale. La démonstration que je propose n'est certainement pas la plus élégante possible[27], mais elle donne les bons arguments physiques qui conduisent au fait que seule la composante du champ électrique selon l'axe de croissance est absorbée.

Le potentiel lié au champ électromagnétique est de type dipolaire. Son expression est de la forme $H = e\vec{E}.\vec{r}\, e^{i\vec{q}\vec{r}}$, avec q l'impulsion du photon. La transition qui intervient dans un QWIP est celle entre le niveau fondamental d'un puits et le premier niveau excité de ce même puits. Les fonctions d'onde sont décrites via le formalisme de la fonction enveloppe soit $\psi = \frac{1}{\sqrt{A}} e^{i\vec{q}''\vec{r}''}\xi(z)$, où $r''$ est la composante dans le plan du vecteur position, et A l'aire du pixel. Compte tenu de ces deux expressions l'élément de matrice qui intervient dans la règle d'or de Fermi est donné par :

$$M = \left\langle \frac{1}{\sqrt{A}} e^{i\vec{q}i''\vec{r}''}\xi_i(z) \middle| e\vec{E}.\vec{r}\, e^{i\vec{q}\vec{r}} \middle| \frac{1}{\sqrt{A}} e^{i\vec{q}j''\vec{r}''}\xi_j(z) \right\rangle \qquad (1\text{-}3)$$

$$M = \frac{e}{A}\iiint d^2r''dz\, e^{i(k''_j+q''-k_i'')r''}\vec{E}.\vec{r}\,\zeta_i(z)\zeta_j(z)e^{iq_z z} \qquad (1\text{-}4)$$

Soit encore en décomposant en deux le terme $\vec{E}.\vec{r}$ :

$$M = \frac{e}{A}\int \zeta_i(z)\zeta_j(z)e^{iq_z z}dz \iint d^2r''e^{i(k''_j+q''-k_i'')r''}E''.\vec{r}'' + \frac{e}{A}\int \zeta_i(z)\zeta_j(z)e^{iq_z z}E_z.z\,dz \iint d^2r''e^{i(k''_j+q''-k_i'')r''}$$

(1-5)

Compte tenu du fait que $L_W$<<λ, il s'ensuit que $q_z z << 1$ et donc $e^{iq_z z} \sim 1$. Notons également que l'impulsion d'un électron est grande devant celle d'un photon, ce qui a pour conséquence $k''_j + q'' - k''_i \approx k''_j - k''_i$ (transition verticale). Alors le premier terme de l'expression de l'élément de matrice s'annule car les deux fonctions d'onde sont orthogonales. Finalement,

$$M = \frac{e}{A}E_z.\delta(k''_i - k''_j)\int \zeta_i(z)\zeta_j(z)z\,dz \qquad (1\text{-}6)$$

Ce qui montre que seule la composante parallèle à la direction de croissance est absorbée par un puits quantique. Le QWIP se comporte donc comme un matériau absorbant et anisotrope.

#### *1.4.2.2. Réseau de couplage*

Comme nous l'avons vu précédemment l'utilisation de transitions inter-sousbandes impose que seule la composante du champ électrique dans la direction de croissance peut être absorbée. Le champ électromagnétique étant transverse, le QWIP est théoriquement aveugle en incidence normale. En réalité il existe une absorption résiduelle[28] due à la diffusion par les défauts ou encore aux effets de taille finie du pixel. Il est toutefois nécessaire afin d'augmenter l'absorption d'ajouter un système de couplage.

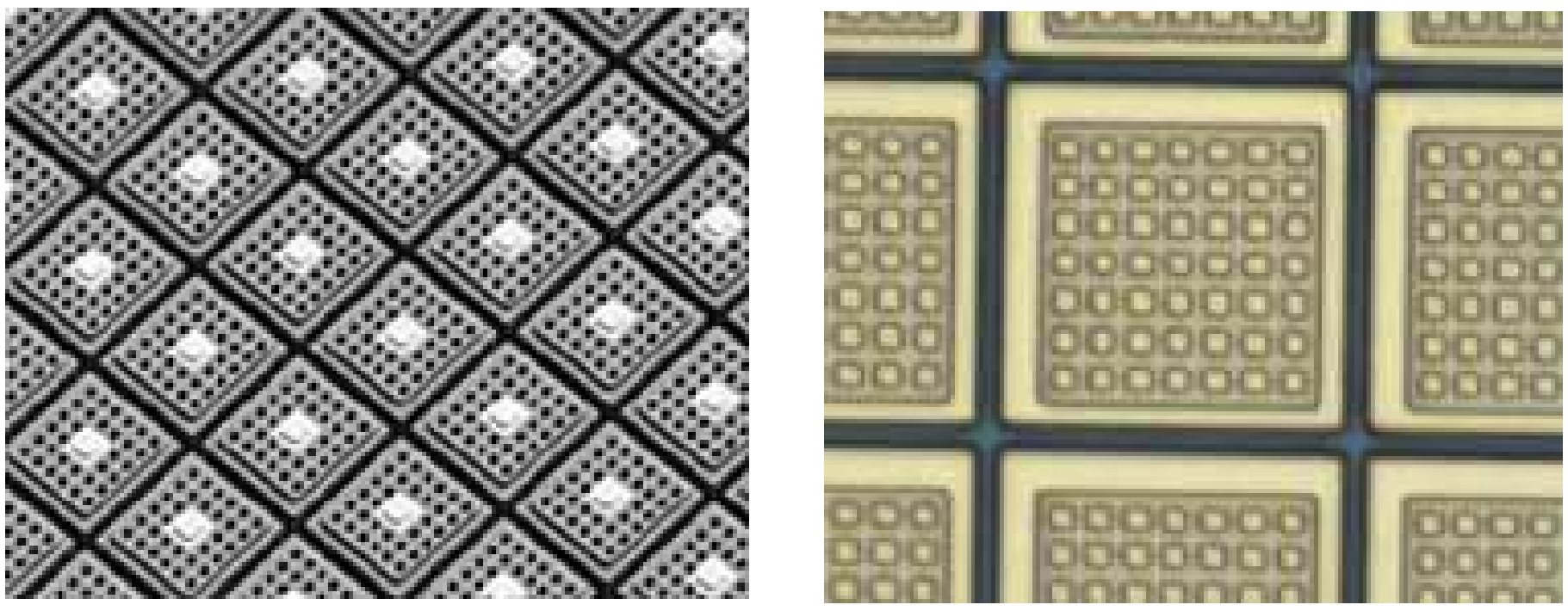

*FIG. 1-7 Image de matrice de pixels QWIP avec le réseau de couplage gravé à la surface.*

Dans la pratique, la solution la plus souvent retenue est la gravure d'un réseau de couplage, voir FIG. 1-7. Sur cette figure nous pouvons voir une matrice de pixels ainsi qu'un zoom sur l'un d'eux qui permet de mieux visualiser le réseau gravé à leur surface. Le flux de photons traverse donc le pixel (éclairage face arrière) avant de rencontrer le réseau qui le réfléchit dans une autre direction. Expérimentalement le couplage est optimal[29] quand le pas du réseau est égal à la longueur d'onde dans le matériau. Comme pour des raisons de gain en résolution, la technologie s'oriente vers des pixels de plus en plus petits, typiquement une vingtaine de microns actuellement, il n'y a pas plus de quelques périodes de réseau sur un pixel. L'intérêt du réseau de couplage est de concentrer l'énergie au niveau des puits, voir FIG. 1-8. Le réseau s'étend sur une vingtaine de microns tandis que le champ doit être concentré à une distance de un ou deux micromètres, le couplage se fait donc en champ proche. Le couplage électro-optique dans les QWIP est l'objet d'intense recherche[30]. La solution du réseau bidimensionnel régulier est un bon compromis entre facilité de fabrication et performance. En pratique ces réseaux permettent d'obtenir des réponses angulaires stables sur une plage de +/- 20° [31].

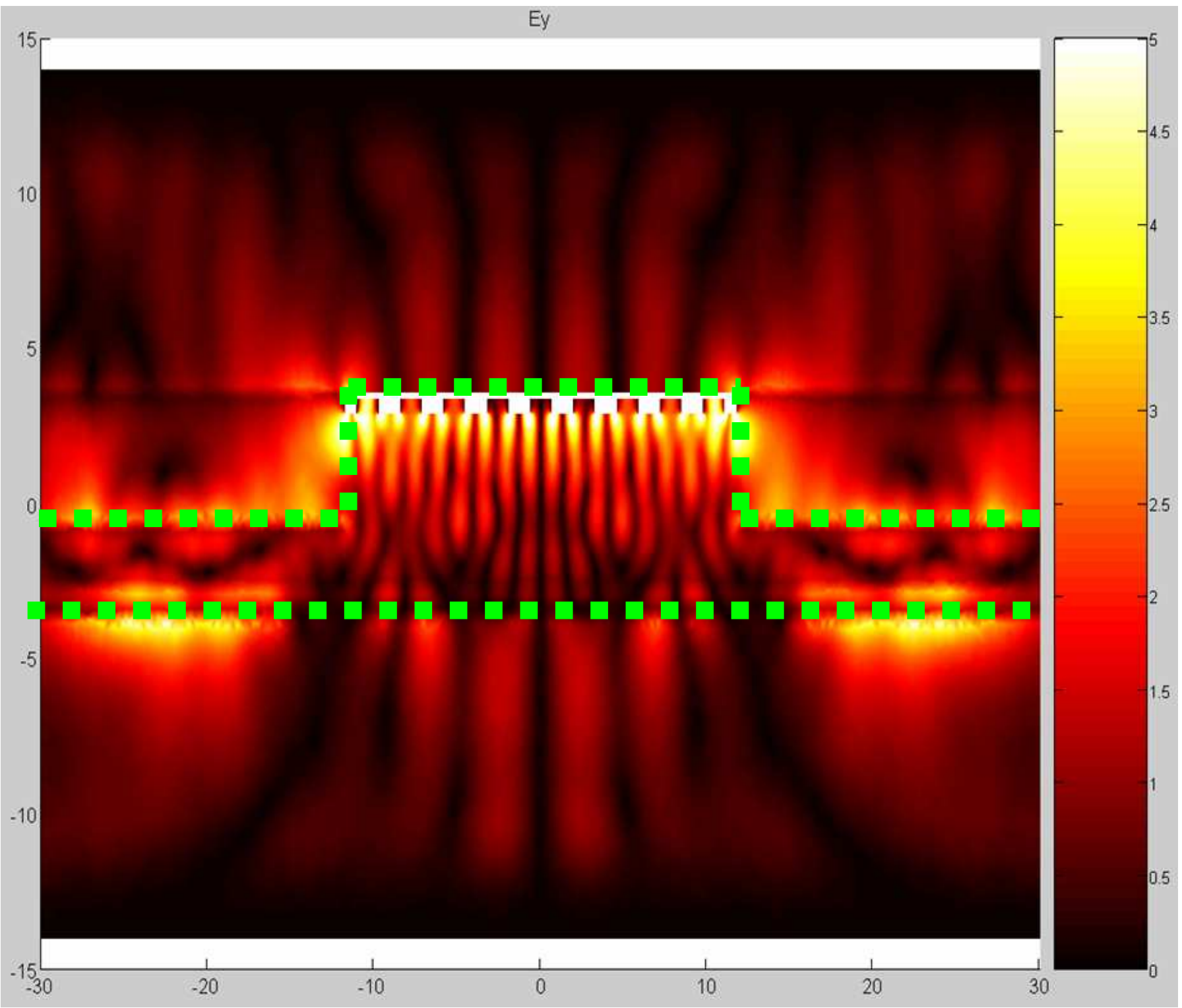

*FIG. 1-8 Cartographie de champ électromagnétique dans un pixel QWIP dans la direction verticale, pour un pixel de taille finie avec réseau. Les points les plus clairs représentent les champs électriques les plus forts. Les traits verts et pointillés symbolisent le pixel et son substrat. Calcul réalisé à Alcatel Thales 3-5Lab par une méthode type éléments finis*

### 1.4.3. Aspect technologique

#### *1.4.3.1. Cryogénie et température d'utilisation*

Dans les QWIP la température de fonctionnement est déterminée par le niveau acceptable de courant d'obscurité. L'intégralité des QWIP commerciaux fonctionne dans un régime de température où le courant d'obscurité est dans son régime thermoionique. L'utilisation à plus basse température que nous envisageons est donc relativement inédite. En bande III, la température d'utilisation est située entre 50K et 77K[32,33] selon la qualité du détecteur et l'utilisation visée. De façon générale la température d'utilisation décroît avec la longueur d'onde (au delà de 100K en bande II[25], entre 35 et 55K vers 15µm[32,34], à peine au dessus de l'hélium pour le THz[35]). L'augmentation de la température d'utilisation des QWIP est un enjeu majeur. En effet plus la température d'utilisation est importante plus la cryogénie pourra être compacte (essentiel pour des applications embarquées), bon marché et présenter une durée de vie importante.

#### *1.4.3.2. Circuit de lecture*

Le circuit de lecture sert d'interface entre le bloc détecteur et l'utilisateur. Il se présente sous forme d'une matrice dont le rôle est de collecter le courant provenant de chaque pixel et de le traduire en terme d'image. Il serait hors de propos de s'intéresser au fonctionnement des circuits de lecture. Toutefois, dans la mesure où il impose certaines contraintes sur le détecteur nous allons présenter plusieurs de leurs spécificités. Le circuit de lecture est hybridé au détecteur par des billes d'indium. Ces billes servent de contact électrique avec le circuit e de lecture. Ce dernier contient en particulier une capacité (0.1pF à 0.5pF) qui intègre les charges. Leur taille est déterminée par des facteurs géométriques tels que la taille pixel et ils peuvent typiquement contenir quelques millions d'électrons. Le signal en sortie de circuit est proportionnel à cette charge. A chaque temps d'intégration les billes sont déchargées afin de permettre une nouvelle charge lors de la trame suivante. Le circuit de lecture a également pour rôle de polariser le détecteur. Les tensions typiques que peuvent appliquer les circuits sont typiquement dans la gamme de -1V à -3V. Le signal obtenu est numérisé sur une dizaine de bits (8 à 14 bits). Au signal provenant du détecteur s'ajoute le bruit propre au circuit de lecture. Celui-ci doit bien entendu être minimisé.

#### 1.4.3.3. Bruit dans les QWIP

Nous détaillerons au chapitre suivant plus précisément le bruit dans les QWIP. Notons tout d'abord que le QWIP présente peu de bruit en 1/f[36]. Cela permet dans une caméra de faire une table de correction deux points à la fabrication et de ne plus avoir à la modifier par la suite. Le bruit thermique dans les QWIP est faible grâce à leur grande résistance différentielle. La principale source de bruit est donc le bruit de génération recombinaison résultant du passage du courant dans la structure.

#### 1.4.3.4. Fonction de mérite

Afin de comparer les performances de différents détecteurs des fonctions de mérite ont été définies. Leur but est de définir une grandeur qui s'affranchit des spécificités propres à un détecteur (taille pixel, température fonctionnement…). Il en existe de nombreuses, je me limiterai donc à trois d'entre elles que sont la réponse (R), la détectivité spécifique (D*) et la NETD (Noise Equivalent Temperature Difference). Le tableau suivant détaille chacun des critères définis précédemment et en donne une valeur typique pour les QWIP.

| Grandeur | Formule | Définition | Valeur typique |
|---|---|---|---|
| R | $R = \frac{\eta g e}{h\upsilon}$ | Traduit la capacité du détecteur à transformer un flux d'énergie lumineuse en courant. | $1mA\cdot W^{-1} \rightarrow qq\ A\cdot W^{-1}$ |
| D* | $D^* = \frac{R\sqrt{A}}{DSB}$ | Inverse du flux pour lequel le rapport signal sur bruit vaut 1, multiplié par la racine de la surface de détection et la racine de la bande passante. | $10^9 \rightarrow 10^{13}$jones ($W^{-1}\cdot cm\cdot\sqrt{Hz}$) |
| NETD | | Contraste thermique d'une scène pour laquelle le rapport signal à bruit vaut 1. | 10mK→300mK |

*tab. 1-2 Fonction de mérite des QWIP. $\eta g$ est le rendement quantique externe, $h\upsilon$ l'énergie de la transition, DSB la densité spectrale de bruit et A l'air d'un pixel.*

Les valeurs qui sont donnés dan le tableau tab. 1-2 ne sont données qu'à titre indicatif, car elles dépendent de l'ouverture numérique du système, de la température de scène, de la bande spectrale observée ainsi que du temps d'intégration

Il est également courant d'utiliser la notion de régime BLIP. Le mode de fonctionnement d'un détecteur est dit BLIP (Background Limited Infrared Detector) quand le bruit du détecteur est limité par le bruit photonique de la scène observée. Le régime BLIP est une notion propre à l'infrarouge car le détecteur, dans le visible, n'est jamais limité par le bruit de fond de la scène.

### 1.4.4. Communauté et état de l'art

#### *1.4.4.1. La communauté QWIP*

La communauté QWIP est un monde assez restreint qui se compose d'une quinzaine de laboratoires essentiellement concentrés aux Etats-unis et en Europe. La carte FIG. 1-9 montre la répartition des principaux laboratoires, et le tableau tab. 1-3 les décrit.

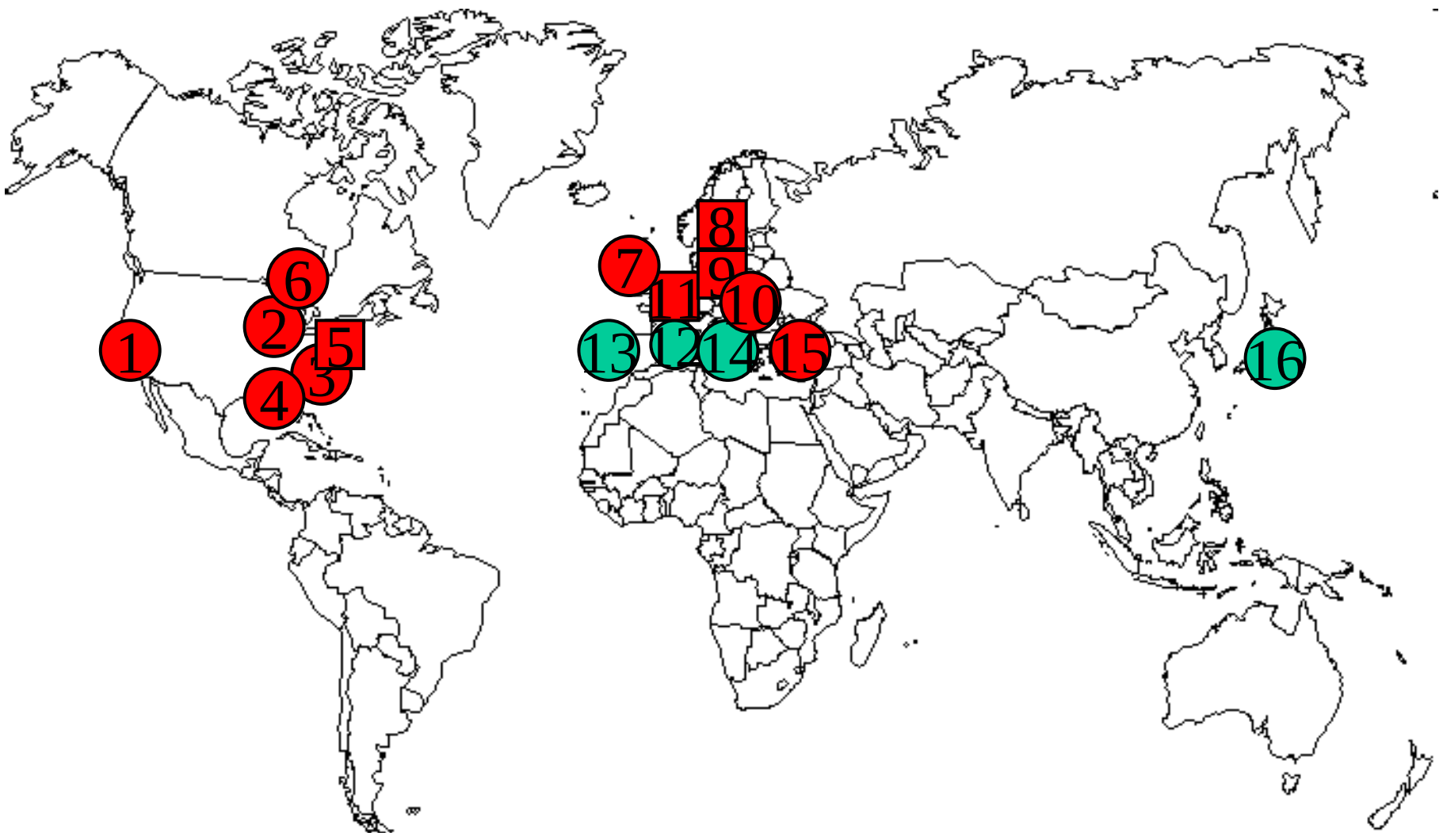

*FIG. 1-9 Localisation des centres d'étude des QWIP. En rouge les lieux disposant de moyens de croissance. Dans un carré les industriels.*

| N° sur la carte | NOM | Chef d'équipe | Statut | Moyen de croissance | Gamme spectrale | QWIP 12-18µm | Commentaires |
|---|---|---|---|---|---|---|---|
| 1 | Jet Propulsion Lab (JPL) | S. Gunapala | Laboratoire de recherche | × | MWIR, LWIR, VLWIR | × | Spécialiste grandes matrices |
| 2 | Center for Quantum devices | M. Razeghi | Universitaire | × | MWIR, LWIR, VLWIR | | |
| 3 | Naval Research Lab | | Laboratoire de recherche | × | MWIR, LWIR, VLWIR | | |
| 4 | Geogia state university | A.G.U. Perera | Universitaire | × | MWIR, LWIR, VLWIR | | |
| 5 | QmagiQ | | Industriel | × | MWIR, LWIR | | |
| 6 | NRC | H.C. Liu | Universitaire | × | MWIR, THz | | |
| 7 | Sheffield university | | Universitaire | × | | | |
| 8 | IR nova/ACREO | J.Y. Anderson | Industriel | × | MWIR, LWIR | × | |
| 9 | AIM | | Industriel | × | MWIR, LWIR | × | Se sont tournés vers les super réseaux de type II |
| 10 | ETH | | Universitaire | × | | × | Plutôt QCD |
| 11 | Alcatel thales 3-5 lab | P. Bois | Industriel | × | MWIR, LWIR, VLWIR, THz | × | |
| 12 | ONERA | | Laboratoire de recherche | | | × | Caractérisation electro-optique |
| 13 | Universidad politecnica de Madrid | E. Munoz | Universitaire | × | MWIR | | |
| 14 | Politecnico di Torino | A. Carbone | Universitaire | | | | Spécialiste du bruit |
| 15 | Middle East Technical University | C. Besikci | Universitaire | × | MWIR, LWIR | | |
| 16 | Aizu university | V. Ryzhii | Universitaire | | | | Thématique arrétée |

*tab. 1-3 Description des laboratoires dont l'une des spécialités est le QWIP.*

#### 1.4.4.2. Etat de l'art des détecteurs QWIP autour de 15µm

Les applications de détection infrarouge autour de 15µm sont exigeantes et le nombre de laboratoire capable de produire une caméra complète est très limité. Ce sont principalement le JPL, IR nova et le 3-5 lab. Le tableau ci-dessous dresse le niveau de performance de différents QWIP donc la longueur d'onde est située dans la gamme 12-18µm. Les valeurs que je rapporte sont celles données par les références associées, mais celles-ci sont généralement avares de détails (car généralement associé à des applications de défense). Une fois encore il est difficile de comparer ces performances entre elles, en particulier la NETD et la détectivité, tant elles dépendent des conditions dans lesquelles les mesures ont été faites.

| $\lambda_{pic}$ ($\Delta\lambda/\lambda$) | $V_{pol}$ | $T_{det}$ | Performances $J_{dark}$, ηg, réponse, NETD, D* | Structure $L_w$, $L_b$, %Al, dopage , | Matrice/pixel | Laboratoire | Année | Référence |
|---|---|---|---|---|---|---|---|---|
| | | | | réseau | | | | |
| 14.6µm<br>(15%) | 3V | 45K | $R=4.3\times10^{-2}$A/W<br>$D^*\approx10^{10}$jones | | 256×1 | National Laboratory for Infrared Physics, Shanghai Institute of Technical Physics | 2007 | 37 |
| 13.5µm<br>(42%) | -<br>2.5V | 55K<br><br>35K | $J_{dark}=5\times10^{-7}$A, pixel $25\times25µm^2$<br>R=0.25A/W<br>$D^*\approx10^{10}$jones<br>NETD≈20mK<br><br>NETD =55mK | Structure à 3 puits bande large | Pixel seul<br><br>FPA<br>650×512<br>25×25µm2 | JPL | 2001 | 38 |
| 9.1µm<br>(16%)<br><br>et<br>14.4µm<br>(10%) | -2V | 40K | LWIR<br>NETD=36mK, R=0.5A/W<br>Absorption QE=6.4%<br><br>VLWIR<br>NETD=44mK, R=0.38A/W<br>Absorption QE=11.6% | bicolore | | JPL | 2000 | 39, 40 |
| 15µm<br>(5%) | | 55K | $D^*\approx1.1\times10^{11}$jones<br>NETD≈350mK | | 2×130 | JPL | 1998 | 41 |
| 9.3µm<br>(19%)<br><br>et<br>15.9µm<br>(12%) | -<br>2.5V | | LWIR<br>R=0.26A/W<br>$\eta_g$ =17.5%<br><br>VLWIR<br>R=0.34A/W<br>peak QE=16.4%<br>$D^*=10^{12}$jones (35K), $2\times10^{11}$(40K), $2\times10^{10}$(50K) | bicolore | | JPL | 1998 | 42 |
| 14.2µm<br>(10%) | | | η=3%<br>$D^*=1.6\times10^{10}$jones (55K)<br>NETD =30mK (45K) | | 128×128<br>Pas 38µm | JPL | 1999 | 43 |
| 15 | -<br>0.8V | 45K | Abs≈40%<br>$J_{dark}=1.5\times10^{-4}A.cm^{-2}$ | | 100µm | 3-5Lab | 2007 | 44 |

| | | | | | | | |
|---|---|---|---|---|---|---|---|
| 14.7 | 2V | 40K | D*=1.1×10$^{10}$jones<br>R=0.09A/W | 20<br>périodes<br>inGaAs<br>$L_b$=500Å | US army<br>research lab | 1998 | 45 |

*tab. 1-4 : Bibliographie des détecteurs QWIP répondant autour de 15µm. En l'absence de consensus sur la définition du rendement quantique, il est plus intéressant de comparer les réponses.*

Le nombre de sources reste limité, les données proviennent majoritairement du JPL. La publication de données liées au niveau de performance reste un sujet sensible, que ce soit pour des questions de confidentialité militaire ou industrielle. Du tableau précédent il nous est toutefois possible de dresser la carte d'identité du QWIP à l'état de l'art actuel. Les résultats seront présentés au chapitre suivant pour comparaison avec le détecteur qui a servi à cette thèse.

### 1.4.5. Détecteurs dérivés

Les QWIP se voient généralement reprocher deux défauts majeurs : le besoin d'un réseau de couplage et le nécessaire compromis entre absorption et courant d'obscurité. Cela a déclenché une recherche importante autour des dispositifs à transitions inter-sousbandes pour remédier à ces deux contraintes. Ces recherches ont abouti à plusieurs types de détecteurs dérivés dont voici trois grandes filières le QCD, le QDIP et le QWISP.

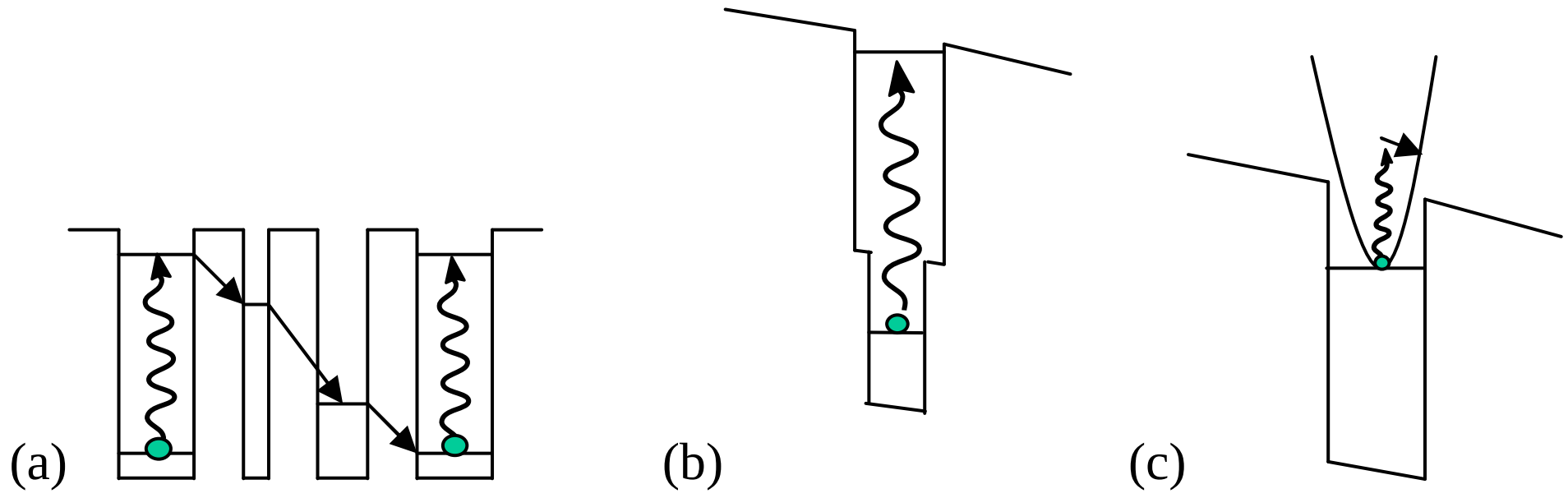

*FIG. 1-10 (a) Détection dans un QCD. (b) QDIP en version DWELL. (c) QWISP.*

#### *1.4.5.1. Quantum Cascade Detector*

Le QCD, pour Quantum Cascade Detector, est la version photovoltaïque du QWIP. Sa structure s'inspire de celle des lasers à cascades quantiques. Ils ont été développés à peu près simultanément en France (par Thales sous l'impulsion de Vincent Berger[46]) et en Suisse par Hofstetter[47]. Leur principe est le suivant : une transition photonique entre un niveau fondamental et un niveau excité est obtenue dans un puits assez large, voir FIG. 1-10(a). Le niveau excité est ensuite couplé à une cascade de niveaux au sein de laquelle l'électron diffuse jusqu'à la période suivante. D'abord conçu en bande II et III[48,49], leur développement dans le THz est plus récent[47,50]. Leur utilisation ne se limite plus à GaAs, mais est possible dans d'autres matériaux comme GaN[51] (~2µm), dans ce dernier de complexes effets piézoélectriques sont alors à prendre en compte. Enfin très récemment les premiers détecteurs à cascade quantique interbandes[52,53], faisant intervenir des hétérostructures de type II, ont été présentés pour le MWIR.

#### *1.4.5.2. Quantum Dot Infrared Photodetector*

Les Quantum Dots Infrared Photodetectors (QDIP), reposent non plus sur des puits quantiques, mais sur des boîtes quantiques. Leur développement est plus récent[54], mais le principe reste très proche de celui des QWIP, au fait près que le confinement est à présent 0D. Premier avantage de ce confinement 0D : la disparition de la règle de sélection qui imposait une composante non nulle du champ électrique selon la direction de croissance. Toutefois le caractère plus pyramidal que sphérique des dots obtenus par croissance Stranski Krastanov[55] fait qu'une structure de couplage peut encore être nécessaire. Des améliorations du point de vue du courant d'obscurité sont également reportées[18]. Ces détecteurs souffrent toutefois d'une absorption faible due à la trop faible concentration de dot. Le QDIP se décline en de nombreuses variantes mais la plus populaire est le « dot in the well » (DWELL), dans laquelle le dot est lui-même piégé au sein d'un puits, voir FIG. 1-10(b). Cette version permet un contrôle plus aisé du profil de bande tout en facilitant le repeuplement du niveau fondamental[18].

#### *1.4.5.3. Quantum Well Intra-Subband Photodetector*

Le QWISP, pour Quantum Well IntraSubband Photodetector, a été développé au JPL et est une alternative au QWIP pour les plus hautes longueurs d'onde[56,57]. Son invention est extrêmement récente puisqu'elle date de 2007. Il se base non plus sur des transitions entre sous bandes, mais au sein d'une même sous-bande, voir FIG. 1-10(c). Ce type de transition, normalement interdit par le caractère vertical des transitions photoniques, est rendu possible par la diffusion importante que subissent les électrons. Cette diffusion est la conséquence directe des forts taux de dopage de ces structures. Ce fort taux de dopage (dix à cent fois plus fort qu'un QWIP de même longueur d'onde de détection) permet également une forte absorption. Le fonctionnement de ce détecteur ressemble donc beaucoup à celui d'un QWIP possédant une transition liée-libre, la réponse spectrale s'en trouve élargie par rapport au QWIP classique. Autre avantage il ne nécessite pas de structure de couplage. Les performances restent toutefois modestes.

## 1.5. Conclusion

J'ai dans ce chapitre présenté les grandes étapes technologiques qui ont conduit à l'utilisation des hétérostructures de semi-conducteurs dans la détection infrarouge. Je me suis focalisé sur la filière des détecteurs à puits quantiques (QWIP) et j'ai présenté ses spécificités, avantages et points faibles. Dans le chapitre suivant je vais présenter la caractérisation d'un composant QWIP haute longueur d'onde à l'état de l'art actuel. Puis j'utiliserai les résultats obtenus pour évaluer les performances d'une caméra basée sur ce détecteur dans le cadre d'un scénario faible flux.

# 2. CARACTERISATIONS ELECTRO-OPTIQUES d'UN QWIP A 15µm.

## Sommaire

**Publications associées à ce chapitre :**

- E. Lhuillier, I. Ribet-Mohamed, M. Tauvy, A. Nedelcu, V. Berger, E. Rosencher, Ultimate performance of Quantum Well Infrared Photodetectors in the tunneling regime, Infrared. Phys. Techn. **52**, 132 (2009).
- E. Lhuillier, I. Ribet-Mohamed, N. Péré-Laperne, M. Tauvy, J. Deschamps, A. Nedelcu, E. Rosencher, 15µm Quantum well infrared photodetector for thermometric imagery in cryogenic windtunnel, accepté dans Infrared Phys. Techn. (2010).

J’ai au chapitre précédent décrit le fonctionnement des QWIP, je vais à présent m’intéresser à la caractérisation d’un composant QWIP dont la réponse est située autour de 15µm. Cette caractérisation va en particulier se focaliser sur les performances du détecteur quand celui-ci est éclairé par un faible flux de photons infrarouges. Je présenterai en particulier le banc de mesure dédié que j’ai pu réaliser, avec un intérêt particulier pour la linéarité de la réponse sur une large gamme de flux de photons. Je montrerai que le composant étudié atteint un niveau de performance plus que comparable à l’état de l’art actuel pour cette longueur d’onde. Enfin je présenterai les résultats d’une simulation qui vise à estimer les performances d’un instrument optique basé sur ce détecteur. Et je ferai la démonstration que le QWIP, grâce à l’amélioration de ses performances, est aujourd’hui un candidat viable pour des applications très bas flux.

# 2.1. Composant

## 2.1.1. Croissance et technologie

Dans tout ce chapitre je vais m’intéresser au niveau de performance atteint par un composant provenant d’Alcatel Thales 3-5lab. C’est un composant que l’on pourrait qualifier de typique du niveau de performance atteint par les QWIP en début de cette thèse (2007).

Le pic de la réponse est situé un peu en dessous de 15µm. Nous allons utiliser ce composant dans des applications de faibles flux de photons et embarquées sur porteur (avion, satellite). Toutefois sa structure n’ pas été optimisé pour de telles applications, car les barrières ne sont pas très épaisses. Sa croissance et sa fabrication technologique ont été réalisées par le 3-5 lab. Deux détecteurs ont été obtenus à partir d’une même croissance. Le premier est fait de mono-éléments dont la taille des pixels est de 50µm et 100µm. Le second est une matrice avec deux tailles pixels (23.5µm et 18.5µm) dont il est possible d’adresser des pixels uniques. Le fait d’utiliser une matrice démontre sa faisabilité et permet d’accéder aux performances du détecteur final. Le tableau ci-dessous (tab. 2-1) donne les paramètres de croissance du composant.

| Grandeur | Valeur |
|---|---|
| $L_w$ | 73 Å |
| $L_b$ | 350 Å |
| %Al | 15.2 |
| $V_b$ | 128 meV |
| $E_1$ | 40 meV |
| $E_2$ | 127 mev |
| $\lambda_{\det}^{theorique}$ | 14.3 µm |
| $N_{2D}$ | $3\times10^{11}$ cm$^{-2}$ |
| $E_f$ | 10.6 meV |

*tab. 2-1 Grandeurs nominales propres au détecteur caractérisé.*

### 2.1.2. Préliminaires expérimentaux

#### *2.1.2.1. Difficultés rencontrées*

La philosophie générale de la caractérisation de ce composant n'est pas du tout celle de mesures de routine. L'objectif est de relaxer au maximum les contraintes sur le point de fonctionnement du détecteur, particulièrement en terme de température, pour en évaluer les meilleurs performances. Autrement dit nous sommes prêt à faire fonctionner le composant à plus basse température, que se qui se fait usuellement, pour en obtenir de meilleures performances.

Nous allons chercher en particulier à atteindre des températures assez basses vis-à-vis de l'utilisation courante de ces détecteurs, typiquement 4K au lieu de 40K. En physique du solide travailler à de telles températures n'a rien de particulièrement difficile. Toutefois la difficulté dans notre cas provient du fait qu'il faille maintenir une température faible, alors que la masse à refroidir est importante. En effet, il ne faut pas se contenter de refroidir l'échantillon mais également l'écran froid qui l'accompagne. Au problème de la masse à refroidir peut s'ajouter l'échauffement dû à l'éclairement de l'échantillon. Il nous faut donc disposer d'une interface mécanique qui soit bonne conductrice de chaleur, et avec de faibles pertes vis-à-vis de l'extérieur.

Conséquence directe de l'utilisation de ce détecteur à basse température, les signaux à mesurer, que ce soit le courant ou le bruit, sont très faibles. Il nous faut donc disposer d'appareils de mesure et d'amplification performants et bas bruits. Et il faut que la connectique soit aussi pensée avec cette philosophie de réduction de bruit.

Dernière difficulté essentielle, il nous faut gérer avec précision la photométrie du dispositif. La caractérisation du composant doit se faire sous faible flux. Or à 15µm pour obtenir de tel flux il faut disposer d'un corps noir à des températures cryogéniques. Il est donc nécessaire de gérer un double système de cryogénie, l'un pour le détecteur, l'autre pour la source de photons. Afin de gérer correctement le flux de photons entre les deux, il faut être capable de réaliser des écrans froids adaptés. L'expérience de l'ONERA sur ce point a été essentielle.

#### *2.1.2.2. Interface opto-électro-mécanique*

Les composants nous sont fournis par le 3-5 Lab sur une céramique, sur laquelle le composant est mécaniquement fixé à l'aide d'une colle assurant aussi la conductivité thermique. La connectique électrique est assurée par des fils d'or (bonding). Nous avons choisi de faire les mesures sur le doigt froid d'un cryostat Janis à circulation d'hélium. Ce cryostat a été préféré à une machine à froid afin de limiter le bruit pouvant provenir du piston nécessaire à la compression de l'hélium.

La céramique qui sert de support à l'échantillon est elle-même fixée sur une pièce en cuivre en contact avec le doigt froid, voir FIG. 2-1. La connectique électrique de l'échantillon est assurée par un nombre réduit de fils. Nous avons préféré limiter le nombre de pixels accessibles afin de limiter le pont thermique en provenance des câbles. Dans le même esprit nous nous contentons d'une mesure deux fils, plutôt qu'un montage quatre fils, ceci étant rendu possible par la grande résistivité des QWIP. Les câbles électriques sont pris relativement longs, afin d'augmenter leur résistance thermique, et sont thermalisés en amont de leur connexion à l'échantillon. Chaque passage de câble est peint en noir, pour éviter un éventuel flux parasite qui pourrait atteindre le détecteur. Enfin plusieurs sondes de température sont disposées sur le montage pour en suivre précisément la température.

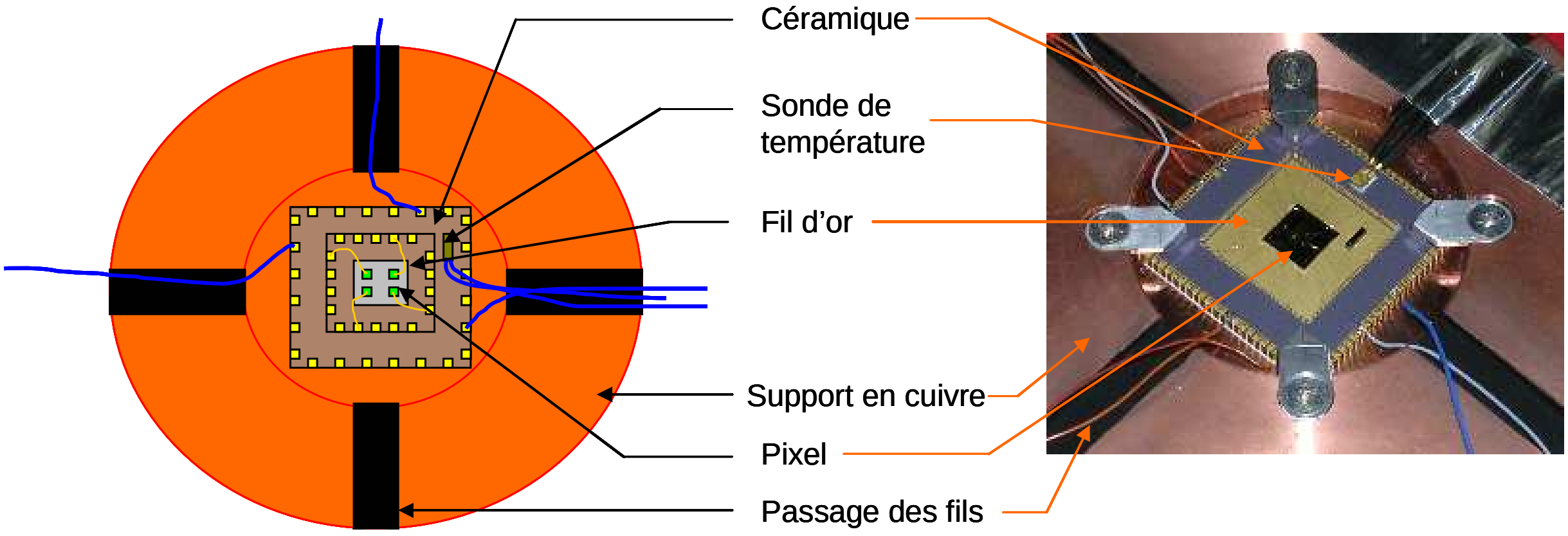

*FIG. 2-1 Photographie et schéma d'une puce et de son interface*

## 2.2. Caractérisation électro-optique classique

Avant de présenter les mesures sous faibles flux qui ont nécessité la réalisation d'un banc spécifique, je vais décrire des mesures de caractérisations électro-optiques classiques de nos détecteurs. Il s'agit de mesure de courant d'obscurité, de mesure de réponse spectrale et de mesure de bruit.

### 2.2.1. Courant d'obscurité

Les applications grandes longueurs d'onde sont particulièrement exigeantes en terme de courant d'obscurité, dans la mesure où les flux visés sont faibles et que les photo-courants sont donc faibles. Il nous faut être capable de mesurer des courants aussi petits que la dizaine de femto ampères.

#### *2.2.1.1. Montage*

Les niveaux de courant mesurés dans nos structures sont compris entre 50fA et 500µA selon la tension appliquée, la température et la taille du pixel à mesurer. Ces mesures de courant d'obscurité nécessitent un appareil de grande résolution avec une bonne dynamique. Notre choix s'est porté sur un sub femtoampèremètre Keithley 6430 qui polarise et mesure le courant simultanément. La régulation thermique est assurée par un Lakeshore 330. Trois sondes de températures sont utilisées sur l'ensemble du montage, voir FIG. 2-2.

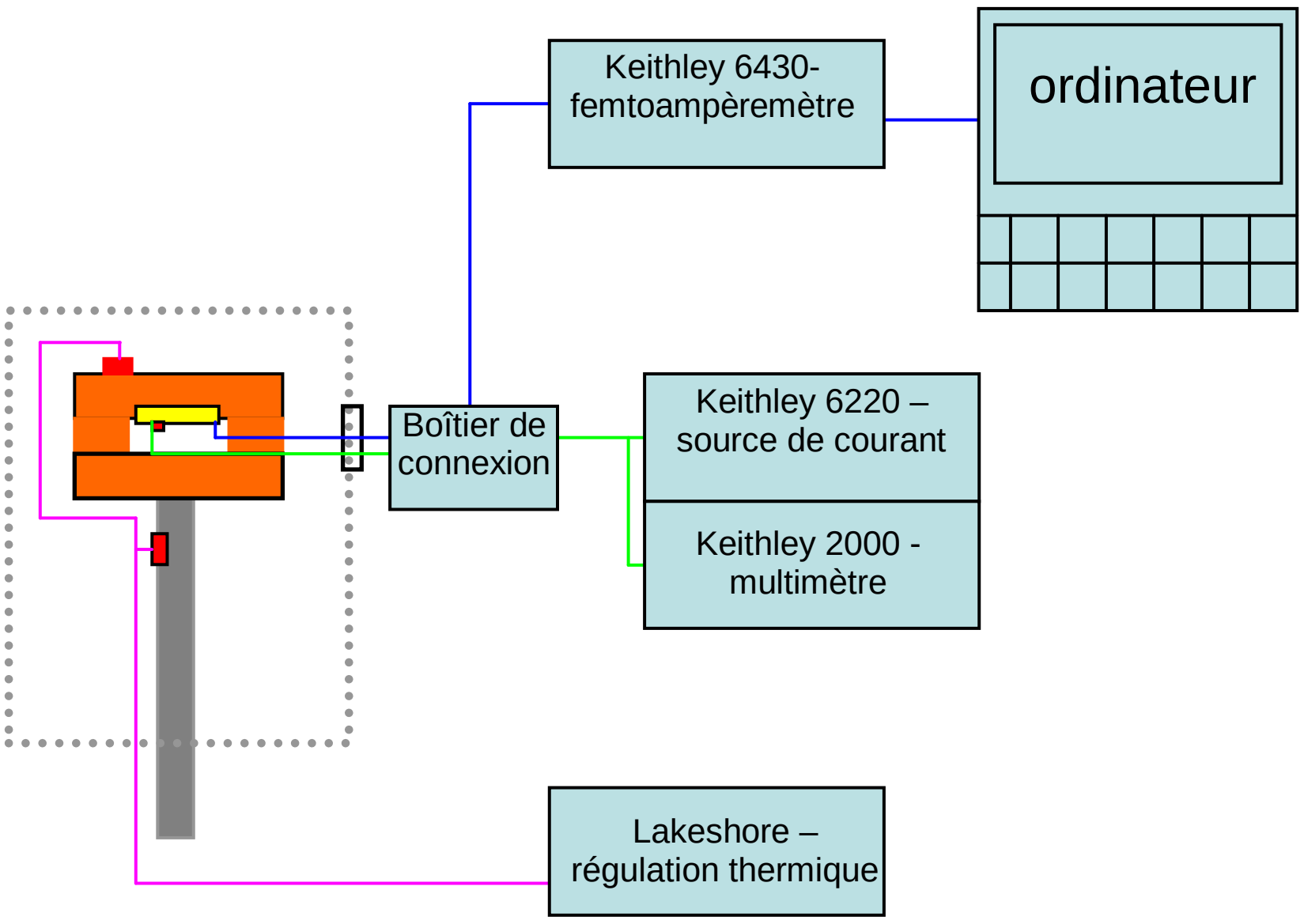

*FIG. 2-2 Schéma du dispositif de mesure du courant d'obscurité.*

#### 2.2.1.2. Mesure I(V)

Trois régimes de transport se distinguent dans le faisceau de courbe *I(V)*, voir FIG. 2-3. Le régime haute température au dessus de 30K, pour lequel les courbes *I(V)* sont strictement monotones avec la tension. Nous verrons au chapitre suivant que ces courbes suivent une loi d'Arrhenius avec la température. A basse température (T<30K) et basse tension, la courbe *I(V)* présente un plateau (résistance différentielle très grande) dont on cherchera à comprendre l'origine au chapitre 3 et 4. Les courbes à différentes températures se superposent, preuve que le transport est alors en régime tunnel. A haute tension (V>2V) le courant s'envole avec la tension. Le transport change alors de régime, selon la zone considérée sur la structure il y a du transport résonant vers le niveau excité[58] (au coeur du QWIP) et des phénomènes d'ionisation par impact[59] (au niveau de l'injecteur).

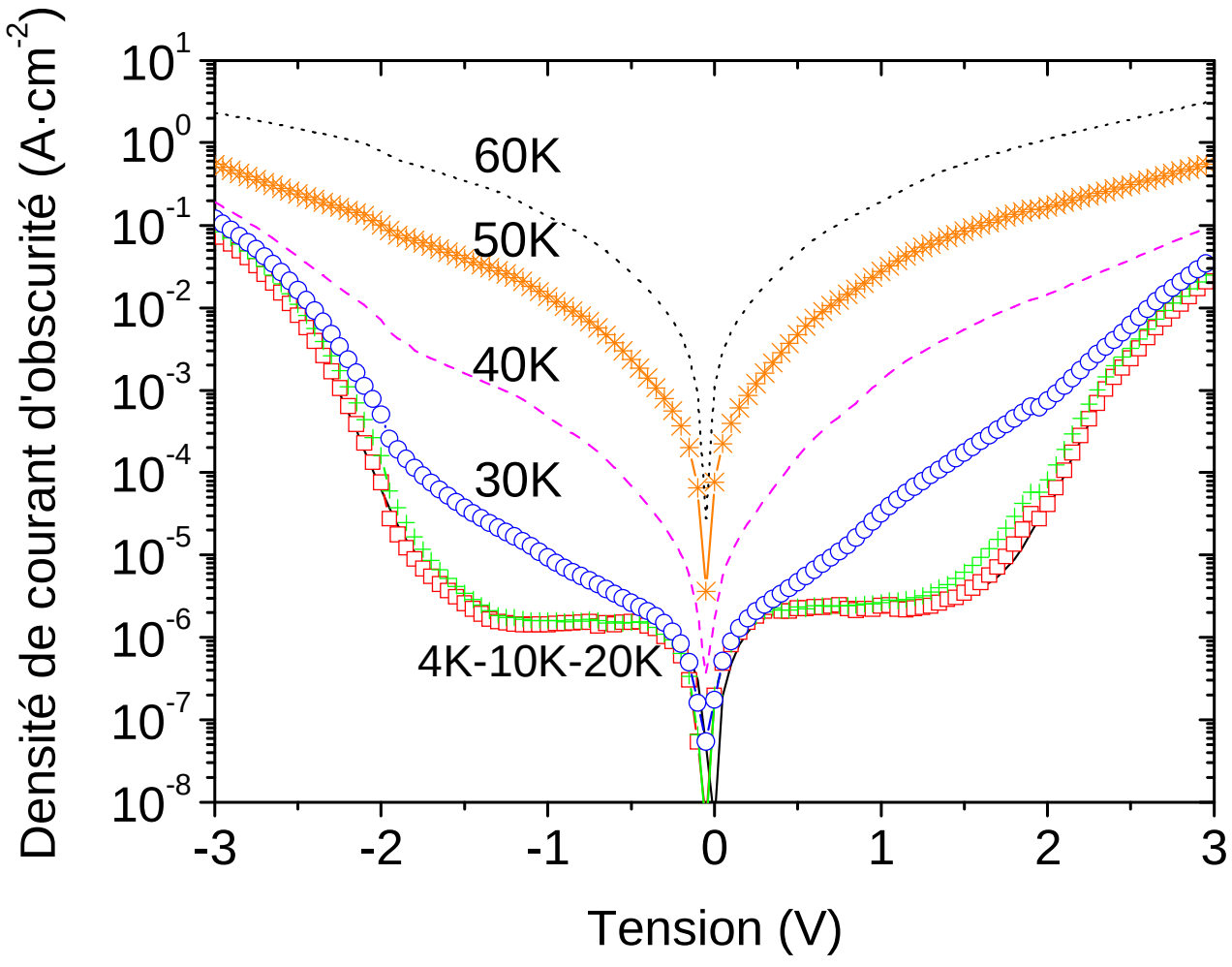

*FIG. 2-3 Courbes I(V) pour des températures comprises entre 4K et 60K*

Les courbes *I(V)* sont légèrement asymétriques, le courant de plateau pour les tensions négatives est deux fois plus faible que pour les tensions positives. L'asymétrie des courbes *I(V)* reflète une asymétrie au niveau de la structure du super réseau. Elle est due à une diffusion superficielle d'espèces (ségrégation). Cette asymétrie peut affecter les barrières (ségrégation de l'aluminium) ou le dopage[60,61] (ségrégation du silicium).

La stratégie qui consiste à abaisser la température pour diminuer le courant d'obscurité atteint sa limite pour T<30K quand le courant passe en régime tunnel. C'est pourquoi il est indispensable de comprendre ce qui se passe en régime tunnel et de proposer de nouvelles structures dont le courant tunnel résiduel est plus faible.

### 2.2.2. Réponse spectrale

La caractérisation spectrale des composants est une étape essentielle. Ces mesures permettent à la fois un contrôle de la bonne qualité de la croissance et une étude des phénomènes physiques liés au photocourant. Dans ce paragraphe nous allons en particulier mettre en évidence le phénomène d'effet tunnel assisté par photon et champ électrique.

#### *2.2.2.1. Choix du montage*

Nous utilisons la source d'un spectromètre à transformée de Fourrier (Bruker Equinox 55), afin d'éclairer notre composant, voir FIG. 2-4. Le pixel est polarisé à l'aide de l'ampli (Femto – DLCPA 200), qui récupère et amplifie le courant de sortie. Le signal est alors renvoyé vers le FTIR qui en fait la transformée de Fourrier afin d'en extraire le spectre. La régulation thermique du montage est assurée par un Lakeshore 330. Un hublot en ZnSe, non traité, sert d'ouverture optique au cryostat. Le choix de ce matériau est justifié par une faible absorption sur la plage 5-20µm. La résolution du FTIR est de 4 cm$^{-1}$, soit à 15 µm, une résolution de 90 nm en longueur d'onde.

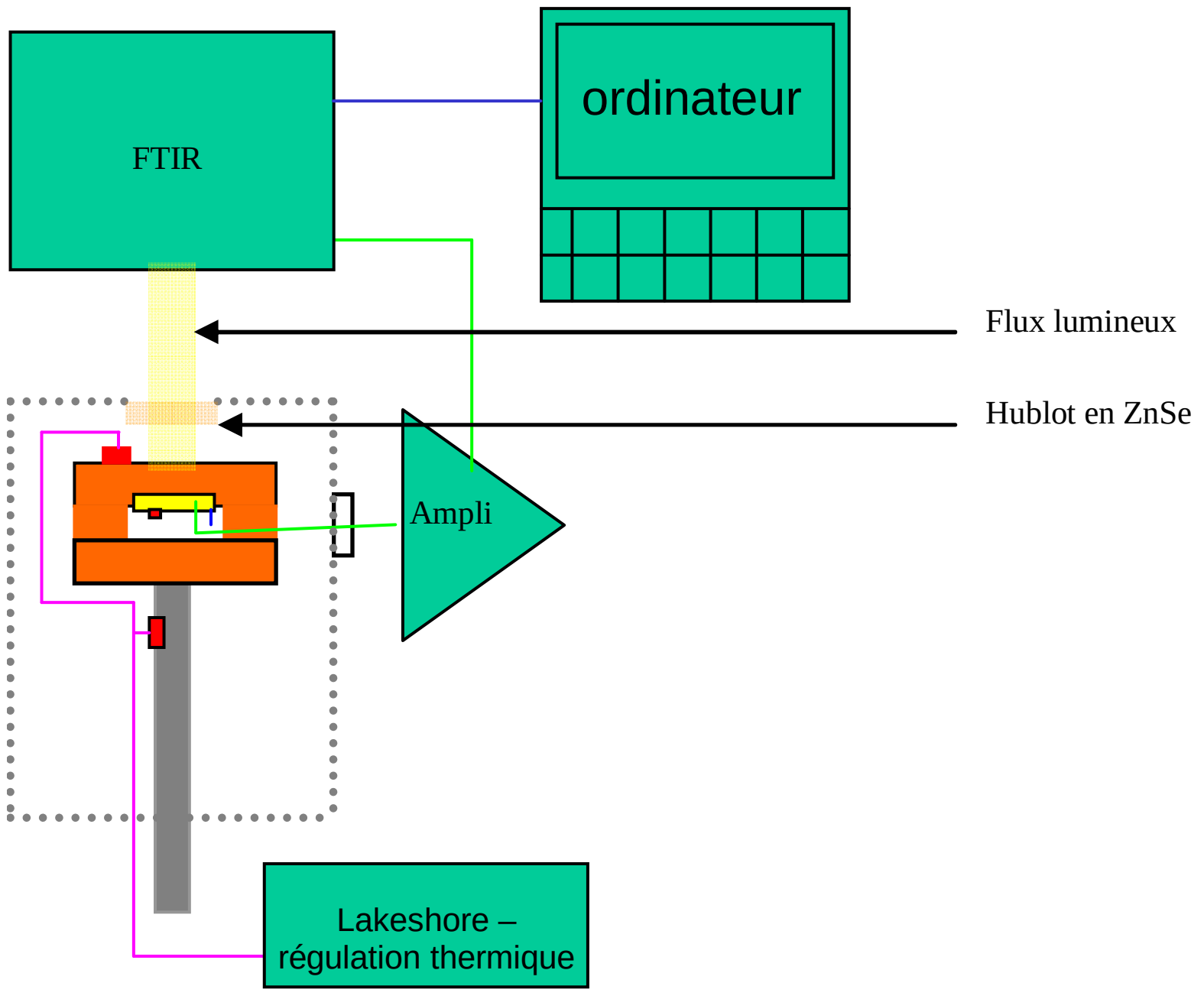

*FIG. 2-4 Montage électrique nécessaire à l'obtention de la réponse spectrale*

La FIG. 2-5 montre la réponse spectrale du composant. La réponse pic de ce dernier est située autour de 14.7µm avec un élargissement à mi hauteur de 2.1µm, ce qui est classique pour une transition lié-quasi liée.

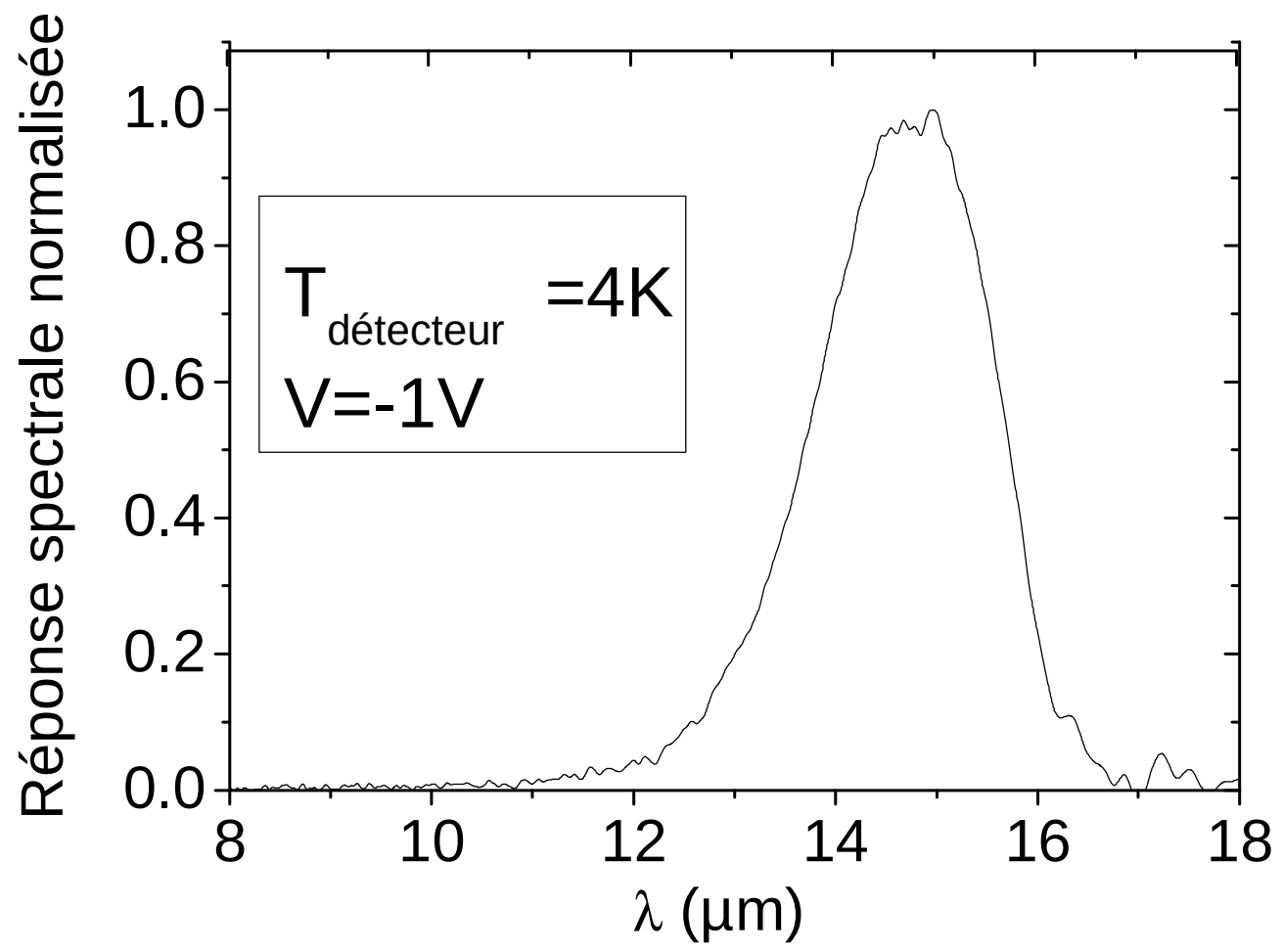

*FIG. 2-5 Réponse spectrale, pour un pixel de100×100 µm².*

### *2.2.2.2. Effet de la polarisation*

Il est possible d'observer dans la zone de basse énergie du spectre un déplacement de celui-ci avec la tension, voir FIG. 2-6. Un zoom sur cette zone permet de mieux appréhender le phénomène.

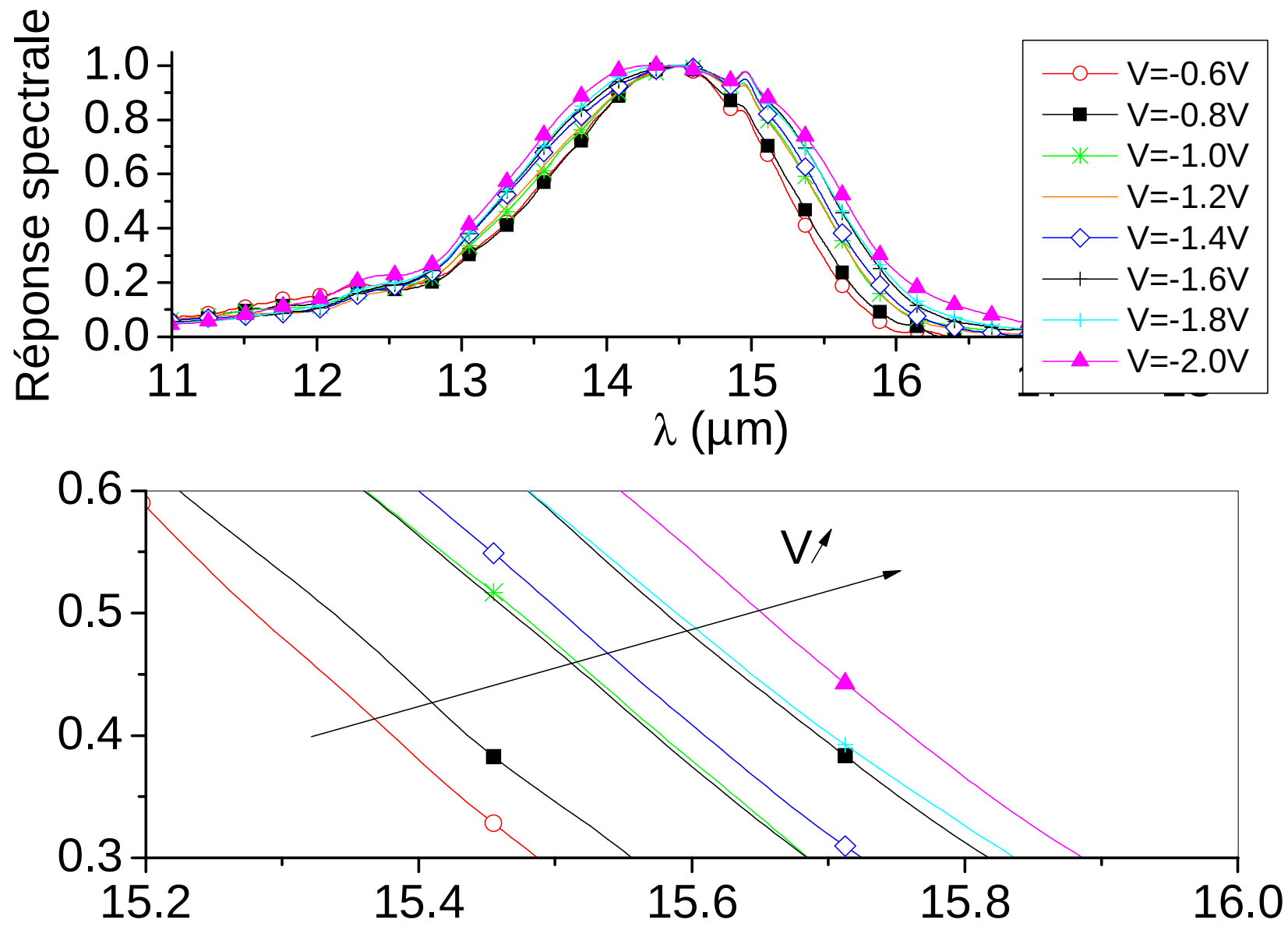

*FIG. 2-6 Réponse spectrale en fonction de la longueur d'onde à différentes tensions de polarisation et son zoom sur la zone ou l'effet du champ électrique est le plus fort*

Nous observons un élargissement du spectre vers les grandes longueurs d'onde quand la tension augmente, voir FIG. 2-6. Nous attribuons cet effet au phénomène d'effet tunnel assisté par photon et par champ électrique. Les photons qui sont absorbés dans cette partie du spectre ont une énergie juste inférieure à celle de la transition vers le niveau excité. L'électron qui absorbe un tel photon est toujours piégé dans le puits sur un niveau virtuel. Toutefois à cette énergie la barrière est relativement fine et sera d'autant plus fine que le champ électrique sur la barrière sera fort. L'électron peut donc facilement la traverser par effet tunnel. Le chemin suivi par l'électron est montré dans l'encart de la FIG. 2-7. Pour valider cette hypothèse nous avons tracé l'évolution du photo-courrant avec la tension et nous l'avons comparée avec celle de la transparence tunnel. Pour une énergie de photon fixée, la réponse mesurée, en fonction de la tension, doit donc être dans le rapport des transparences tunnels[i] : $\frac{I(V)}{I(V_0)} = \frac{T(V)}{T(V_0)}$ avec $V_0$ une tension référence. La FIG. 2-7 montre l'excellent accord obtenu entre les données expérimentales et la courbe théorique, si l'on admet un léger ajustement de la hauteur de barrière.

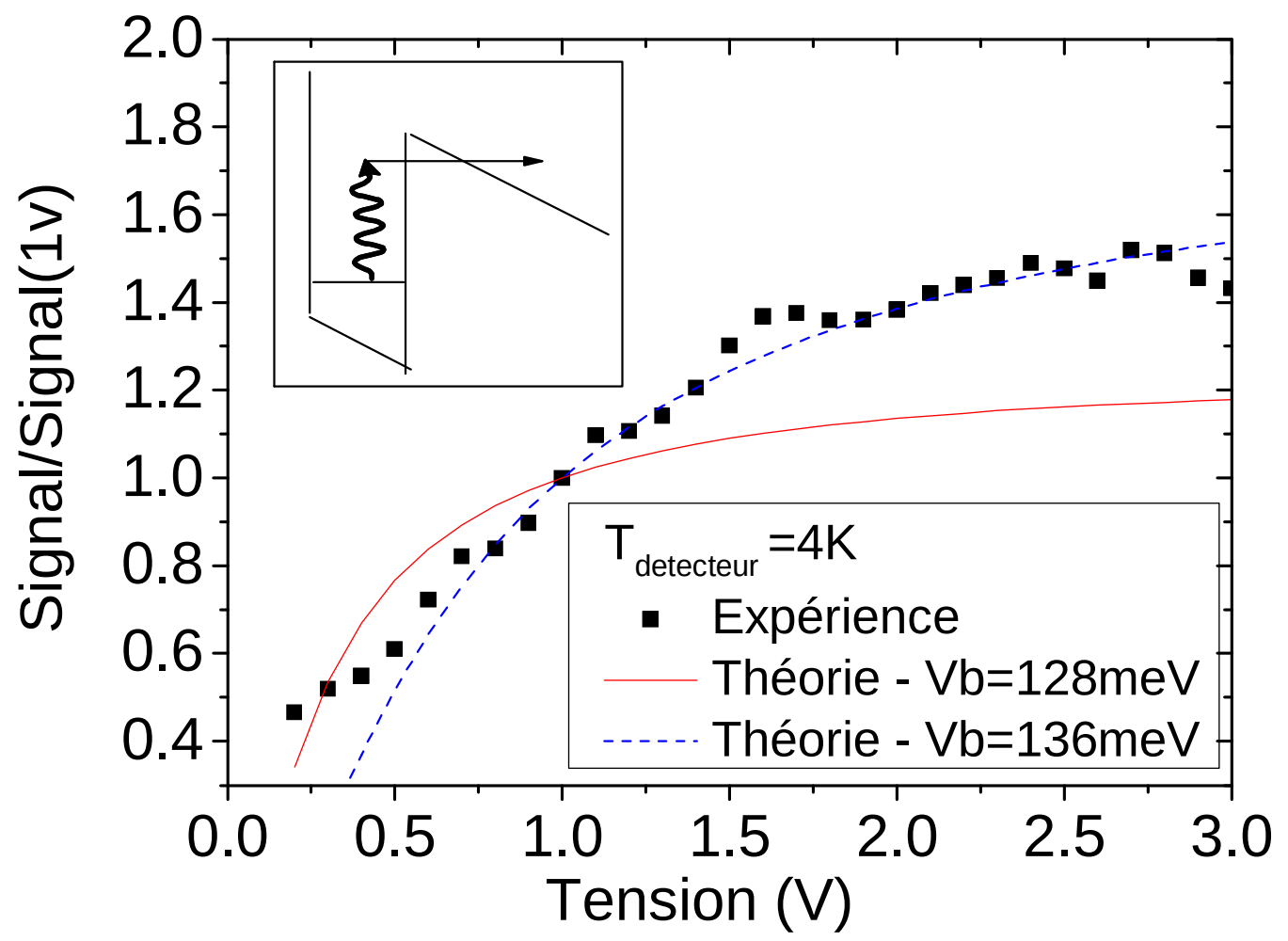

*FIG. 2-7 Evolution des rapports des courants $\frac{I(V)}{I(V_0)}$ et des transparences tunnel $\frac{T(V)}{T(V_0)}$ en fonction de la tension pour une longueur d'onde de 15.5$\mu m$ et $V_0$=1V. Encart : schéma du chemin suivi par l'électron lors du processus d'effet tunnel assisté par photon et champ électrique.*

### 2.2.3. Mesure du bruit

La mesure du bruit est une mesure essentielle pour estimer le niveau de performance d'un détecteur. Je présente ici les mesures qui ont été faites sous obscurité.

---

[i] La transparence tunnel est ici calculée en utilisant l'approximation WKB qui sera plus longuement discutée au chapitre suivant.

Le résultat des mesures de bruit sous faible flux seront présentées dans la seconde partie de ce chapitre.

Dans les QWIP, il est généralement admis que le bruit[29] provient de deux sources que sont le bruit thermique et le bruit de génération recombinaison. Cette écriture du bruit s'inspire du modèle de Beck[62], la densité spectrale de bruit (DSB) peut se mettre sous la forme :

$$DSB^2 = 4eIg_{noise} + \frac{4k_bT}{R_{dyn}} \qquad (2\text{-}1)$$

Avec I le courant dans la structure, $R_{dyn}$ sa résistance différentielle et $g_{noise}$ le gain de bruit qui *a priori* est différent du gain de photoconduction. Le premier terme provient du bruit de génération recombinaison tandis que le second résulte du bruit thermique. La résistance des QWIP étant importante, c'est le bruit de génération recombinaison qui domine dans ces structures.

#### *2.2.3.1. Dispositif de mesure*

La difficulté des mesures de bruit réside dans la faiblesse du signal mesuré. De nombreuses précautions ont été prises afin d'obtenir des mesures propres : établissement d'un plan de masse sous le cryostat, développement d'une connectique blindée appropriée, limitation des boucles de masse. La résolution de notre dispositif se situe autour de 3 fA·Hz$^{-1/2}$. Le montage électrique finalement retenu pour la mesure de bruit est présenté sur la FIG. 2-8.

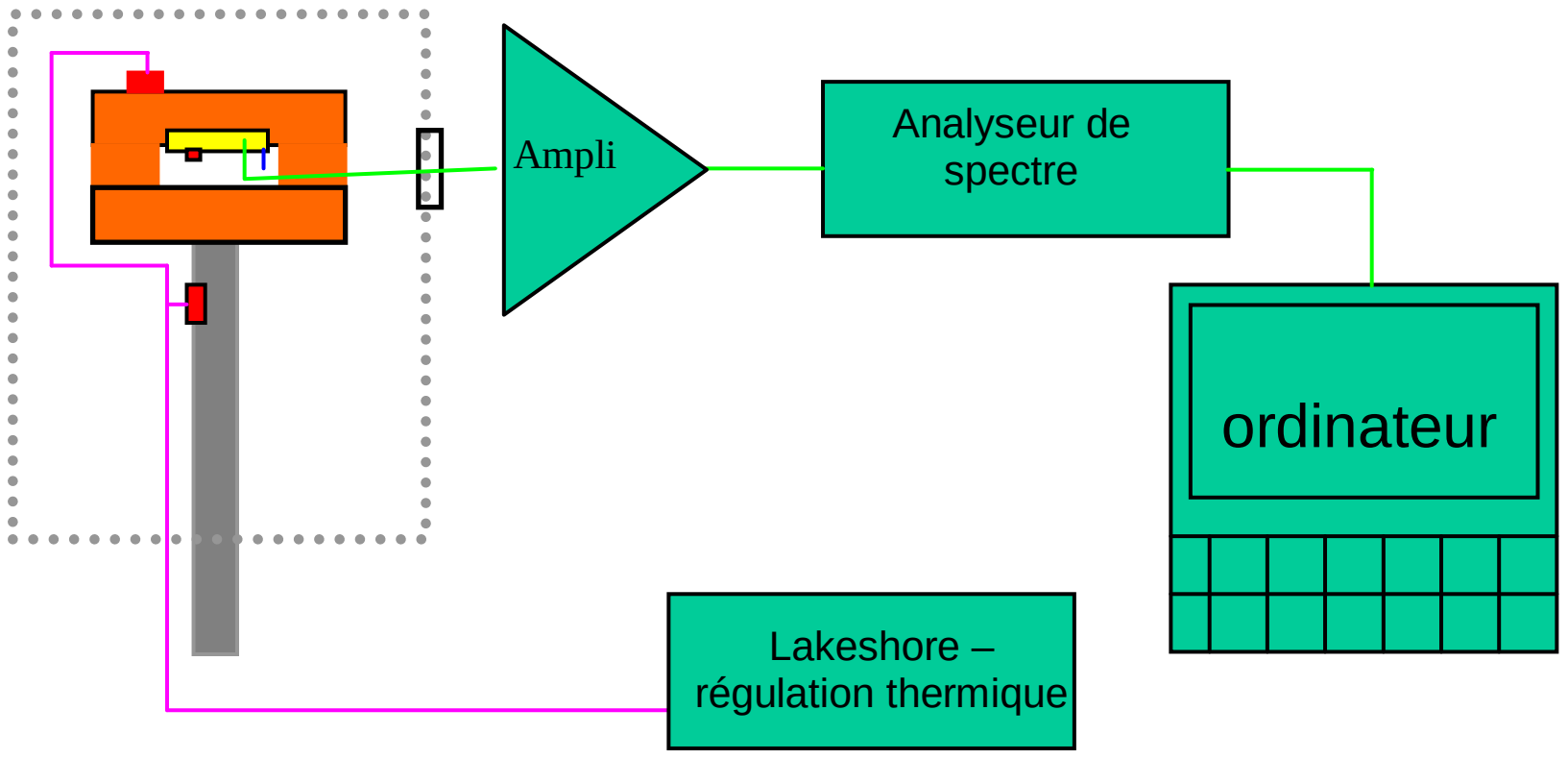

*FIG. 2-8 Schéma du montage de l'expérience de mesure de bruit*

Les pixels sont polarisés via la source de tension d'un ampli à trans-impédance (FEMTO – DLCPA 200), dont le gain varie typiquement de $10^5$ à $10^9$. Le signal est analysé par un analyseur de spectre (Ono Sokki – CF5220Z). La régulation thermique est assurée par un Lakeshore 330, dont les impulsions de courant nécessaires à la régulation sont filtrées afin de limiter le couplage thermo-électrique. L'acquisition du bruit suit un protocole définit par le 3-5Lab.

### 2.2.3.2. Résultats

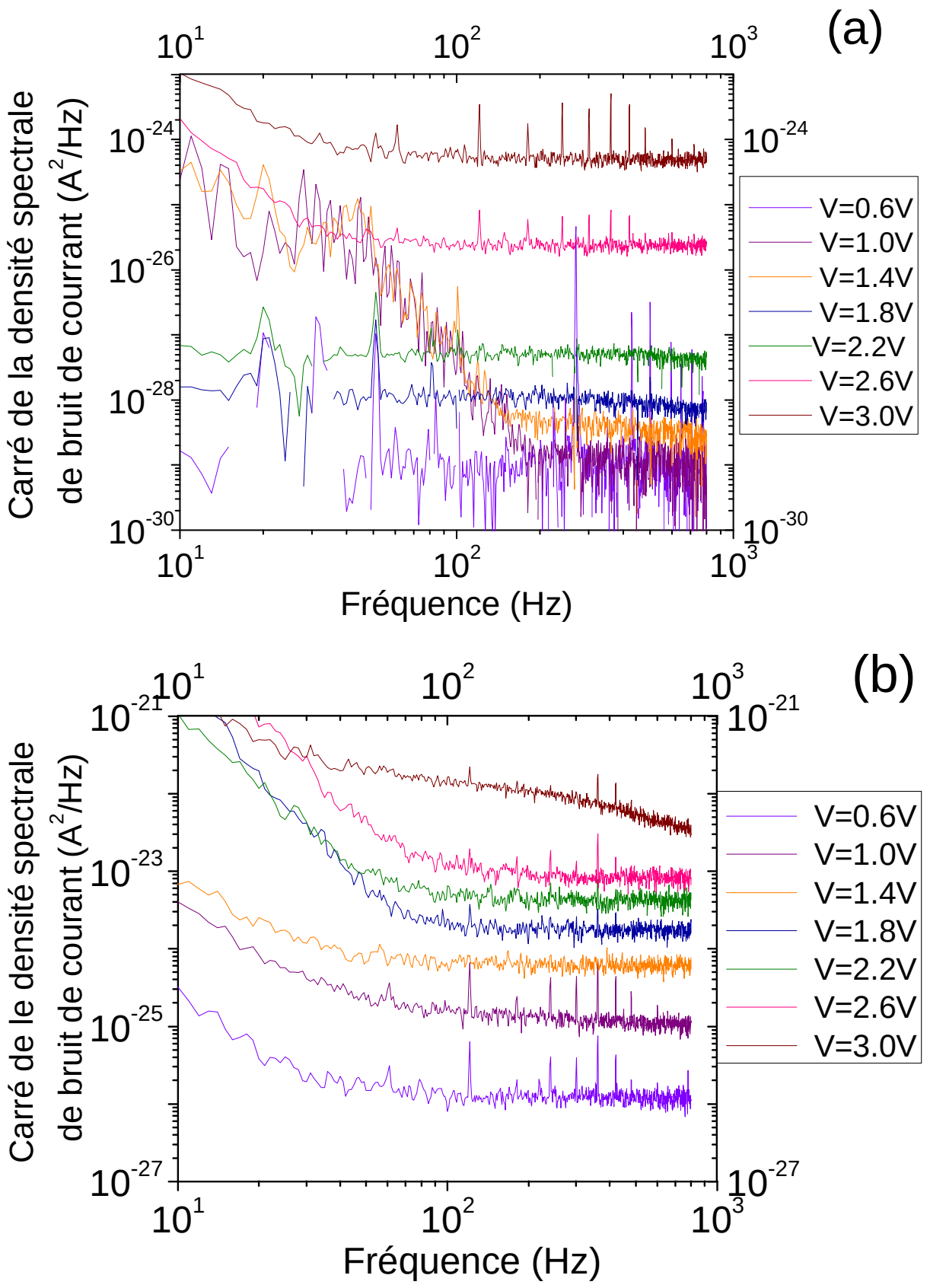

*FIG. 2-9 Carré de la densité spectrale de bruit à 4K (a) et 50K (b) pixel 100×100µm².*

La FIG. 2-9 présente des spectres de mesure de bruit pour deux températures différentes. Au delà de 100Hz chacune des courbes présente un plateau de bruit dont la valeur en fonction de la tension est rapportée par la FIG. 2-10.

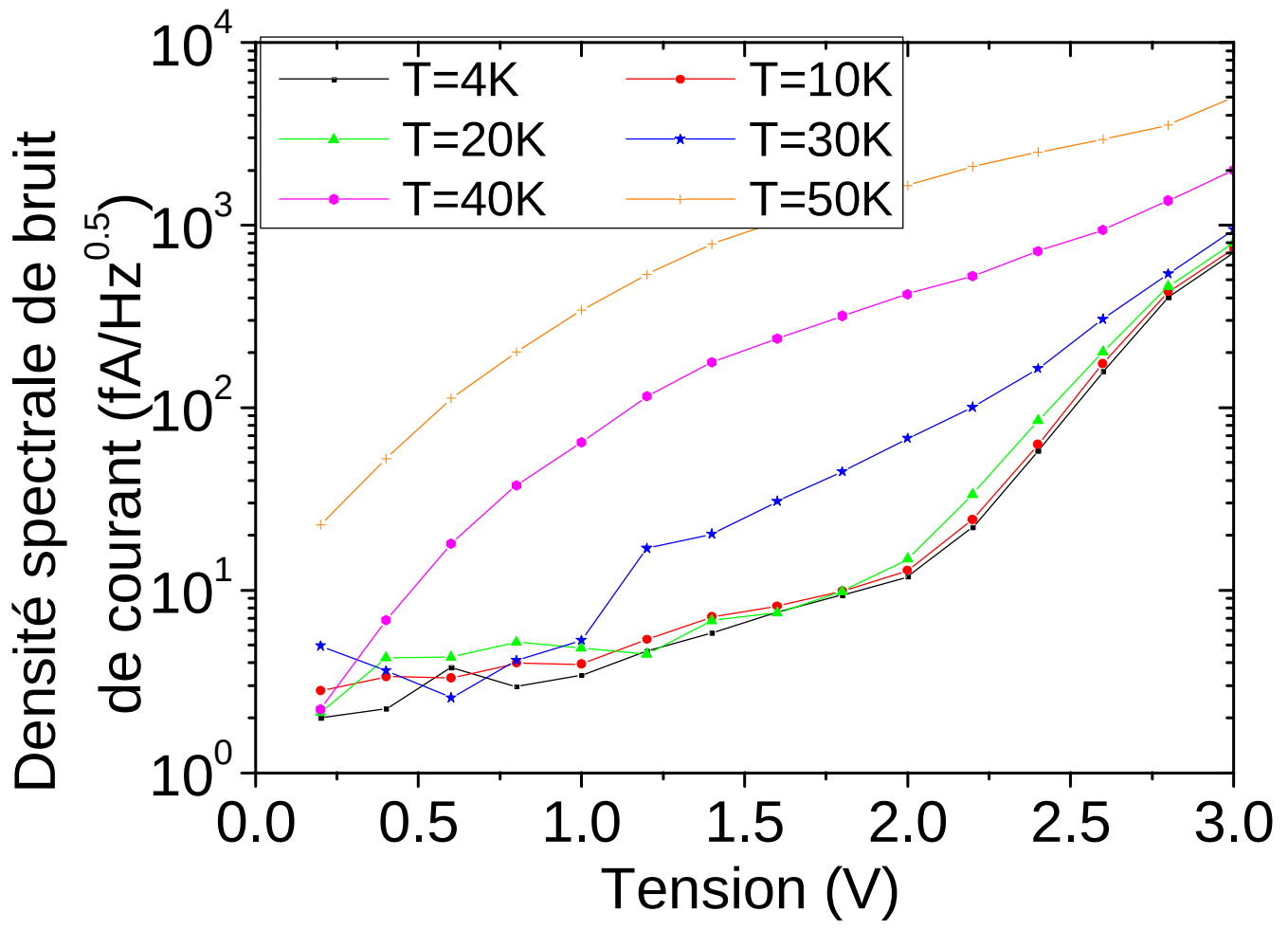

*FIG. 2-10 Densité spectrale de bruit au niveau du plateau de courant tunnel, en fonction de la tension de polarisation pour différentes températures - pixel 100×100$\mu$m$^2$.*

A partir du graphe précédent (FIG. 2-10) il est possible d'extraire la valeur du gain, qui rappelons le, est donné par :

$$g_{noise}(V,T) = \frac{DSB^2 - \frac{4k_bT}{R_{dyn}}}{4eI} \quad (2\text{-}2)$$

Ou DSB est la densité spectrale de bruit, I est le courant préalablement mesuré et $R_{dyn}$ est obtenue en dérivant les courbes de courant. C'est cette procédure qui sera utilisée par la suite pour calculer le gain. Seules les mesures de bruit sous flux seront présentées dans la partie suivante de ce chapitre

## 2.3. Mesures sous faible flux

Les mesures sous flux répondent à un triple objectif : premièrement, caractériser le détecteur au plus proche de son utilisation finale. Deuxièmement, les mesures doivent permettre une évaluation de la linéarité de la réponse du composant. Enfin, les données obtenues doivent servir d'entrée à notre code de simulation de performance d'instrument optique basé sur ce détecteur.

### 2.3.1. Dispositif de mesure

#### *2.3.1.1. Besoins*

Le banc de mesure qui doit servir aux mesures faibles flux doit répondre à trois contraintes principales :

- Afin que la caractérisation du détecteur se fasse au plus proche des conditions réelles d'utilisation, nous souhaitons soumettre le détecteur à un flux large spectralement plutôt qu'à une excitation piquée en longueur d'onde. Cela passe donc par l'illumination du détecteur par un corps noir plutôt qu'un laser.

- Les flux visés par nos applications sont comme nous le verrons à la fin de ce chapitre de l'ordre de $10^{13}$-$10^{14}$photons·$s^{-1}$·$cm^{-2}$, ce qui compte tenu de la forme de la réponse spectrale, est un flux très faible. Le corps noir doit donc atteindre des températures cryogéniques.

- Enfin et afin de tester la linéarité du composant nous souhaitons que le dispositif de caractérisation permette de soumettre le détecteur à une large gamme de flux. La température du corps noir doit donc pouvoir varier sur une plage importante.

#### *2.3.1.2. Solutions retenues*

Nous avons donc choisi un banc de mesure basé sur l'utilisation de deux cryostats qui se font face. Le premier, opérant à l'hélium, est utilisé pour refroidir le détecteur, tandis que le second, fonctionnant à l'azote, refroidit le corps noir. Ce dernier nous permet alors d'explorer des températures comprises entre 77K et 340K (le corps noir est alors chauffé), au sein de la même enceinte. Je me contenterai ici de présenter des résultats pour des températures comprises entre 80 et 290 K, plus représentatives des besoins réels.

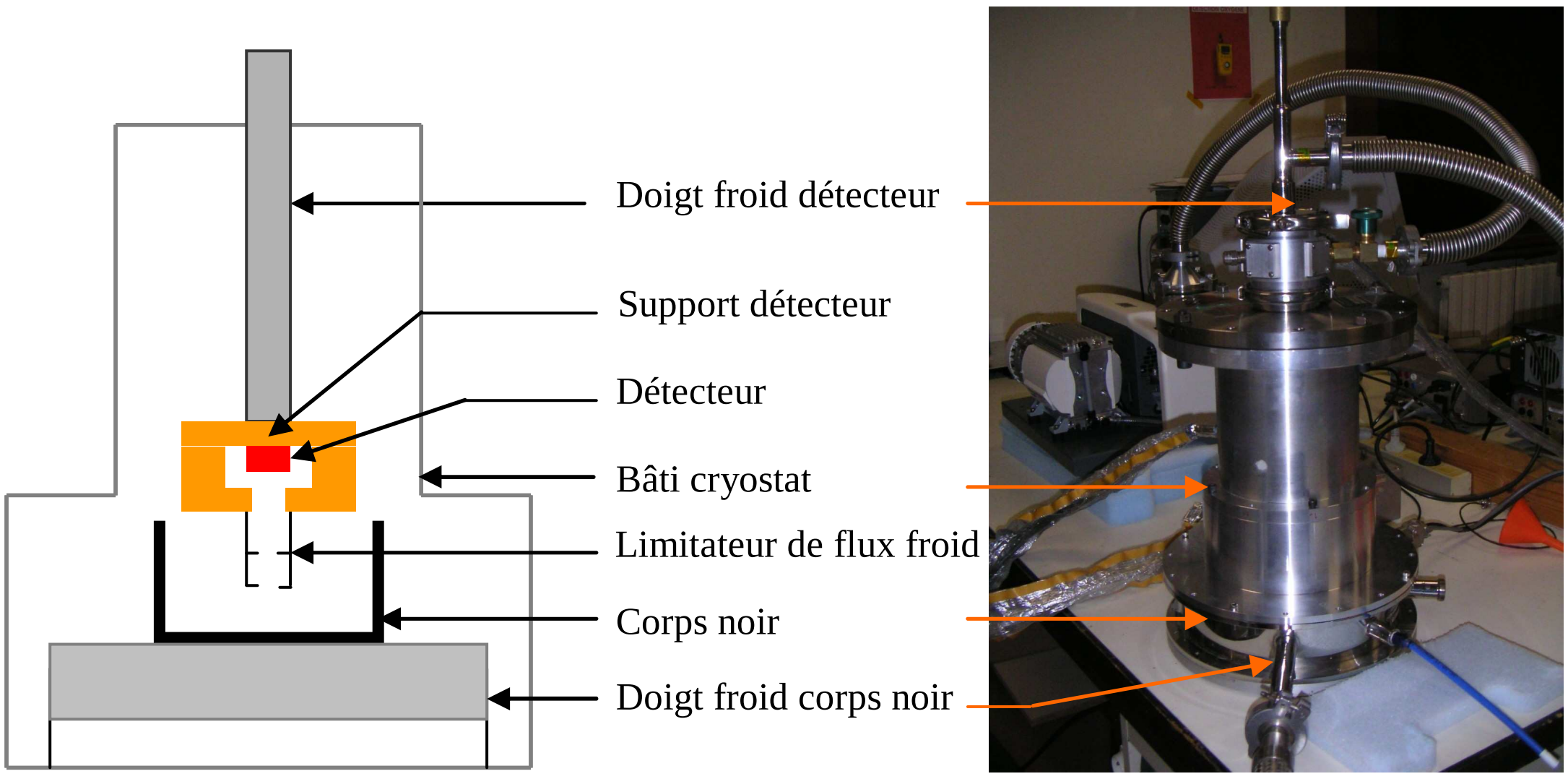

*FIG. 2-11 Schéma et photographie du banc de mesure sous faible flux.*

La maîtrise de la photométrie du dispositif est essentielle au calcul du niveau de performance. L'élaboration de l'écran froid doit donc être soignée. Le corps noir se compose d'une pièce métallique peinte en noir avec de la peinture Nextel (Primer 5523-7525 Anthrazit de 3M) de haute émissivité ($\varepsilon \approx 1$), voir FIG. 2-11. La bonne conductivité thermique du métal doit assurer l'uniformité de la luminance du corps noir. Le haut rebord de la pièce limite le flux, en provenance du bâti, reçu par le détecteur. Une seconde pièce métallique fixée sur le support du détecteur sert à définir l'ouverture numérique du banc de mesure. Il est important que cette pièce ne soit pas trop proche du corps noir afin que sa température ne varie pas trop quand celle du corps noir varie. Notons que les diaphragmes intermédiaires de cette pièce servent à limiter les réflexions à sa surface, et cela en plus de la peinture noire qui en recouvre la surface. L'ouverture de ce banc est de f/2.8[ii].

Finalement le graphe FIG. 2-12 donne le flux de photons émis par notre dispositif ainsi que l'éclairement en fonction de la température. Nous réussissons à explorer près de quatre décades de flux situées entre $10^{12}$ photons·s$^{-1}$·cm$^{-2}$ et $10^{16}$photons·s$^{-1}$·cm$^{-2}$, ce qui permet d'encadrer largement la gamme de flux d'intérêt. En terme de puissance émise ces flux se situent entre 20nW·cm$^{-2}$ et 200µW·cm$^{-2}$, ce qui est particulièrement intéressant comparativement à ce qui est rapporté dans la littérature. En effet les gammes de flux précédemment rapportées pour l'étude de la linéarités des QWIP se situent entre 1µW·cm$^{-2}$et 1GW·cm$^{-2}$, voir les références 63, 64 et 65.

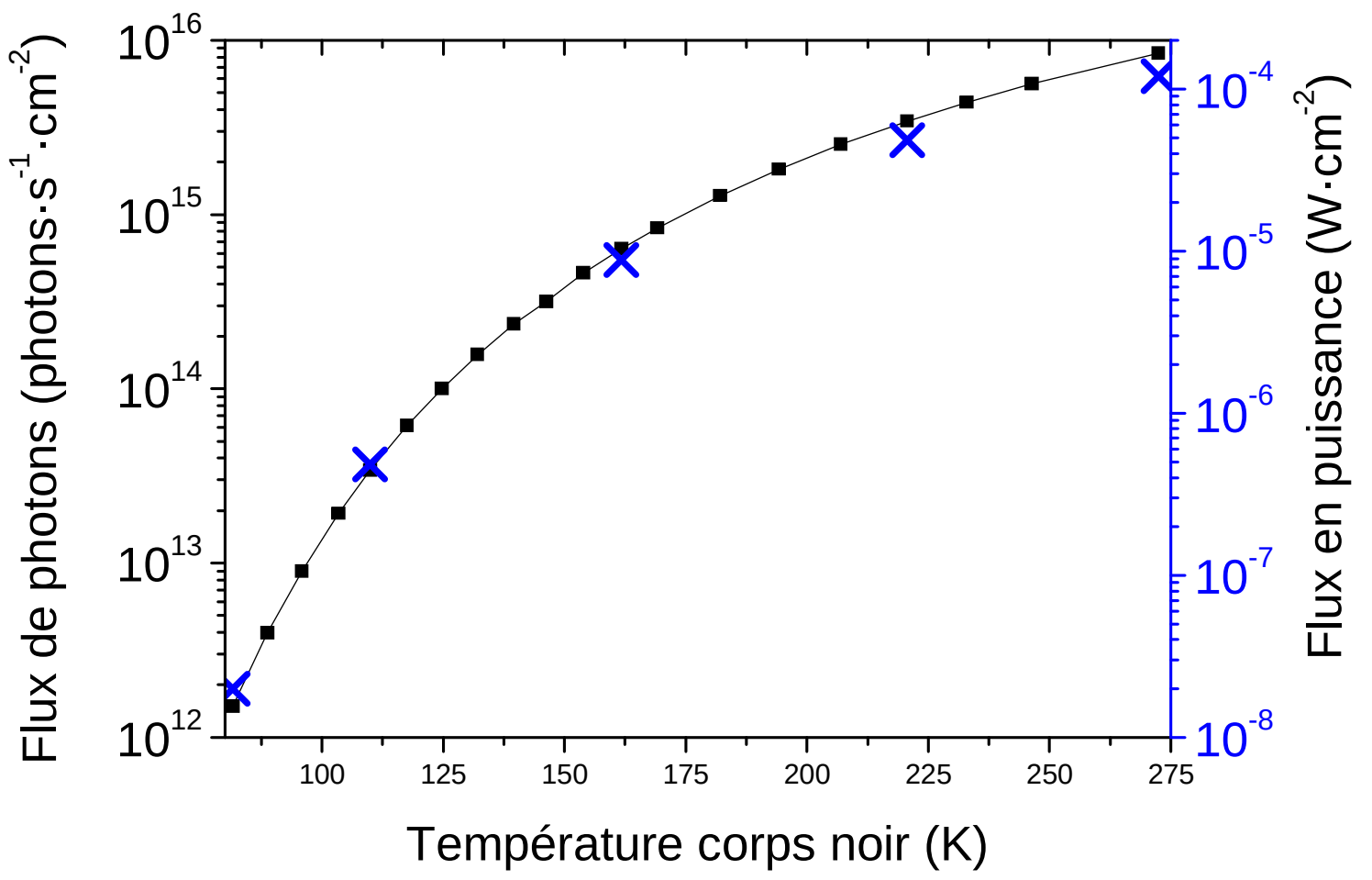

*FIG. 2-12 Flux de photons et éclairement en fonction de la température du corps noir*

## 2.3.2. Mesures

### *2.3.2.1. Mesures réalisées*

Nous avons réalisé deux types de mesures :

- Faisceau de courbe *I(V)* en fonction de la température du corps noir et de la température du détecteur avec pour objectif d'en extraire le rendement quantique externe et la résistance différentielle du composant.

- Mesure de bruit sous flux pour différentes températures de corps noir et de détecteur, l'objectif étant d'extraire le gain de bruit du composant.

Les résultats obtenus seront présentés à la fin de cette partie afin de comparer les performances de ce détecteur avec l'état de l'art actuel.

---

[ii] Les ouvertures classiques des dispositifs hautes longueurs d'onde sont généralement situé entre f/2 et f/9.

#### 2.3.2.2. Rendement quantique

Dans la suite nous définirons le rendement quantique externe comme la dérivée de la densité de courant par rapport au flux divisée par la charge du proton. La relation qui lie le courant total dans la structure est alors donné par :

$$J = J_{dark} + e{\cdot}\eta{\cdot}g{\cdot}\phi \qquad \text{(2-3)}$$

Avec η le rendement quantique interne, g le gain de photoconduction et $\phi$ le flux de photons par unité de surface. La quantité $\eta{\cdot}g$ est appelée rendement quantique externe. Nous avons choisi de suivre l'hypothèse de linéarité locale de la réponse. En faisant l'hypothèse que sur un petit intervalle de flux le rendement quantique est constant, il est possible de déduire le rendement quantique externe.

$$\eta g = \frac{J_{total}(\phi_1) - J_{total}(\phi_2)}{e(\phi_1 - \phi_2)} \qquad \text{(2-4)}$$

Ou $\phi_1$ et $\phi_2$ sont deux valeurs de flux proches. Cette méthode est préférable à celle qui consisterait à évaluer de la façon suivante $\eta{\cdot}g = \frac{(J - J_{dark})}{e{\cdot}\phi}$. En effet cette approche peut conduire à des aberrations si le courant d'obscurité n'est pas exactement le même entre la mesure sous flux et la mesure de courtant d'obscurité. De telles erreurs peuvent par exemple se produire en régime thermique si la température du détecteur n'est pas exactement la même. Cette erreur est d'autant plus grande que le flux est faible, ce qui est notre cas. Voila pourquoi nous n'avons pas retenu cette procédure.

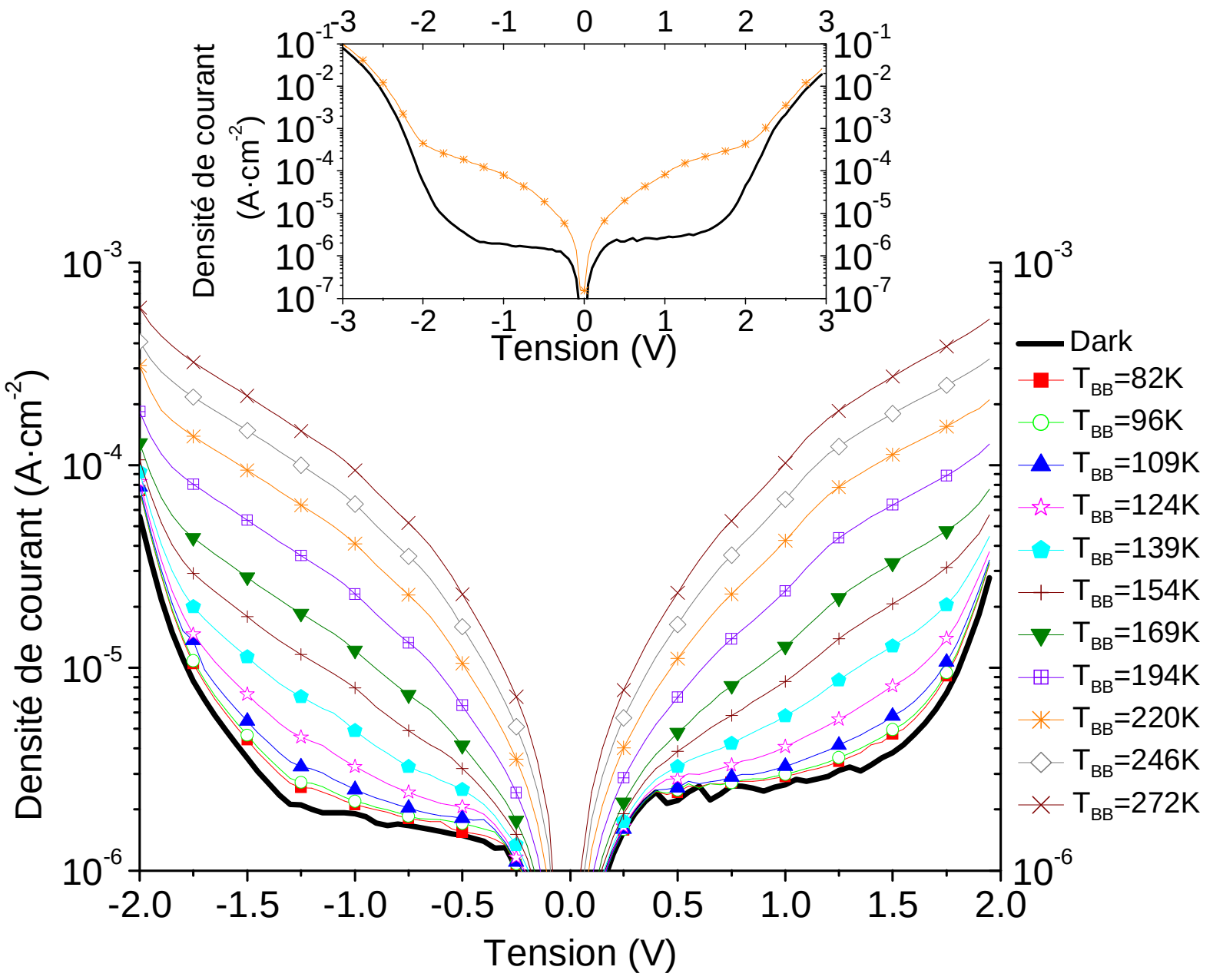

*FIG. 2-13 Faisceau de courbes I(V) réalisé à 4K sous différentes températures de corps noirs. Encart : courbe I(V) sur une plage plus étendue de tension, la température de corps noir est de 220K.*

La figure FIG. 2-13 montre un faisceau de courbes *I(V)* pour différentes valeurs de température du corps noir. Dès 80K le courant total se distingue du courant d'obscurité. A haute tension (V>2V,) le signal est essentiellement composé de courant d'obscurité, voir l'encart de FIG. 2-13. Le point de fonctionnement du détecteur sera donc à chercher dans une plage de tension inférieure à ce seuil.

#### 2.3.2.3. Linéarité du composant

En suivant la procédure décrite précédemment nous pouvons déduire de ce faisceau de courbes le rendement quantique du détecteur en fonction du flux. Les résultats sont présentés pour différentes tensions sur le graphe FIG. 2-14. Ce graphe montre une très faible dépendance du rendement quantique avec le flux auquel le détecteur est exposé. Le rendement quantique du détecteur vaut 7% sous -1V et 30% sous -2V.

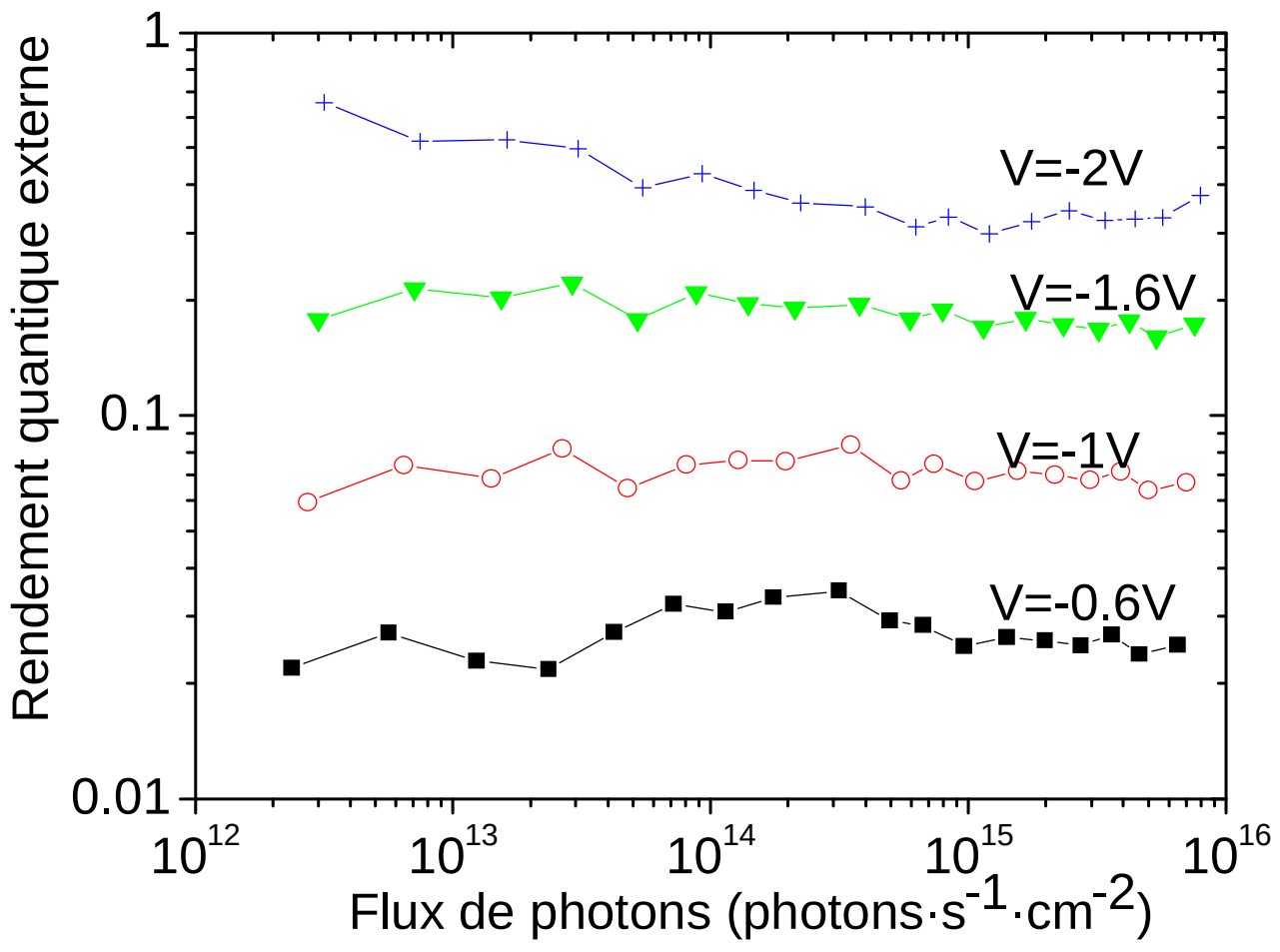

*FIG. 2-14 Rendement quantique externe en fonction du flux de photons pour différentes tensions de polarisation du détecteur.*

La linéarité des QWIP peut paraître de prime abord surprenante dans la mesure où l'exposition à un flux du détecteur implique une réorganisation de la structure de champ. Comme, de plus, le comportement en courant avec le champ n'est généralement pas linéaire, il est étonnant que ces différents processus conduisent à une réponse du détecteur quasi indépendante du flux auquel il est exposé. Néanmoins la linéarité des QWIP est excellente comme le rapporte les références 63, 64 et 65. Toutefois dans notre cas cette linéarité est moins surprenante puisque le flux appliqué est faible. Ainsi il faut s'attendre à ce que cette réorganisation de la distribution de champ soit modérée.

Au-delà de 2V le transport rentre dans une zone de champ où intervenir le phénomène d'ionisation par impact[59] qui peut provoquer une multiplication des porteurs. Toutefois il ne faut pas y voir la possibilité d'un point de fonctionnement intéressant dans la mesure où ce processus s'accompagne d'une explosion du bruit.

#### 2.3.2.4. Fiabilité des mesures

Sous -1V le rendement quantique moyen est de 7.1% avec un écart type des données, sur plus de trois ordres de grandeur de flux, de 0.6%. La réponse associée est de l'ordre de $0.8 A \cdot W^{1}$.

Dans la mesure où le rapport signal sur bruit visé reste assez faible, il est important de valider la précision des mesures. Les sources d'imprécision sur la mesure du rendement quantique externe sont multiples :

- Méconnaissance du flux, celle-ci peut résulter d'une imprécision sur la température du corps noir ou d'une inhomogénéité de température de la surface du corps noir.
- Méconnaissance du flux, liée à une imprécision sur la valeur des paramètres géométriques de l'écran froid.
- Méconnaissance du flux, liée à la présence de flux parasite.
- Bruit de mesure sur le courant.

La FIG. 2-15 redonne le rendement quantique externe sous -1V du composant avec les barres d'erreur sur le flux et le rendement quantique. Notons que l'imprécision est généralement plus forte aux faibles flux. L'incertitude associée à ces mesures est de l'ordre de quelques dizaines de pourcent.

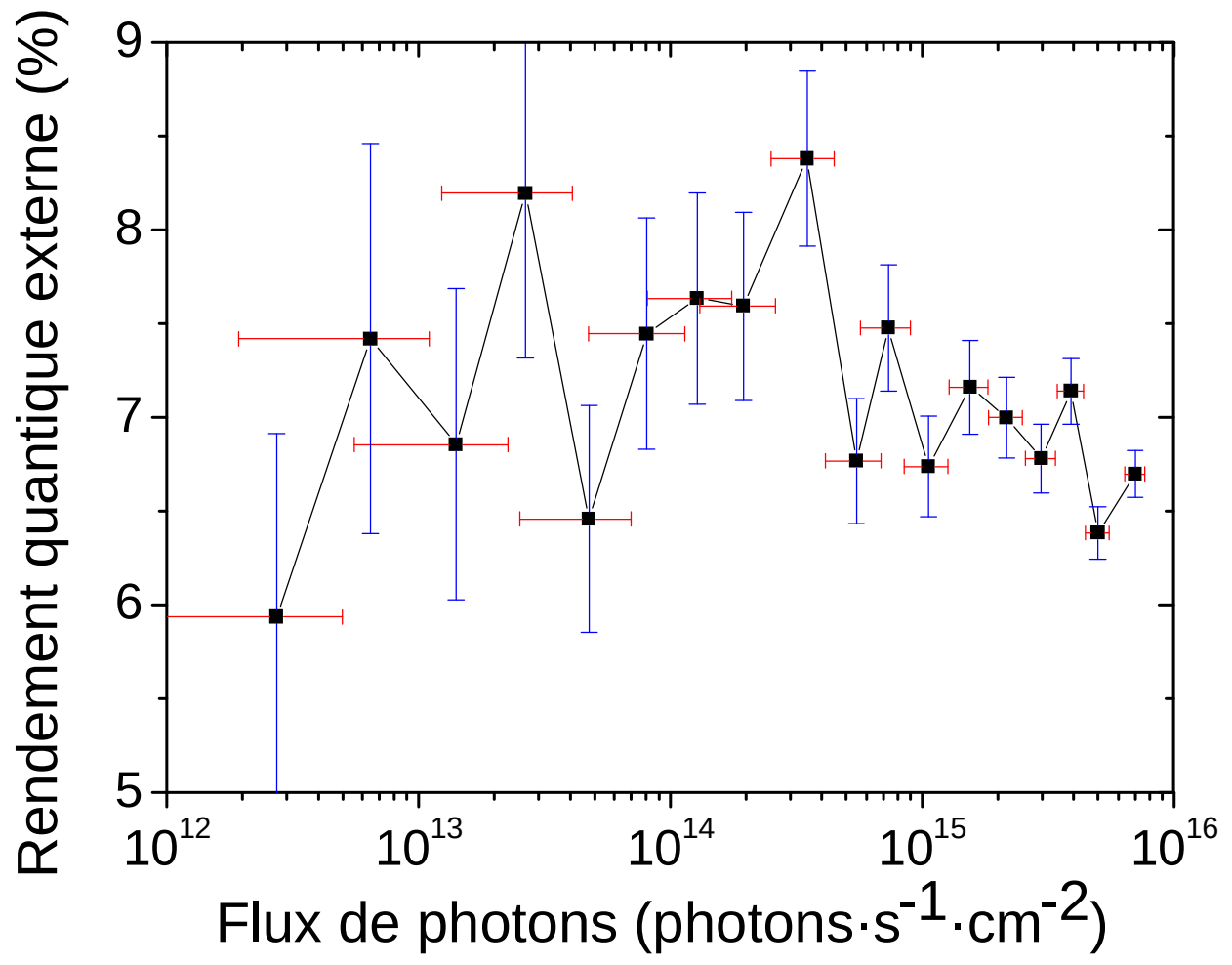

*FIG. 2-15 Rendement quantique externe en fonction du flux de photons sous -1V, T=4K. Les barres d'erreurs donnent l'imprécision sur le flux et le rendement quantique externe.*

#### *2.3.2.5. Résultats et comparaison à l'état de l'art actuel*

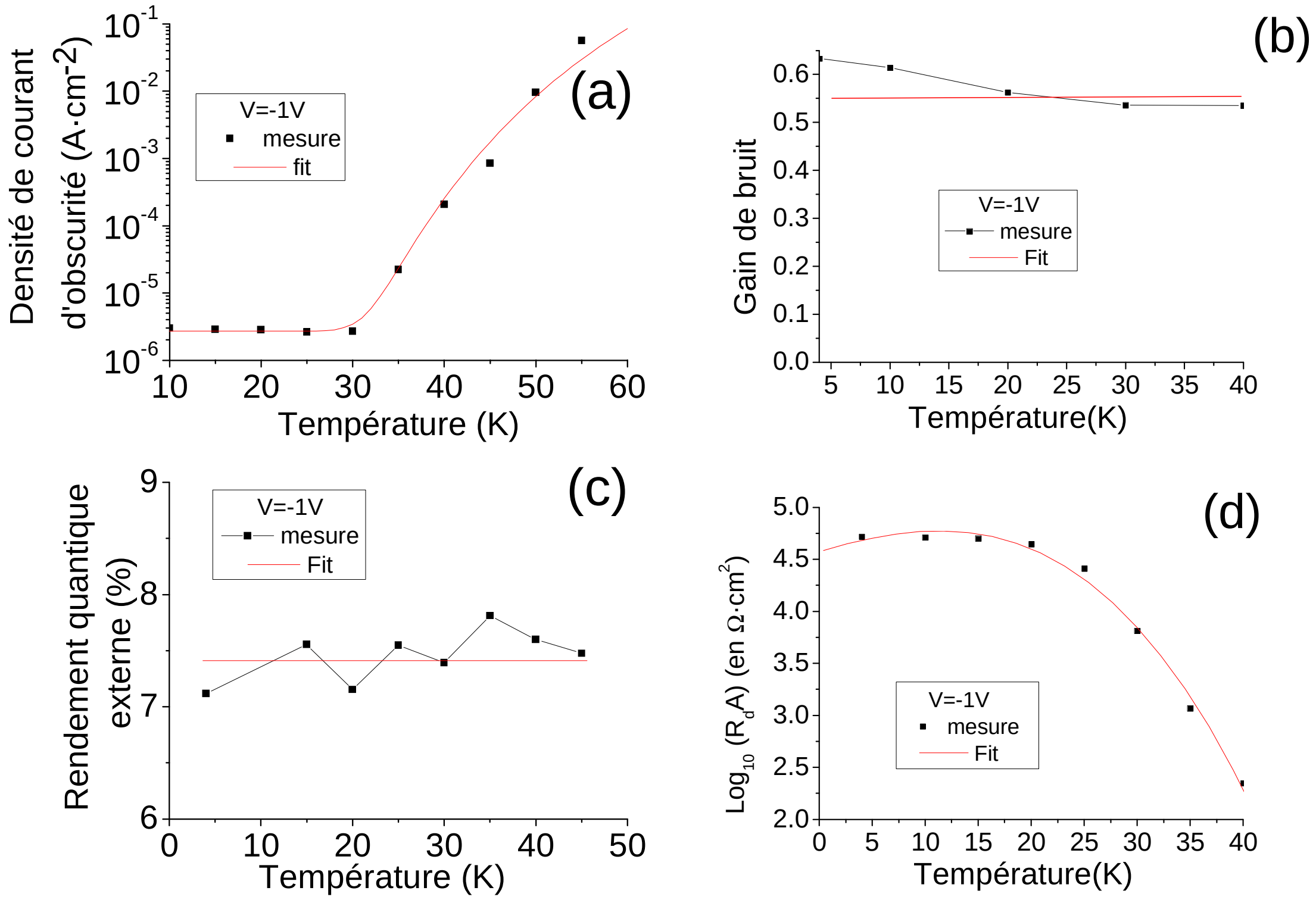

*FIG. 2-16 Tracé des valeurs expérimentales et des modélisations associés en fonction de la température : de la densité de courant (a), du gain de bruit (b), du rendement quantique externe (c) et de la résistance dynamique(d).*

La FIG. 2-16 donne la dépendance du courant d'obscurité, du gain de bruit, du rendement quantique externe et de la résistance dynamique en fonction de la température du détecteur quand ce dernier est exposé à un corps noir froid ($T_{CN}$≈150K). Nous constatons que le gain de bruit et le rendement quantique externe sont des grandeurs très faiblement dépendante de la température du détecteur. Ces données vont à présent pouvoir servir d'entrée à notre outil de simulation, qui va nous permettre de prédire les performances d'une caméra basée sur ce détecteur.

Avant d'aller plus loin dans l'utilisation de ce détecteur il est primordial de s'assurer que son utilisation est bien pertinente, c'est-à-dire que son niveau de performance est au moins acceptable vis-à-vis d'autres composants QWIP opérant à cette longueur d'onde. Une comparaison entre notre composant et l'état de l'art actuel est proposée dans le tableau tab. 2-2. Ces donnés proviennent du tableau tab. 1-4, mais une fois encore il faut bien comprendre les limites de cette comparaison. Plusieurs de ces valeurs sont dépendantes des conditions de la mesure qui ne sont pas toujours connues !

| Grandeurs | QWIP typique de la gamme 12-18µm | Notre QWIP |
|---|---|---|
| λ=15µm | λ=15µm | λ=14.5µm |
| Δ λ/ λ | 10-15% (transition lié quasi lié) | 13% |
| $T_{det}$ | 35-55K | $\leq 30K$ |
| $V_{pol}$ | -2.5V → -2V | -1.5V → -1V |
| $J_{dark}$ | $10^{-3}$-$10^{-1}$A·cm$^{-2}$ | $10^{-6}$ A·cm$^{-2}$ |
| Réponse | 0.5A·W$^{-1}$ | 0.8A·W$^{-1}$ |
| Rendement quantique | 10% | 7% |
| NETD | 20-60mK sous 300K ($10^{15}$photons·s$^{-1}$·cm$^{-2}$) | |
| Détectivité spécifique | $10^{10}$-$10^{11}$jones sous 300K | 8.4×$10^{10}$ jones sous $T_{CN}$=300K<br>5.3×$10^{11}$ jones sous $T_{CN}$=120K |

*tab. 2-2 Comparaison du niveau de performance entre le composant caractérisé et l'état de l'art QWIP à la même longueur d'onde*

En terme de point de fonctionnement la température d'utilisation est plus faible que ce qui est généralement rapporté sur cette longueur d'onde. Cela permet essentiellement d'abaisser le courant d'obscurité de trois à cinq décades. La tension de polarisation est également un peu plus faible et cela également dans le but de limiter le courant d'obscurité. Les performances mesurées sont tout à fait comparables à ce qui se fait de mieux autour de 15µm. La détectivité spécifique obtenue égale, voir dépasse les meilleurs performances de la concurrence.

## 2.4. Application à un scénario faible flux

La première partie de ce chapitre s'intéressait à la caractérisation ($J_{dark}$, réponse spectrale, bruit, réponse sous faible flux) du composant 15µm sous faible flux. Nous avons détaillé les dispositifs de mesure et les performances à l'état de l'art de notre composant QWIP haute longueur d'onde. La dernière partie de ce chapitre se propose d'évaluer les performances d'un instrument fabriqué à partir de ce détecteur. En effet les fonctions de mérite propres au QWIP, qui ont pu être définies précédemment, sont des grandeurs importantes pour comparer les détecteurs au sein d'une même filière mais elles restent insuffisantes pour statuer quant à la capacité d'un détecteur à imager une scène donnée.

Cette dernière partie sera donc dédiée à la conception et à l'évaluation d'une caméra infrarouge basée sur le composant QWIP 15µm. Nous avons fait le choix d'une application faible flux civile, ce qui nous permet de publier tous les paramètres et besoins applicatifs. Néanmoins cette application se rapproche suffisamment des autres besoins pour représenter à la fois un défi en terme de performance tout en gardant la possibilité d'extrapoler les résultats à des applications plus militaires. Au final nous souhaitons faire un bilan des performances des QWIP actuels et identifier quels sont les points qui restent encore à améliorer.

### 2.4.1. Choix du scénario

#### *2.4.1.1. Soufflerie ETW*

Le scénario retenu pour tester les performances de notre QWIP est celui de la soufflerie ETW (European Transsonic Windtunnel[4]), voir FIG. 2-17. C’est une soufflerie transsonique cryogénique construite en Allemagne au milieu des années 90. La particularité de cette soufflerie est sa capacité à simuler des écoulements avec un grand nombre de Reynolds ($Re>10^7$). De tels écoulements servent à simuler les phases de décollage et d’atterrissage des avions. Pour obtenir de tels nombres de Reynolds, la solution technique retenue consiste à diminuer la température de l’écoulement. La température de l’air dans la soufflerie est typiquement comprise entre 110K et 300K. Le refroidissement est obtenu en vaporisant de l’azote liquide dans l’écoulement.

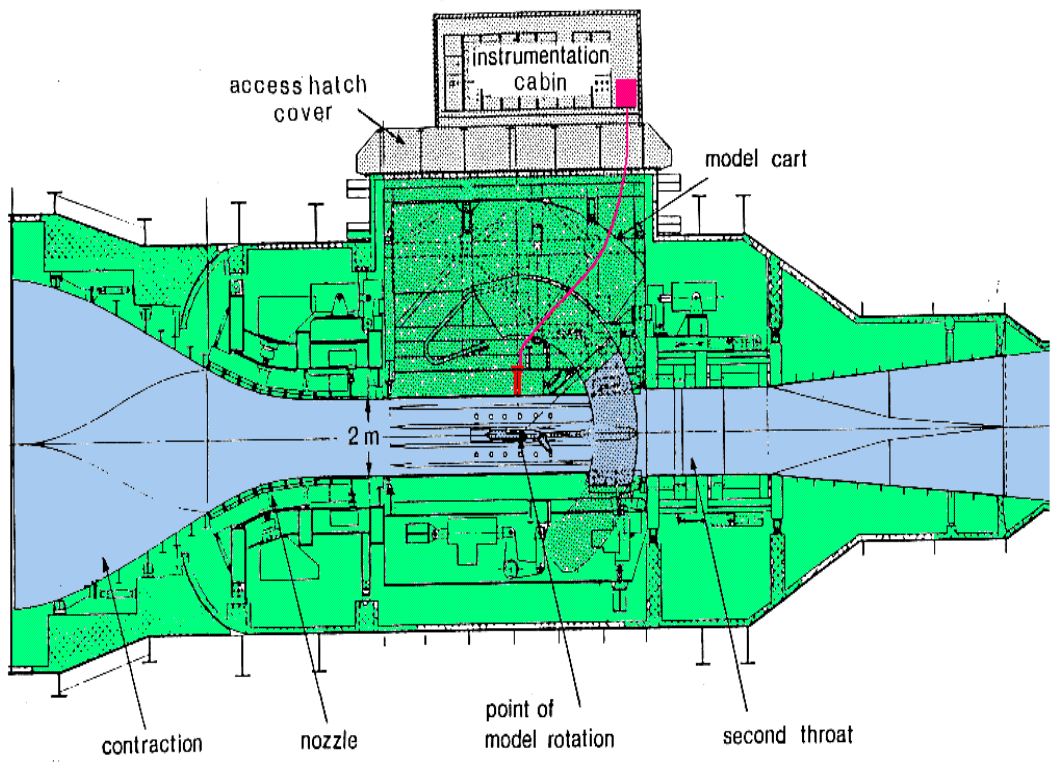

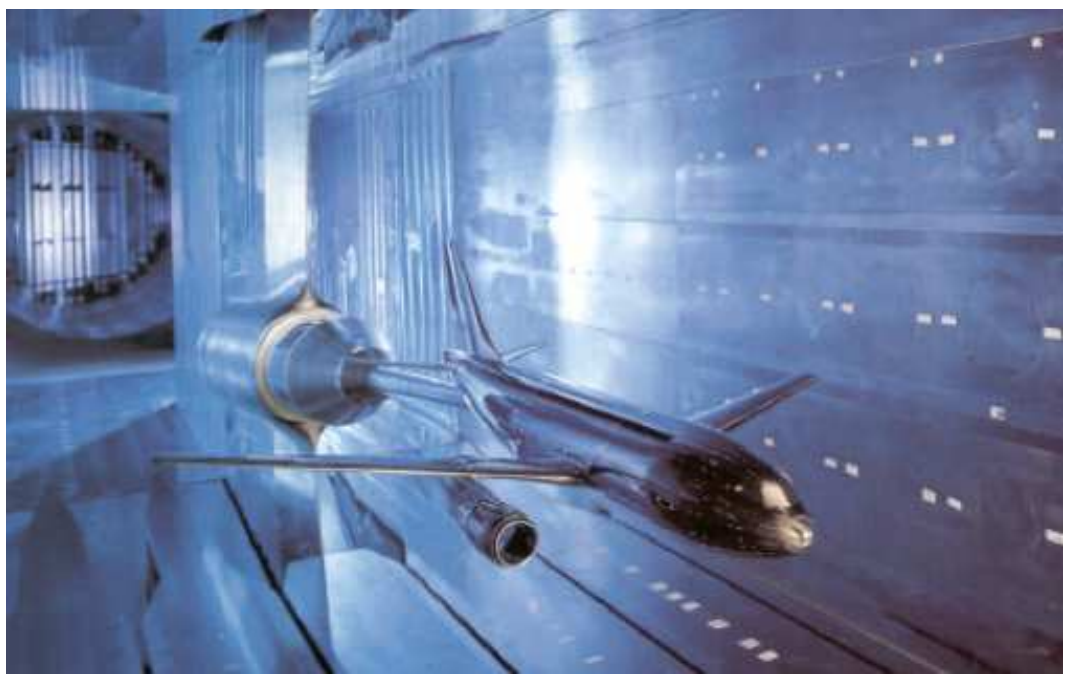

*FIG. 2-17 A gauche, schéma de la soufflerie ETW. En rouge la position du détecteur. A droite, photo de l’intérieur de la soufflerie.*

Les phénomènes qui doivent être observés sont des transitions d’écoulement laminaires/turbulent en surface d’une maquette. Cela nécessite une thermométrie adaptée.

#### *2.4.1.2. Besoin photométrique*

L’ONERA, de par sa double compétence en soufflerie et en optique, est un partenaire fondateur de cette soufflerie. Notre département s’est vu charger à la fin des années 90 de la réalisation d’une caméra destinée à l’observation des écoulements froids d’ETW[66]. L’objectif était alors d’observer l’intégralité d’une maquette de quarante centimètres avec une résolution millimétrique, tout en visualisant des différences de température à sa surface de 200mK (avec un objectif de 100mK). Le tab. 2-3 donne quelques données chiffrées sur les dimensions de la maquette et les besoins optiques nécessaires aux observations physiques.

| Grandeurs | Valeurs | Commentaires |
|---|---|---|
| Température du fond | 120K | |
| Température de l’objet | 121K | |
| Température maximale de scène | 140K | Donne le temps d’intégration maximal |
| Taille de l’objet | 40 cm | |
| Résolution spatiale souhaitée | 1mm | 1 cm correspond à la taille des marqueurs |
| | | thermiques en bordure d'aile |
| Distance caméra objet | 1 m | |
| Temps intégration | <20 ms | Lié au temps caractéristique de l'écoulement |
| NETD | <0.2K | Correspond à un rapport signal sur bruit de 5, nécessaire pour une cible résolue quand on visualise une variation de température de 1K |

*tab. 2-3 Tableau récapitulatif des besoins pour l'imagerie de la soufflerie ETW.*

#### *2.4.1.3. Détecteur retenu*

La filière retenue dans les années 90 pour répondre à ce besoin est celle du Si:Ga. Le détecteur se présente sous la forme d'une matrice de 192×128 pixels au pas de 75µm. Le système fonctionne à 10K pour un rendement quantique de l'ordre de 30%. La faible taille matricielle empêchait d'obtenir à la fois la résolution spatiale sur la maquette et le champ nécessaire à la visualisation de la maquette complète. Un double système d'objectif était donc nécessaire, ce qui est particulièrement contraignant vu que le système complet devait être froid. La masse à refroidir est telle qu'elle oblige la présence d'un double étage de cryogénie. La NETD finale se situe autour de 100mK. Au final, bien que le détecteur réponde aux besoins de thermométrie requis pour cette expérience, deux principaux défauts persistent : sa faible température de fonctionnement et sa faible taille matricielle. Le QWIP est au minimum est un bon candidat pour résoudre ce second point et nous allons voir qu'il s'avère complètement adapté à notre problématique.

### 2.4.2. Outil de simulation

L'ONERA, de par son rôle d'expert, est amené à évaluer la performance de différentes filières de détecteurs et réalise également des instruments d'optique infrarouge. Le DOTA a donc développé des outils de simulation nécessaires au dimensionnement des ces instruments. C'est un modèle ingénieur, qui comprend de nombreux paramètres euristiques et donc l'objectif est de lier les performances de briques élémentaires à celle du système complet. L'outil de simulation *Planfoc* prend en compte trois principaux blocs que sont la cible, l'atmosphère et le bloc instrumental. Le détecteur est un élément du dernier bloc, au même titre que l'optique et le circuit de lecture. L'ensemble « cible plus atmosphère » définit un scénario. Ce code est purement analytique, ce qui en fait un instrument de simulation particulièrement rapide. Notons enfin que cet outil a été comparé et validé par rapport à d'autres codes. Cela reste toutefois un outil de dimensionnement dont la précision est de l'ordre de la dizaine de pourcent.

Ce code aura donc entre autre pour entrée les performances du détecteur que nous avons mesuré précédemment. La sortie du code sera le rapport signal sur bruit en sortie d'instrument.

### 2.4.3. Utilisation du QWIP 15µm

#### *2.4.3.1. Version du QWIP retenue*

Le but de cet exercice est essentiellement de tester la capacité du détecteur QWIP dans une application faible flux. Nous avons donc fait en sorte que tous les composants optiques et électroniques aient les performances de composants sur étagère. Ce qui signifie que les paramètres de la matrice ont été pris identiques à ceux du détecteur caractérisé quand ils étaient connus (taille pixel, pas de la matrice). Pour les autres caractéristiques nous avons pris des valeurs réalistes qui ont été validé par les technologues. Le tab. 2-4 donne les principales informations relatives à la matrice QWIP et à son circuit de lecture.

| Grandeurs | Valeurs |
|---|---|
| Tension de sortie du circuit de lecture | 2V |
| Capacité de lecture | 0.3pF |
| Dimension matrice | 512×512 |
| Pas matrice | 25µm |
| Taille pixel | 23.5µm |

*tab. 2-4 Caractéristique de la matrice de détecteur et de son circuit de lecture.*

Pour cette étude nous avons choisi de fixer le point de fonctionnement du détecteur en terme de tension de polarisation. L'étude précédente a mis en évidence que le meilleur rapport courant photonique sur courant d'obscurité se trouvait un peu au delà de -1V. Ce point offre un bon rapport entre le courant photonique et le courant d'obscurité, voir FIG. 2-18. Notre étude porte donc sur l'influence de la température du détecteur. Le but étant d'estimer la température maximale d'utilisation de ce détecteur qui permet d'obtenir le niveau de performance requis.

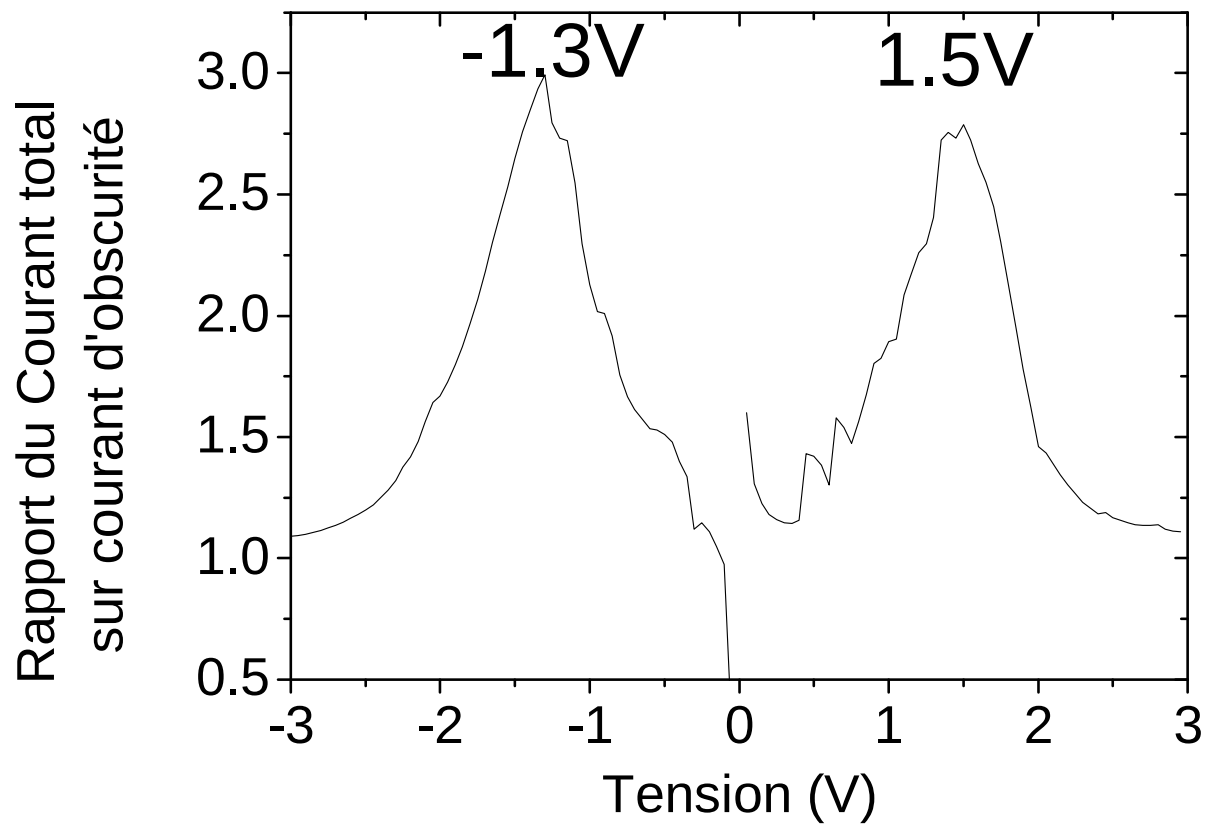

*FIG. 2-18 Evolution du rapport entre le courant total et le courant d'obscurité en fonction de la tension. Ce graphe est utile à la détermination du point de fonctionnement en tension du détecteur.*

*Planfoc* n'est pas un code qui vise à décrire les propriétés physiques microscopiques des détecteurs. Il fonctionne plutôt selon un mode « boîte noir ». Le logiciel utilise des formules analytiques qui n'ont pas nécessairement d'origine physique. Ainsi les paramètres propres à nos détecteurs ($J_{dark}$, bruit, résistance dynamique, rendement quantique, …) sont modélisés par des fonctions analytiques, typiquement des polynômes. Ces formules analytiques nous permettent d'évaluer le rapport signal sur bruit de sortie sur la plage des paramètres d'entrée, y compris entre deux points expérimentaux.

Certaines grandeurs, telles que le bruit, n'ont pas pu être mesurées exactement dans la gamme de flux envisagée par notre scénario, dans ce cas nous avons fait le choix d'utiliser le jeu de données qui se rapprochait le plus.

De nos mesures expérimentales, FIG. 2-16, nous extrayons la dépendance en température pour une tension de fonctionnement de -1V :

- Le courant d'obscurité est modélisé par un plateau (indépendance du courant tunnel avec la température) suivi d'une croissance exponentielle. $J_{dark}(T) = A_0 + A_1 \cdot e^{(-A_2/T)}$ avec $A_0$=2.7×$10^{-6}$·A·$cm^{-2}$, $A_1$=$10^4$ A·$cm^{-2}$ et $A_2$=700K.
- La résistance dynamique est modélisé par une loi de puissance polynomiale telle que $R_{dyn}(T) = 10^{(C_0 + C_1 \cdot T + C_2 \cdot T^2 + C_3 \cdot T^3)}$ avec $C_0$=4.57193, $C_1$=0.03124·$K^{-1}$, $C_2$=7.81212×$10^{-4}$·$K^{-2}$ et $C_3$=3.56613×$10^{-5}$·$K^{-3}$.
- Le gain de bruit et le rendement quantique ont été pris constants avec la température dans la mesure ou leurs variations sont du même ordre que la résolution de mesure.

#### 2.4.3.2. Discussion sur l'optique

En raison du changement du format matriciel par rapport au détecteur Si:Ga, une complète refonte de l'optique était nécessaire afin de profiter pleinement des performances du détecteur. La procédure qui a été suivie pour dimensionner l'optique est la suivante. Connaissant les dimensions du détecteur et de la maquette à observer, nous pouvons calculer la focale équivalente du système pour que la maquette soit entièrement imagée. Par ailleurs, il existe une plage acceptable d'ouverture numérique qui est raisonnable techniquement et financièrement à atteindre pour les systèmes infrarouges : typiquement entre f/2 et f/4. Cela conditionne le diamètre de la pupille d'entrée. La valeur de transmission de l'ensemble de la chaîne optique a été prise proche de celle de la caméra existante. Finalement le tableau tab. 2-5 récapitule les propriétés optiques de la caméra dessinée.

| Grandeurs | Valeurs |
|---|---|
| Longueur d'onde coupure optique | 20µm |
| Transmission des optiques | 25% |
| Focale du système | 15mm |
| Diamètre de la pupille entrée | 6mm |

*tab. 2-5 Caractéristiques de l'optique de l'instrument.*

#### 2.4.3.3. Performance de la caméra basée sur le QWIP

Le rapport signal sur bruit que nous avons choisi d'évaluer est la NETD. Cette NETD est globalement constante en dessous de 30K (autour de 130mK, pour 200mK nécessaire), puis s'envole au-delà, voir FIG. 2-19 (a). Cette envolée est la conséquence directe de l'augmentation du courant d'obscurité et s'accompagne d'une chute du temps d'intégration. A basse température notre caméra remplit donc bien sa mission.

Le bruit du détecteur est toujours limité par celui lié au courant d'obscurité, voir FIG. 2-19 (b). Cette caméra n'atteint donc jamais le régime BLIP dans lequel le bruit de la caméra est limité par celui de la scène. Le régime BLIP n'est donc absolument pas une nécessité pour remplir une mission d'observation bas flux.

Enfin notons que la moitié de la capacité de lecture est remplie par du courant d'obscurité à basse température, voir FIG. 2-19 (c). Une dizaine de pourcent est remplie par le signal utile dans la gamme de mesure. A haute température et en dépit de la diminution du temps d'intégration, la capacité de lecture est majoritairement remplie par le courant d'obscurité, ce qui réduit à peau de chagrin le signal utile.

La caméra conduit à une résolution au niveau de la cible inférieure à 2mm pour 1 cm requis. La caméra basée sur le QWIP répond au besoin en terme de champ et de résolution.

En conclusion ce QWIP répond parfaitement à la mission qui lui est demandé à condition de le faire fonctionner à basse température. Il conduit à des NETD de l'ordre de 130mK. La NETD maximale requise de 200mK est atteinte pour une température de détecteur de 32K. Le principal défaut de composant est sans aucun doute son courant d'obscurité qui remplit les capacités de lecture (ce qui limite le temps d'intégration) et ajoute un bruit excessif.

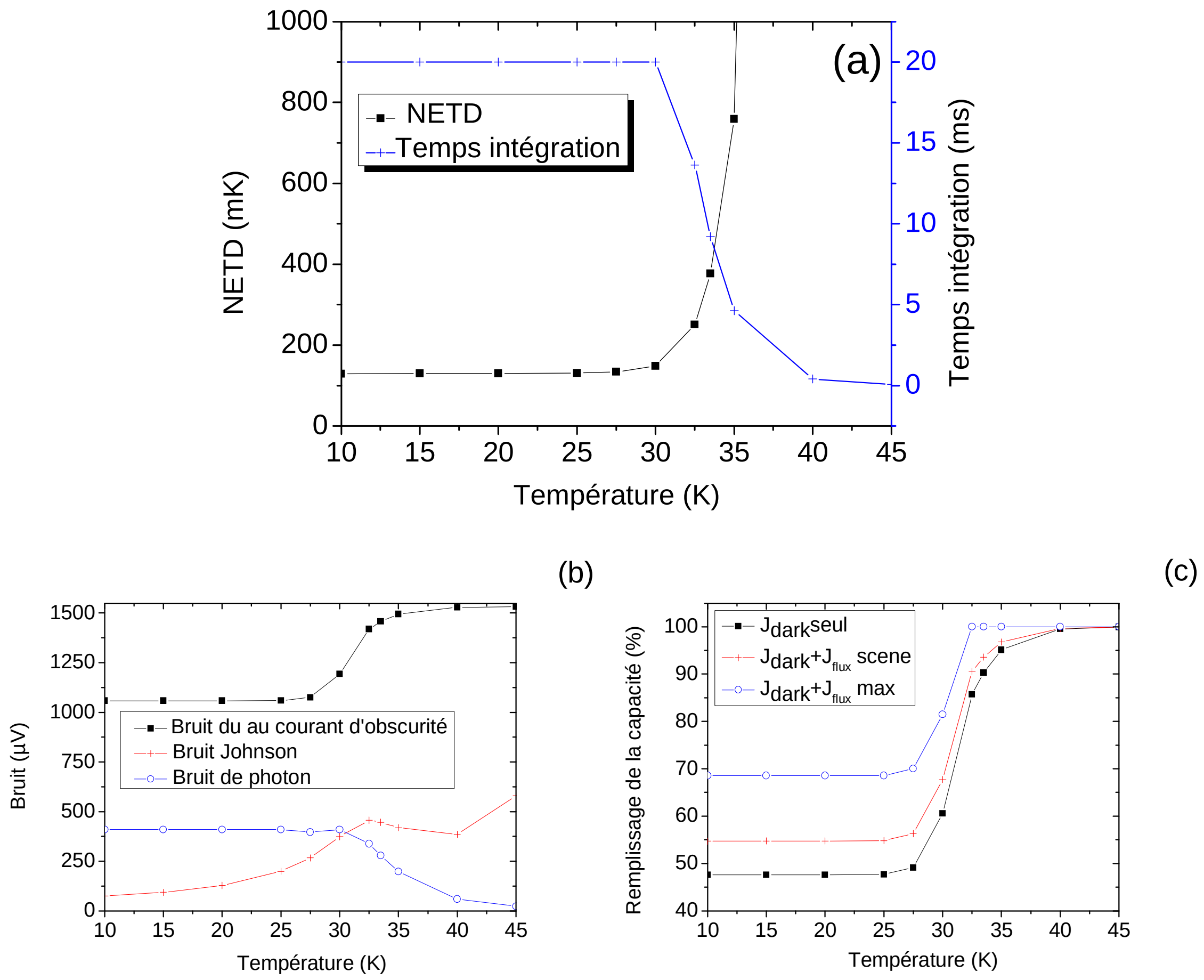

*FIG. 2-19 (a) NETD et temps d'intégration en fonction de la température. (b) Bruit en fonction de la température. (c) Remplissage de la capacité en fonction de la température.*

#### 2.4.3.4. Comparaison aux autres filières

Nous venons de démontrer que le QWIP réussirait parfaitement cette mission de thermographie pour la soufflerie ETW. Avec une NETD similaire à celle de la caméra Si:Ga autour de 100mK, le QWIP autorise de plus une forte augmentation de la température de fonctionnement, puisqu'elle passe de 10K à 30K. Autre avantage de ce QWIP, sa grande taille matricielle permet de ne plus avoir à recourir au système de double objectif dans la mesure où il permet à la fois d'obtenir un champ important et une bonne résolution spatiale. Cela devrait donc conduire à une grande simplification de l'optique et également à une diminution de la masse nécessitant d'être refroidie, ce qui devrait encore alléger les contraintes sur l'appareil cryogénique.

Le recours à la filière MCT aurait été une autre solution pour répondre à notre besoin. Cette filière propose de bons rendements quantiques et de grandes tailles matricielles. Bien que la démonstration de MCT haute longueur d'onde ($\lambda$>12µm) ait été apportée[67, 68], aucun fabricant français ne commercialise de tel produit à ce jour. Or le MCT avec une longueur d'onde de coupure trop basse ($\lambda$<12µm) ne peut répondre à cette mission !

## 2.5. Conclusion

J'ai démontré la capacité du QWIP 15µm à répondre à une application faible flux telle que celle définie par la soufflerie ETW. Tout en offrant une meilleure NETD il permet de multiplier par trois la température de fonctionnement du détecteur. Sa grande taille matricielle permet également de simplifier l'étage d'optique par rapport au détecteur existant actuellement. La diminution du courant d'obscurité en régime tunnel est la priorité pour améliorer les performances de ce détecteur.

# 3. TRANSPORT ELECTRONIQUE DANS LES HETEROSTRUCTURES DE TYPE I

## Sommaire

**Publications associées à ce chapitre :**

- F. Castellano, F. Rossi, J. Faist, E. Lhuillier and V. Berger, Modeling of dark current in midinfrared quantum well infrared photodetectors, Phys. Rev. B **79**, 205304 (2009).
- E. Lhuillier, I. Ribet-Mohamed, M. Tauvy, A. Nedelcu, V. Berger, E. Rosencher, Ultimate performance of Quantum Well Infrared Photodetectors in the tunneling regime, Infrared. Phys. Techn. **52**, 132 (2009).

J’ai établi au chapitre précédent que même si les QWIP présentent de nombreux atouts pour une utilisation dans des applications faibles flux, la diminution du courant d’obscurité est nécessaire en vue de leur optimisation. Ce sera donc l’objectif que je chercherai à remplir au cours des trois prochains chapitres de cette thèse.

Ce troisième chapitre va s’attacher à des généralités sur le transport électronique dans les super-réseaux de type I. L’objectif est essentiellement de présenter les formalismes de transport que nous développerons par la suite. Je vais tacher de présenter, classer, d’exhiber les hypothèses et les limites mais aussi de montrer les liens qui peuvent exister entre les différents modèles de transport vertical. La littérature sur ce chapitre est abondante, ce chapitre est donc un mélange entre un rapport bibliographique et une application de ces résultats au transport dans les QWIP hautes longueurs d’onde.

Cette partie peut également s’avérer assez calculatoire. Afin de ne pas trop en alourdir la lecture, j’ai choisi de reporter en annexe les aspects les plus mathématiques du problème, pour ne laisser dans le cœur du texte que les formules finales et leurs interprétations physiques.

Ce chapitre s’organise de la façon suivante. Je commencerai par quelques mots sur le transport à haute température dans les super réseaux de type I. Puis dans toute la suite nous nous focaliserons sur le transport tunnel basse température et ses différents traitements possibles. L’idée est de commencer par présenter les différents modèles simplifiés modélisant ce transport. Ces modèles sont très riches en physique et je tacherai de décortiquer les hypothèses qui leur sont nécessaires et leur domaine d’application. J’ai tenté de classer ces modèles selon trois grandes catégories : matrice de transmission, modèle hydrodynamique et modèle de hopping. Ce chapitre ne serait pas complet si je ne disais pas quelques mots sur les modèles plus complexes existant pour traiter ce transport que sont les formalismes de la matrice densité, de la matrice de Wigner, de la fonction de Green hors équilibre et la méthode Monte-Carlo. Pour finir je tacherais d’unifier ces différentes théories en présentant leurs points communs et ceux qui les opposent afin de donner une vision cohérente de leur utilisation.

Sauf exception, qui serai précisée, toutes les applications numériques de ce chapitre sont appliquées à la structure QWIP décrite au chapitre précédent.

## 3.1. Transport à haute température

Tous les QWIP utilisés dans des caméras commerciales opèrent à des températures pour lesquelles le courant d’obscurité est dans le régime thermionique. Il n’est pas possible, pour la plupart des applications, pour des raisons de masse et de durée de vie de l’appareil frigorifique, de faire fonctionner le détecteur à très basse température. Mais nos applications exigent davantage de performances quittes à en payer le prix sur la cryogénie. Voila pourquoi je ne vais m’attacher à la modélisation du régime haute température que très brièvement.

Par régime haute température, j’entends le régime de transport dans lequel les électrons sont thermiquement activés. Ces derniers passent alors du niveau fondamental d’une période vers le continuum où ils dérivent alors sous champ. Ce régime est particulièrement bien modélisé par une loi d’activation de type Arrhenius.

Différentes expressions peuvent être trouvées dans la littérature[29,74] :

$$J_{th\ Levine} = e.\frac{m^*}{\pi\hbar^2}.\frac{v_d}{L_w}.eFL_b.e^{-\frac{V_b - eFL_w - E_f}{k_b T}} \qquad (3\text{-}1)$$

$$J_{th\,SchneiderLiu} = e\frac{m^*}{\pi\hbar^2}\frac{v_d \tau_c}{\tau_{scatt}}\frac{k_b T}{L_d} e^{-\frac{V_b - \frac{eFL_w}{2} - E_f + E_{ex}}{k_b T}} \qquad (3\text{-}2)$$

$$J_{th\ Kane} = 2ev_d\left(\frac{m^* k_b T}{2\pi\hbar^2}\right)^{3/2} e^{-\frac{V_b - \frac{eFL_w}{2} - E_f + E_{ex}}{k_b T}} \qquad (3\text{-}3)$$

Où

$$E^{ex} \approx -\frac{e^2}{4\pi\varepsilon} k_F (1 - 0.32\frac{k_F L_w}{\pi}) \qquad (3\text{-}4)$$

Avec $v_d = \frac{\mu F}{\sqrt{1 + \left(\frac{\mu F}{v_{sat}}\right)^2}}$, m* la masse effective de GaAs, $v_d$ la vitesse de dérive, $\mu$ et $v_{sat}$ sont respectivement la mobilité des électrons et leur vitesse de saturation, $\tau_c$ le temps de capture dans le puits, $\tau_{scatt}$ le temps de diffusion du puits vers le continuum, $k_F$ le vecteur d'onde de Fermi, $\varepsilon$ la permittivité de GaAs et $E_{ex}$ l'énergie d'échange[69]. La valeur de $E_{ex}$~-13.6meV est obtenue via l'expression approximée donnée par (3-4)[70]. Les équations *(3-1)* à *(3-3)* peuvent toutes se résumer sous la forme :

$$I = I_0(T,V)\exp(-\frac{E_a(V)}{k_b T}) \qquad (3\text{-}5)$$

Il faut voir une telle expression comme le courant dans le continuum ($I_0(T,V)$) multiplié par la probabilité de l'avoir peuplé ($\exp(-\frac{E_a(V)}{k_b T})$). Le pré facteur est généralement pauvre en élément géométrique propre à la structure. Suivant le modèle, une dépendance en tension (via un terme de vitesse de dérive sous champ) et en température peut être présente, mais cette dépendance reste modérée dans la mesure où elle se fait via des lois de puissances. La dépendance en température se fait essentiellement via le terme d'activation thermique. Il est alors nécessaire de définir convenablement l'énergie d'activation. La version la plus naïve consiste à la prendre égale à la différence d'énergie entre le continuum et le niveau fondamental. Il faut alors ajouter à cette quantité une dépendance en champ électrique : $E_a(V) = V_b - E_1 + \eta e \cdot F \cdot L_W$ où la quantité $\eta$ est un paramètre compris entre zéro et un[58]. Cette dépendance en champ traduit simplement l'abaissement de la barrière de potentiel avec le champ électrique. Il est possible de complexifier ce modèle à l'infini en lui ajoutant par exemple l'énergie d'échange (terme de confinement supplémentaire dû à l'interaction coulombienne[29,70]) ou en soustrayant l'énergie de Fermi (afin de considérer la distribution en énergie des électrons). Mais bien que ces quantités aient une origine physique, ils ont tendance à devenir des paramètres d'ajustement qui font perdre au modèle de sa généralité.

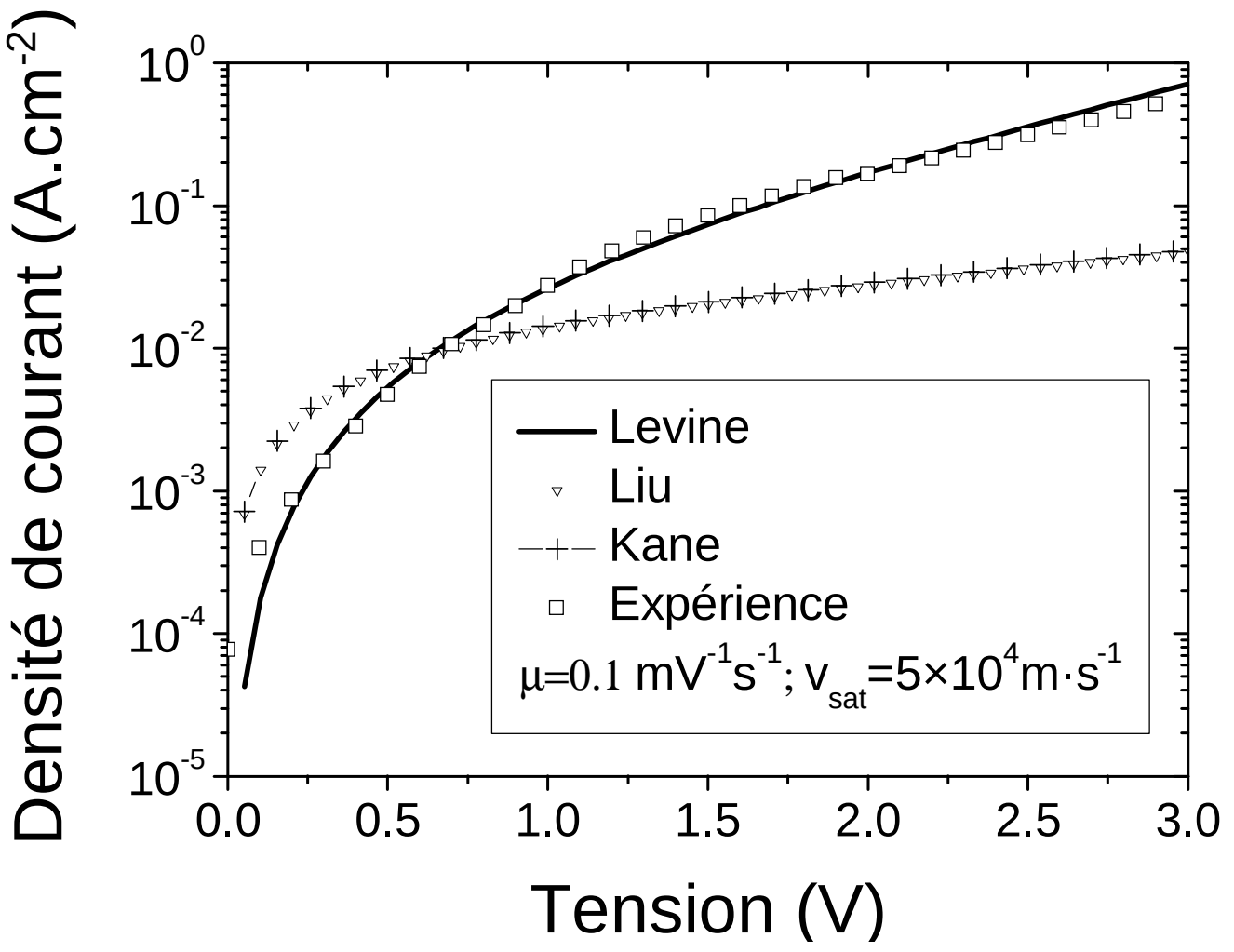

*FIG. 3-1 Comparaison du courant expérimental et de celui donné par les trois expressions théoriques précédentes ((3-1) (3-2) et (3-3)) à 50K. Le paramètre* $\frac{\tau_c}{\tau_{scatt}}$ *a été pris égal à un comme proposé par la référence 29.*

Notons que cette physique se rapproche beaucoup de celle du semi-conducteur massif (utilisation de mobilité et de vitesse de saturation), c'est donc une physique empirique qui a été longuement validée[71]. Pour notre composant c'est l'expression la plus simple, voir équation *(3-1)*, qui conduit au meilleur accord, celui ci est d'ailleurs excellent. Dans la mesure où la physique de ce régime est bien maîtrisée par des formules analytiques simples et que ce régime n'est pas celui qui nous intéresse il n'est pas utile de chercher à complexifier ces modèles. La suite de ce chapitre se focalisera donc sur le transport à basse température et l'effet tunnel qui l'accompagne.

## 3.2. Transport à basse température

Le transport vertical à basse température dans les super-réseaux de type I fait intervenir deux éléments de physique quantique essentiels que sont le confinement quantique et l'effet tunnel. Il nous faut donc mettre en place une modélisation qui contienne la physique de ces deux effets. Il est également important de tenir compte du particularisme des super-réseaux des QWIP qui sont très faiblement couplés.

La modélisation que nous mettrons en place devra en particulier rendre compte du phénomène original que nous avons observé sur le composant QWIP 15μm, à savoir le plateau des courbes *I(V)*. Ce phénomène de résistance différentielle infinie a été observé sur plusieurs structures et restait avant cette thèse incompris. Nous devrons toutefois aller au delà de la simple compréhension qualitative de la forme de ces courbes dans la mesure où l'objectif que nous devons garder en tête est la réduction du courant en régime tunnel. Nous nous fixons également pour objectif la prise en compte réaliste du dopage des structures. Cela sera justifié par la suite puisque nous verrons que le dopage joue un rôle déterminant sur les propriétés de transport.

Cette partie du chapitre va se subdiviser en trois sous parties :

- La première s'intéressera aux modèles simplifiés de transport que sont les approches de type matrice de transmission, le modèle hydrodynamique qui aboutit au modèle de transport d'Esaki-Tsu et enfin le modèle de transport par saut entre états localisés.

- La seconde sous-partie traitera des approches complètes de transport hors équilibre, cette partie sera surtout bibliographique. J'y aborderai le traitement du transport par des approches telles que la matrice densité particulièrement utile dasn le cas du transport résonant, et les fonctions de Green ou de Wigner.

- Enfin la dernière partie s'attachera à donner une vision cohérente de ces différentes approches.

### 3.2.1. Matrice de transmission

L'approche du transport à travers une hétérostructure via un terme de transmission dérive directement du calcul des fonctions d'onde dans ces structures. Il existe en effet deux approches non *ab initio* de calcul des structures électroniques de super réseaux[72]. La première qui porte le nom de « supercell approach » utilise directement le fait qu'un super-réseau est un « super » réseau. Il est alors possible d'utiliser toute « l'artillerie » qui existe pour le calcul des fonctions d'onde du réseau atomique : fonction de Bloch, condition de Born von Karman... La version la plus utilisée de cette approche est la méthode des liaisons fortes. L'autre approche, dite des conditions aux limites, est celle qui consiste à connaître les fonctions d'onde par zone, puis à les rabouter en utilisant les conditions aux limites. Pour le détail je renvois le lecteur à un cours de mécanique quantique au chapitre de calcul de fonction d'onde dans un puits carré[73] La méthode $k \cdot p$ dérive également de cette seconde approche.

Finalement l'objectif de ce formalisme de transmission est de relier le courant en amont de la structure à celui en aval via des considérations sur la structure électronique du dispositif.

L'approche la plus classique pour traiter le transport tunnel dans un QWIP est justement basée sur une version simple et analytique de la matrice de transfert. Elle est très bien adaptée au traitement des structures faiblement couplées. Dans ce cas il faut alors parler d'effet tunnel séquentiel. L'électron traverse la structure par sauts successifs. Il traverse une barrière de potentiel avant d'être diffusé, et ainsi de suite jusqu'au contact.

#### *3.2.1.1. Le transport tunnel séquentiel et son traitement par l'approximation WKB*

La méthode WKB pour Wentzel-Kramers-Brillouin permet de calculer la probabilité de transfert à travers une barrière de potentiel. Elle est basée sur la résolution d'une version lentement variable de l'équation de Schrödinger, qui conduit à une probabilité de passage à travers la barrière de potentiel U (voir annexe 3-B) :

$$T(E) = \exp\left[-2.\sqrt{\frac{2m^*}{\hbar^2}}.\int_{x_1}^{x_2}\sqrt{U(x)-E}.dx\right] \qquad (3\text{-}6)$$

Avec m* la masse effective des électrons, E leur énergie et où $x_1$ et $x_2$ sont respectivement les positions d'entrée et de sortie de la barrière tunnel. Cette approximation repose sur deux hypothèses majeures (i) la probabilité de passage vers l'état final doit être faible et (ii) le potentiel ne doit pas varier trop vite devant la longueur d'onde électronique. La première hypothèse impose ainsi l'absence d'onde réfléchie dans la barrière, ce que certains auteurs traduisent par le couplage entre un état discret et un continuum. Ce qui est une façon classique de traiter un réservoir en physique mésoscopique. La seconde hypothèse, malgré sa formulation très physique, est juste là pour permettre un calcul analytique du problème. L'approximation WKB est très souvent utilisée pour modéliser le régime tunnel des QWIP[29,74]. Elle a l'avantage de grossièrement reproduire le bon ordre de grandeur du courant tout en étant analytique. Elle rend également simplement compte du phénomène d'abaissement de barrière avec le champ électrique, généralement non considéré dans la majorité des approches alternatives. Dans le cas d'une barrière trapézoïdale ou triangulaire, l'expression précédente se calcule explicitement, voir l'équation (3-7) et le graphe FIG. 3-2.

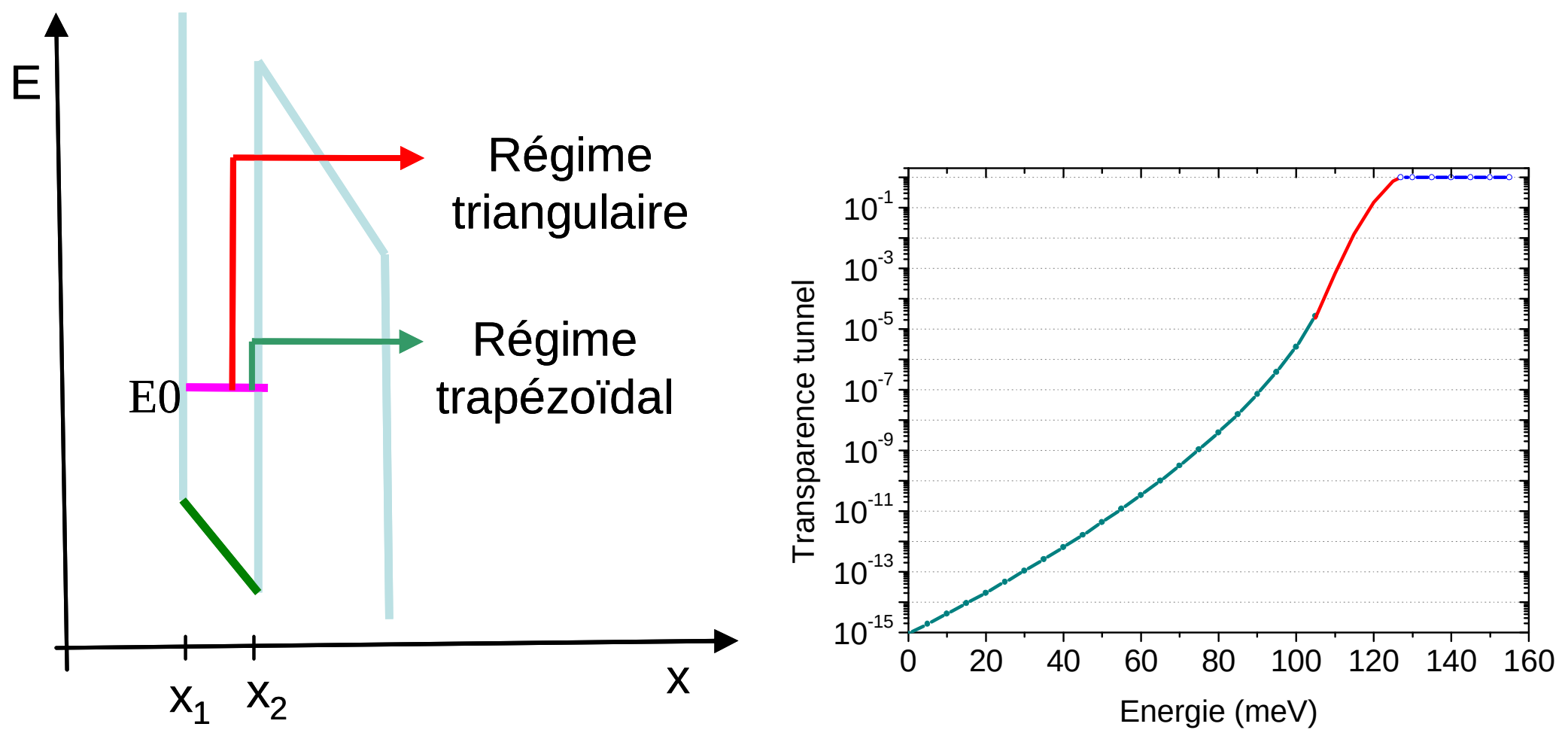

*FIG. 3-2 A gauche : Schéma des différents chemins tunnels envisageables pour l'électron. A droite : probabilité de passage tunnel à travers la barrière pour l'électron (calcul effectué dans la structure étudiée au chapitre deux).*

$$T(E,F)=\begin{cases} 1 \quad si \quad E>V_b \\ \exp\left(-\dfrac{4\sqrt{2m_b*}}{3.e.F.\hbar}(Vb-E)^{3/2}\right) \quad si \quad V_b-e.F.L_b<E<V_b : régime \quad triangulaire \\ \exp\left(-\dfrac{4\sqrt{2m_b*}}{3.e.F.\hbar}[(Vb-E)^{3/2}-(Vb-E-e.F.L_b)^{3/2}]\right) \quad si \quad E<V_b-e.F.L_b : régime \quad trapézoidal \end{cases}$$

(3-7)

La probabilité de passage tunnel peut facilement être reliée au courant via les différentes expressions que l'on peut trouver dans la littérature[29,74,75].

$$J_{ST \quad Levine}=\frac{ek_bT}{\hbar L_w^{\,2}}.T(E_1+eFL_w)\times\ln\left(\frac{1+e^{\frac{E_{fw}-E_1}{k_bT}}}{1+e^{\frac{E_{fw}-eFL_b-E_1}{k_bT}}}\right) \qquad (3\text{-}8)$$

$$J_{ST \quad Gomez}=\frac{ek_bT}{2\hbar L_w^{\,2}}.T(E_1)\times\ln\left(\frac{1+e^{\frac{E_{fw}-E_1}{k_bT}}}{1+e^{\frac{E_{fw}-eFL_b-E_1}{k_bT}}}\right) \qquad (3\text{-}9)$$

$$J_{ST \quad SchneiderLiu}=e\frac{m*}{\pi\hbar^2}\int_{E1}^{\infty}\tau^{-1}(E)f_{FD}(E)dE-e\frac{m*}{\pi\hbar^2}\int_{E1-eFL_b}^{\infty}\tau^{-1}(E)f_{FD}(E)dE \qquad (3\text{-}10)$$

Dans cette dernière expression le taux de diffusion tunnel est donné par $\tau=\dfrac{2L_w}{v_1}T^{-1}$, où $v_1$ est la vitesse de l'électron dans le puits donnée par $E=1/2.m*v_1^{\,2}$. Ces expressions ont en commun l'utilisation d'une probabilité tunnel de passage à travers une barrière et d'un terme de population, sous forme d'une fonction de Fermi Dirac ou d'un log quand l'expression est déjà intégrée.

Ce modèle conduit à une dépendance strictement monotone du courant avec la tension, voir FIG. 3-3. Un tel comportement est en accord avec les observations habituelles dans la majorité des QWIP[75] et structures dérivées[76], mais est en désaccord avec la présence de plateau sur la courbe expérimentale de notre composant 15µm.

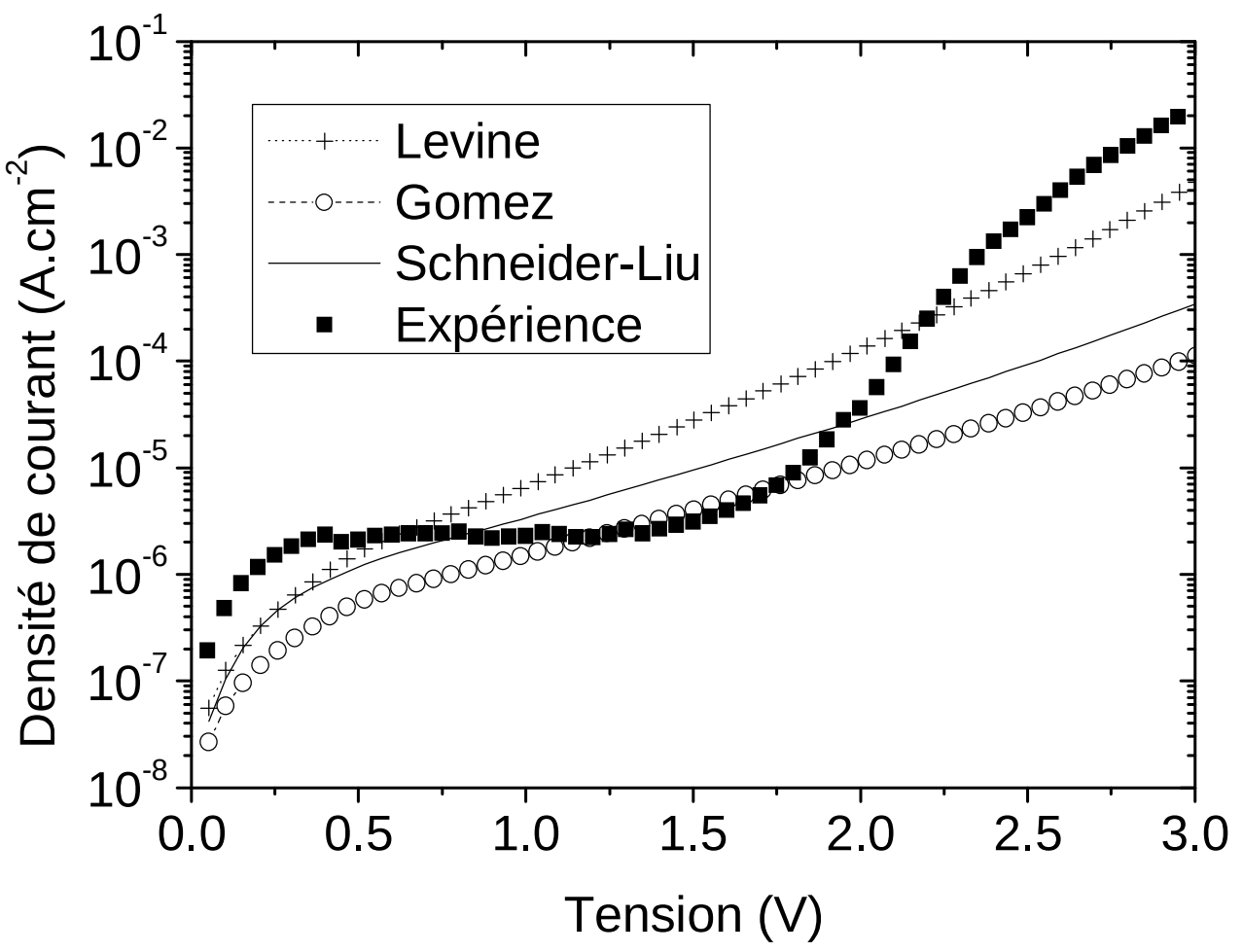

*FIG. 3-3 : Comparaison des courbes I(V) expérimentales et théoriques issues du modèle WKB*

Cette approximation WKB conduit donc grossièrement au bon ordre de grandeur de courant mais s'avère incapable de reproduire la forme de la courbe *I(V).* Cette incapacité résulte du grand nombre d'approximations qui sont contenues dans ce modèle.

### 3.2.1.2. Matrice de transfert

L'approche WKB est la version la plus simple de cette approche de transfert dans la mesure où c'est une version scalaire d'un problème matriciel plus complexe. L'approche en terme de matrice de transfert se propose de faire disparaître certaines limitations qui résultent des hypothèses fortes nécessaires à WKB.

Un traitement non plus scalaire mais matriciel permet de réintroduire les ondes réfléchies. Toutefois dans la mesure où les QWIP sont très peu couplés, il ne faut pas s'attendre à un rôle essentiel de la part de ces ondes réfléchies. Ce traitement peut se faire soit sur une base d'onde planes (pour une structure à tension nulle), ce qui conduit alors au modèle de Kronig-Penney ; soit, si la structure est polarisée, sur la base des fonctions d'Airy. Un tel calcul permet par exemple de faire ressortir le caractère résonant du transport à travers une diode à double barrière[77]. Ce calcul a été mené dans notre structure par F. Castelanno[58] en suivant la procédure de Rakityansky[78]. L'accord obtenu est assez bon, voir FIG. 3-4, et permet en particulier de rendre compte de la remonté des courbes I(V) à basse température aux fortes tensions (V>2V ou F>12kV.cm$^{-1}$)

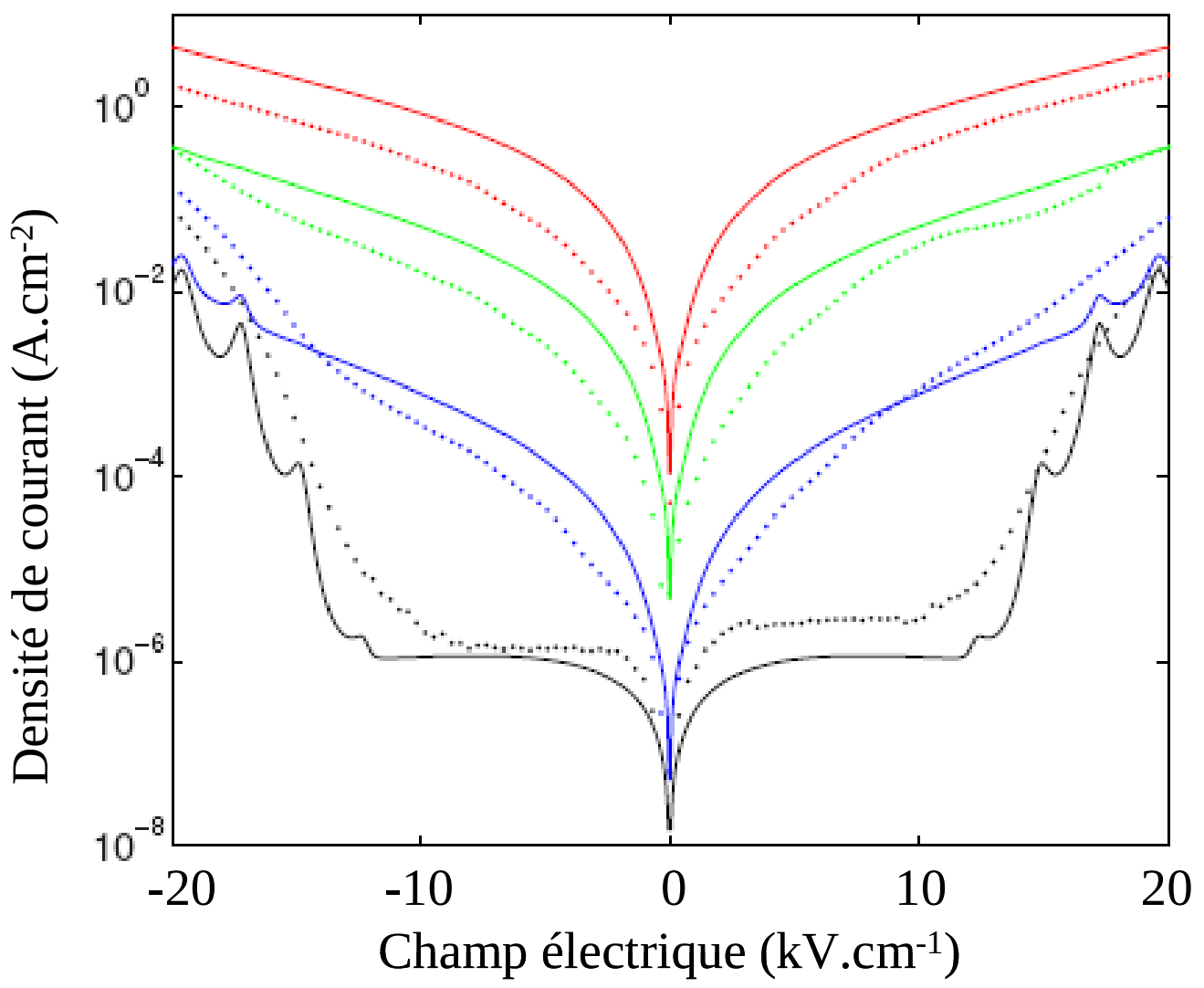

*FIG. 3-4 : Courbes J(F,) à 10K (noir), 40K (bleu), 50K (vert) et 60K (rouge), obtenues par F. Castelanno, en utilisant une approche de type mini-bande et matrice de transfert.*

Un des défauts de ces approches est le traitement très basique du dopage (uniquement via un niveau de Fermi). Par ailleurs d’un point de vue numérique un traitement très rigoureux de la précision des calculs est nécessaire, à cause de l’utilisation des fonctions de Bessel.

### 3.2.2. Modèle hydrodynamique

Ce modèle simple de transport a connu un certain succès pour traiter le cas des semiconducteurs massifs. Cette approche, basée sur une équation de conservation, correspond à une vision intuitive du transport dans laquelle les électrons coulent tel un liquide.

Dans un tel modèle,[79,80,81] le courant est vu comme la somme d'un terme de dérive sous champ ($\frac{nv}{L_d}$, avec $n$ la concentration de porteurs et $v$ leur vitesse) et d'un terme de diffusion ($D\nabla n$, avec $D$ le cœfficient de diffusion). Le courant prend alors la forme :

$$J(t) = \frac{\varepsilon}{e}.\frac{dF}{dt} + \frac{nv(F)}{L_d} - D(F)\nabla n \qquad (3\text{-}11)$$

Toutefois l'utilisation d'une telle équation dans une hétérostructure est assez contestable. La prise en compte des effets de confinement des niveaux électroniques n'est pas explicite : v et D cachant beaucoup de la physique du problème. Il faut également s'interroger quant à la pertinence de l'utilisation d'une vitesse de dérive quand l'électron est en régime tunnel. Notons également que les cœfficients de diffusion n'ont aucune raison d'être égaux à ceux du matériau massif et sont donc introduits de façon empirique. Ces différentes raisons font que ce type de modèle de transport peut convenir pour décrire un électron dans le continuum mais *stricto senso* beaucoup moins pour un électron en régime tunnel. Néanmoins une équation de transport basée sur une équation de conservation de la charge reste une idée pertinente, couramment utilisée dans les lasers à cascade.

#### *3.2.2.1. Equation de Boltzmann*

L'équation de Boltzmann, d'origine classique, est très utilisée pour décrire le transport dans les hétérostructures[82]. Elle reprend les principaux éléments du modèle de diffusion-dérive mais y ajoute un terme de relaxation. L'équation de Boltzmann est aussi une équation de conservation de la charge[83] :

$$\frac{\partial f}{\partial t} = \frac{f(r,k,t+dt) - f(r,k,t)}{dt} = \frac{f(r - vdt, k - \frac{dk}{dt}dt, t) - f(r,k,t)}{dt} = -v(k)\nabla_r f - \frac{dk}{dt}\nabla_k f$$
(3-12)

Avec f le facteur de population d'un puits. Comme précédemment on y retrouve un terme de diffusion $v(k)\nabla_r f$ et un terme de dérive $\frac{dk}{dt}\nabla_k f$. Equation à laquelle il faut ajouter un terme source qui représente les électrons gagnés par diffusion entre états électroniques :

$$\frac{\partial f}{\partial t} + v(k)\nabla_r f + \frac{eF}{\hbar}\nabla_k f = \left.\frac{\partial f}{\partial t}\right|_{diff} \qquad (3\text{-}13)$$

#### 3.2.2.2. Equation d'Esaki-Tsu

Esaki, en plus d'être à l'origine des premières hétérostructures, a également initié le premier modèle de transport dans les super réseaux[84]. Esaki et Tsu ont pour cela établi une loi de vitesse qui rend compte du phénomène de résistance négative différentielle. Cette loi de vitesse dérive de l'équation de Boltzmann. L'avantage de ce modèle par rapport au calcul complet de Boltzmann est de donner de la physique du super-réseau dans un modèle qui en a peu. Le détail des calculs pour passer de l'une à l'autre est donné en annexe (3-A) de ce chapitre mais nous pouvons ici rappeler les principales hypothèses ainsi que quelques grandes étapes du calcul. Trois principales hypothèses sont nécessaires : (i) la solution est supposée homogène spatialement ce qui supprime le terme de diffusion. (ii) Le terme source ou d'injection d'électrons par diffusion est traité comme un terme de relaxation avec un temps $\tau$ de relaxation constant. Il est alors possible de montrer que la loi de vitesse est donnée par :

$$\langle v \rangle = \int_{-\pi/L_d}^{\pi/L_d} dk_0 \, f_0(k) \int \frac{\partial v(k(t))}{\partial t} e^{-\frac{t}{\tau}} dt \qquad (3\text{-}14)$$

Enfin (iii) le super-réseau est supposé infini et faiblement peuplé ce qui se traduit mathématiquement par le fait que la population à l'équilibre est modélisé par à un simple delta dans l'espace des énergies. En injectant de plus la relation de dispersion sinusoïdale du super-réseau[130] dans l'équation précèdente il est possible d'obtenir la loi d'Esaki Tsu (voir annexe 3-A) :

$$\langle v \rangle = \frac{eF{L_d}^2\Delta}{2\hbar^2} \frac{\frac{1}{\tau}}{\left(\frac{1}{\tau}\right)^2 + \left(\frac{eFd}{\hbar}\right)^2} \qquad (3\text{-}15)$$

Cette expression peut s'exprimer avec des variables réduites $F_c = \frac{\hbar}{e\tau d}$ et $v_m = \frac{\hbar}{dm(0)}$ où $m(0)$ est la masse en bas de bande donnée par $m(0) = \frac{\hbar^2}{\left.\frac{\partial^2 E}{\partial {k_z}^2}\right|_{k_z=0}} = \frac{2\hbar^2}{\Delta d^2}$ ; pour conduire à :

$$\langle v \rangle_{ET} = v_m \frac{F/F_c}{1+\left(\dfrac{F}{F_c}\right)^2} \qquad (3\text{-}16)$$

Tsu, en collaboration avec Döhler, a également apporté une interprétation quantique à cette loi de vitesse et où cette fois-ci la RDN résulte de la localisation par le champ des fonctions d'onde[85].

Notons enfin que ce type de transport est particulièrement adapté au cas des super-réseaux fortement couplés (minibande large) ou des champs électriques faibles et cela de manière à ce que l'électron soit bien délocalisé sur la structure.

#### 3.2.2.3. Extension au cas des super-réseaux non faiblement dopés

L'un des défauts de l'approche précédente vient de l'hypothèse que la fonction de distribution des électrons en énergie doit être un simple Dirac, ce qui physiquement doit se traduire par une population faible en électron et donc un dopage faible. Or ce n'est pas le cas de nos structures. Il était donc intéressant d'étendre le modèle pour ne plus avoir recours à cette hypothèse.

Ce travail a été fait en collaboration avec le groupe de Jérome Faist (ETH Zürich) et en particulier Fabrizio Castellano (lui-même détaché du groupe de Fausto Rossi, Politecnico di Torino). Toute la partie modélisation des super-réseaux fortement dopés, traitée via le formalisme de minibande, est leur travail. Ce travail a donné lieu à une publication, voir la référence[58]. En voici les principales étapes :

Afin de s'affranchir de cette hypothèse sur la fonction de distribution il devient nécessaire de modifier la relation de dispersion qui garde son caractère sinusoïdal selon l'axe de croissance et devient parabolique dans la direction des couches.

$$E = E_0 + \frac{\Delta}{2}(1-\cos(k_z d)) + \frac{\hbar^2 k''^2}{2m^*} \qquad (3\text{-}17)$$

Par ailleurs, l'effet du champ électrique est de seulement déplacer la distribution électronique dans l'espace des moments. Castellano a alors montré que la nouvelle vitesse moyenne se déduit de l'expression non corrigée d'Esaki-Tsu via la relation[58]

$$\langle v \rangle = \langle vg \rangle_{ET} \frac{\Delta}{4E_f} \qquad (3\text{-}18)$$

$$j = en \frac{\Delta \langle vg \rangle_{ET}}{4E_f} \qquad (3\text{-}19)$$

Où n est la densité surfacique de porteurs, $E_f$ le niveau de Fermi. Le terme $\frac{\Delta}{E_f}$ vient du fait que seule une petite fraction des porteurs contribue au courant. Ceux pour lesquels le niveau de Fermi est dans la minibande. Leur nombre est proportionnel à la largeur de minibande. A l'opposé le nombre d'électrons qui ne contribuent pas est lui proportionnel au nombre d'électrons $E_f$. Le facteur ¼ résulte quant à lui du fait que l'électron ne doit pas être appareillé pour contribuer au transport.

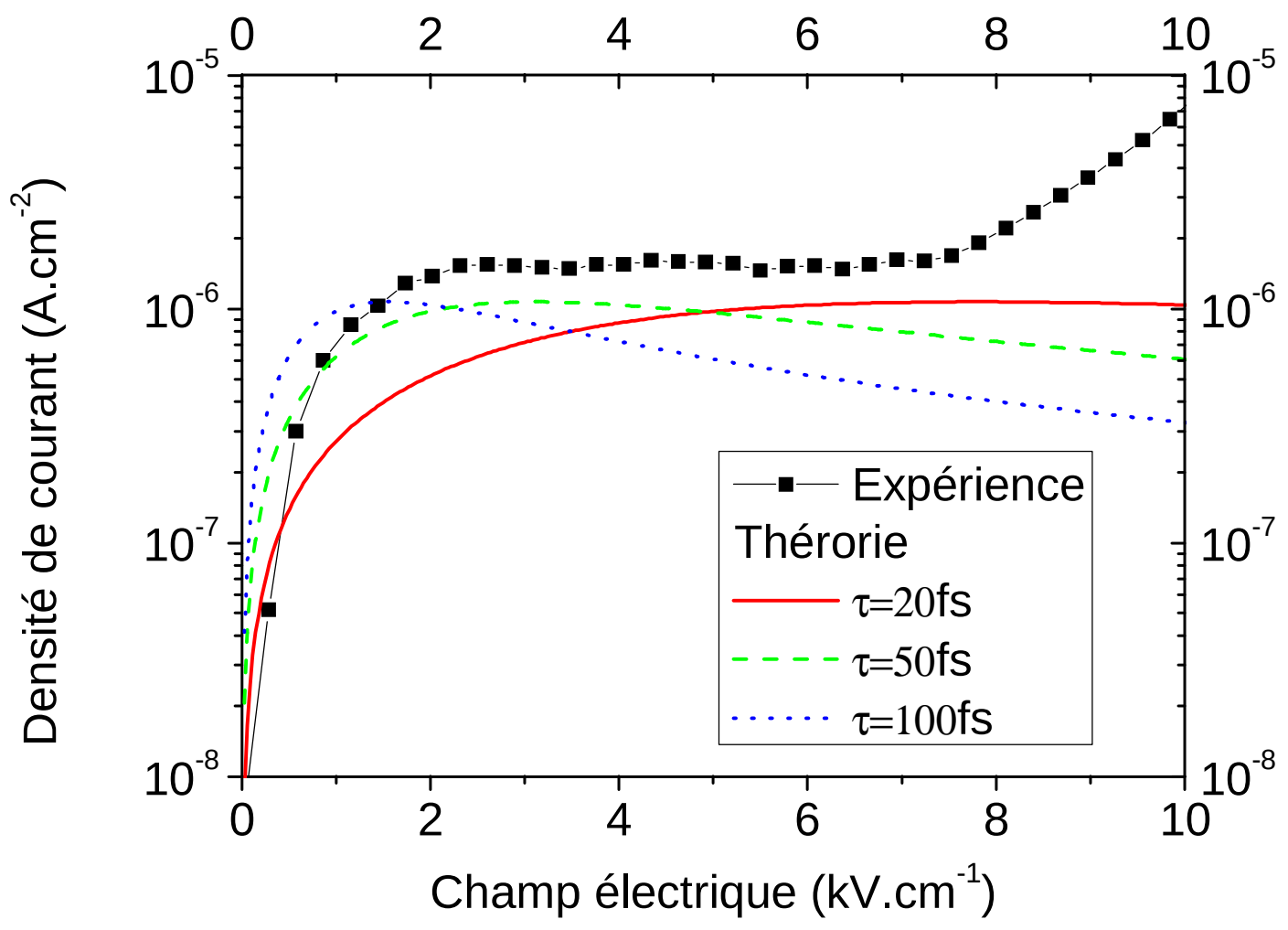

*FIG. 3-5 Courbe J(F) obtenue par le modèle de mini-bande étendu au cas des structures non faiblement dopées.*

Ce modèle très simple et analytique conduit à d'assez bons résultats. Il arrive de plus, en ajustant le temps de relaxation dans une gamme de valeurs réaliste, à reproduire la forme du plateau des courbes *I(V)*, voir FIG. 3-5. Cette forme de plateau s'interprète alors comme la saturation de la vitesse des électrons avec le champ électrique. La corrélation entre théorie et expérience est assez bonne, ce qui est assez surprenant dans la mesure où l'utilisation d'un modèle basé sur des mini bandes peut paraître saugrenue dans une gamme de champ telle que la notre : $eFL_d >> \Delta$. Néanmoins malgré ce résultat, ce modèle se voit reprocher une dépendance du courant avec le dopage irréaliste. En effet le courant est indépendant du dopage en raison du facteur $\frac{n}{E_f}$, ce qui n'est pas du tout en accord avec l'expérience. Paradoxalement c'est en voulant faire disparaître une hypothèse sur la distribution des électrons que l'on aboutit à un résultat étrange vis-à-vis de la population électronique.

### 3.2.3. Hopping : transport par saut entre états localisés

Le transport par hopping a connu ses heures de gloire avec l'étude des transitions de Mott[iii] et Anderson[iv]. Dans ces transitions, il existe un certain seuil (de dopage et de désordre respectivement) où le transport ne résulte plus d'un état délocalisé comme dans un métal ou dans une minibande mais de sauts entre états localisés. Bien qu'initialement ce type de transport soit inélastique, le qualificatif a été étendu à tout transport entre états localisés qui convient bien à nos super réseaux quand la chute de potentiel par période est grande devant la taille de mini-bande.

Toutefois le passage du transport par hopping du bulk[86] vers le super-réseau a jusqu'au début des années 90 été sujet à débat. En effet jusqu'en 1988, l'existence des états localisés (dits de Wannier-Stark) ne faisait pas l'objet d'un consensus dans la communauté des théoriciens[87]. Mais en 1988, le débat a pris fin avec la confirmation expérimentale de l'existence de ces niveaux[88,89] ce qui a repopularisé cette approche du transport dans les hétérostructures.

Dans cette approche le transport est évalué via le calcul d'un temps de diffusion entre deux états localisés[90]. Ce temps diffère fortement de ceux qui sont introduits dans les approches de type minibande ou Kazarinov-Suris (cf section 3.2.4.1 sur la matrice densité). En effet dans ces approches c'est un temps de relaxation (de la quantité de mouvement ou de l'énergie) qui ne s'accompagne pas de transport de particule. Le transport est traité de façon cohérente[91], même s'il se trouve affecté par la relaxation. Au contraire dans une approche de type hopping, la diffusion contribue au transport.

L'utilisation de ce formalisme dans le cas des QWIP est surtout le résultat de l'équipe de P. Harrison à Leeds[92,93,94,95]. Ce formalisme est aussi abondamment utilisé pour les lasers à cascade quantique[96]. Il présente également le grand avantage d'être facilement utilisable en présence d'un champ magnétique[97]. J'entrerai plus dans le détail de ce formalisme au chapitre suivant car il va s'avérer tout à fait adapté au traitement du transport dans nos structures.

---

[iii] La transition de Mott consiste en une transition conducteur/isolant induite par l'interaction coulombienne. Dans un semi conducteur dopé il existe une valeur de dopage en deçà de laquelle les électrons deviennent localisés sur les impuretés.

[iv] La localisation d'Anderson est un phénomène de localisation par le désordre.

### 3.2.4. Théories complètes

Je viens de décrire les trois formalismes simplifiés de transport dans les super réseaux. Nous avons vu que le formalisme de la matrice de transfert pouvait s'appliquer aux super-réseaux et conduire à l'approximation WKB pour traiter du transport tunnel séquentiel. J'ai présenté ensuite le modèle hydrodynamique et son application au transport qu'est l'équation de Boltzmann. Esaki et Tsu ont dérivé ce formalisme pour traiter du transport dans les mini-bandes. Enfin j'ai brièvement abordé le transport par sauts entre états localisés, mais nous y reviendrons par la suite. Pour finir sur ces différentes façons d'aborder le transport, je vais présenter quelques formalismes plus généraux liés au traitement des systèmes ouverts.

#### *3.2.4.1. Matrice densité*

Je vais m'attarde un peu longuement sur ce formalisme car il permet d'introduire une notion essentielle qui sera reprise par la suite à savoir la notion de transition à énergie complexe.

- *Formalisme*

La matrice densité est un formalisme qui opère la jonction entre la mécanique quantique et la thermodynamique. Ce formalisme est couramment utilisé pour décrire le transport dans les hétérostructures[90,98,99,100]. Il est en particulier utilisé pour traiter le cas du transport résonant, ce qui en fait l'un des outils les plus utilisés pour le transport dans les lasers à cascade quantique[101].

L'évolution temporelle du système est gérée par son hamiltonien $H$. Ce dernier peut se décomposer en deux parties $H = H_0 + V$ ou $H_0$ est l'hamiltonien du système sans diffusion, qui est diagonal en impulsion de par la périodicité du super-réseau. $V$ est l'hamiltonien associé à la diffusion et qui permet donc l'évolution du système entre états d'impulsions différentes.

L'opérateur matrice densité[102] est utilisé pour calculer les valeurs moyennes d'une observable, ainsi l'opérateur $eTr(\rho v)$ se confond avec l'opérateur courant. Reste à donner une expression quantique de la vitesse, pour cela il suffit de considérer que cette dernière est la dérivée de l'opérateur position. Son évolution est donnée par $v = \frac{1}{i\hbar}[z, H_0]$ [91]. En compilant les deux expressions précédentes, l'opérateur courant se déduit de la matrice densité et de l'hamiltonien via :

$$J = \frac{e}{i\hbar} Tr(\rho \ [z, H_0]) \quad (3\text{-}20)$$

Afin de le calculer nous suivrons la procédure décrite par Kazarinov et Suris[103] puis reprise et détaillée par Willenberg et al[104]. Mon travail a donc consisté à adapter leur calcul à notre cas particulier du système à deux niveaux, mais il permet d'introduire des notions essentielles pour la suite telle que la notion de transition à énergie complexe.

Appliquons le formalisme de la matrice densité au transport entre deux sous-bandes (dans le cas qui nous intéresse les deux sous-bandes associées à l'état fondamental de deux puits consécutifs[105]).

$$H_0 = \begin{pmatrix} \varepsilon_k^n & \Delta \\ \Delta & \varepsilon_k^{n+1} \end{pmatrix} \text{ Avec } \varepsilon_k^n = E_1 + n \cdot eFL_d + \frac{\hbar^2 k^2}{2m^*} \quad (3\text{-}21)$$

L'hamiltonien de diffusion $V$ est généralement pris diagonal, c'est-à-dire que seule la diffusion intra bande est considérée. $V = \begin{pmatrix} V_{kk'}^{nn} & 0 \\ 0 & V_{kk'}^{n+1n+1} \end{pmatrix}$. Il est également possible d'expliciter l'opérateur position $Z = \begin{pmatrix} Z^{nn} & Z^{nn+1} \\ Z^{n+1n} & Z^{n+1n+1} \end{pmatrix}$ avec $z^{n+1n+1} - z^{nn} = L_d$.

Les éléments non diagonaux de l'opérateur vitesse qui interviennent dans l'expression du courant s'écrivent :

$$v^{nn+1} = \frac{1}{i\hbar}\left[\Delta(z^{n+1n+1} - z^{nn}) + z^{nn+1}(\varepsilon_k^n - \varepsilon_k^{n+1})\right] \quad (3\text{-}22)$$

Or $z^{nn+1} \approx \dfrac{\Delta}{\varepsilon_k^{n+1} - \varepsilon_k^n}(z^{n+1n+1} - z^{nn}) \approx \dfrac{\Delta}{eF}$ ce qui est faible si le couplage l'est, ce qui permet de limiter l'expression de la vitesse à $v^{nn+1} = \dfrac{\Delta(z^{n+1n+1} - z^{nn})}{i\hbar}$. L'expression du courant est alors donnée par

$$J = \frac{e}{i\hbar}\sum_k \Delta(\rho_k^{nn+1} - \rho_k^{n+1n}) \quad (3\text{-}23)$$

Il faut donc connaître l'expression des termes de la forme $\rho_k^{nn+1}$, le détail du calcul est donné en annexe 3-C de ce chapitre. L'expression finale donne :

$$\underbrace{\left(\Delta E - i(\gamma_k^n + \gamma_k^{n+1})\right)}_{\text{Energie complexe}} f_k^{nn+1} = \underbrace{\Delta(f_k^{nn} - f_k^{n+1n+1})}_{1\ :\ \text{Processus tunnel}} - \underbrace{i\frac{\Delta}{\Delta E}\left((\gamma_k^{n+1}(f_{q_{n+1}}^{n+1n+1} - f_k^{n+1n+1}) - \gamma_k^n(f_{q_n}^{nn} - f_k^{nn})\right)}_{2\ :\ \text{Diffusion intrabande}}$$

(3-24)

L'interprétation de cette dernière expression est la suivante : la transition entre l'état n et n+1 se fait en deux étapes, la première consiste en un processus tunnel à impulsion constante vers un état virtuel, voir FIG. 3-6. Un second processus, de diffusion, permet à l'électron de relaxer son impulsion. Il faut noter que l'énergie de la transition est alors complexe : à la différence d'énergie classique vient s'ajouter l'élargissement des niveaux. Dans une telle approche le couplage tunnel est celui qui permet le déplacement des charges, alors que la diffusion n'y contribue pas[91]. Par contre elle permet de faire relaxer l'impulsion et l'énergie des électrons.

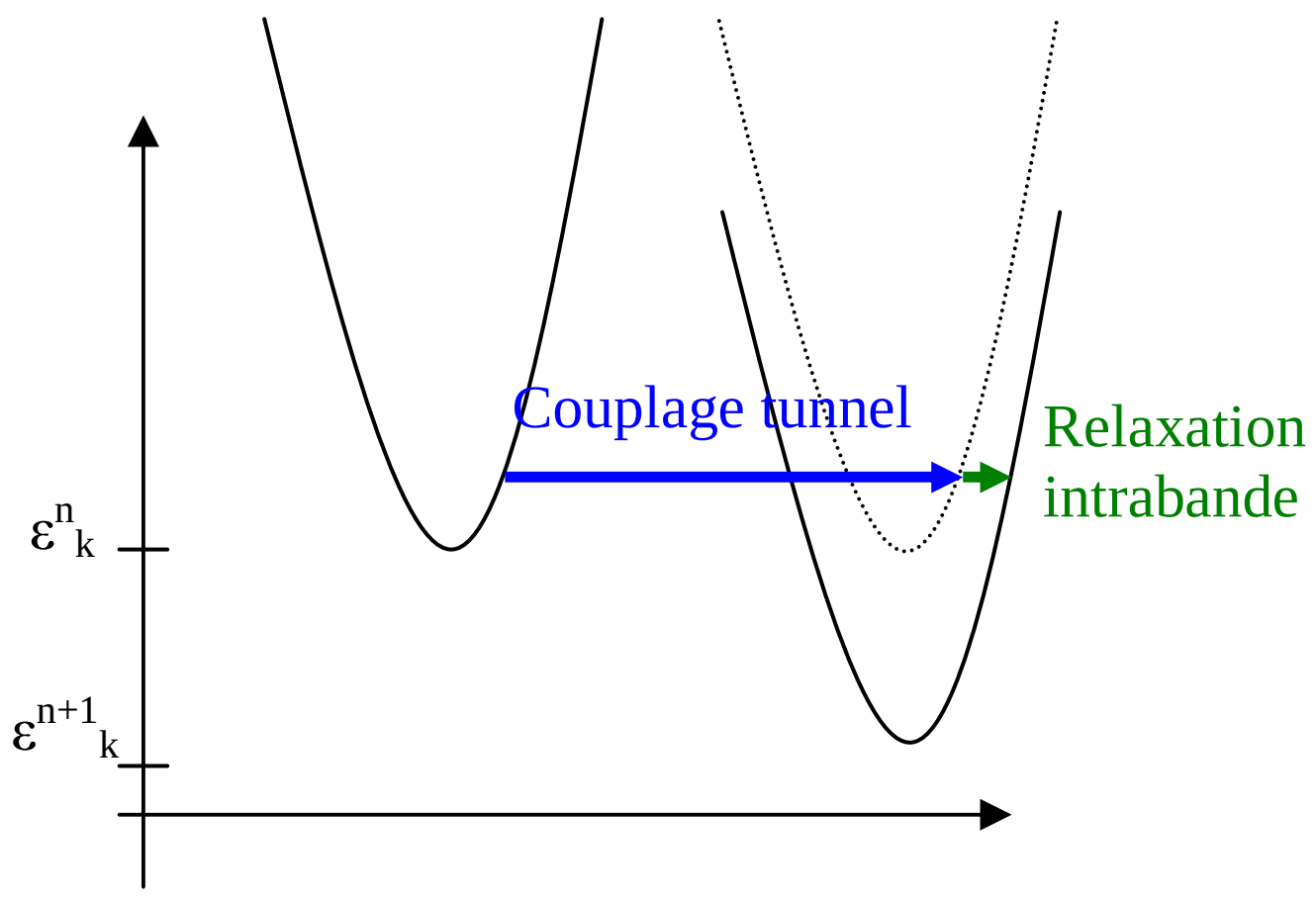

*FIG. 3-6 Principe du transport tunnel assisté par diffusion.*

Il ne reste plus alors qu'à injecter (3-24) dans l'expression (3-23) du courant :

$$J = \frac{e\Delta^2 L_d}{\hbar} \sum_k \frac{\gamma_k^n \left( f_k^{n+1n+1} - f_{q_n}^{nn} \right) + \gamma_k^{n+1} \left( f_{q_{n+1}}^{n+1n+1} - f_k^{nn} \right)}{\Delta E^2 + \left( \gamma_k^{n+1} + \gamma_k^n \right)^2} \qquad (3\text{-}25)$$

Où $q$ est l'impulsion de l'état d'arrivée. Dans ce processus en deux étapes (parfois appelé processus tunnel du second ordre) le transport se fait avec conservation de l'énergie. Kazarinov et Suris ont poussé plus loin l'approximation en utilisant un taux de diffusion constant et en prenant l'impulsion de l'état d'arrivée égale à k, ce qui simplifie encore l'expression du courant

$$J = \frac{e\Delta^2 L_d}{\hbar} \frac{\gamma}{\Delta E^2 + \gamma^2} \Delta n \qquad (3\text{-}26)$$

Dans ce cas le transport se fait à impulsion constante : seul la première étape du processus décrit sur la figure FIG. 3-6 est considérée. Cette dernière expression fonctionne efficacement quand les états en présence sont de population différente, c'est par exemple le cas lors de la mise en résonance d'un niveau fondamental avec un état excité de la période suivante. Par contre elle n'est pas valable entre états équivalents.

Pour ce qui est de la forme des courbes, nous retrouvons la forme de Lorentzienne vue dans le modèle de minibande. Notons par ailleurs que les éléments de matrice, et donc les taux de diffusion, ne dépendent pas du champ électrique de par l'utilisation de fonctions de Wannier pour leurs calculs. Cette approche de matrice densité conduit donc aux mêmes formes de courbes. Par contre $\gamma$ représente une véritable quantité physique, un taux de diffusion, alors que ce n'était qu'un paramètre servant à introduire la diffusion dans l'approche de minibande.

- *Cas du transport résonant et inversion de population*

Cette approche du transport est à l'origine du laser à cascade quantique. Cette structure, inventée au milieu des années 90 par Jérôme Faist et al[10], est sans conteste le plus grand succès des dispositifs intrabandes. Le caractère résonant du transport est utilisé pour créer l'inversion de population dans un laser dont il devient alors possible d'ajuster la longueur d'onde d'émission uniquement en jouant sur des paramètres géométriques de croissance. A présent ces lasers réussissent à fonctionner à température ambiante[106] et la longueur d'onde peut varier du proche infrarouge (2.6µm[107]) jusqu'au THz[108]. Ce dispositif a son pendant en terme de détecteur avec le QCD.

#### 3.2.4.2. Fonction de Wigner

La fonction de Wigner est l'analogue quantique de la fonction de distribution qui intervient dans l'équation de Boltzmann. Son expression dérive directement de celle de la matrice densité.

$$f_W(r,k) = \int dr' e^{-ik\cdot r'} \rho(r + r'/2; r - r'/2) \qquad (3\text{-}27)$$

Elle a l'avantage par rapport à celle-ci de toujours rester réelle, mais elle peut être de signe négatif. Cette négativité se manifeste lorsque l'état considéré n'est pas classique. Ce formalisme est bien adapté au traitement des systèmes ouverts que sont les super-réseaux.

L'évolution de la fonction de Wigner est régie par une équation similaire à celle de Boltzmann :

$$\frac{\partial f_W(r,k)}{\partial t} = -\Gamma_{out} \cdot f_W(r,k) + \sum_{k'} \Gamma_{in}(r,k') \cdot f_W(r,k') \qquad (3\text{-}28)$$

ou $\Gamma_{out}, \Gamma_{in}$ sont respectivement les taux de transitions vers des états extérieurs et internes au système. Ce formalisme est très utilisé en physique quantique (étude des chats de Schrödinger par exemple), mais son utilisation pour le transport dans les super-réseaux[109,110] est plus marginale que celui de la matrice densité ou de la fonction de Green hors équilibre que nous allons voir juste après.

#### 3.2.4.3. Fonction de Green hors équilibre

La fonction de Green dans son formalisme hors équilibre offre une belle alternative au traitement du transport par la matrice densité. Je vais dans ce paragraphe me contenter de donner les bases de ce formalisme et tenter d'en faire un parallèle avec celui de la matrice densité[100]. Je finirais par quelques mots sur son utilisation dans les hétérostructures de semi-conducteur.

Je passe volontairement l'aspect mathématique de l'objet fonction de Green pour ne parler que de physique. Toutefois le lecteur intéressé pourra se reporter, par exemple, à l'introduction du livre de S. Datta[111]. La fonction de Green a commencé à être utilisée dans le domaine de la matière condensée, dans les années 60 par Keldish et Kadanoff[112]. Elle est définie de la façon suivante[111,113] :

$$G(r_1, r_2, t_1, t_2) = \frac{1}{i\hbar} \left\langle T \cdot \Psi(r_1, t_1) \cdot \Psi^+(r_2, t_2) \right\rangle \qquad (3\text{-}29)$$

Avec $\Psi$ et $\Psi^+$ les opérateurs champs qui se définissent comme $\Psi(r,t) = \sum_s a_s(t) \cdot \varphi_s(r)$ avec s un état propre, a l'opérateur d'annihilation électronique et $\varphi$ la fonction d'onde associée à ce même état. $\Psi^+$ est l'hermitien conjugué de $\Psi$. Enfin T est l’opérateur d’ordonnancement de Wick, qui assure la causalité de la fonction de Green.

La fonction de Green traduit la réponse de la mer d’électrons à l’addition ou à la destruction d’une particule supplémentaire. C’est donc un formalisme qui inclut facilement les effets d’injection ainsi que les interactions entre porteurs.

Il est courant de définir deux opérateurs de corrélation qui dérivent de G[114] :

$$G^{>}(r_1,r_2,t_1,t_2)=\frac{1}{i\hbar}\left\langle \Psi(r_1,t_1)\cdot\Psi^{+}(r_2,t_2)\right\rangle \qquad (3\text{-}30)$$

$$G^{<}(r_1,r_2,t_1,t_2)=-\frac{1}{i\hbar}\left\langle \Psi^{+}(r_2,t_2)\cdot\Psi(r_1,t_1)\right\rangle \qquad (3\text{-}31)$$

Dont il est facile de trouver une interprétation physique puisque $G^{<}$ se confond avec la densité moyenne de particules et que $G^{>}$ correspond à la densité d’états. Le nom de fonction de corrélation n’a pas été choisi par hasard dans la mesure ou il est possible de relier $G^{<}$ aux termes non diagonaux de la matrice densité[100,115] via :

$$\rho_{ij}=\frac{1}{2i\pi}\int G_{ij}^{<}(E)\cdot dE \qquad (3\text{-}32)$$

$G_{ij}^{<}$ traduit donc la cohérence des états i et j entre eux. Pour finir cette fonction peut être reliée au courant[116] par

$$J=\frac{1}{2im^{*}}\lim_{r'\to r}\left[(\nabla r-\nabla r')G^{<}\right] \qquad (3\text{-}33)$$

Ce formalisme a été utilisé pour modéliser un certain nombres de dispositifs[117] basés sur des hétérostructures de semiconducteurs, tels que la diode à double barrières tunnel,[116] les super-réseaux[118] et les lasers à cascade[91]. Par contre rien à ma connaissance qui soit spécifique au QWIP.

#### ***3.2.4.4. Monte Carlo***

L’approche Monte Carlo est couramment utilisée pour décrire le transport dans les détecteurs[119] et autres nanostructures[79]. Cette méthode consiste à suivre temporellement l’évolution d’une population d’électrons qui suit un transport balistique interrompu par des événements de diffusion. Leur population s’en trouve réorganisée en énergie et en impulsion.

A chaque pas de temps, un nombre est tiré au sort dont la valeur va déterminer si le vol libre se poursuit ou si au contraire il y a diffusion[120]. La distribution de probabilité de ces tirages au sort est liée au taux de diffusion total. Ensuite un processus de diffusion possible est tiré au sort, tout en respectant le rapport des taux de diffusion entre eux.

Contrairement à la plupart des approches précédentes où l’on cherchait à obtenir des grandeurs microscopiques par un moyennage temporel (recherche de solution stationnaire), les grandeurs macroscopiques sont obtenues par des moyennes d'ensemble sur la population d'électrons considérée.

Ce genre d'approche a été principalement utilisé dans les QWIP par l'équipe de Ryzhii[121,122,123] pour estimer la vitesse de réponse des QWIP à une impulsion brève.

## 3.3. Vers une unification des théories

Le but de cette partie est de montrer que bien que ces différentes théories aient l'air parfois extrêmement éloignées, il existe une seule et même physique et qu'il est donc possible d'établir des relations entre les modèles. L'idée est de montrer qu'il est souvent possible d'obtenir ces différentes théories d'une autre façon que celle présentée dans la première partie de ce chapitre.

### 3.3.1. Choix des fonctions d'onde

Les formalismes développés dans ce chapitre utilisent abondamment la notion d'état quantique et de fonctions d'onde. Ces dernières interviennent souvent à travers le calcul d'éléments de matrice. Les propriétés physiques ne doivent pas dépendre de la base de fonctions d'onde qui est utilisée pour les évaluer. Mais certaines hypothèses simplificatrices qu'utilisent les modèles peuvent en dépendre, ce qui aboutit généralement à des modèles dépendants du choix de fonctions d'onde. Trois grandes bases de fonctions d'ondes sont généralement utilisées.

- Les ondes de Bloch qui sont solutions de l'hamiltonien associé à la structure du super-réseau quand celui-ci n'est pas biaisé. Les fonctions d'onde sont délocalisées sur l'ensemble de la structure. Ce sont elles qui sont utilisées dans l'approche de mini-bande.

- Les fonctions de Wannier-Stark sont les solutions de l'hamiltonien associé à la structure du super-réseau en présence de champ électrique. Elles sont plus localisées que les ondes de Bloch. Elles sont généralement traitées comme des états stationnaires, même si elles deviennent métastables quand leur énergie devient complexe. Ce dernier point sera discuté de façon plus poussée au chapitre 4. Ce type d'onde est utilisé dans l'approche de hopping.

- Enfin les fonctions de Wannier, ce sont des états extrêmement localisés puisque l'on peut les calculer comme des états propres de l'opérateur position[118]. Elles ont l'avantage de ne pas dépendre du champ électrique appliqué, ce qui fait que les éléments de matrice associés peuvent être calculés une fois pour toute. C'est ce type de fonctions d'onde qui est généralement utilisé dans le formalisme de la matrice densité.

### 3.3.2. Minibande vs Hopping : choix de jauge

Nous allons dans ce paragraphe montrer que la théorie de transport basée sur les minibandes ou celle de Wannier-Stark (hopping) ne diffèrent, en principe, que par un choix de jauge[58]. Pour comprendre cela rappelons que les champs électriques (F) et magnétiques (B) dans la structure sont reliés à leur potentiel scalaire $\varphi$ et à leur potentiel vecteur (A) par :

$$\begin{cases} \vec{F} = \vec{\nabla}\varphi + \dfrac{\partial \vec{A}}{\partial t} \\ \vec{B} = r\vec{o}t(\vec{A}) \end{cases} \qquad (3\text{-}34)$$

Dans l'approche de Wannier-Stark le choix de jauge qui est fait est $\begin{cases} \varphi = \vec{F}.\vec{r} \\ \vec{A} = \vec{0} \end{cases}$, tandis que pour l'approche minibande il est $\begin{cases} \varphi = 0 \\ \left|\vec{A}\right| = Ft \end{cases}$. Dans tous les cas l'hamiltonien du système s'écrit $H = \dfrac{(p + eA)^2}{2m*} + \varphi + V$. Dans une telle écriture $\varphi$ est la partie du potentiel liée à la présence du champ électrique tandis que V est le potentiel lié au profil de bande des différents matériaux qui composent l'hétérostructure.

Dans l'approche de type minibande le potentiel est donc un créneau non biaisé ($\varphi = 0$) mais l'hamiltonien dépend du temps, ce qui conduit à une dépendance temporelle du vecteur d'onde $k(t) = k_0 + \dfrac{Ft}{\hbar}$. Cela traduit simplement le fait que les électrons sont accélérés par le champ électrique. Tandis que dans l'approche de Wannier-Stark le potentiel $\varphi$ biaise la structure, mais l'hamiltonien ne dépend pas du temps. Donc formellement la même physique est incluse dans les deux approches, elle ne se distinguent[124] donc que par les approximations successives qui sont faites dans chacune d'elles.

### 3.3.3. Cohérence du transport : choix de la base de fonctions d'onde

Je voudrais à présent faire un parallèle entre l'approche de hopping entre états de Wannier Stark et le processus de diffusion au second ordre résultant de la matrice densité. A priori on pourrait voir ses deux formalismes comme antagonistes dans la mesure où la diffusion ne contribue pas au transport dans l'approche de matrice densité alors qu'au contraire elle est le mécanisme de transport dans l'approche de hopping[91]. Pourtant ces deux processus se ressemblent assez, les deux font en particulier du transport à énergie constante, mais sans conservation de l'impulsion. Dans l'approche de matrice densité, les états quantiques sont développés sur la base des fonctions de Wannier qui sont des fonctions relativement localisées. Le transport consiste en une première étape de couplage tunnel entre deux fonctions d'onde de puits consécutifs (transport cohérent), voir FIG. 3-7, en haut. Dans l'approche de hopping, cette première étape n'est pas nécessaire car les fonctions incluent ce couplage tunnel (le transport est incohérent, mais la cohérence est incluse dans les fonctions d'onde). Reste ensuite le processus de diffusion qui vient ensuite relaxer l'impulsion gagnée lors de ce changement d'état quantique, (voir FIG. 3-7, en bas).

C'est donc essentiellement le choix de la base de fonctions d'onde qui différencie ces deux processus, mais le transport que subit l'électron est bien le même dans les deux cas.

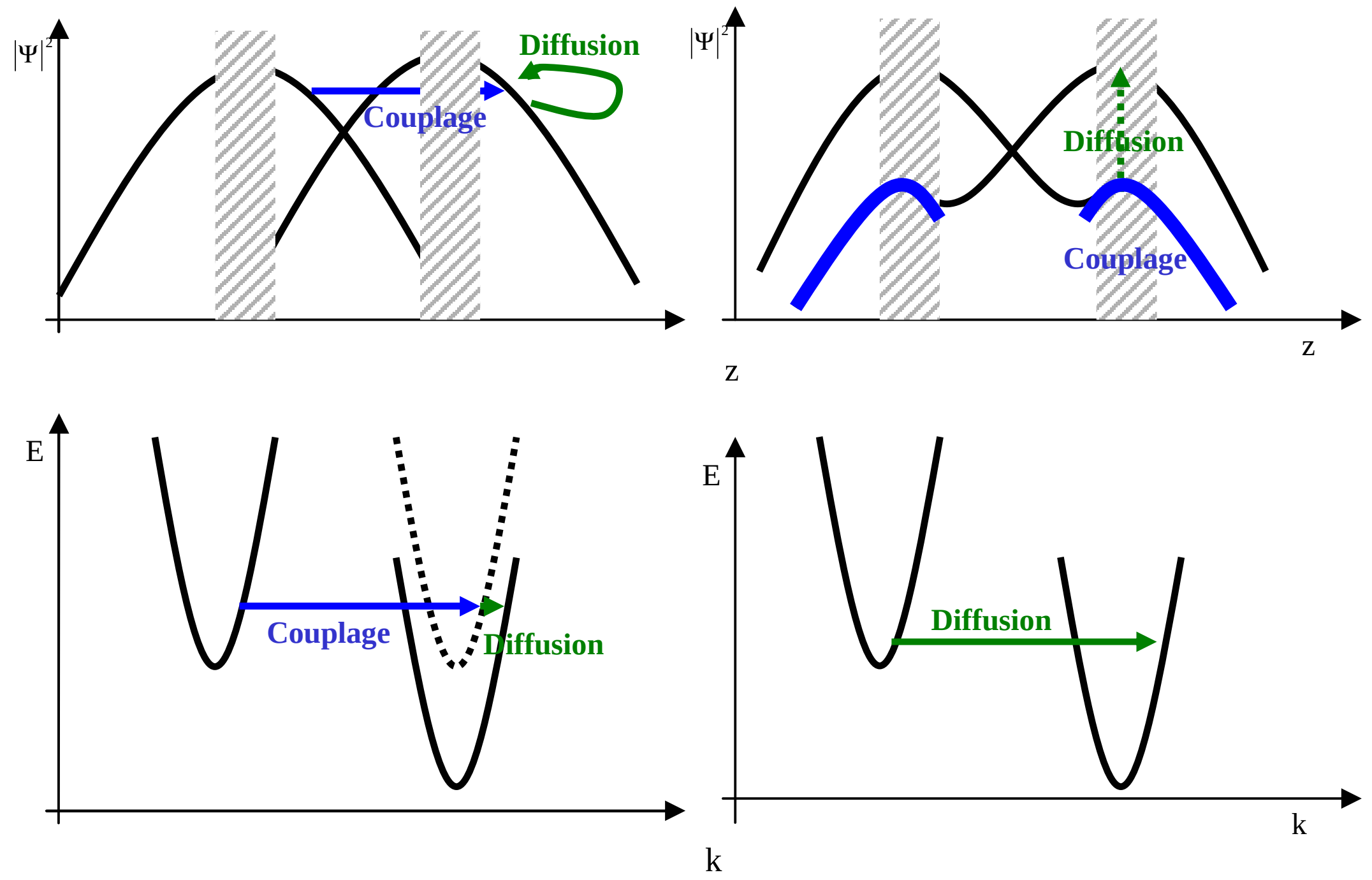

*FIG. 3-7 En haut à gauche : fonctions d'onde et déplacement des charges associées au processus de diffusion du second ordre dans l'approche de matrice densité. En haut à droite fonctions d'onde et déplacement des charges associés au processus de hopping. En bas : déplacement des charges associées au processus de diffusion du second ordre dans l'approche de matrice densité dans l'espace énergie impulsion. En bas à droite : déplacement des charges associées au processus de hopping dans l'espace énergie impulsion. Les zones grisées représentent les puits*

### 3.3.4. Effet du champ électrique

L'effet du champ électrique sur le courant s'avère parfois assez différent entre ces différentes approches. Mais cela résulte essentiellement du choix de la base de fonctions d'onde. Les approches de matrice densité et de minibande semblent s'accorder pour que la courbe J(F) présente une forme de Lorentzienne. Le phénomène de résistance différentielle négative apparaît alors à fort champ. Cela résulte du fait que le champ électrique tend à localiser les fonctions d'onde.

Au contraire dans l'approche WKB (tunnel séquentiel), la probabilité de passage à travers la barrière est calculée en utilisant des ondes planes indépendantes du champ électrique. Le phénomène de RDN n'est alors pas décrit. Mais cette approche rend compte de l'abaissement des barrières à fort champ électrique, c'est ce qui explique le caractère strictement monotone de l'*I(V)* associée à ce processus. Ce dernier processus n'est par contre pas pris en compte dans une approche comme la matrice densité qui utilise les fonctions de Wannier, indépendantes du champ électrique.

### 3.3.5. Diagramme de validité

Il est important de comprendre que chacune des ces approches simplifiées repose sur au moins une hypothèse forte[118]. L'approche de minibande traite mal la localisation par le champ du fait de l'utilisation de fonctions d'ondes délocalisées. L'approche par saut entre états localisés (Wannier Stark hopping) traite la diffusion au premier ordre, car elle n'inclut pas l'élargissement des niveaux. Enfin l'approche de transport tunnel séquentiel repose sur un traitement du couplage inter-puits à l'ordre le plus bas. Le tableau tab. 3-1 récapitule la façon dont chacune des approches simplifiées traite le couplage inter-puits, le champ électrique et l'élargissement des niveaux.

| | Couplage | Champ électrique | Elargissement des niveaux |
|---|---|---|---|
| 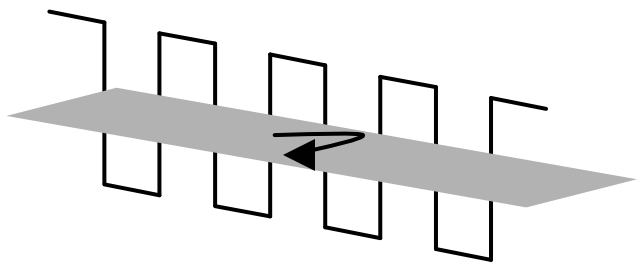 Minibande | Traitement exact | Accélération des électrons | Règle d'or (1$^{er}$ ordre) |
| 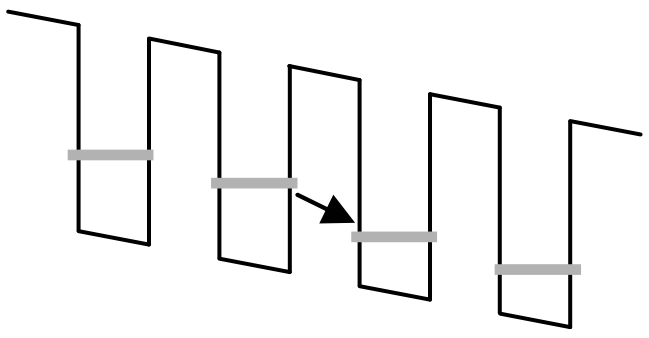 Wannier Stark hopping | Traitement exact | Traitement exact | Règle d'or (1$^{er}$ ordre) |
| 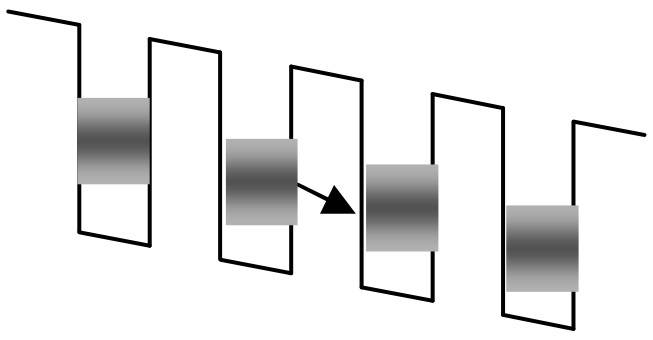 Tunnel séquentiel | 1$^{er}$ ordre | Ecart en énergie | Traitement exact |

*tab. 3-1 Traitement par les différentes approches simplifiées du couplage tunnel, du champ électrique et de la diffusion, d'après Wacker[118].*

Du tableau précédent, Wacker a déduit un diagramme de validité de chacune de ces approximations (FIG. 3-8). Comme nous pouvions nous y attendre l'approche de minibande fonctionne quand le couplage entre puits est important. Alors qu'au contraire l'approche par effet tunnel séquentiel est plutôt adaptée au couplage faible. L'approche par sauts entre états localisés est, elle, adaptée au cas où le champ électrique appliqué est fort. Il persiste des zones ou aucune de ces théories simplifiées n'est valide, dans ce cas une approche complète (matrice densité ou fonction de Green…) est alors nécessaire. Wacker a également montré comment il était possible de retrouver chacune des ces approches simplifiées à partir du formalisme de fonction de Green.

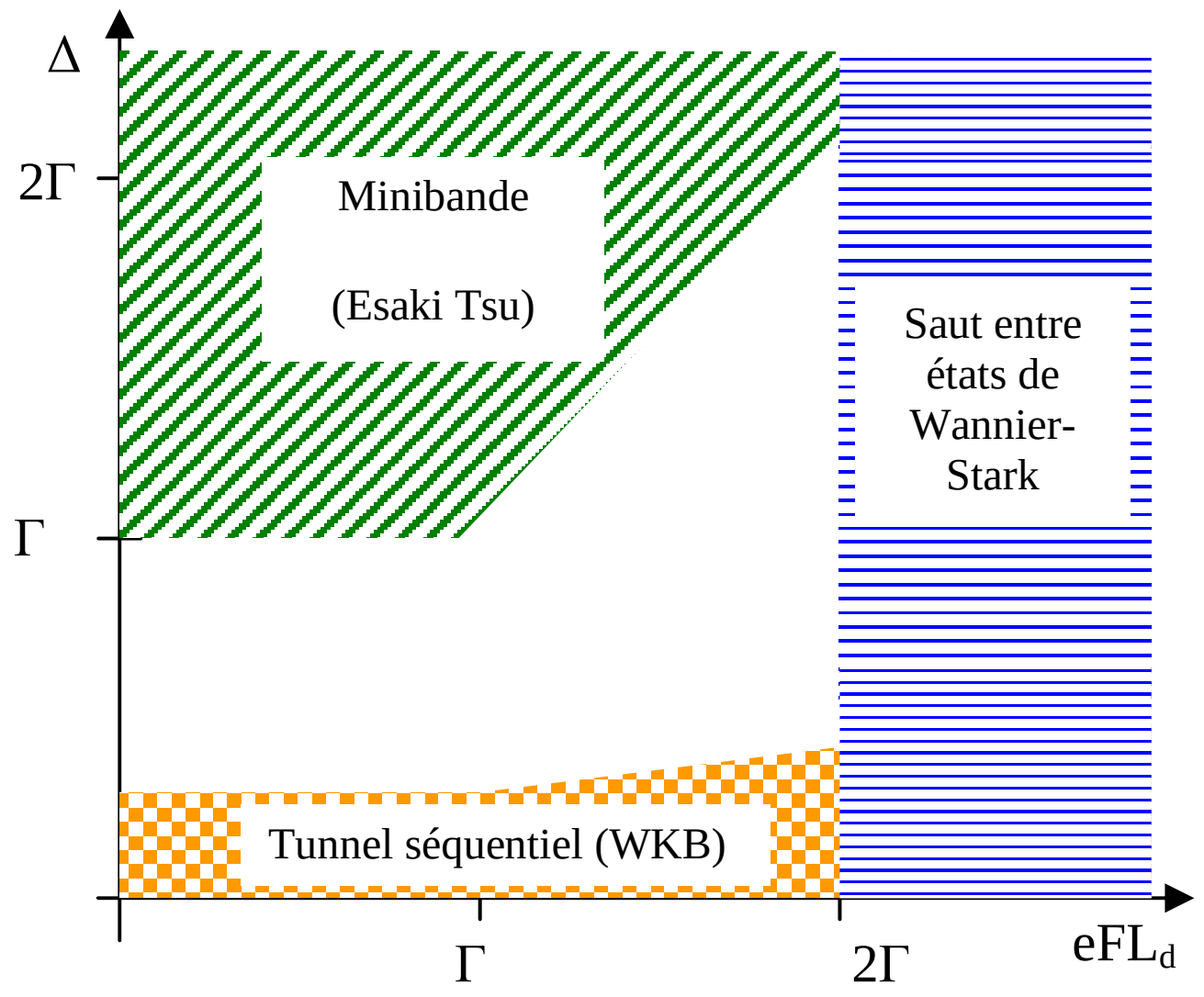

*FIG. 3-8 Diagramme de validité des différentes approches simplifiées, Γ est l'élargissement en énergie des niveaux. D'après Wacker*[118].

## 3.4. Conclusion

A la vue de ce diagramme de validité, voir FIG. 3-8, il devient légitime de se demander ou se situe le point de fonctionnement d'un QWIP haute longueur d'onde. Le couplage dans ces structures étant faible, la zone de fonctionnement se situe plutôt le long de l'axe des abscisses. Reste maintenant à savoir si les QWIP se situent plutôt dans la zone des forts champs électriques ou dans celle des faibles champs. Pour cela il nous faut estimer l'élargissement des niveaux. Or justement il est très facile de l'estimer en utilisant un code de diffusion basée sur une approche de type Wannier-Stark. Il me faudrait donc disposer d'un tel outil de simulation et en vérifier *a posteriori* la validité. Mais ce n'est pas la seule raison qui me pousse à vouloir développer ce formalisme. L'approche WKB s'est en effet montrée incapable de rendre compte de la forme de plateau que nous avons pu observer sur les *I(V)*. De l'autre côté l'utilisation d'une approche de minibande s'est montrée capable d'apporter une première explication à cette forme de plateau via l'introduction d'une vitesse de saturation, mais son utilisation vis-à-vis du diagramme de validité reste discutable. Et plus grave, cette approche a conduit à des résultats tout à fait incompatibles avec l'expérience en terme de dépendance vis-à-vis du dopage. Voila donc un certains nombres d'arguments qui nous pousse à développer un outil de simulation basé sur l'approche de hopping entre états de Wannier-Stark. C'est donc cet outil de simulation et les résultats de transport qui lui sont associés que je vais présenter dans les deux prochains chapitres.

# ANNEXE 3-A : De l'équation de Boltzmann à celle d'Esaki-Tsu

Afin d'établir la loi de vitesse d'un électron dans un super-réseau, Esaki et Tsu sont partis de l'équation de Boltzmann, voir (3-13), qu'ils ont utilisée dans une version simplifiée.

## 3.5.1. De l'équation de Boltzmann à celle de Chambers

Deux hypothèses préliminaires ont été faites :

- La solution recherchée est supposée homogène, le terme de dépendance spatiale s'annule donc. Cette hypothèse ne serait pas justifiée dans un système présentant des inhomogénéités (en particulier de champ électrique).
- Le terme de diffusion $\left.\frac{\partial f}{\partial t}\right|_{diff}$ est réduit à sa plus simple expression en remplaçant la dérivée, par un simple taux de croissance : $\left.\frac{\partial f}{\partial t}\right|_{diff} = -\frac{f - f_0}{\tau}$ avec $f_0$ la fonction de distribution à l'équilibre.

Compte tenu de ces deux hypothèses l'équation de Boltzmann se réécrit $\frac{\partial f}{\partial t} + \frac{eF}{\hbar}\nabla_k f = -\frac{f - f_0}{\tau}$. Cette équation est soluble analytiquement[125], et donne pour solution stationnaire $f(k_0) = \int_{-\infty}^{0} \frac{\exp(t/\tau)}{\tau} f_0(t)dt$, ce qui est exactement l'équation de Chambers $f = \int_{-\infty}^{0} f_0 \frac{\partial P}{\partial t} dt$, dans l'hypothèse ou la fonction P qui régit la probabilité de choc par unité de temps est donnée par $P(t) = e^{\frac{t}{\tau}}$ (t<0), ce que nous montrerons dans le paragraphe suivant.

## 3.5.2. Modèle de Chambers

L'approche développée par Esaki et Tsu, en vue d'établir leur loi de vitesse, dérive de lois plus générales, cette approche porte le nom de « Path integration method »[126].

Soit P(t), la probabilité de ne pas subir de choc par unité de temps et soit τ le temps entre deux diffusions. Il est alors possible d'établir l'équation différentielle qui régit P(t). La probabilité pour que à t+dt l'électron n'ait pas encore subi de diffusion est $P(t + dt) = P(t)[1 - \frac{dt}{\tau}]$. Compte tenu du fait que l'on suppose qu'à t=0, la particule n'a pas encore subi de choc, soit $P(0) = 1$, nous obtenons finalement $P(t) = e^{-\frac{t}{\tau}}$ (t>0). Il s'agit alors d'établir la fonction de distribution f des électrons qui prend en compte la diffusion et connaissant la fonction de distribution à l'équilibre $f_0$.

f est alors donnée par $f(k_0)=\int_{-\infty}^{0} f_0 \frac{\partial P}{\partial t}dt$. Par ailleurs l'évolution du vecteur d'onde k est donnée par les équations du mouvement $\hbar\dot{k}=eF$, et cela dans l'hypothèse ou seul un champ électrique est appliqué. Il est immédiat d'intégrer cette équation $k=\frac{eFt}{\hbar}+k_0$. L'expression de f peut alors s'intégrer par partie pour donner $f=f_0(k_0)-\int_{-\infty}^{0}\frac{\partial f(k(t))}{\partial t}P(t)dt$. L'expression de la vitesse moyenne est alors donnée par :

$$\langle v\rangle=\int_{-\pi/L_d}^{\pi/L_d} dk_0 v(k_0) f(k_0)=\int_{-\pi/L_d}^{\pi/L_d} dk_0 v(k_0) f_0(k_0)-\int_{-\pi/L_d}^{\pi/L_d} dk_0 v(k_0)\int_{-\infty}^{0}\frac{\partial f_0(k(t))}{\partial t}P(t)dt$$

Le premier terme est simplement la vitesse moyenne à l'équilibre et est donc nul. Le second peut se calculer et compte tenu de l'expression obtenue pour P donne $\langle v\rangle=\int_{-\pi/L_d}^{\pi/L_d} dk_0 f_0(k)\int_{0}^{\infty}\frac{\partial v(k(t))}{\partial t}e^{-\frac{t}{\tau}}dt$, voir la référence 125 pour plus de détail.

### 3.5.3. Loi d'Esaki-Tsu

Il est alors question de particulariser la loi de vitesse dans le cas d'un super-réseau.

#### *3.5.3.1. Relation de dispersion*

Pour un super-réseau dans lequel on se limite au couplage premier voisin, la relation de dispersion est donnée par $E=\frac{\Delta}{2}(1-\cos(k_z d))$, ou d est la période du super réseau et $\Delta$ la taille de mini bande. Cette relation est obtenue par une approche de type liaison forte. Dans cette hypothèse, la vitesse d'un état $k_z$ est donnée par $v(k_z)=\frac{1}{\hbar}\frac{\partial E}{\partial k_z}$. Il est également possible de définir la masse effective en bas de bande $m(0)=\frac{\hbar^2}{\left.\frac{\partial^2 E}{\partial {k_z}^2}\right|_{k_z=0}}=\frac{2\hbar^2}{\Delta d^2}$

#### *3.5.3.2. Hypothèse fondamentale*

Afin de calculer la vitesse moyenne il suffit donc d'injecter dans l'expression obtenue par Chambers celle de la vitesse tout en tenant compte de la relation de dispersion sinusoïdale. Il faut néanmoins pour cela faire une hypothèse sur la fonction de distribution des électrons à l'équilibre. L'hypothèse retenue par Esaki et Tsu est très forte et consiste à dire que $f_0=\delta(k)$. Cela a pour conséquence immédiate de supposer que le super-réseau n'est que faiblement peuplé/dopé. Ils supposent de plus que le système est à température nulle.

#### 3.5.3.3. Loi de vitesse

Compte tenu de cette hypothèse l'expression de Chambers se simplifie grandement $\langle v\rangle = \int \frac{\partial v(k(t))}{\partial t} e^{-\frac{t}{\tau}} dt$. Sachant que $v(k_z) = \frac{1}{\hbar}\frac{\partial E}{\partial k_z}$ et que $\hbar\dot{k} = eF$, il vient immédiatement que $\frac{\partial v}{\partial t} = \dot{k}\frac{\partial v}{\partial k_z} = \dot{k}\frac{\partial \frac{1}{\hbar}\frac{\partial E}{\partial k_z}}{\partial k_z} = \frac{eF}{\hbar^2}\frac{\partial^2 E}{\partial k_z{}^2} = \frac{eF}{\hbar^2}\frac{\Delta d^2}{2}\cos(k_z d)$, d'où finalement $\langle v\rangle = \frac{eFd^2\Delta}{2\hbar^2}\int \cos(\frac{eFd}{\hbar}t)\exp(-\frac{t}{\tau})dt$. Dans cette dernière expression il est possible de reconnaître la transformé de Laplace de cosinus et donc $\langle v\rangle = \frac{eFd^2\Delta}{2\hbar^2}\frac{\frac{1}{\tau}}{\left(\frac{1}{\tau}\right)^2 + \left(\frac{eFd}{\hbar}\right)^2}$. Expression qui se réécrit

$$\langle v\rangle = \frac{\hbar}{dm(0)}\frac{\frac{eFd\hbar}{\tau}}{\left(\frac{\hbar}{\tau}\right)^2 + (eFd)^2} \quad (3\text{-}35)$$

Cette loi de vitesse[9] conduit au phénomène de Résistance Différentielle Négative (RDN). Il est possible de donner l'interprétation suivante à ce phénomène : aux champs faibles les électrons sont accélérés par le champ au sein de la minibande. Mais si le champ devient trop important, les réflexions de Bloch aux interfaces du super-réseau ne permettent plus aux électrons d'atteindre leur pleine vitesse et alors leur vitesse moyenne diminue jusqu'à changer de signe.

#### 3.5.3.4. Influence des paramètres

Afin d'évaluer l'importance des différents paramètres intervenant dans la structure, nous introduisons le champ $F_c = \frac{\hbar}{e\tau d}$. on définit également $v_m = \frac{\hbar}{dm(0)}$ et il est alors possible de réécrire la loi de vitesse réduite[127]

$$\langle v\rangle_{ET} = v_m \frac{F/F_c}{1+\left(\frac{F}{F_c}\right)^2} \quad (3\text{-}36)$$

Cette courbe passe par un maximum en $F_c$, la valeur du maximum est $\frac{v_m}{2}$. $\tau$ peut donc jouer sur la position (en champ) du maximum, mais pas sur sa valeur. Plus le temps de diffusion est court, plus la RDN est faible en module. La diffusion retarde l'apparition de la RDN et en limite la valeur, voir FIG. 3-9.

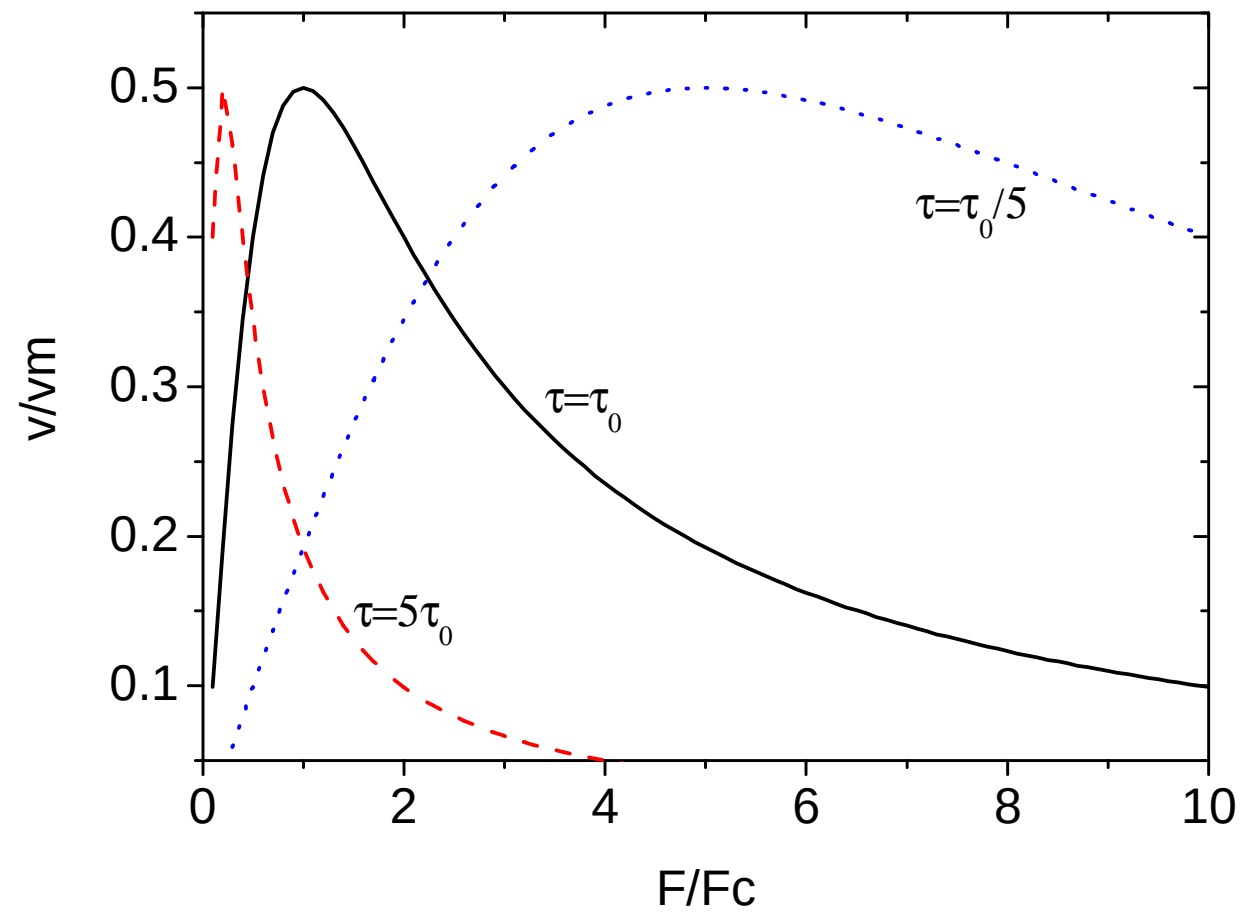

*FIG. 3-9 Loi de vitesse d'Esaki-Tsu, pour trois valeurs du temps de diffusion.*

#### 3.5.3.5. Possibilité d'amélioration

Deux améliorations immédiates peuvent être faites à ce premier modèle :

- *Prise en compte de la température*

Si plutôt que de prendre une fonction de distribution de type delta, nous choisissons une distribution de type Maxwell-Boltzmann[125], la loi de vitesse obtenue est légèrement modifiée par rapport à la loi précédente[125,127]

$$\langle v \rangle = \langle v \rangle_{ET} \frac{I_1\left(\frac{\Delta}{k_b T}\right)}{I_0\left(\frac{\Delta}{k_b T}\right)} \quad (3\text{-}37)$$

Avec $I_n$ la fonction de Bessel de seconde espèce et d'ordre n.

- *Modification de la relation de dispersion*

Esaki et Tsu ont dans leur article fondateur[9] envisagé la possibilité d'une relation de dispersion non sinusoïdale. Ils envisagent une relation de dispersion composée de deux paraboles juxtaposées. La loi de vitesse qui s'en suit est plus complexe du point de vue analytique, mais la forme de la courbe n'est que faiblement affectée, voir FIG. 3-10.

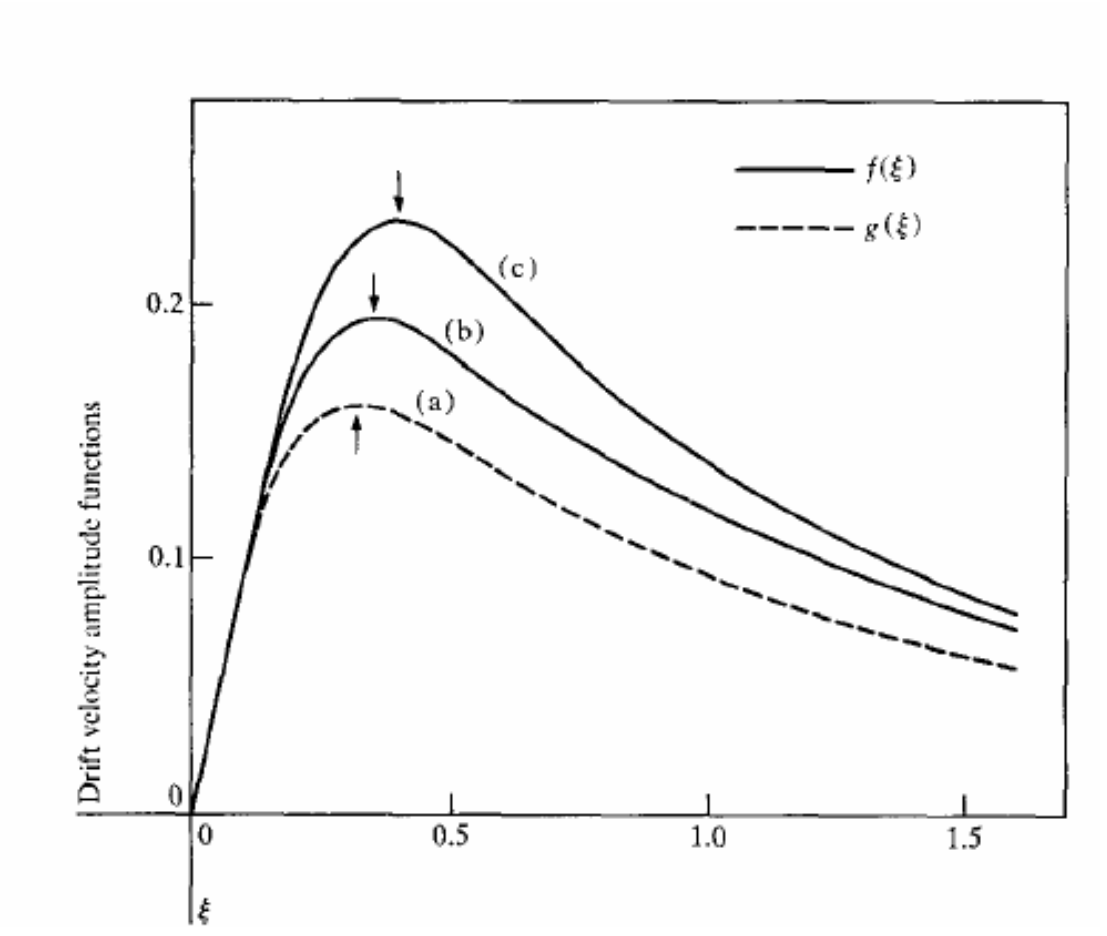

*FIG. 3-10 En trait pointillé la loi de vitesse originale, en traits pleins la loi de vitesse pour une relation de dispersion faite de paraboles, d'après la référence 9.*

# ANNEXE 3-B : L'approximation WKB[128]

Pour établir l'expression WKB, considérons un électron d'énergie $E$ qui doit passer par effet tunnel à travers une barrière de potentiel $U(x) = -eV(x)$. L'équation de Schrödinger régissant le mouvement de l'électron s'écrit alors

$$\left[-\frac{\hbar^2}{m^*}\frac{d^2}{dx^2} + U(x)\right]\psi(x) = E\psi(x) \quad (3\text{-}38)$$

La quantité $K(x) = \sqrt{\frac{2m^*}{\hbar^2}(U(x) - E)}$ est définie, ce qui permet de réécrire l'équation (3-38) :

$$\frac{d^2}{dx^2}\psi(x) - K^2(x)\psi(x) = 0 \quad (3\text{-}39)$$

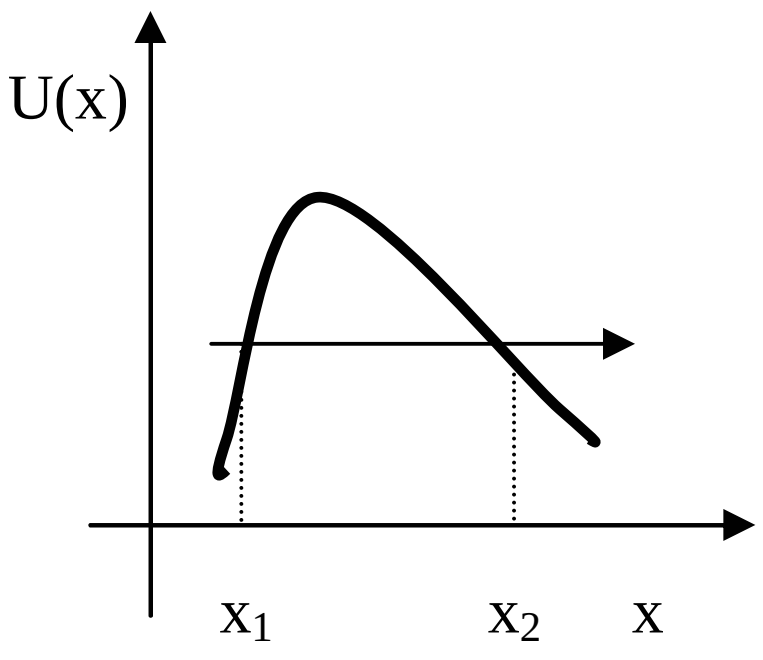

*FIG. 3-11 Profil de potentiel à travers lequel l'électron passe par effet tunnel*

Nous cherchons alors la solution à cette équation sous la forme $\psi(x) = e^{\alpha(x)}$, où $\alpha(x)$ est négatif de manière à ce que $\psi(x)$ soit décroissante. Il s'agit alors d'obtenir à partir de (3-39) une équation dans laquelle l'inconnue sera $\alpha(x)$ :

$$\frac{d^2}{dx^2}\alpha(x) + [\frac{d}{dx}\alpha(x)]^2 - K^2(x) = 0 \quad (3\text{-}40)$$

On néglige dans l'équation (3-40) le premier terme devant le second $\frac{d^2\alpha(x)}{dx^2} << [\frac{d}{dx}\alpha(x)]^2$, l'hypothèse simplificatrice porte donc sur $\alpha(x)$ qui doit être assez douce, notons que la fonction d'onde elle-même n'est pas nécessairement douce en raison de l'exponentielle. Il est intéressant de noter que contrairement à une idée très répandue l'hypothèse porte sur $\alpha(x)$ et pas sur *K(x)* (donc pas sur le profil). L'équation (3-40) est alors considérablement simplifiée : $[\frac{d}{dx}\alpha(x)]^2 - K^2(x) = 0$. On en déduit immédiatement une expression pour $\alpha(x) = -\int_{x_1}^{x} K(x)dx$ et pour $\psi(x) = \exp(-\int_{x_1}^{x} K(x)dx)$. Le cœfficient de transmission de la barrière s'obtient simplement en évaluant $T = \psi^*(x_2)\psi(x_2) = \exp(-2\int_{x_1}^{x_2} K(x)dx)$, soit encore :

$$T = \exp[-2.\sqrt{\frac{2m}{\hbar^2}}.\int_{x_1}^{x_2}\sqrt{U(x) - E}.dx] \quad (3\text{-}41).$$

## ANNEXE 3-C : Calcul lié à l'approche de matrice densité

Cette annexe s'inspire largement des calculs des références 9 et 104. Et vient détailler les calculs qui sont fait dans la partie 3.2.4.1.

Il nous faut connaître l'expression des termes de la forme $\rho_k^{nn+1}$ qui interviennent dans l'expression du courant (3-23), pour cela écrivons leur évolution temporelle

$$i\hbar \frac{d}{dt}\left[\rho_k^{nn+1}\right] = \left[H,\rho\right]_k^{nn+1} \qquad (3\text{-}42)$$

qui peut se réécrire

$$i\hbar \frac{d}{dt}\left[\rho_k^{nn+1}\right] = (\varepsilon_k^n - \varepsilon_k^{n+1})\rho_k^{nn+1} + \Delta(\rho_k^{n+1n+1} - \rho_k^{nn}) + \sum_{k'} V_{kk'}^{nn}\rho_{k'k}^{nn+1} - \rho_{kk'}^{nn+1}V_{k'k}^{n+1n+1} \qquad (3\text{-}43)$$

Expression dont on prend la transformée de Laplace et dont on cherche les solutions stationnaires (s→0).

$$(\varepsilon_k^n - \varepsilon_k^{n+1}) f_k^{nn+1} = \Delta( f_k^{nn} - f_k^{n+1n+1}) - \sum_{k'} V_{kk'}^{nn} f_{k'k}^{nn+1} - f_{kk'}^{nn+1}V_{k'k}^{n+1n+1} \qquad (3\text{-}44)$$

Dans cette expression interviennent des termes de la forme $f_{k'k}^{nn+1}$ dont il convient de connaître l'expression, à leur tour leur évolution est donnée par

$$i\hbar \frac{d}{dt}\left[\rho_{kk'}^{nn+1}\right] = \left[H,\rho\right]_{kk'}^{nn+1} \qquad (3\text{-}45)$$

Définissons $\rho_k$ la partie diagonale en impulsion de $\rho$ et nous négligeons le commutateur $\left[V;(\rho-\rho_k)\right]$

$$i\hbar \frac{d}{dt}\left[\rho_{kk'}^{nn+1}\right] \approx (\varepsilon_k^n - \varepsilon_{k'}^{n+1})\rho_{kk'}^{nn+1} + \Delta(\rho_{kk'}^{n+1n+1} - \rho_{kk'}^{nn}) + V_{kk'}^{nn}\rho_{k'}^{nn+1} - \rho_k^{nn+1}V_{kk'}^{n+1n+1} \qquad (3\text{-}46)$$

Expression dont nous prenons à nouveau la transformée de Laplace avant d'en chercher les solutions stationnaires.

$$-(\varepsilon_k^n - \varepsilon_{k'}^{n+1}) f_{kk'}^{nn+1} = \Delta( f_{kk'}^{n+1n+1} - f_{kk'}^{nn}) + V_{kk'}^{nn} f_{k'}^{nn+1} - f_k^{nn+1}V_{kk'}^{n+1n+1} \qquad (3\text{-}47)$$

En remarquant que si $xT = A(x)$ avec T la distribution inconnue alors $T = A(x)\left[i\pi\delta(x) + P(1/x)\right]$, dans laquelle $\delta$ et $P$ sont respectivement la distribution delta et la partie principale. Nous pouvons écrire :

$$f_{kk'}^{nn+1} = -\left[ i\pi\delta(\varepsilon_k^n - \varepsilon_{k'}^{n+1}) + P\left( \frac{1}{(\varepsilon_k^n - \varepsilon_{k'}^{n+1})} \right) \right] \left( \Delta( f_{kk'}^{n+1n+1} - f_{kk'}^{nn}) + V_{kk'}^{nn} f_{k'}^{nn+1} - f_k^{nn+1} V_{kk'}^{n+1n+1} \right)$$

(3-48)

A présent il nous faut connaître l'expression des termes de la forme $f_{kk'}^{nn}$, en repartant de l'expression (3-45) appliquée au cas ou n=n+1

$$-(\varepsilon_k^n - \varepsilon_{k'}^n) f_{kk'}^{nn} = \Delta( f_{kk'}^{n+1n} - f_{kk'}^{nn+1}) + V_{kk'}^{nn} ( f_{k'}^{nn} - f_k^{nn}) \qquad (3\text{-}49)$$

Comme par ailleurs nous cherchons à évaluer non pas $f_{kk'}^{nn}$ seul, mais $\Delta( f_{kk'}^{n+1n+1} - f_{kk'}^{nn})$, nous négligerons dans l'expression de $f_{kk'}^{nn}$ les termes d'ordre un ou plus en $\Delta$, ce qui physiquement interdit les allers-retours multiples dans une barrière tunnel, ce qui semble raisonnable dans la mesure ou le couplage est faible. L'expression (3-49) se calcule comme précédemment :

$$f_{kk'}^{nn} \approx \underbrace{- i\pi\delta(\varepsilon_k^n - \varepsilon_{k'}^n) V_{kk'}^{nn} ( f_{k'}^{nn} - f_k^{nn})}_{=0} - P\left( \frac{1}{(\varepsilon_k^n - \varepsilon_{k'}^n)} \right) V_{kk'}^{nn} ( f_{k'}^{nn} - f_k^{nn}) \qquad (3\text{-}50)$$

d'où

$$f_{kk'}^{nn} \approx \frac{V_{kk'}^{nn} ( f_{k'}^{nn} - f_k^{nn})}{(\varepsilon_k^n - \varepsilon_{k'}^n)} \qquad (3\text{-}51)$$

Reste à réinjecter cette expression dans (3-48). Si de plus nous négligeons la partie principale

$$f_{kk'}^{nn+1} = -i\pi\delta(\varepsilon_k^n - \varepsilon_{k'}^{n+1}) \left( \Delta( \frac{V_{kk'}^{nn} ( f_k^{n+1n+1} - f_{k'}^{n+1n+1})}{(\varepsilon_k^{n+1} - \varepsilon_{k'}^{n+1})} - \frac{V_{kk'}^{nn} ( f_k^{nn} - f_{k'}^{nn})}{(\varepsilon_k^n - \varepsilon_{k'}^n)} ) + V_{kk'}^{nn} f_{k'}^{nn+1} - f_k^{nn+1} V_{kk'}^{n+1n+1} \right)$$

(3-52)

Cette dernière expression est à présent réinjectée dans (3-44). Nous faisons l'hypothèse de négliger toute corrélation entre éléments de diffusion, c'est-à-dire qu'un terme de la forme $V^{ij}$ ne peut se coupler qu'à un terme de la forme $V^{ij} / V^{ji}$ mais pas à des termes de la forme $V^{ii}, V^{jj}$...Dans cette expression les termes du type $f_{k'k}^{nn+1}$ interviennent sou la forme $V_{kk'}^{nn} f_{k'k}^{nn+1}$ que l'on peut calculer

$$V_{kk'}^{nn} f_{k'k}^{nn+1} = i\pi\delta(\varepsilon_{k'}^{n} - \varepsilon_{k}^{n+1})\left|V_{kk'}^{nn}\right|^2 \left( \frac{\Delta(f_{k'}^{nn} - f_{k}^{nn})}{\varepsilon_{k'}^{n} - \varepsilon_{k}^{n}} - f_{k}^{nn+1} \right) \qquad (3\text{-}53)$$

$$f_{kk'}^{nn+1} V_{k'k}^{n+1n+1} = i\pi\delta(\varepsilon_{k}^{n} - \varepsilon_{k'}^{n+1})\left|V_{kk'}^{n+1n+1}\right|^2 \left( f_{k}^{nn+1} - \frac{\Delta(f_{k}^{n+1n+1} - f_{k'}^{n+1n+1})}{\varepsilon_{k}^{n+1} - \varepsilon_{k'}^{n+1}} \right) \qquad (3\text{-}54)$$

Ce qui une fois remis dans (3-44) conduit à :

$$(\varepsilon_{k}^{n} - \varepsilon_{k}^{n+1}) f_{k}^{nn+1} - i\pi f_{k}^{nn+1} \sum_{k'} \delta(\varepsilon_{k'}^{n} - \varepsilon_{k}^{n+1})\left|V_{kk'}^{nn}\right|^2 + \delta(\varepsilon_{k}^{n} - \varepsilon_{k'}^{n+1})\left|V_{kk'}^{n+1n+1}\right|^2 = \Delta(f_{k}^{nn} - f_{k}^{n+1n+1})$$
$$- i\pi \sum_{k'} \delta(\varepsilon_{k'}^{n} - \varepsilon_{k}^{n+1})\left|V_{kk'}^{nn}\right|^2 \frac{\Delta(f_{k'}^{nn} - f_{k}^{nn})}{\varepsilon_{k'}^{n} - \varepsilon_{k}^{n}} - i\pi \sum_{k'} \delta(\varepsilon_{k}^{n} - \varepsilon_{k'}^{n+1})\left|V_{kk'}^{n+1n+1}\right|^2 \frac{\Delta(f_{k}^{n+1n+1} - f_{k'}^{n+1n+1})}{\varepsilon_{k}^{n+1} - \varepsilon_{k'}^{n+1}}$$

(3-55)

Cette expression à l'allure barbare se simplifie en définissant des taux de diffusion $\gamma_{k}^{n} = \pi \sum_{k'} \delta(\varepsilon_{k'}^{n} - \varepsilon_{k}^{n+1})\left|V_{kk'}^{nn}\right|^2$, la différence d'énergie associée à la transition $\Delta E = \varepsilon_{k}^{n} - \varepsilon_{k}^{n+1}$ et $q_n$ l'impulsion de l'état final. Ce qui permet de réécrire (3-55) :

$$\left(\Delta E - i(\gamma_{k}^{n} + \gamma_{k}^{n+1})\right) f_{k}^{nn+1} = \Delta(f_{k}^{nn} - f_{k}^{n+1n+1}) - i\frac{\Delta}{\Delta E}\left((\gamma_{k}^{n+1}(f_{q_{n+1}}^{n+1n+1} - f_{k}^{n+1n+1}) - \gamma_{k}^{n}(f_{q_n}^{nn} - f_{k}^{nn})\right)$$

(3-56)

# 4. MODELE DE TRANSPORT PAR SAUTS ENTRE ETATS LOCALISES[v]

## Sommaire

**Publication associée à ce chapitre :**

- E. Lhuillier, I. Ribet-Mohamed, A. Nedelcu, V. Berger, E. Rosencher, Quantum transport in weakly coupled superlattices at low temperature, Phys. Rev. B 81, 155305 (2010).

[v] Dans ce chapitre je vais utiliser indifféremment le terme « transport entre états localisés » et « hopping ».

Dans le chapitre précédent j'ai montré les limites des approches simplifiées basées sur l'hypothèse WKB (transport tunnel séquentiel) et l'approche de minibande. En particulier ces approches n'ont pas permis de donner une interprétation quantique du phénomène de plateau observé sur les courbes *I(V)* du composant étudié au chapitre deux. Par ailleurs le traitement de la dépendance du courant avec le dopage reste insatisfaisant. L'approche de hopping offre une bonne alternative à ces deux précédentes approches de plus, elle permet en théorie de traiter du cas des structures faiblement couplées que sont les QWIP. La validité de l'utilisation de cette approche semble donc en accord avec le diagramme (voir FIG. 3-8) présenté au chapitre précédent. Il faut toutefois s'attendre à de possibles difficultés à bas champ électrique.

L'approche par sauts entre états localisés (hopping) est couramment utilisée dans la modélisation du transport dans les lasers à cascade quantique[96], mais son utilisation dans les QWIP reste marginale, seul le groupe de Harrison à Leeds semble s'y être intéressé. Ils se sont toutefois heurtés à des difficultés liées à des effets de localisation de fonctions d'onde auxquels ils n'ont pas apporté de réponses.

Cette approche de hopping présente un autre avantage : elle permet de dissocier aisément les différents mécanismes à l'origine du transport. Cette dissociation facilite grandement la possibilité d'identifier des pistes d'optimisation des détecteurs comparativement aux approches qui utilisent des paramètres effectifs et qui ainsi dissimulent une partie de la physique du problème. L'un des objectifs de ce chapitre est donc de mettre en place une modélisation qui repart de principes fondamentaux de la physique et qui va, sans introduire de paramètres, nous permettre d'évaluer des grandeurs macroscopiques.

Néanmoins, comme nous pouvions nous y attendre, notre modélisation s'est heurtée à des difficultés pour reproduire le régime ohmique de transport présent aux basses tensions. Nous analyserons donc dans ce chapitre l'origine de ce problème et je présenterai une modification de l'approche de hopping qui permet d'étendre son domaine de validité aux faibles tensions.

Ce chapitre va s'organiser de la manière suivante, je commencerai par présenter l'outil de simulation de transport que j'ai développé. Je présenterai la physique qui y ait incluse et les différentes hypothèses que nous avons pu faire. Ensuite j'engagerai une discussion sur la façon dont il convient de modéliser le régime ohmique de transport dans une approche de hopping. Cela nous conduira à une extension du modèle classique du courant. Je finirai ce chapitre en appliquant notre outil de simulation au QWIP étudié au chapitre deux. J'apporterai alors une explication quantique au phénomène de plateau des *I(V)* ainsi que des pistes qui doivent nous permettre de diminuer l'amplitude du courant d'obscurité en régime tunnel.

## 4.1. Modèle

Par transport par hopping il faut ici entendre transport tunnel à travers une barrière de potentiel entre deux états localisés. Il est vrai que certains puristes[129] se limitent à un couplage inélastique dans le cas du transport par hopping. Mais j'ai pris la liberté, couramment utilisée, de généraliser ce terme à n'importe quelle type d'interaction, qu'elle soit élastique ou non.

En pratique, à chaque état est associé une fonction d'onde et une énergie, il est alors possible via la règle d'or de Fermi de calculer un temps de passage entre ces deux états, voir l'équation (4-1) et le schéma FIG. 4-1.

$$\Gamma(K_i) = \tau_{tunnel}^{-1}(E,F) = \frac{2\pi}{\hbar}\sum_{Kf}\left|\langle f|H|i\rangle\right|^2 \delta(\varepsilon_i - \varepsilon_f) \qquad (4\text{-}1)$$

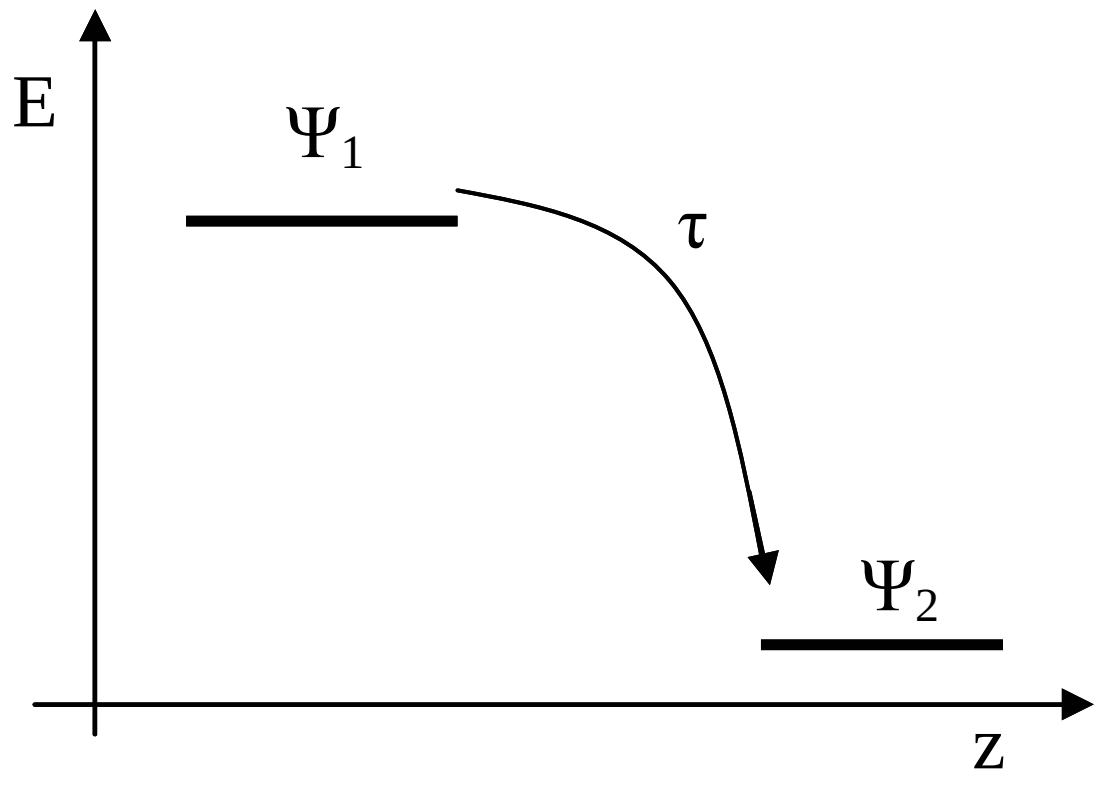

*FIG. 4-1 Schéma du transport par hopping.*

A partir de ce temps de diffusion il va ensuite nous être possible de remonter au courant dans la structure.

### 4.1.1. Evaluation des états localisés

#### *4.1.1.1. Notation*

Avant tout calcul donnons une convention sur la dénomination des états $|i\rangle = |K_i, k_{iz}\rangle \propto \xi_{kiz}.e^{i\vec{K}\vec{R}}$, $K_i$ est la composante dans le plan de la fonction d'onde et $k_{iz}$ celle selon la direction de croissance z. $|i\rangle$ signalera l'état initial et $|f\rangle$ l'état final. Les états de phonons seront notés quant à eux $|q\rangle = |Q, q_z\rangle$. Les vecteurs dans l'espace direct seront notés $\vec{r} = (r'', z)$. Enfin l'énergie totale sera décrite de la façon suivante $\varepsilon_i = E_i + \frac{\hbar^2 K_i^{\ 2}}{2m^*}$ avec $E_i$ l'énergie du niveau considéré. Au final la quantité $E_f - E_i$ vaut $e.F.L_d$ dans le cas de deux puits consécutifs avec $L_d$ la période du QWIP.

#### *4.1.1.2. Equation de Schrödinger*

L'utilisation de la règle d'or de Fermi de manière à évaluer $\tau_{tunnel}^{-1}(E,F)$ nécessitera la connaissance des fonctions d'onde dans nos structures. Pour cela nous nous plaçons dans le formalisme de la fonction enveloppe $\psi(r'',z) = \xi(z)e^{i\vec{K}\vec{r}''}$, dans lequel on considère le mouvement dans le plan assimilable à celui d'un électron libre. L'équation de Schrödinger indépendante du temps qui permet d'évaluer $\xi(z)$ est donnée par l'hamiltonien de Ben-Daniel-Duke[130] :

$$\left[-\frac{\hbar^2}{2}\frac{d}{dz}\frac{1}{m^*(z)}\frac{d}{dz}+V(z)-eFz\right]\xi(z)=E\xi(z) \qquad (4\text{-}2)$$

Afin de résoudre cette équation deux codes ont été utilisés :

- Un code k·p à deux bandes, très pratique pour évaluer les états dans une hétérostructure complexe à grand nombre de puits.
- Un code maison basé sur la « shooting method »[131], utile pour traiter le cas des profils de bandes non idéaux. Ce code a également l'avantage de permettre une résolution autoconsistante du système Schrödinger-Poisson.

La FIG. 4-2 montre le profil de bande et les états propres obtenus dans la structure QWIP étudiée au chapitre deux.

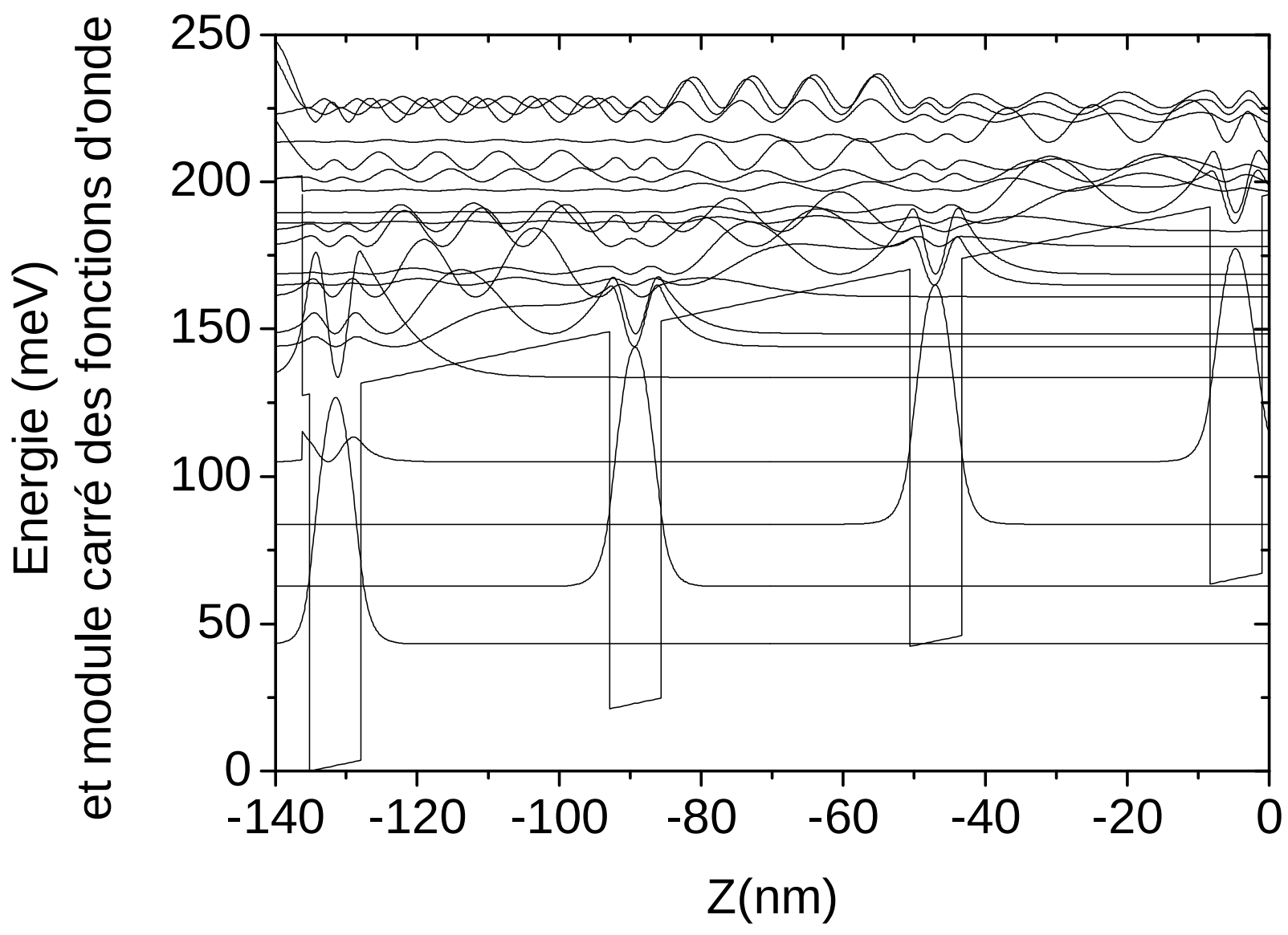

*FIG. 4-2 : Fonctions d'onde et potentiel d'un QWIP.*

Il est intéressant de souligner l'une des difficultés de modélisation propre au QWIP à savoir la difficulté numérique d'obtenir des fonctions d'onde correctes dans des structures pour lesquelles les barrières sont très larges.

### 4.1.2. Traitement du transport dans un système à deux puits

#### *4.1.2.1. Equation de Schrödinger*

Dans la mesure où les QWIP sont des super réseaux extrêmement faiblement couplés, il semble raisonnable de dire que le transport se fait entre états plus proches voisins. Il n'est donc pas nécessaire pour modéliser le transport de décrire l'ensemble de la structure composée de quelques dizaines de puits. L'ensemble de ce chapitre s'intéressera donc à la modélisation du courant dans une structure à deux puits quantiques couplés[132].

Néanmoins cette hypothèse peut être lourde de conséquences puisqu'elle néglige toute éventuelle inhomogénéité de champ[29,133] ou effet de contact[134]. Nos simulations vont donc nous conduire à des courbes *J(F)* que nous ne pourrons comparer aux courbes expérimentales J(V) qu'en supposant le champ uniforme. Toutefois pour ne pas laisser ce point en suspend le dernier chapitre de cette thèse s'intéressera aux effets non locaux du transport.

Dans le cas des QWIP la modélisation va donc être la suivante, nous avons une structure à deux puits couplés, dont on ne considère que l'état fondamental. Cela signifie donc que je limite volontairement la validité de ma modélisation, aux tensions pas trop fortes, pour lesquelles la chute de potentiel par période ne permet pas la mise en résonance d'un état fondamental avec le niveau excité de la période suivante. Nous considérerons le courant qui va vers l'aval (*J+*) et celui qui remonte vers l'amont (*J-*) et qui joue principalement aux tensions faibles, voir FIG. 4-3.

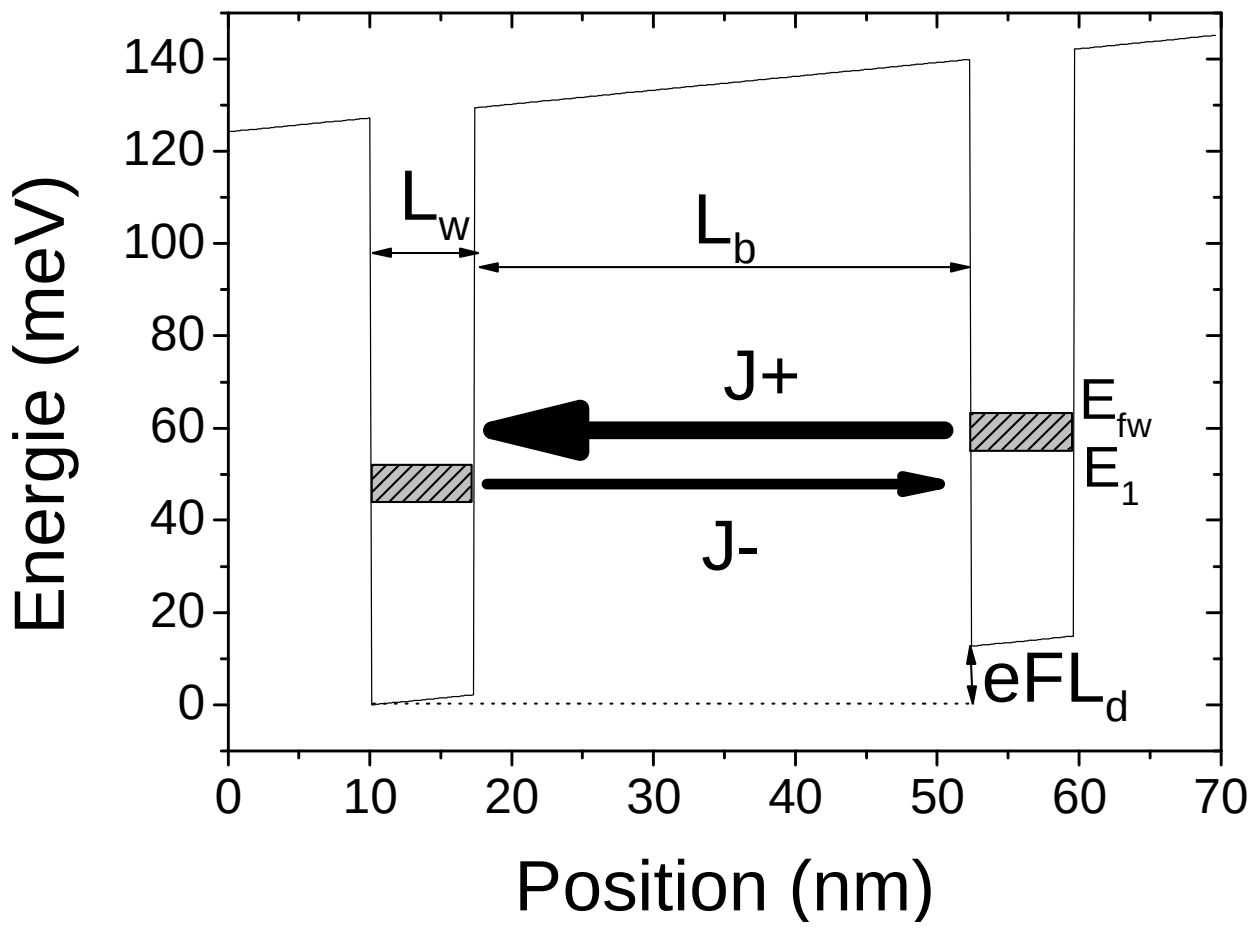

*FIG. 4-3 Schéma du transport dans une structure à deux puits couplés*

#### 4.1.2.2. Analogie mécanique

Il est intéressant de faire une analogie mécanique entre nos systèmes de puits quantiques couplés et un ensemble d'oscillateurs. Il est en effet possible de modéliser le mouvement d'un électron au sein du super-réseau par celui d'un pendule qui serait couplé au pendule de la période suivante.

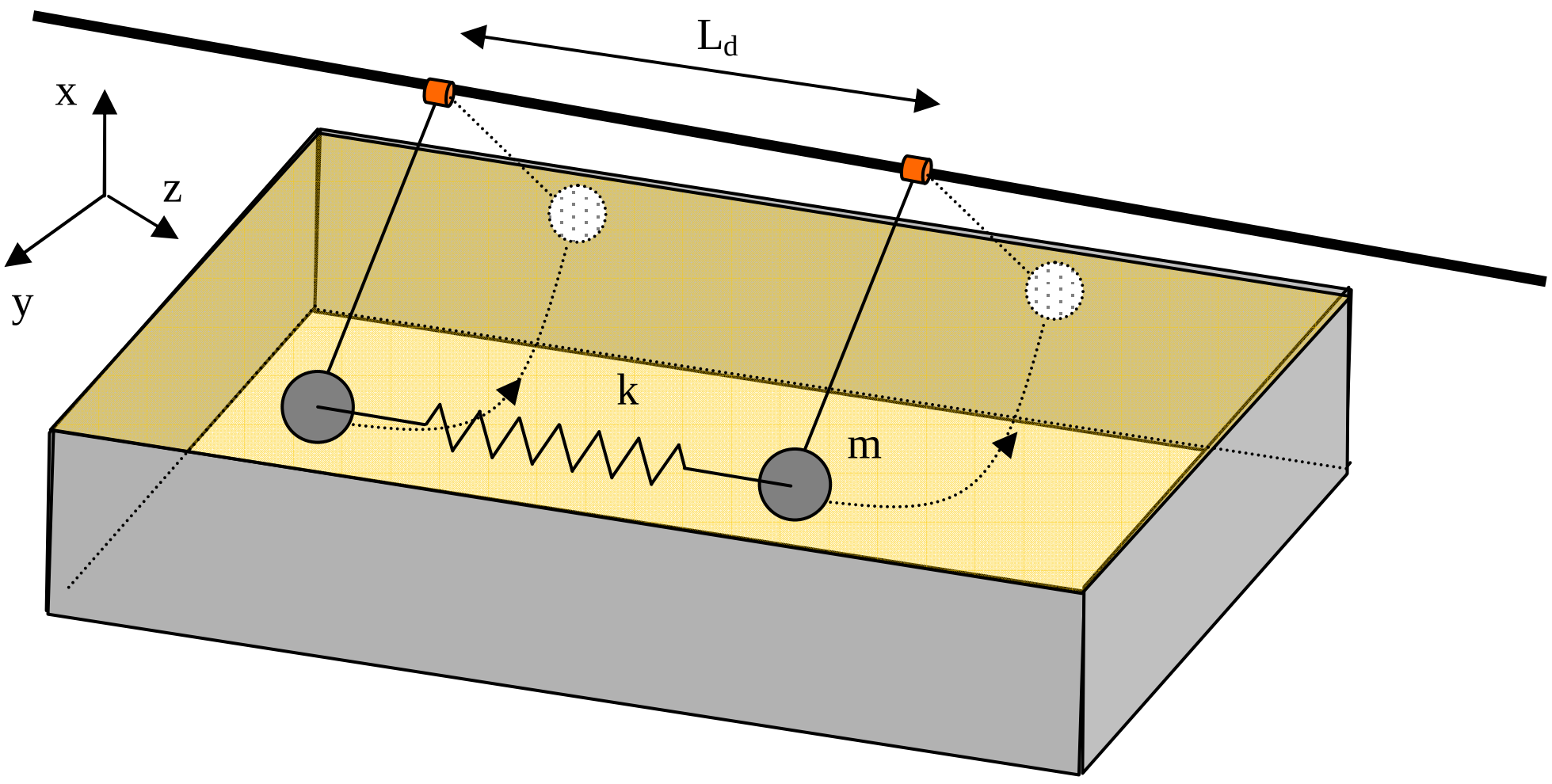

*FIG. 4-4 : Schéma d'un couple de pendules oscillant dans un bain d'huile et couplé dans la direction z par un ressort de faible raideur.*

Chaque pendule oscille dans le plan Oxy et est couplé au pendule suivant par un ressort de faible raideur. La diffusion intra sous-bande peut être modélisée par un terme dissipatif : le pendule oscille alors dans un bain d'huile. Ce système peut facilement se mettre en équation :

$$m\frac{d^2}{dt^2}\vec{r}_n = J\ddot{\theta}\cdot\vec{e}_\theta - f\dot{\theta}\cdot\vec{e}_\theta + k\left[z_{n+1} + z_{n-1} - 2\cdot z_n\right]\vec{e}_z \qquad (4\text{-}3)$$

Avec $J$ le moment d'inertie du pendule, $\theta$ l'angle formé par le pendule et l'axe Ox, f un terme de frottement visqueux. Le tableau ci-dessous donne l'équivalence entre les grandeurs du système mécanique et celle du super-réseau.

| Système de pendule | Superréseaux |
|---|---|
| $J\ddot{\theta}$ | Déplacement dans le plan |
| $f\dot{\theta}$ | $\frac{n_{2D}}{\tau_{\text{intra}}}$ Diffusion intrabande |
| $k$ | $\Delta$ Couplage inter puits |
| mgz | eFz |

Il est même possible d'imaginer un équivalent à l'application de la tension. Il suffit pour cela d'incliner le système de pendule. La gravité joue alors le rôle du potentiel électrostatique.

Dans un tel système d'oscillateur, l'un des comportements usuels auquel nous pourrions nous attendre est la synchronisation des oscillateurs. Toutefois le couplage entre période est extrêmement faible. Le déplacement hors du plan des pendules (équivalent du passage d'un électron d'une période à l'autre) est donc très vite atténué. Dans les QWIP les passages des electrons d'un puits à l'autre ne sont donc pas corrélés.

### 4.1.3. Interactions prises en compte

Dans notre modélisation le temps de diffusion entre les deux états localisés est calculé en utilisant la règle d'or de Fermi dont l'expression est donnée par :

$$\Gamma(K_i) = \frac{2\pi}{\hbar} \sum_{Kf} \left| \langle f | H | i \rangle \right|^2 \delta(\varepsilon_i - \varepsilon_f)$$

Avec $|i\rangle$ et $|f\rangle$ les états initiaux et finaux respectivement et $H$ l'hamiltonien de couplage. La règle d'or de Fermi provient de la théorie des perturbations au premier ordre, ce qui signifie que les fonctions d'onde sont évaluées dans la structure en l'absence de perturbation.

Dans toute cette thèse le matériau utilisé pour les épitaxies est l'alliage GaAs/AlGaAs obtenu par MBE, la qualité du matériau est donc supposée être excellente (faible concentration en dislocation, faible niveau d'impuretés résiduels). Cela signifie que notre modélisation va pouvoir se contenter d'interactions très fondamentales pour le couplage entre états. En particulier nous négligerons la présence d'éventuelles dislocations.

Notre modélisation prend en compte la diffusion des électrons par les phonons LO et LA, le désordre d'alliage, la rugosité d'interface, et les diffusions coulombiennes avec les impuretés et entre électrons. Le schéma FIG. 4-5 rappelle les différents processus considérés par notre modélisation. L'intérêt de prendre en compte un spectre assez large d'interactions réside dans le fait que nous n'avons pas à faire à priori d'hypothèse quand à l'amplitude de chacune d'elle. Le modèle pourra donc être utilisé dans une large gamme d'hétérostructures[135]. Les commentaires que je vais faire dans la suite de cette présentation quant à l'amplitude relative de ces interactions sont destinés à la structure QWIP 15µm. elles ne sont donnés que pour se faire un avis à priori sur leur amplitude mais le modèle les prend toutes en compte qu'elle que soit leur amplitude.

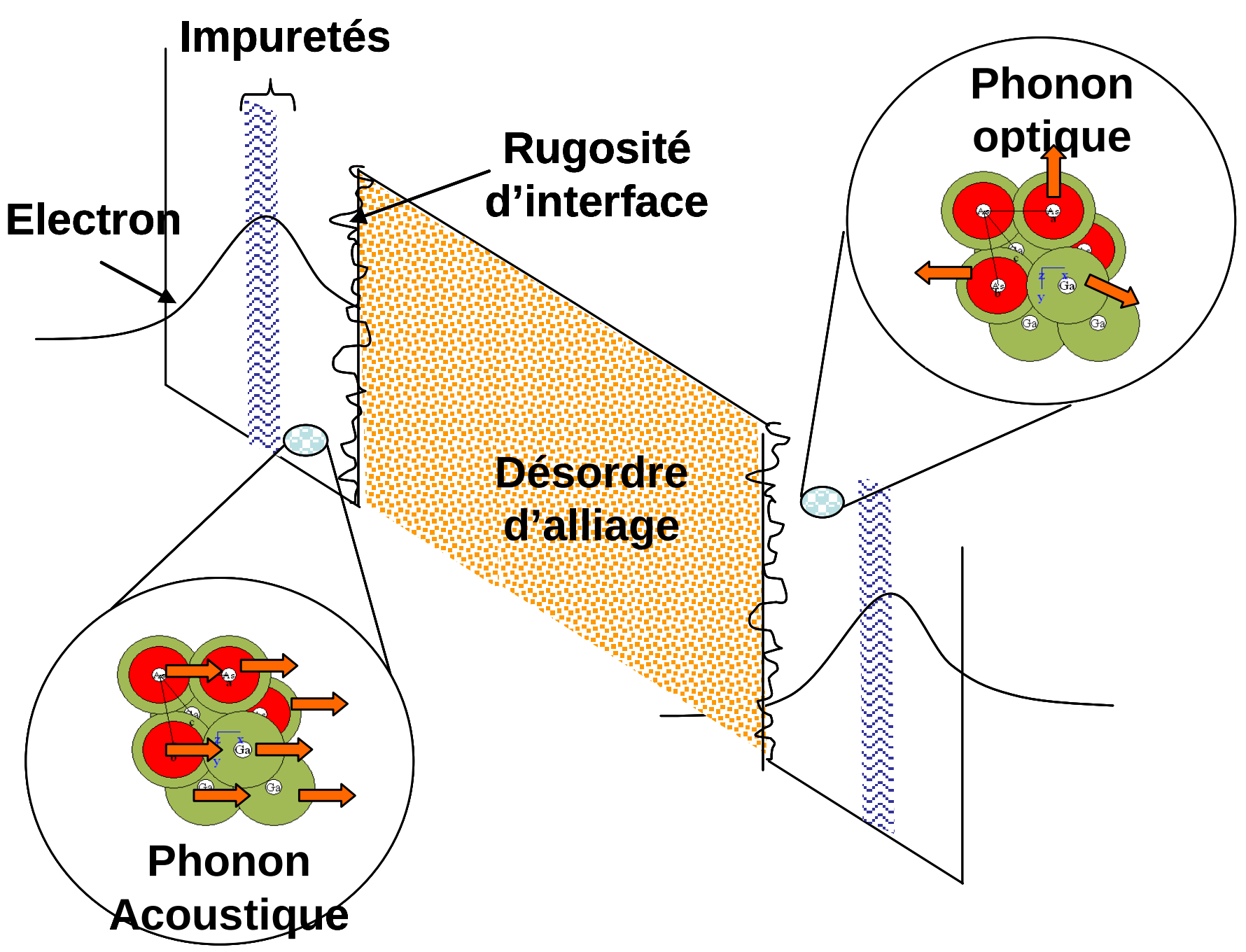

*FIG. 4-5 : Schéma des interactions prises en compte par notre modélisation*

Passons à présent plus en détail les six interactions considérées. Afin de ne pas alourdir cette partie je renverrai le lecteur intéressé par le détail des calculs à la référence[136]. De plus je vais faire ici un certain nombre de commentaires sur l'amplitude attendue de chacune des interactions. Ceux-ci ne sont destinés qu'au QWIP 15µm et ne s'appliquent *a priori* pas à d'autres structures (tel que le QCD THz dans le chapitre suivant).

#### *4.1.3.1. Phonons LO*

Deux éléments jouent en la défaveur de ce processus. Premièrement la chute de potentiel aux bornes d'une barrière est comprise, dans la gamme de tension du plateau, entre 12 et 35meV, là ou le phonon optique de GaAs a une énergie de 36meV. Deuxièmement la température très basse nuit fortement à la présence de tel phonon, il faut s'attendre à une absorption des phonons négligeable.

L'hamiltonien de Frölich[83] qui régit cette interaction s'écrit :

$$H_{e^- / phonon} = \sum_q \sqrt{\frac{e^2 \hbar w_{LO}}{2\Omega \varepsilon_p} \frac{1}{q^2 + q_0{}^2}} e^{i\vec{q}\vec{r}} b_q^+ + hc \qquad (4\text{-}4)$$

Où $\hbar w_{LO}$ est l'énergie des phonons optiques dans GaAs supposé non dispersif[137], $\Omega$ le volume de l'échantillon, $\frac{1}{\varepsilon_p} = \frac{1}{\varepsilon_\infty} - \frac{1}{\varepsilon_s}$ et $\varepsilon_\infty$ et $\varepsilon_s$ sont respectivement les constantes diélectriques à fréquence infinie et à fréquence nulle, $q_0$ est l'inverse de la longueur de Debye et $b_q^+$ l'opérateur création de phonon. Enfin *hc* désigne la quantité conjuguée. Dans cette écriture de l'hamiltonien, j'ai supposé que les phonons étaient ceux du matériau massif et qu'ils n'étaient pas affectés par le passage d'un matériau à l'autre. D'autres auteurs ont inclus les effets de phonons sur les interfaces, voir les références[138,139]. Pour plus de commodité je définis la quantité $\alpha_{LO}{}^2 = \frac{e^2\hbar w_{LO}}{2\Omega\varepsilon_p}$, le taux de diffusion en émission est alors donné par[131,132] :

$$\Gamma_{LO}(K_i) = \frac{\alpha_{LO}{}^2 m^*}{(2\pi)^2\hbar^3}\int d\theta \int dq_z \frac{F_{LO}(q_z)}{q^2+q_0{}^2}(1+n_{BE}(\hbar w_{LO})) \qquad (4\text{-}5)$$

Ou $q^2 = Q^2 + q_z{}^2$, $Q^2 = K_i{}^2 + K_f{}^2 - 2K_iK_f\cos(\theta)$ $K_f{}^2 = K_i{}^2 + \frac{2m^*}{\hbar^2}(E_i - E_f - \hbar w_{LO})$ et le facteur de forme de l'interaction qui relie l'amplitude du taux de diffusion aux paramètres géométriques de la structure s'écrit :

$$F_{LO}(q_z) = \left|\left\langle \xi_f \left| e^{iq_z z} \right| \xi_i \right\rangle\right|^2 \qquad (4\text{-}6)$$

Dans le cas de l'absorption de phonon le facteur $(1+n_{BE}(\hbar w_{LO}))$ est simplement remplacé par $n_{BE}(\hbar w_{LO})$. La loi de conservation en énergie s'en trouve également modifiée.

#### 4.1.3.2. Phonons LA

Dans le cas des phonons acoustiques l'hamiltonien de couplage aux électrons prend la même forme que celui imputable au phonon optique et seule l'expression du préfacteur change :

$$H_{e^-/phonon} = \sum_q \sqrt{\frac{\hbar D_c{}^2}{2\Omega\rho.c_s}}\, q e^{i\vec{q}\vec{r}} b_q{}^+ + hc \qquad (4\text{-}7)$$

Dans une telle expression les phonons acoustiques ont une relation de dispersion linéaire qui suit le modèle de Debye[137]. $D_c$ est le potentiel de déformation acoustique, $\Omega$ le volume de l'échantillon, $\rho$ la densité de GaAs, $c_s$ la vitesse du son dans le matériau, $q$ le vecteur d'onde du phonon et $b_q{}^+$ l'opérateur création de phonon. Définissons alors $\alpha_{AC}{}^2 = \frac{\hbar D_c{}^2}{2\rho.c_s}$, qui permet une écriture du taux de diffusion sous la forme[131] :

$$\Gamma_{AC} = \frac{\alpha_{AC}{}^2}{4\pi^2\hbar^2 c_s}\int_0^\infty K_f dK_f \int_0^{2\pi} d\theta \frac{q^2}{q_z}(1+n_{BE}(\hbar w)).F_{AC}(q_z) \qquad (4\text{-}8)$$

Avec $q = \dfrac{E_i - E_f + \dfrac{\hbar^2}{2m^*}(K_i{}^2 - K_f{}^2)}{\hbar c_s}$, $Q^2 = K_i{}^2 + K_f{}^2 - 2K_i K_f \cos(\theta)$,

$q_z = \sqrt{q^2 - Q^2}$

Le facteur de forme de l'interaction qui relie l'amplitude du taux de diffusion aux paramètres géométriques de la structure s'écrit :

$$F_{AC} = \left|\left\langle \xi_f \left| e^{iq_z z} \right| \xi_i \right\rangle\right|^2 \quad (4\text{-}9)$$

Dans le cas de l'absorption de phonon le facteur $(1+n_{BE}(\hbar w))$ est simplement remplacé par $n_{BE}(\hbar w)$.

#### *4.1.3.3. Désordre d'alliage (AL)*

Le désordre d'alliage résulte de la répartition aléatoire de l'aluminium dans la matrice de GaAs au sein de l'alliage $Al_xGa_{1-x}As$. L'électron trouve lors de son passage dans l'alliage un potentiel qui est localement désordonné. Ces fluctuations locales d'énergie peuvent à leur tour engendrer de la diffusion.

Dans notre structure en GaAs/AlGaAs il ne faut pas s'attendre à un rôle important de ce processus diffusif dans la mesure où les fonctions d'onde sont surtout localisées dans les puits qui sont constitués par un alliage binaire. Ce ne serait pas forcement la même chose pour une structure en GaInAs/AlInAs.

Le potentiel lié à à la présence d'un atome d'aluminium s'écrit $V = \Delta V.\delta x(r)$ où $\Delta V$ est l'écart en énergie entre GaAs et AlAs et $\delta x(r)$ est une fonction delta centrée sur un atome d'aluminium. Généralement la procédure consiste ensuite à définir une fonction de corrélation entre les positions des atomes d'aluminium qui suit la loi de Nordheim[137] : $\langle \delta x(r)\delta x(r')\rangle = \Omega_0[x(1-x)]\delta(r-r')$ avec x la concentration relative en aluminium contenu dans l'alliage. Il vient ensuite que le taux de diffusion est donné par[140] :

$$\Gamma_{AL} = \frac{m^*}{\hbar^3}\Delta V^2 \Omega_0 [x(1-x)] F_{AL} \qquad (4\text{-}10)$$

Avec le facteur de forme de l'interaction qui relie l'amplitude du taux de diffusion aux paramètres géométriques de la structure:

$$F_{AL} = \int_{alliage} \left|\xi_f(z)\right|^2 \left|\xi_i(z)\right|^2 dz \qquad (4\text{-}11)$$

#### *4.1.3.4. Rugosité d'interface (IR)*

L'expression du temps de diffusion imputable à la rugosité d'interface fait en particulier intervenir le recouvrement des fonctions d'onde au niveau de l'interface. Or contrairement au laser à cascade quantique où certaines barrières sont très fines de manière à favoriser le processus tunnel, l'épaisseur de la barrière dans nos QWIP (plusieurs centaines d'Angstrom) nuit fortement au recouvrement de la queue des fonctions d'onde. En conséquence il est fort probable que ce processus soit lui aussi négligeable.

La procédure pour le traitement de la rugosité d'interface est assez proche de celle utilisée pour le désordre d'alliage : c'est son équivalent surfacique. Ainsi le potentiel associé à la présence d'un défaut d'interface s'écrit $V_{IR}(r) = V_b \Delta \delta(z - z_i) F(r)$ avec $z_i$ la position des interfaces, $\Delta$ l'amplitude des défauts et F(R) de nouveau une fonction de corrélation entre la position des défauts. Cette fonction de corrélation est souvent choisie gaussienne[140,141] $\langle F(r).F(r')\rangle = \exp(-\frac{|r-r'|^2}{\xi^2})$. $\xi$ s'interprète alors comme une longueur de corrélation c'est à dire comme la distance moyenne entre deux défauts. Finalement l'expression du taux de diffusion est

$$\Gamma_{IR} = \frac{m^* V_b{}^2 \Delta^2 \xi^2 F_{IR}}{2\hbar^3} \int_0^{2\pi} e^{-\frac{Q^2\xi^2}{4}} d\theta \qquad (4\text{-}12)$$

Avec le facteur de forme de l'interaction qui relie l'amplitude du taux de diffusion aux paramètres géométriques de la structure:

$$F_{IR} = \left|\xi_i(z_i)\right|^2 \left|\xi_f(z_i)\right|^2 \qquad (4\text{-}13)$$

#### *4.1.3.5. Impuretés ionisées (II)*

Dans les hétérostructures l'origine des impuretés est double. Il y a bien sûr les impuretés non désirées qui peuvent provenir des sources utilisées lors de l'épitaxie[142]. Ce point sera discuté plus avant au chapitre suivant. Mais l'origine majeure des impuretés est le dopage. L'expression du potentiel écranté est donné par :

$$V(r) = \frac{e^2}{4\pi\varepsilon_0\varepsilon_r} \frac{e^{-q_0 r}}{r} \qquad (4\text{-}14)$$

Qui conduit à un taux de diffusion sous la forme[132,143,144] :

$$\Gamma_{II} = \frac{e^4}{8\pi\hbar\varepsilon_0{}^2\varepsilon_r{}^2}\frac{m^*}{\hbar^2}\int_{impuretés} dz_{ii} N(z_{ii}) \times \int d\theta \frac{F_{II}\left(\sqrt{\left|K_i - K_f\right|^2 + q_0{}^2}\right)}{\left|K_i - K_f\right|^2 + q_0{}^2} \qquad (4\text{-}15)$$

Avec $z_{ii}$ la position des impuretés, $\left|K_i - K_f\right| = \sqrt{K_i{}^2 + K_f{}^2 - 2K_iK_f\cos(\theta)}$, $K_f{}^2 = K_i{}^2 + \frac{2m^*}{\hbar^2}(E_i - E_f)$, $q_0$ le vecteur d'onde de Thomas Fermi donné par $q_0{}^2 = \frac{e^2 n}{\varepsilon_r k_b T}$. Le facteur de forme de l'interaction qui relie l'amplitude du taux de diffusion aux paramètres géométriques de la structure s'écrit :

$$F_{II}(Q) = \left|\int dz \xi_f{}^*(z) e^{-Q|z-zi|}\xi_i(z)\right|^2 \qquad (4\text{-}16)$$

#### *4.1.3.6. Electron-Electron (ee)*

Ce processus va être fortement fonction du niveau de dopage du QWIP. Il présente la difficulté d'être complexe à modéliser tant d'un point de vue théorique que numérique. Un grand nombre d'auteurs traitent cette interaction via le formalisme de la fonction de Green[145,146]. Nous n'utiliserons pas pour notre part ce point de vue, préférant celui développé par Smet et al[147] puis par Harrison et al[131].

Dans le cas de la diffusion entre électrons, le potentiel coulombien est le même que pour les impuretés ionisées, à savoir $V(r) = \frac{e^2}{4\pi\varepsilon_0\varepsilon_r}\frac{e^{-q_0 r}}{r}$. La principale différence vient du fait que deux états initiaux et finaux interviennent dans l'expression du taux de diffusion. En pratique cela se traduit surtout par une forte inflation du temps de calcul nécessaire. L'expression du taux de diffusion est donnée par

$$\Gamma^{ee} = \frac{m^* e^4}{(4\pi\hbar)^3\varepsilon_0{}^2\varepsilon_r{}^2}\int\int_0^{2\pi}\int_0^{2\pi}\left|\frac{F^{ee}{}_{ijfg}(q_{xy})}{q_{xy}}\right|^2 \times P_{j,f,g}(k_j,k_f,k_g)k_j dk_j d\alpha d\theta \qquad (4\text{-}17)$$

Avec $q_{xy}{}^2 = \frac{2k_{ij}{}^2 + \Delta k_o{}^2 - 2k_{ij}\sqrt{k_{ij}{}^2 + \Delta k_o{}^2}\cos\theta}{4}$, $k_{ij}{}^2 = k_i{}^2 + k_j{}^2 - 2k_ik_j\cos\alpha$

et $\Delta k_o{}^2 = \frac{4m^*}{\hbar^2}(E_i + E_j - E_f - E_g)$.

Le facteur de forme de l'interaction qui relie l'amplitude du taux de diffusion aux paramètres géométriques de la structure s'écrit :

$$F^{ee}{}_{ijfg}(q_{xy}) = \iint \xi_f^*(z)\xi_g^*(z')\xi_i(z)\xi_j(z') \times e^{-q_{xy}|z-z'|} dzdz' \qquad (4\text{-}18)$$

### 4.1.4. Outil de simulation

#### *4.1.4.1. Hypothèses*

Avant de calculer les taux de diffusion il reste encore un certain nombre d'hypothèses à faire.

Tout d'abord je n'ai considéré que les électrons dans la vallée Γ, délaissant ceux des vallées L et X, voir schéma de la structure de bande de GaAs (FIG. 4-6). Cette hypothèse semble raisonnable dans la mesure où l'écart en énergie de ces vallées est d'au moins 0.3eV, alors que nous nous intéressons justement à des QWIP hautes longueurs d'onde donc basses énergies. La même hypothèse est faite dans la barrière, malgré la présence d'aluminium qui rapproche encore un peu plus les vallées Γ et X, cela se justifie par la faible teneur en aluminium. Cela pourrait par contre être une limitation de cet outil si nous étions amenés à modéliser des structures en bande II.

Par ailleurs les expressions des taux de diffusion et du courant vont faire intervenir la population associée à chacun des états. Nous avons fait l'hypothèse que les populations de chaque état étaient à l'équilibre. Ce qui se traduit par l'utilisation de populations données par les distributions de Fermi Dirac, pour lesquelles la température électronique est prise égale à celle du réseau. Cette hypothèse de thermalisation des électrons sera justifiée *a posteriori* quand nous verrons que le taux de diffusion inter-puits est près de dix décades plus faible que celui de diffusion intrabande[148].

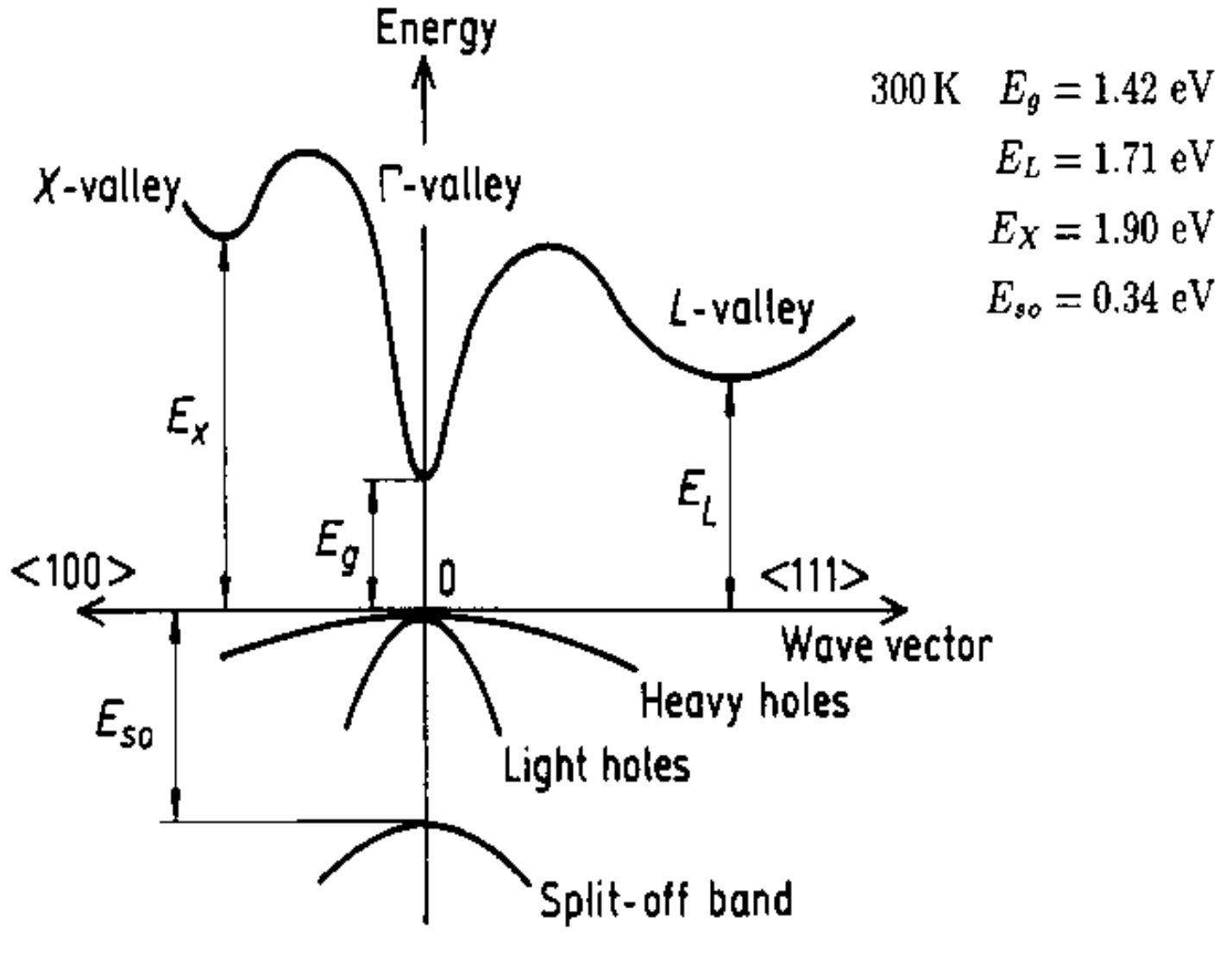

*FIG. 4-6 : Structure de bande de GaAs, d'après 149.*

Par ailleurs le nombre de porteurs est pris égal au dopage, ce qui revient à supposer que tous les dopants sont ionisés. Cette hypothèse semble raisonnable puisque la transition de Mott dans GaAs massif est située autour de $10^{16}$cm$^{-3}$. Il a certes été rapporté que les effets de confinement pouvaient augmenter cette valeur[150] mais de toute façon les concentrations en porteurs, supérieures à $10^{18}$cm$^{-3}$ dans nos structures, font que tous les porteurs peuvent être considérés comme ionisés.

#### 4.1.4.2. Paramètres utiles à la modélisation

Notre modélisation utilise des paramètres (non ajustables) qui sont rapportés dans le tab. 4-1.

| Grandeur | Unité | Valeur | Référence |
|---|---|---|---|
| $m_w^*=(0.067+0.083x)m_0$ | kg | $6.1\ 10^{-32}$ (x=0%) | 132 |
| $\varepsilon_r \approx \varepsilon_s$ | | $12.9\,\varepsilon_0 = 1.14\ 10^{-10}$ | 83 |
| $\varepsilon_\infty$ | | $10.9\,\varepsilon_0 = 9.65\ 10^{-11}$ | 83 |
| $\hbar w_{LO}$ | meV | 36.6 | 83 |
| $\rho$ | $Kg.m^{-3}$ | 5320 | 151 |
| $c_s$ | $ms^{-1}$ | 5220 | 151 |
| $D_c$ | eV | 12 | 151 |
| $\Delta$ | nm | 0.3 | 140, 141, 152, 153 |
| $\xi$ | nm | 6.5 | 140, 141, 152, 153 |
| $\Delta V=V_{AlAs}-V_{GaAs}$ | eV | 0.836 | 140 |
| a | nm | 0.565 | 83 |
| $Vb=\Delta V.x$ | eV | 0.128 (x=15%) | 29 |

*tab. 4-1 Paramètres liés aux interactions et utilisés dans la modélisation.*

#### 4.1.4.3. Le code de simulation

- *Le code*

Le code de simulation du transport a beaucoup évolué au cours de cette thèse, quatre évolutions majeures ont accompagné le développement du programme. La première version a été développée en scilab (simple mais peu rapide). Elle contient alors le calcul des taux de diffusion pour cinq interactions (LO en émission, AC en émission, II, IR, et AL). La version 2 est écrite en C++, ce qui permet un gain de vitesse non négligeable et est nécessaire au calcul de la sixième interaction (ee). S'ajoute à cela la prise en compte du courant qui remonte la structure (J-) et dont le rôle est important à basse tension. La version 3 correspond à une optimisation de la version 2 à laquelle s'ajoute l'absorption des phonons acoustiques et l'écrantage de l'interaction LO. L'effet Stark est également ajouté. La version 4 inclut de nouvelles améliorations en terme de code ainsi que l'inclusion de l'absorption des phonons LO. La non parabolicité des bandes est également ajoutée. La version 5 finale est un programme tout intégré qui va du calcul des fonctions d'onde au courant final. En pratique ce code calcule essentiellement des intégrales multiples, ce qui peut s'avérer fort coûteux numériquement.

- *Validation de l'outil de simulation*

L'outil de simulation a été validé sur des structures déjà publiées. Pour l'ensemble des interactions à l'exception de celles entre électrons, j'ai utilisé la référence[97]. Pour valider le taux de diffusion entre électrons j'ai utilisé l'article de Smet et al[147]. L'accord obtenu avec ces deux papiers est très bon, les écarts mineurs qui peuvent persister sont dus à des paramètres qui ne sont pas toujours égaux.

### 4.1.5. Résultats : taux de diffusion

#### *4.1.5.1. Ordre de grandeur*

Avant de montrer les résultats de simulation il est important d'avoir une idée de l'ordre de grandeur du taux de diffusion inter puits auquel il faut s'attendre. En effet un calcul rapide va nous montrer que ce taux de diffusion est très éloigné du taux de diffusion intrabande, qui est, lui, dans la gamme térahertz. Pour estimer le taux de diffusion inter puits il suffit de considérer une expression simple de la densité de courant dans laquelle le courant peut être vu comme le produit de la densité surfacique de charge, multipliée par le taux de diffusion inter puits. Ce qui conduit à :

$$\tau = \frac{e.n}{J} \approx \frac{1{,}6.10^{-19}\ 3.10^{11}}{10^{-6}} = \text{quelques dizaines de millisecondes.} \qquad (4\text{-}19)$$

Cela signifie donc qu'entre deux puits couplés il y a typiquement un électron qui passe toutes les 10ms. Nous cherchons donc à modéliser des événements extrêmement peu probables.

#### *4.1.5.2. Résultats de simulation*

La simulation des taux de diffusion associés à chacune des interactions est présentée sur la FIG. 4-7. Comme nous pouvions nous y attendre le taux de diffusion dans la gamme de tension du plateau de courant est bien situé dans la gamme entre 10 et 100Hz. Je discuterais plus avant de l'importance relative des différentes interactions à la fin de ce chapitre.

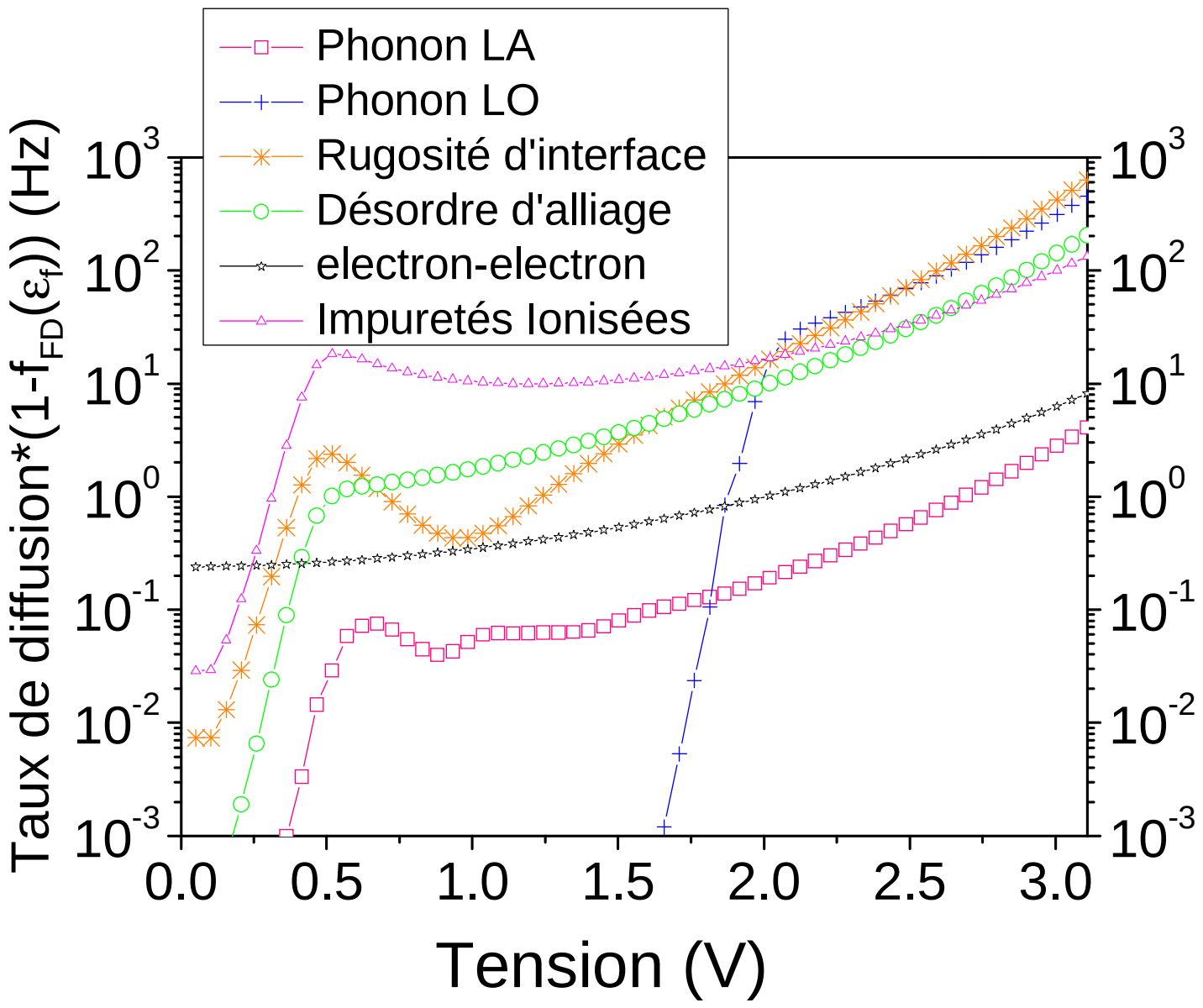

*FIG. 4-7 : Produit du taux de diffusion par la population du niveau d'arrivée, pour $K_i$=0, en fonction de la tension appliquée sur la structure. Calcul effectué pour la structure du chapitre deux. Dans l'hypothèse d'un champ homogène 1V sur la structure correspond à 5.8kV·cm$^{-1}$. Calcul fait à 10K.*

## 4.2. Du taux de diffusion au courant

Notre outil de simulation nous permet à présent d'obtenir les temps de diffusion associés aux différents processus et les résultats sont bien de l'ordre de grandeur espéré. Il serait donc à présent légitime de vouloir non plus se contenter du taux de diffusion en bas de bande ($K_i$=0) mais au contraire de chercher à évaluer le courant qui traverse la structure. Cela devrait en particulier nous permettre de prendre en compte les effets liés à la population de la structure en électrons. Dans l'approche WKB, nous avions construit une expression du courant sous la forme :

$$J = J^+ - J^-$$

$$= \int_{E_1}^{\infty} e\frac{m^*}{\pi\hbar^2}.\tau^{-1}(E,F).(1-f_{FD}(\varepsilon_f)).f_{FD}(\varepsilon_i)d\varepsilon_i - \int_{E_1-eFL_d}^{\infty} e\frac{m^*}{\pi\hbar^2}.\tau^{-1}(E,F).(1-f_{FD}(\varepsilon_f)).f_{FD}(\varepsilon_i)d\varepsilon_i$$

(4-20)

Expression dans laquelle $\tau^{-1}$ est le temps qui sépare deux tentatives pour l'électron de passer à travers la barrière. Il était donc légitime de vouloir utiliser une telle expression pour notre modèle de hopping mais en interprétant cette fois-ci l'expression de $\tau^{-1}$ comme d'un taux de diffusion entre états fondamentaux, somme des différentes interactions considérées plus haut. Le résultat est donné sur la FIG. 4-8.

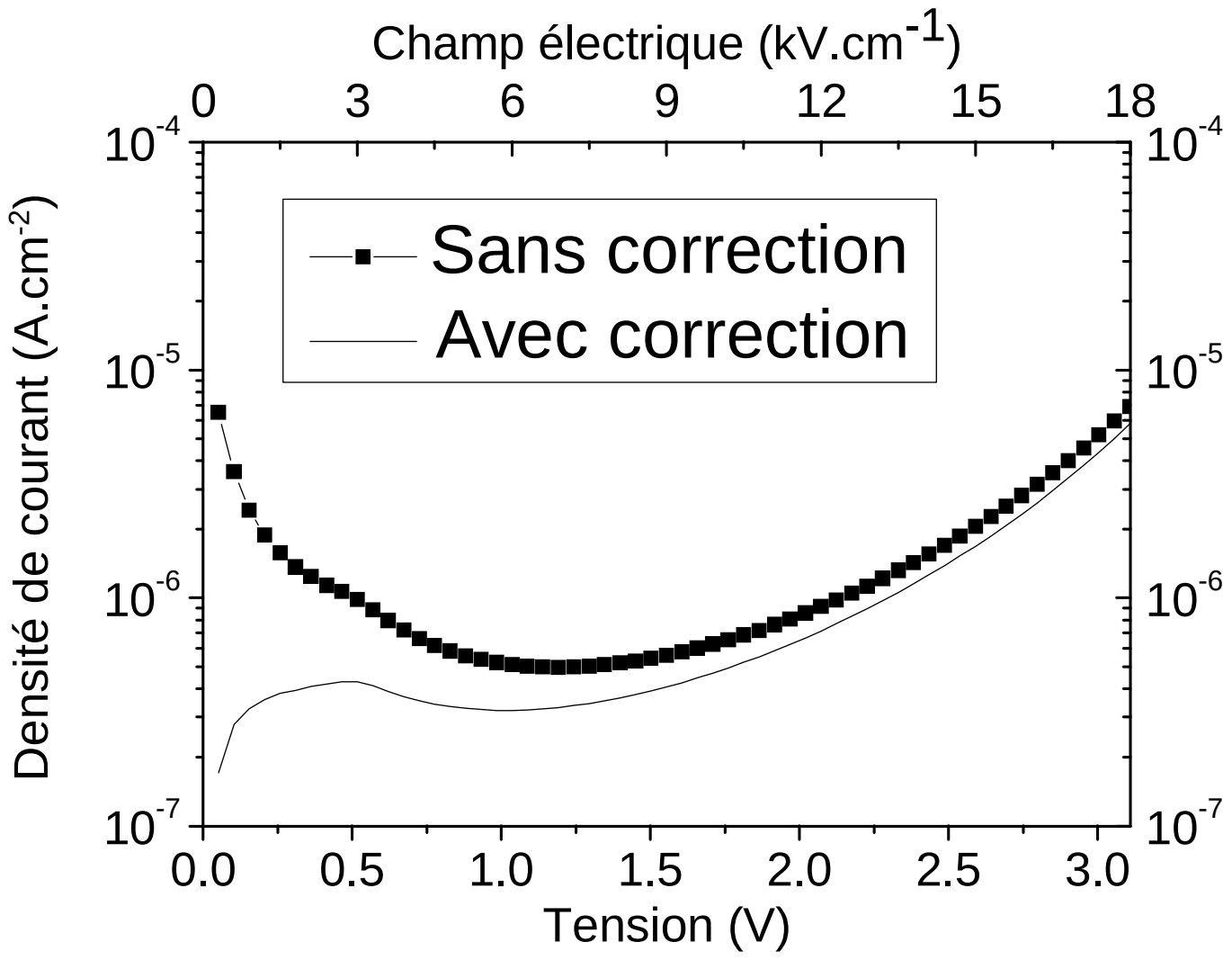

*FIG. 4-8 : Densité de courant total, somme des six processus, en fonction de la tension appliquée sur la structure.*

L'ordre de grandeur est bien le bon, la simulation conduit à un courant de l'ordre de $10^{-6}$A·cm$^{-2}$. Mais la courbe J(V) présente une différence importante avec la courbe expérimentale à savoir le fait que le courant semble diverger à tension nulle. Notons bien que si cette divergence n'apparaît pas sur le taux de diffusion, voir FIG. 4-7, c'est en raison du facteur de population du niveau d'arrivée qui en bas de bande évite cette divergence, mais ce n'est plus le cas autour du niveau de Fermi.

### 4.2.1. Présence d'une divergence à basse tension

Notre modèle présente alors une différence avec l'expérience dont nous souhaitons comprendre l'origine. Voici les différentes hypothèses que nous avons envisagées :

Notre premier réflexe a été de vouloir attribuer cette divergence à une erreur numérique ou à un problème de calcul des taux de diffusion. En particulier les premières versions du code n'incluaient que le courant J+ et pas celui J- qui remonte la structure. Or justement J- joue un rôle important aux basses tensions. Mais la divergence a persisté.

Cet effet pourrait également résulter d'une divergence du taux de diffusion. Nous avons en particulier soupçonné l'interaction coulombienne de pouvoir être à l'origine de cet effet. En effet cette interaction est supposée diverger en *1/q,* avec *q* l'impulsion échangée. Aux basses tensions l'échange d'énergie selon l'axe z (pour de l'impulsion dans le plan) devrait donc être soumis à cette divergence. La solution a consisté à prendre en compte l'écrantage de l'interaction via l'introduction d'un potentiel de Yukawa. Mais cette divergence persistait.

Le problème semble bien être lié à la physique de notre approche et plus précisément à la mise en résonance de niveaux équivalents. Lors de la mise en résonance de deux niveaux équivalents, leurs fonctions d'onde semblent trop délocalisées. Ainsi dans une structure asymétrique à deux puits, cette divergence ne se produit plus à champ nul, mais à champ fini, voir FIG. 4-9. Bien sûr à la résonance on pourrait s'attendre à une délocalisation totale de l'électron, mais ce point de vue est remis en cause si les niveaux ont un temps de vie fini. Dans ce cas l'électron est plus localisé que ne le laisse croire le calcul de Wannier-Stark.

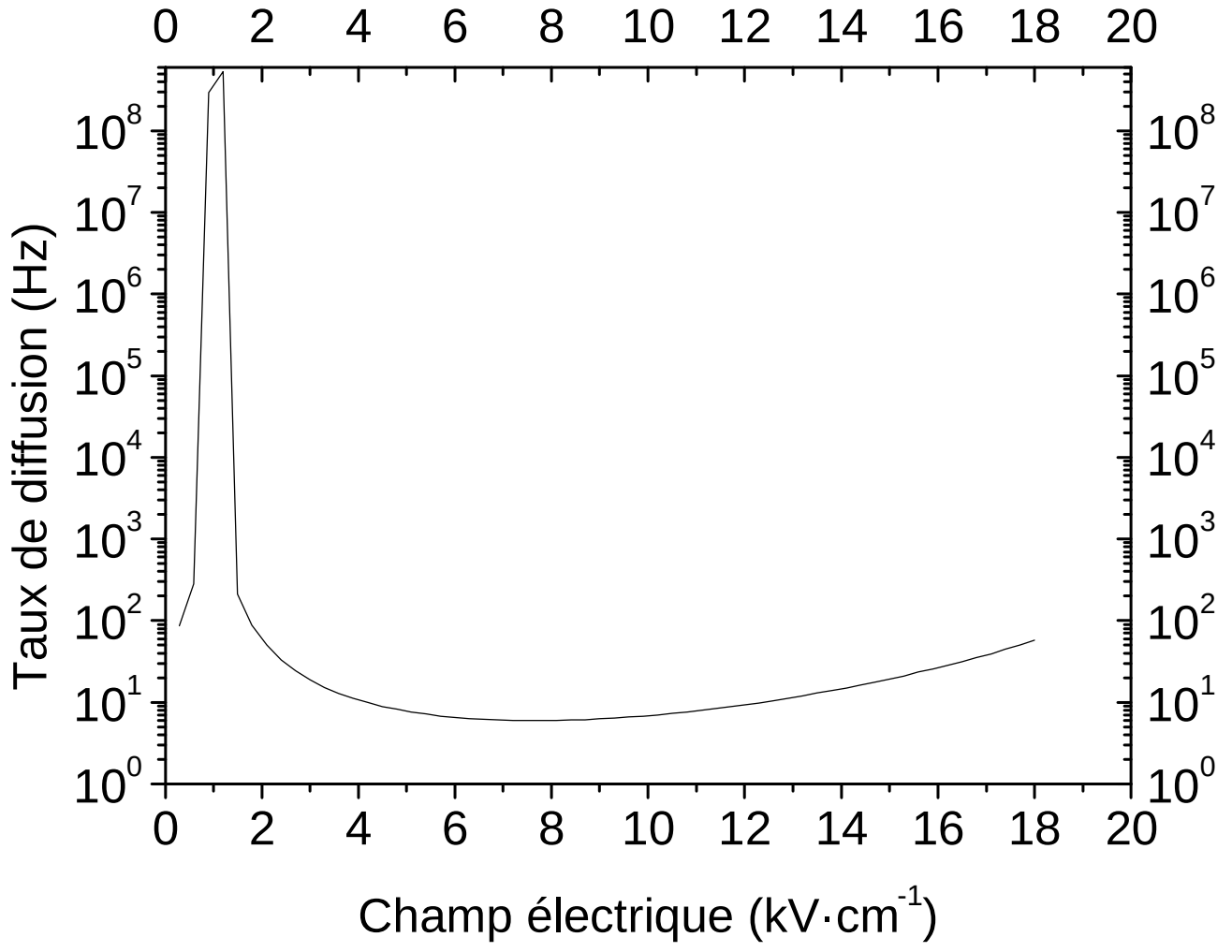

*FIG. 4-9 Taux de diffusion en fonction du champ électrique pour une structure à deux puits asymétriques. Le puits amont passe de 73 Å à 80 Å et le puits aval est inchangé (73Å). La mise en résonance des niveaux se produit pour* $F \approx 1kVcm^{-1}$.

### 4.2.2. Dans la littérature

Cette divergence est également rapportée dans la littérature. Le groupe d'Harrison (Leeds) qui s'intéressait alors au QWIP a observé des effets similaires de divergence liés aux fonctions d'onde[154]. Ils en ont conclu qu'il devait exister des effets de localisation, qu'ils ont attribués à la rugosité d'interface, qui devait localiser les fonctions d'onde dans les structures. Leur investigation n'a pas été au delà. Le groupe de Hu (MIT) a lui aussi constaté des effets similaires dans les lasers à cascade. Ils ont alors proposé une solution drastique qui consiste à supprimer le transport entre niveaux peu couplés[155,156].

Mais cette divergence n'est pas une surprise, il fallait en effet s'attendre depuis le chapitre 3 à ce que cette modélisation du transport par hopping ne soit pas valide aux faibles champs électriques, voir la FIG. 3-8 et la référence 118. Rott[143] a d'ailleurs formalisé cette divergence et montré qu'il fallait s'attendre à une divergence du courant aux faibles champs suivant une loi en $1/F^n$, dans laquelle n dépend de l'interaction considérée.

Mathématiquement cette divergence résulte de la non stabilité des fonctions de Wannier Stark en présence de niveau d'énergie complexe[87,157]. Ce qui physiquement veut dire que notre approche de hopping n'est actuellement pas valide quand la chute de potentiel par période est inférieure à l'élargissement des niveaux[158,159]. L'expression du courant que nous avons, telle qu'elle est donnée par l'équation (4-20), n'est donc pas valide à proximité d'une résonance de niveau équivalent. Il nous revient alors de trouver une solution pour étendre la validité du formalisme de hopping aux bas champs.

### 4.2.3. Extension du formalisme de hopping aux faibles champs électriques

#### *4.2.3.1. Elargissement des niveaux de Wannier Stark*

Dans des structures similaires aux nôtres, l'élargissement des niveaux a pu être évalué expérimentalement[160] ou par des simulations[132], il est généralement admis que celui-ci se situe dans la gamme 1meV à 10meV, soit encore des temps de diffusion entre 0.1 et 1ps. Il est également possible d'utiliser notre propre outil de simulation, ce qui dans notre structure a conduit à un élargissement de 90fs ($\frac{1}{\tau} = 1.1 \times 10^{13}\ Hz$), soit 7.3meV.

Il faut à présent introduire dans cette approche de hopping l'élargissement des niveaux. Pour cela il faut rappeler que le calcul analytique des fonctions de Wannier Stark conduit à des fonctions d'onde sous la forme[118] :

$$\psi_n(z) = \sum_{puits} J_{n-i}\left(\frac{\Delta_0}{2eFL_d}\right)\phi_i(z) \qquad \text{(4-21)}$$

Avec $J_{n-i}$ la n-i$^{ème}$ fonction de Bessel du premier ordre et $\phi_i(z)$ la fonction de Wannier associée au puits i. Dans notre cas, en raison du couplage faible de la structure nous pouvons nous limiter au couplage premier voisin. Cela signifie que dans le puits voisin, la fonction d'onde associée au n$^{ème}$ puits a un « excès » de présence dans le puits voisin donné par $J_1\left(\frac{\Delta_0}{2eFL_d}\right)$. La densité de courant qui est proportionnelle au nombre de porteurs délocalisés, doit donc être proportionnelle à $J_{\text{Wannier-Stark}} \propto J_1\left(\frac{\Delta_0}{2eFL_d}\right)^2$.

Dans cette approche basée sur les fonctions de Wannier, le rôle du champ électrique est de localiser les fonctions d'onde comme cela a été discuté au chapitre trois. L'idée consiste alors à traiter une forme de localisation du à la perte de cohérence par déphasage, par une autre forme de localisation, plus facile à décrire qu'est la localisation par le champ électrique. Pour cela nous utilisons la notion de transition entre état à énergie complexe que j'ai introduit au chapitre précèdent (voir section 3.2.4.1). Dans la mesure où les niveaux sont élargis d'une quantité $\frac{\hbar}{\tau}$, l'énergie de la transition entre les deux états fondamentaux s'écrit $eFL_d + i\frac{\hbar}{\tau}$. Nous introduisons ainsi un champ électrique complexe tel que $eF_{eff}L_d = eFL_d + i\frac{\hbar}{\tau}$. Et nous utilisons à présent ce champ effectif dans le calcul de nos fonctions d'onde, ce qui conduit à une expression du courant sous la forme.

$$J_{corrigé} \propto J_1\left(\frac{\Delta_0/2}{\left|eFL_d + i\frac{\hbar}{\tau}\right|}\right)^2 \qquad (4\text{-}22)$$

Dans la mesure où le couplage est faible $eFL_d >> \Delta_0$, nous pouvons approximer l'expression des fonctions de Bessel $J_1(x) \underset{0}{\propto} \frac{x}{2}$. Finalement l'expression du courant est égale à l'expression précédente multipliée par une fonction correctrice qui prend en compte l'élargissement des niveaux.

$$J_{corrigé} = J_{\text{non corrigé}} \frac{(eFd)^2}{(eFd)^2 + \left(\frac{\hbar}{\tau}\right)^2} \qquad (4\text{-}23)$$

La divergence du courant disparaît alors comme le montre la FIG. 4-8. Avec correction, le modèle est capable de reproduire le régime ohmique de transport qui existe aux basses tensions.

Très récemment Gordon et Majer[161] ont également traité ce problème et lui ont apporté une solution dans le formalisme de la matrice densité. Leur approche introduit la même physique que celle qui est présentée ici, et elle s'avère numériquement très avantageuse pour les structures à grand nombre de niveaux (QCL et QCD).

#### *4.2.3.2. Longueur de cohérence*

Il est possible de montrer que l'introduction d'une chute de potentiel imaginaire est équivalente à l'introduction d'une longueur de cohérence. L'expression des fonctions de Wannier-Stark qui sont solutions de l'hamiltonien de Ben-Daniel-Duke est donnée par des fonctions de Bessel. En particulier, l'excès de présence des fonctions d'onde au niveau du puits premier voisin est donné par la fonction de Bessel du premier ordre $J_1\left(\frac{\Delta_0}{2eFL_d}\right)$, en présence d'un phénomène de localisation cette amplitude serait modifiée et deviendrait $J_1\left(\frac{\Delta_0}{2eFL_d}\right).\exp(-\frac{L_d}{\lambda_c})$. Quantité qui, dans notre approche basée sur l'introduction d'un temps de vie s'écrit également $J_1\left(\frac{\Delta_0}{2eF_{eff}L_d}\right)$.

Comme dans notre cas le couplage est faible, il est raisonnable de faire l'approximation suivante $J_1(x) \underset{0}{\propto} \frac{x}{2}$, qui conduit à $\left|F_{eff}\right| = Fe^{\frac{L_d}{\lambda_c}}$. Cette dernière expression peut donner l'expression de la longueur de localisation :

$$\lambda_c = \frac{L_d}{\ln\left(\frac{\left|eFL_d + i\frac{\hbar}{\tau}\right|}{eFL_d}\right)} \quad (4\text{-}24)$$

Cette longueur de cohérence est de l'ordre de 1à 2nm dans la structure 15µm et présente une très faible dépendance avec le champ électrique.

## 4.3. Résultats pour les QWIP hautes longueur d'onde

### 4.3.1. Simulation des courbes J(V)

Nous pouvons à présent présenter les courbes de densité de courant en fonction de la tension pour la structure du chapitre deux, voir FIG. 4-10.

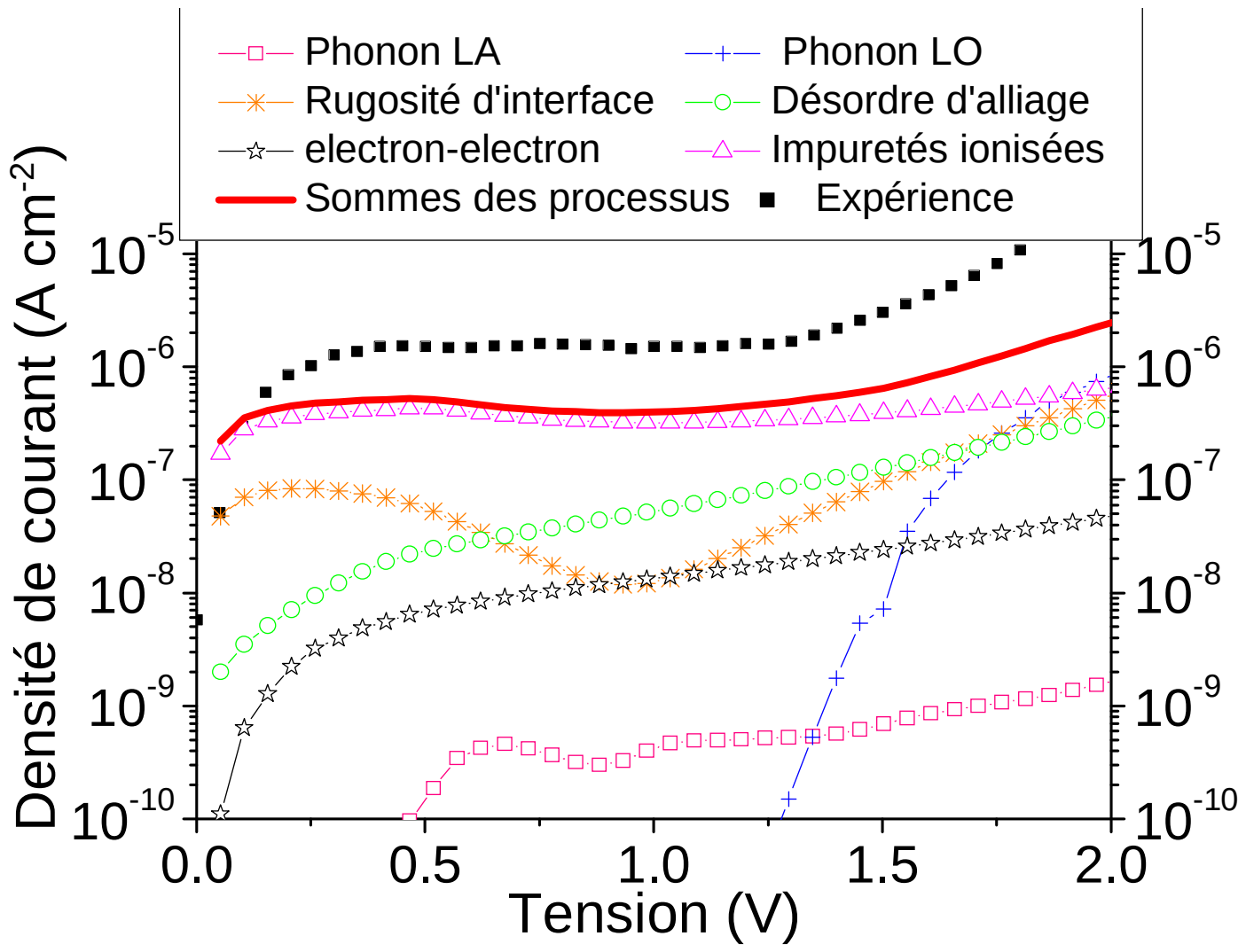

*FIG. 4-10 : Densité de courant en fonction de la tension appliquée pour les six processus inclus dans nos simulations. La courbe théorique (en rouge épais), somme de ces processus, peut être comparée à la courbe expérimentale (en noir et pointillé). Calculs et mesures faits à 10K. Dans l'hypothèse d'un champ homogène 1V sur la structure correspond à 5.8kV·cm$^{-1}$*

Commentons, à présent, les résultats de ces simulations :

- La reproduction du phénomène de plateau sur ces courbes est assez bonne. A plus haute tension (V>2V) la courbe expérimentale s'envole avec la tension, cela résulte de la mise en résonance du niveau fondamental avec le niveau excité du puits suivant.

- Dans notre plage d'intérêt en tension, l'interaction qui domine est celle entre les électrons et les impuretés ionisées. Cela signifie donc que le profil de dopage va avoir une incidence majeure sur le niveau de courant. La réduction du courant d'obscurité en régime tunnel passera nécessairement par la compréhension de l'influence de ce profil de dopage. Ce point sera étudié largement au chapitre suivant.

- La valeur plateau est obtenue avec un désaccord correspondant à un facteur entre deux et trois. Comme nous le verrons au chapitre suivant ce désaccord pourrait être attribué à une méconnaissance du niveau de dopage de l'ordre de 40 à 70 %. Un tel écart semble toutefois au delà de l'imprécision de la MBE sur le niveau de dopage.

### 4.3.2. Origine du plateau

Trois phénomènes entrent en compétition et conduisent à cette forme des courbes *I(V)*, voir FIG. 4-11 :

Notre approche contient l'effet de résistance différentielle négative, qui conduit dans notre plage de champ électrique à une localisation des fonctions d'onde avec le champ électrique. Cela se traduit sur le courant par une décroissance de la densité de courant avec le champ.

D'autre part, notre modèle contient également l'effet d'abaissement de barrière avec le champ électrique. Cette physique, qui est incluse dans l'approximation WKB, conduit au contraire à une augmentation du courant avec le champ.

Enfin l'élargissement des niveaux est à l'origine du transport ohmique aux basses tensions.

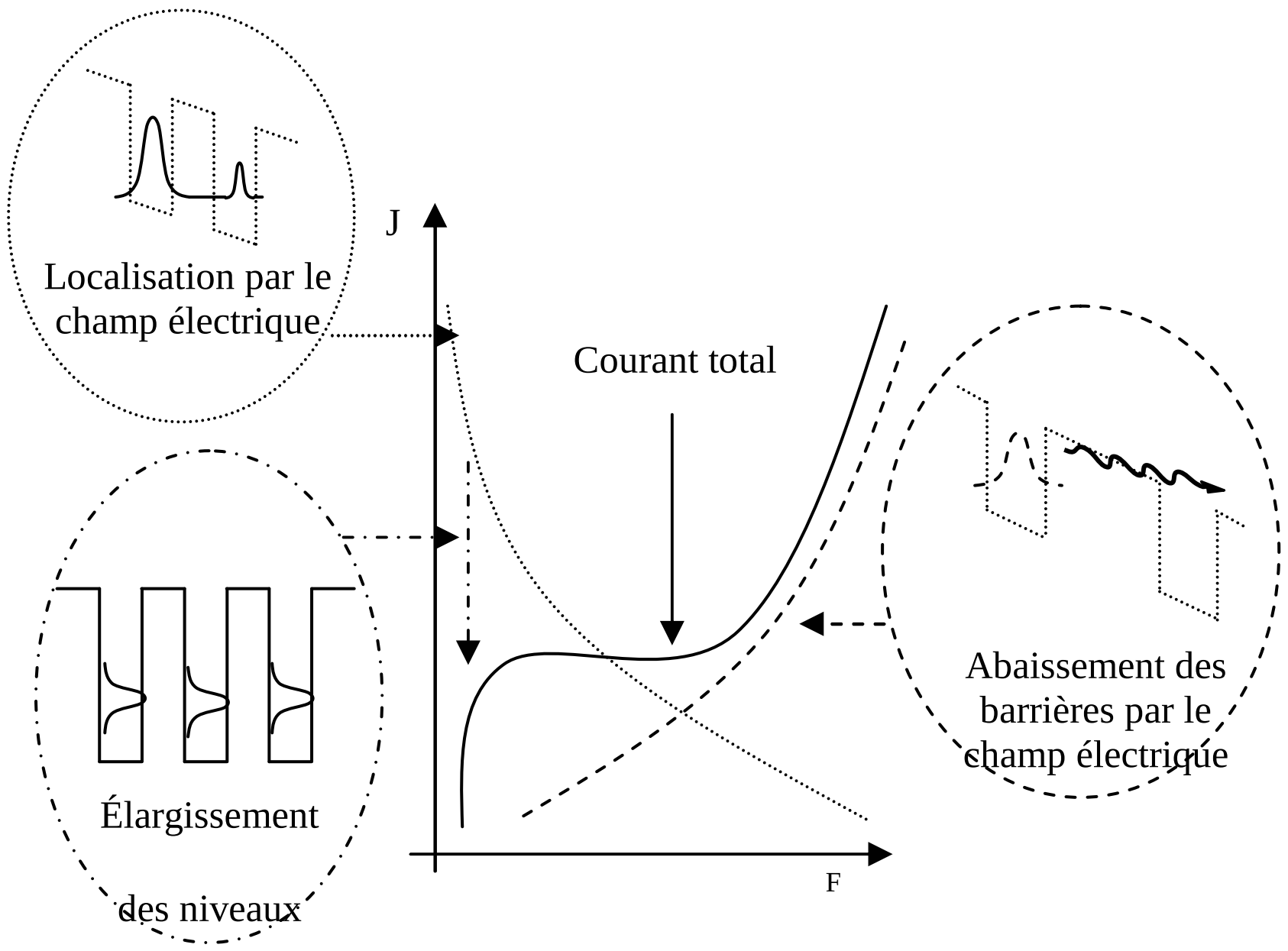

*FIG. 4-11Principaux phénomènes à l'origine de la forme des courbes I(V).*

Ce sont en particulier ces deux premiers processus (localisation des fonctions d'onde et abaissement des barrières) qui sont à l'origine de cette forme plateau. En effet ces deux effets antagonistes viennent se compenser pour donner une zone de résistance infinie. La localisation par le champ joue surtout sur la fonction d'onde aval en diminuant son amplitude dans le puits amont, voir FIG. 4-12 (a). Au contraire l'abaissement de barrière joue sur la fonction d'onde amont et augmente sa probabilité de présence dans le puit aval, voir FIG. 4-12 (b). Au final le recouvrement des fonctions d'onde est donc très faiblement dépendant du champ électrique. Il est intéressant de noter que ces deux effets ne sont pas inclus simultanément dans les autres approches de transport des super réseaux. Cette approche de hopping qui inclut à la fois la physique du modèle de minibande (RDN) et celle de l'effet tunnel séquentiel (abaissement de barrière) est donc la seule qui puisse expliquer ce plateau.

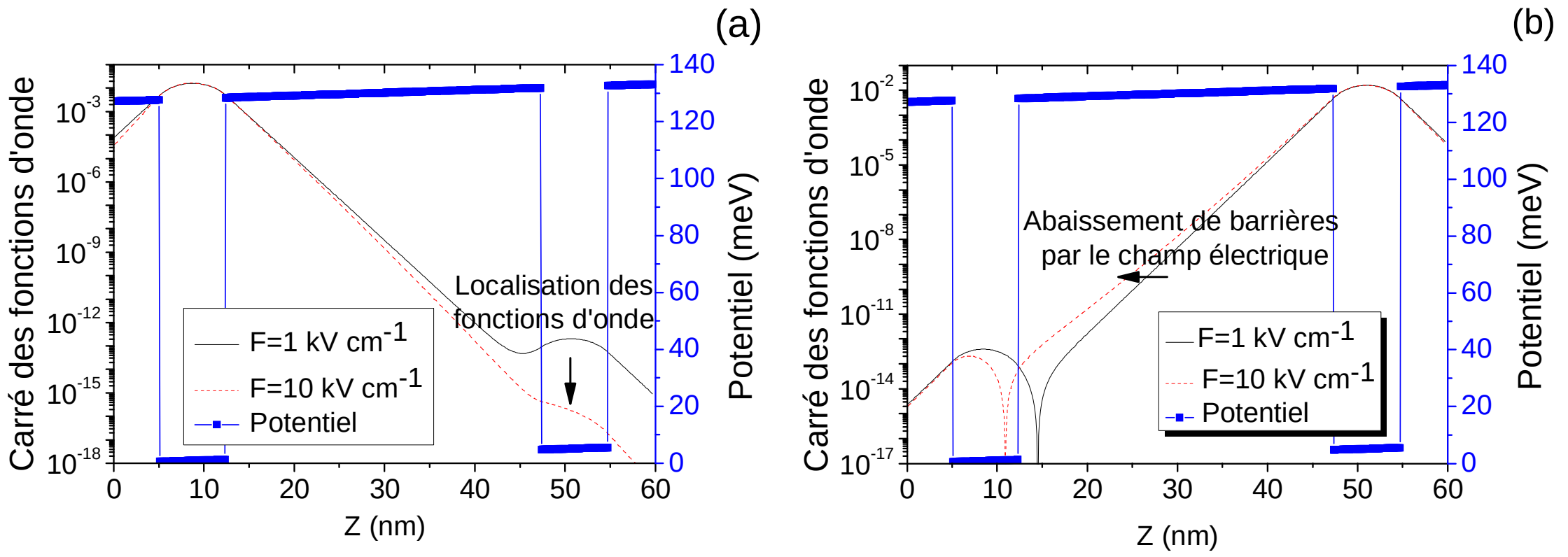

*FIG. 4-12(a) Effet du champ électrique sur les fonctions d'onde aval, la flèche indique l'effet de la localisation des fonctions d'onde. (b) Effet du champ électrique sur les fonctions d'onde amont, la flèche indique l'effet de l'abaissement de barrière. Le profil de bande n'est trace que pour un champ de 1kV.cm$^{-1}$.*

Il faut bien noter que ce plateau peut ne pas se retrouver sur d'autres structures QWIP. En particulier si la structure est beaucoup plus couplée, le processus de RDN peut devenir dominant, ce qui se traduit pas un plateau avec des dents de scie, voir le dernier chapitre de cette thèse. Et au contraire si l'on réduit encore le couplage, l'abaissement de barrière devient prédominant et conduit à des courbes *I(V)* strictement croissantes.

### 4.3.3. Effet de la taille de barrière

L'un des paramètres essentiels pour contrôler le niveau de courant dans les QWIP est la taille des barrières. L'atténuation exponentielle avec la taille de barrière constitue la principale dépendance du couplage dans les QWIP, loin devant les autres paramètres telle que le type de matériau utilisé, le niveau du dopage ou encore la taille des puits. Il était donc important de vérifier que notre modèle conduisait bien à une dépendance réaliste avec ce paramètre. La FIG. 4-13 compare le résultat de simulation avec les donnés expérimentales pour des structures LWIR (8.4µm). Comme pour notre structure à 15µm les courbes I(V) de ces composants présentent un plateau avec la tension dont la valeur en courant est comparée aux simulations.

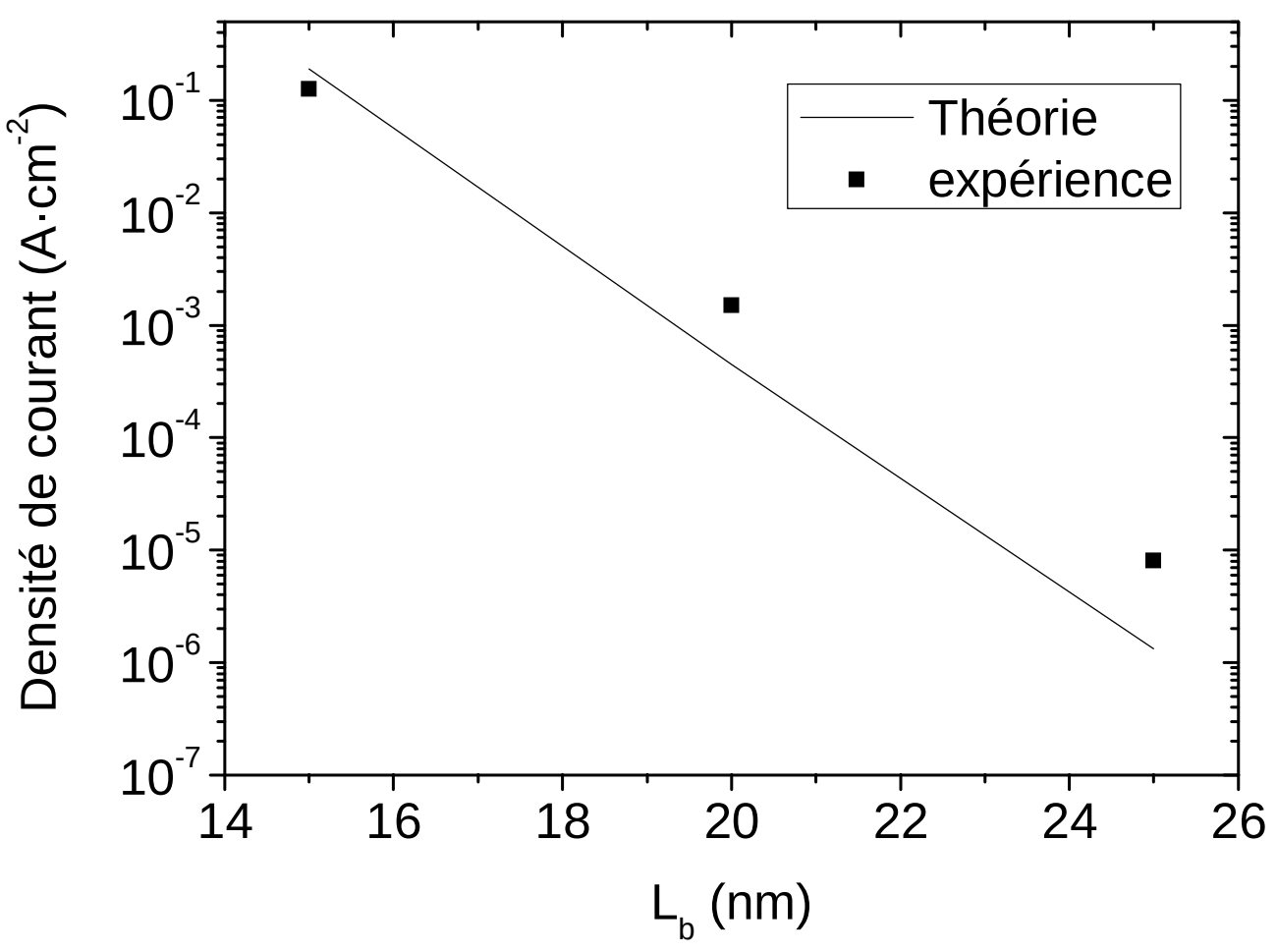

*FIG. 4-13 Effet de ta taille de barrière sur un QWIP 8.5µm ($L_w$=49Å, dopage $2\times10^{11}cm^{-2}$), sous un champ de $10kVcm^{-1}$ et T~40K. Points expérimentaux fournis par Alcatel Thales 3-5 Lab.*

L'accord théorie expérience est assez correct, notre modèle conduit à une légère surestimation de l'atténuation du courant avec la barrière. Une surestimation similaire avait été obtenue avec le modèle basé sur les minibandes, voir références 58.

## 4.4. Conclusion

Plus qu'un outil pouvant prédire avec un bon accord les courbes expérimentales, notre code de simulation est une boîte à outils quantique qui permet de relier la structure microscopique à ses propriétés de transport. Dans ce chapitre j'ai montré comment l'approche de hopping connaissait des difficultés de modélisation à bas champ électrique. L'introduction d'un temps de vie fini pour les états de Wannier-Stark est une nécessité pour reproduire le régime ohmique. Et nous avons apporté une extension de cette approche de transport qui nous permet de l'utiliser y compris aux basses tensions. Grâce à ce modèle j'ai trouvé une origine quantique au phénomène de plateau des courbes *I(V)*, celui-ci résulte de la compétition entre la localisation des fonctions d'onde par le champ électrique et l'effet d'abaissement de barrière par ce même champ. Enfin j'ai identifié que le courant dans cette structure résulte essentiellement de l'interaction entre électrons et impuretés ionisées. J'ai fait la démonstration que ce code était capable de rendre compte efficacement des effets liés à la diffusion. L'étape suivante est donc d'utiliser ce code afin de dessiner de meilleurs QWIP destinés aux faibles flux. C'est ce travail qui va constituer le chapitre suivant.

# 5. OPTIMISATION DES DETECTEURS ET APPLICATIONS AUX STRUCTURES COMPLEXES.

## Sommaire

**Publications associées à ce chapitre :**

- E. Lhuillier, E. Rosencher, I. Ribet-Mohamed, A.Nedelcu, L. Doyennette and V. Berger, Quantum scattering enginnering for the reduction of dark current in high wavelength quantum well infrared photodetectors, accepté dans J. Appl. Phys.
- E. Lhuillier, N. Péré-Laperne, I. Ribet-Mohamed, E. Rosencher, G. Patriarche, A. Buffaz, V. Berger, A. Nedelcu and M. Carras, Quantum well infrared photodetectors hardiness to the non ideality of the energy band profile, J. Appl. Phys 107, 123110 (2010).
- E. Lhuillier, I. Ribet-Mohamed, E. Rosencher, G. Patriarche, A. Buffaz, V. Berger and M. Carras, Interface roughness transport in terahertz quantum cascade detectors, Appl. Phys. Lett. 96, 061111 (2010)

J'ai montré au chapitre deux, l'intérêt d'une diminution du courant tunnel dans les QWIP. J'ai ensuite, au cours des deux chapitres suivants, présenté les outils nécessaires à la modélisation du transport électronique dans les structures très faiblement couplées. J'ai en particulier développé un modèle de transport par sauts entre états localisés qui a donné une interprétation quantique du phénomène de plateau observé expérimentalement sur les courbes *I(V)*.

Ce chapitre ne s'intéresse plus à la compréhension de la physique du transport d'un point de vue fondamental mais va plutôt s'attacher aux résultats qui ont découlé de nos simulations. Nous pouvons classer ces résultats en trois grandes parties. La première est dans la continuité directe de notre objectif de réduction du courant d'obscurité et va s'attacher à la présentation de nouvelles structures. Ce travail va au-delà de la simple ingénierie de fonction d'onde utilisée dans les QWIP. En effet, nous avons développé une véritable ingénierie de diffusion qui a abouti à une réduction conséquente du courant d'obscurité. Jusque là toutes les simulations qui ont pu être menées ont été faites dans des structures à deux puits idéales. Les deux dernières parties de ce chapitre vont donc s'attacher à montrer que ce code peut être utilisé à des fins beaucoup plus générales. En effet je présenterai l'influence de la non idéalité dans la structure QWIP. Je me focaliserai principalement sur les défauts d'interfaces et sur le désordre que les impuretés peuvent introduire. Enfin la dernière partie de ce chapitre sera dédiée à l'étude du transport dans les QCD THz, dans lesquels j'ai pu établir le rôle déterminant joué par les interfaces sur le transport.

## 5.1. Optimisation du profil de dopage : de l'ingénierie de fonction d'onde à l'ingénierie d'élément de matrice

Nous avons identifié dans la partie précédente le rôle déterminant joué par l'interaction entre les électrons et les impuretés ionisées sur le transport en régime tunnel dans les structures QWIP hautes longueurs d'onde. Or ces impuretés proviennent majoritairement du dopage. Je vais donc m'intéresser à présent à l'influence du profil de dopage sur les propriétés de transport. Deux principaux phénomènes vont être étudiés : l'effet de l'amplitude du dopage et l'effet de la position du dopage. Cette étude va aboutir à la conception de nouveaux composants proposant un courant d'obscurité réduit grâce à une optimisation du profil de dopage.

### 5.1.1. Influence du profil de dopage

#### *5.1.1.1. Amplitude du dopage*

Dans les approches de transport développées au chapitre trois nous avions soulevé l'une des limites des approches de minibandes et WKB, à savoir leur mauvais traitement de la dépendance du courant avec le niveau de dopage. L'approche minibande conduisait à un courant indépendant du niveau de dopage tandis que l'approche WKB attribuait une dépendance linéaire. Or il semble, empiriquement, que le courant suive une dépendance avec le dopage plus forte que cette loi linéaire. Notre approche de hopping semble résoudre ce point. En effet dans la mesure où le courant est dominé par la diffusion liée aux impuretés, le taux de diffusion est lui-même linéairement dépendant du dopage. Il en résulte une dépendance quadratique du courant avec le dopage. Afin de valider expérimentalement cette prédiction nous avons utilisé deux séries d'échantillons (B et C). Dans ces séries les composants différent par leur dopage. La première série est composée de QWIP dont la structure est relativement proche du QWIP 15µm qui a été utilisé jusque là. La seconde est une série de QWIP dont la réponse pic est située un peu au-delà de 17µm. Le détail des structures est donné par le tab. 5-1. Ces échantillons proviennent du III-V Lab.

| Paramètres | Serie B | Serie C |
|---|---|---|
| Taille des puits (Å) | 72 | 80 |
| Taille des barrières (Å) | 340 | 400 |
| Composition en aluminium (%) | 15 | 13 |
| Réponse pic (µm) | 14.5 | 17.2 |
| Dopage ($cm^{-2}$) ($J_{dark}$ en A $cm^{-2}$) | $B_1$ : $1\times10^{11}$ ($2.7\times10^{-7}$) | $C_1$ : $2\times10^{11}$ ($1.5\times10^{-7}$) |
| | $B_2$ : $2\times10^{11}$ ($1.3\times10^{-6}$) | $C_2$ : $4\times10^{11}$ ($6\times10^{-7}$) |

*tab. 5-1 : Structures utilisées pour mesurer l'influence du niveau de dopage sur le courant.*

Les mesures de courant pour la série B sont présentées sur la FIG. 5-1, elles conduisent à un rapport 4.6 sur la valeur du courant, contre quatre attendu. La série C, conduit a un accord encore meilleur sur les valeurs plateau puisque le ratio est très proche de 4. Ce qui valide bien cet aspect de notre modèle.

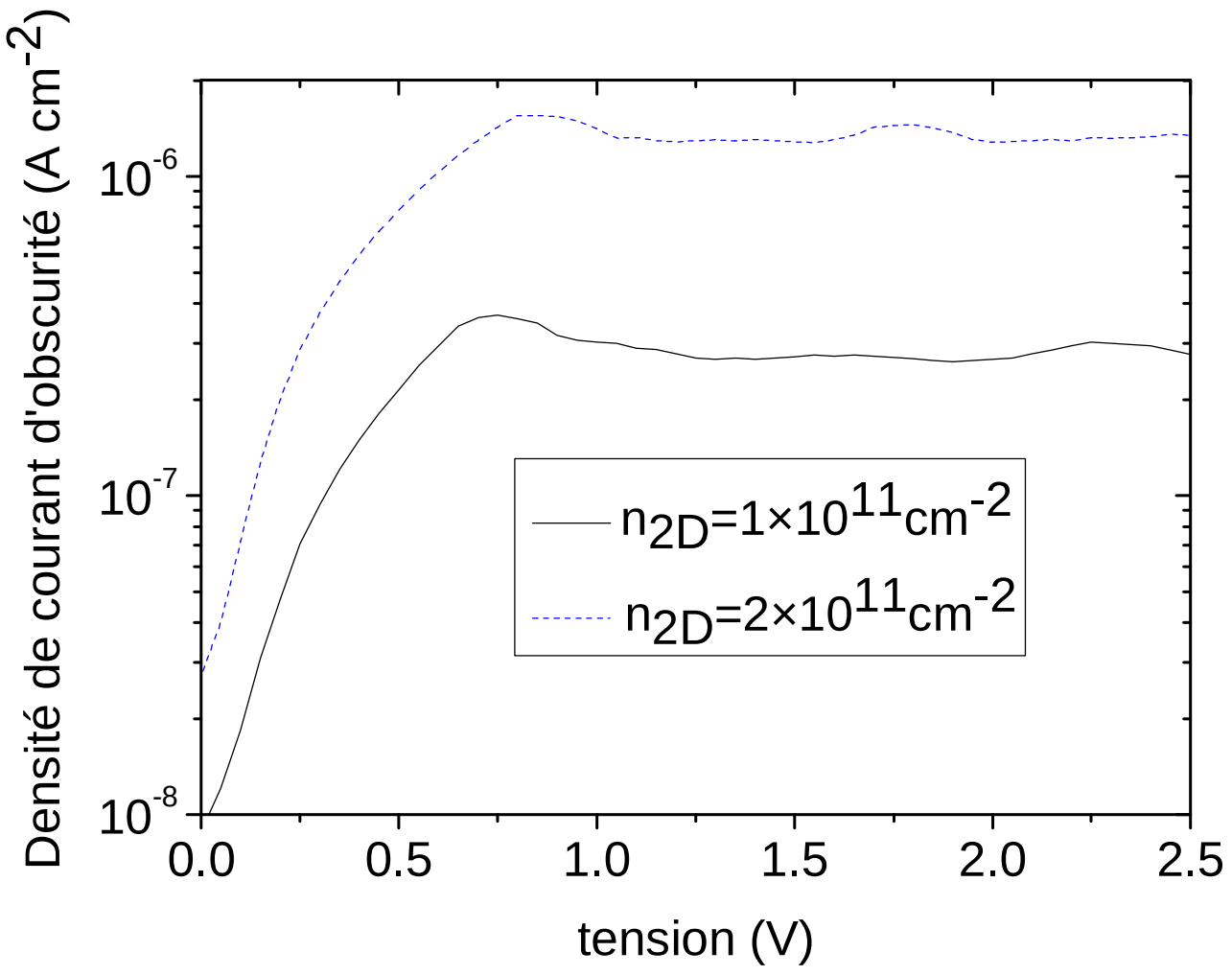

*FIG. 5-1 : Densité de courant d'obscurité en fonction de la tension appliquée pour les composants $B_1$ (dopage $1\times10^{11}cm^{-2}$) et $B_2$ (dopage $2\times10^{11}cm^{-2}$ ). Mesures fournies par le III-V Lab Alcatel-Thales.*

Je tiens toutefois à signaler que la dépendance avec le dopage peut s'avérer beaucoup plus complexe que cette simple loi quadratique dans la mesure où le vecteur d'onde de Thomas-Fermi dépend lui aussi du dopage, d'une façon plus ou moins complexe selon le modèle d'écrantage choisi. Cela signifie concrètement que sur une large plage de dopage, la dépendance du courant avec le dopage peut s'éloigner de cette loi quadratique.

#### 5.1.1.2. Effet de la ségrégation

Nous avons observé au chapitre 2 que les courbes *I(V)* présentent généralement une petite asymétrie qui conduit à un courant tunnel plus faible à tension négative que positive. Cette asymétrie du courant reflète bien sûr une asymétrie de la structure. Une asymétrie du profil de bande[vi] serait envisageable mais un consensus semble se dessiner pour attribuer cet effet à une asymétrie du profil de dopage. Cette asymétrie résulte de la brisure de symétrie qu'est le choix d'une direction de croissance. Le profil de dopage s'étend alors dans la direction de croissance par rapport au créneau idéal, voir les références 162 et 163. Cette ségrégation du dopage se fait préférentiellement dans la direction de croissance, car la ségrégation est un effet superficiel[164] par opposition à la diffusion dans la masse qui se fait à plus haute température[165]. Je n'ai pas cherché à reproduire finement le profil de dopage, en particulier je n'ai pas utilisé un profil de dopage lié à un modèle de ségrégation des espèces. J'ai au contraire préféré modéliser le profil ségrégé par un simple trapèze asymétrique, voir l'encart de la figure

FIG.5-2. La ségrégation conduit à une diminution du courant tunnel typiquement d'un facteur deux, voir FIG.*5-2*. Les valeurs choisies pour l'extension du profil ségrégé (voir FIG.*5-2*) sont en accord avec les références 162 à 165.

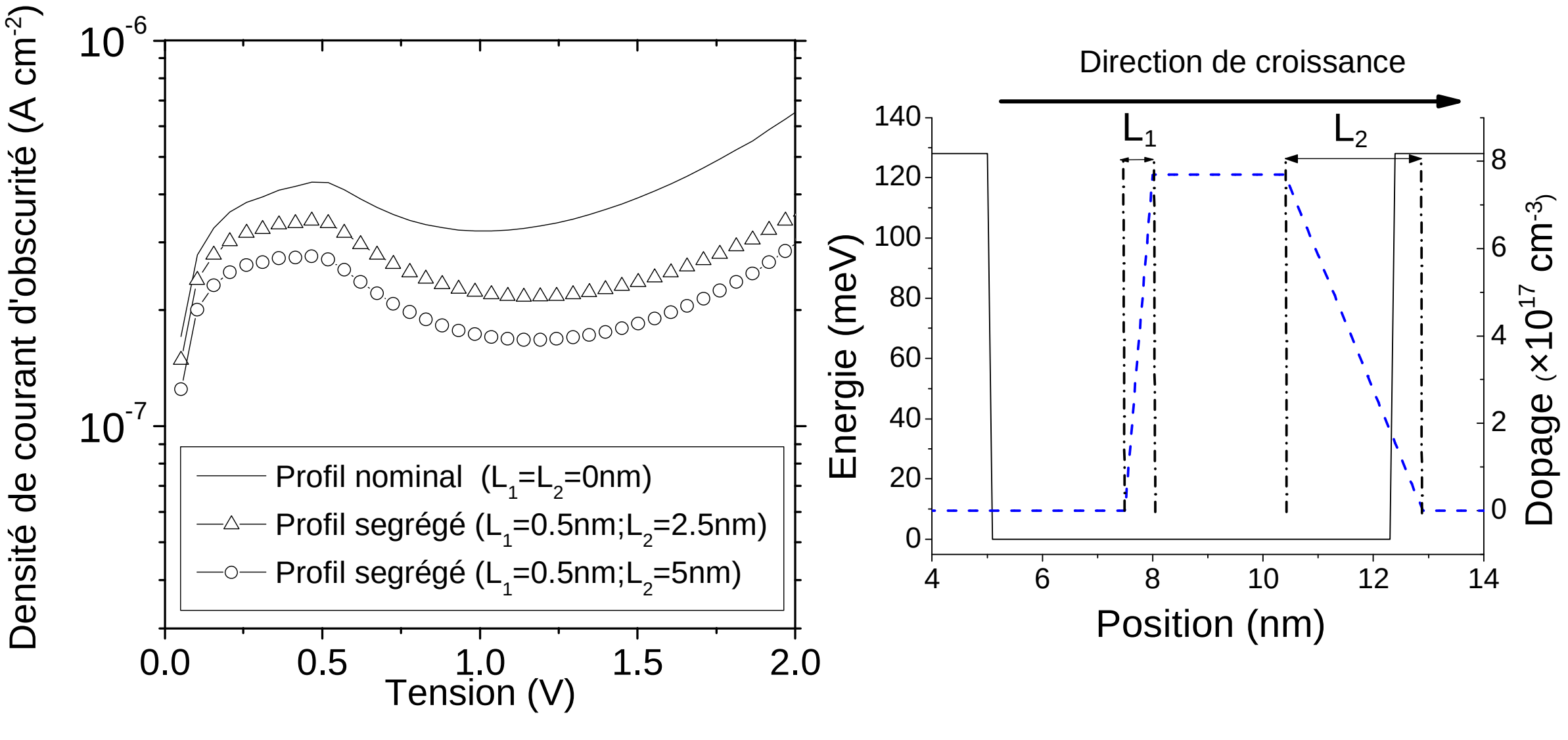

*FIG.5-2 : Densité de courant d'obscurité en fonction de la tension appliquée pour différentes valeurs de la longueur de ségrégation $L_2$. Encart : profil de dopage dans le puits quantique.*

---

[vi] Voir discussion plus loin dans ce chapitre

#### *5.1.1.3. Influence de la technique de croissance*

Jusque là nous n'avons considéré que les impuretés en provenance du dopage. Or il existe dans les hétérostructures des impuretés résiduelles qui peuvent provenir d'une purification imparfaite des matériaux sources lors de la croissance ou de la présence d'impuretés (C, O…) dans le bâti. Ces impuretés ont tendance à se localiser dans les barrières du fait de la grande réactivité de l'aluminium[142]. La quantité d'impuretés contenue dans ces barrières peut en particulier dépendre de la technique de croissance. Il est généralement admis que leur concentration est plus faible en MBE (entre $10^{14}$ et $10^{15}cm^{-3}$) qu'en MOVPE (entre $10^{15}$ et $10^{16}cm^{-3}$). La FIG. 5-3 montre l'influence de l'ajout d'impuretés dans les barrières sur le courant dans la structure. En-dessous de $10^{16}cm^{-3}$ impuretés, le niveau de courant tunnel n'est quasiment pas affecté. Cela signifie que les deux processus de croissance utilisés pour les détecteurs à puits quantiques ne sont pas à l'origine d'effets parasites liés à la présence d'impuretés dans les barrières. Toutefois certaines publications rapportent des valeurs plus élevées du taux d'impuretés en MOVPE[142]. De tels composants ne pourraient alors pas être utilisés lors d'applications bas flux.

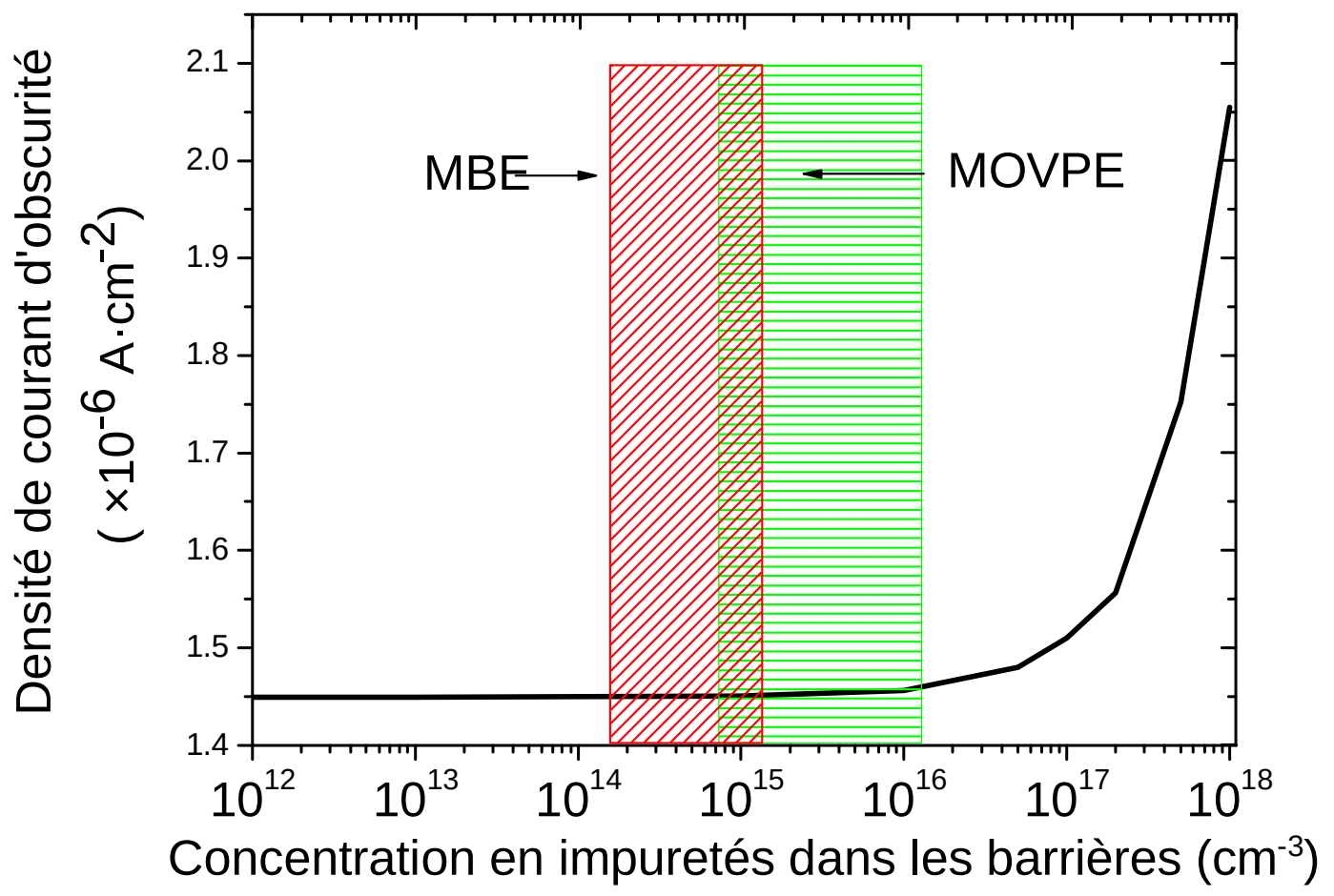

*FIG. 5-3 : Densité de courant d'obscurité en fonction de la concentration en impuretés dans les barrières sous un champ électrique de 10 kVcm$^{-1}$. Les deux rectangles donnent les gammes de concentration en impuretés liées au processus de MBE et de MOVPE.*

#### *5.1.1.4. Position du dopage*

Un dernier point reste à adresser concernant l'influence du profil de dopage sur le courant, à savoir l'effet de la position du dopage. L'interaction coulombienne fait intervenir l'intégrale de recouvrement entre les fonctions d'onde et les positions du dopage. Cette intégrale s'exprime à travers le terme $F_{II}(Q) = \left| \int dz \xi_f^*(z) e^{-Q|z-zi|} \xi_i(z) \right|^2$ ou $\xi_i$ et $\xi_f$ sont les fonctions d'ondes associées aux états finaux et initiaux et $z_i$ la position des impuretés. Le taux de diffusion lié aux impuretés sera donc d'autant plus faible que les fonctions d'onde associées à un puits seront éloignées des impuretés de ce même puits. J’ai donc tracé sur le graphe FIG. 5-4 le niveau de courant en fonction de la position d’un delta dopage dans un puits. Ce graphe montre clairement une influence de la position du dopage. Ainsi en localisant les impuretés d’un coté du puits il est quasiment possible d’annuler la contribution liée aux impuretés, nous reparlerons de cet effet dans la suite de ce chapitre. Comme la structure est polarisée les fonctions d’onde sont légèrement décalées du centre du puits (vers les abscisses négatives), le courant passe par un maximum quand les impuretés sont localisées à l’endroit du maximum des fonctions d’ondes. Nous allons donc pouvoir utiliser cet effet de déplacement du dopage pour réduire le courant d’obscurité dans la structure.

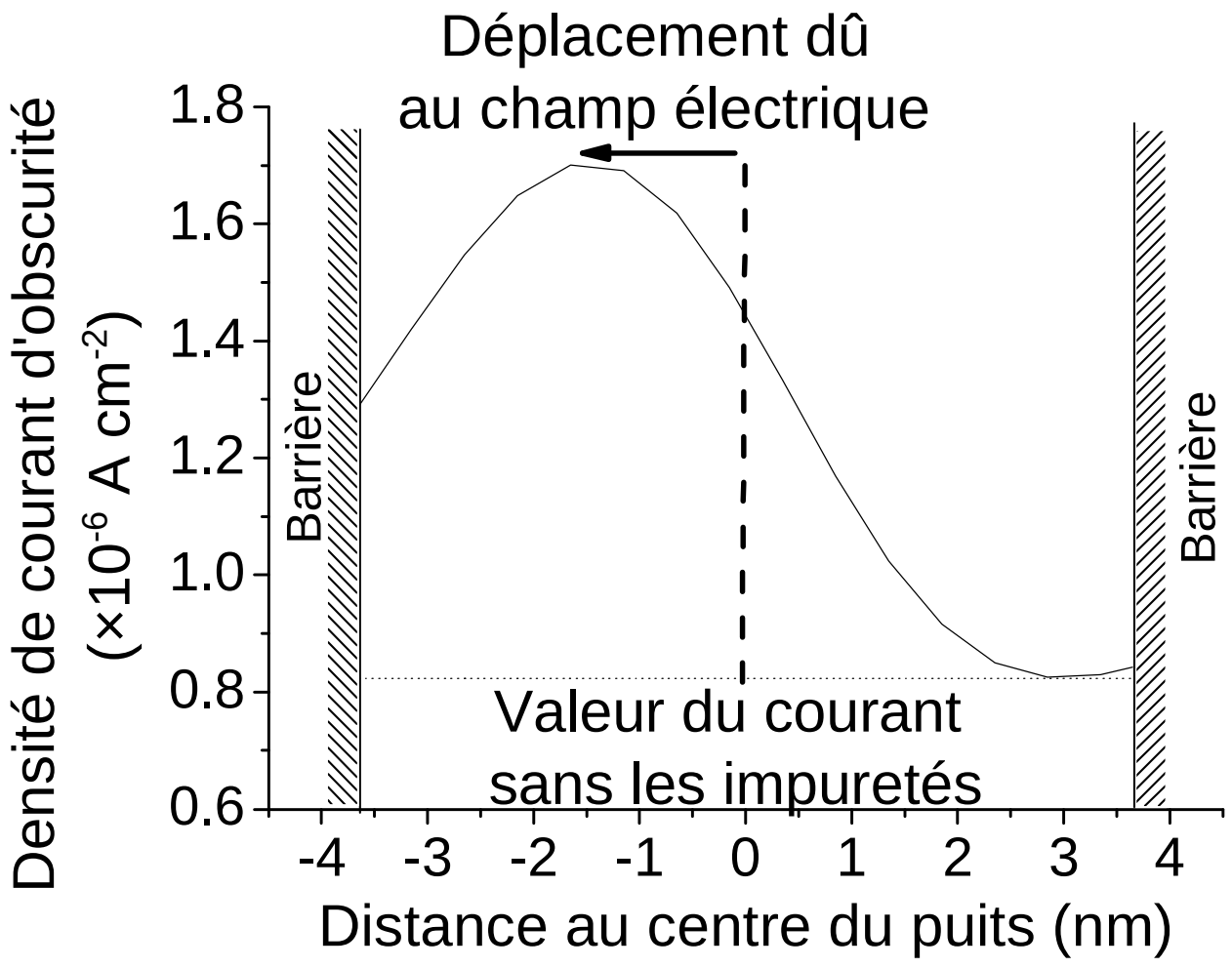

*FIG. 5-4 : Densité de courant d’obscurité (ligne continue) en fonction de la position du dopage dans le puits, sous un champ électrique de 10 kVcm$^{-1}$. La ligne pointillée est le courant au même champ électrique mais sans tenir compte des impuretés.*

### 5.1.2. Réduction du courant d’obscurité

Nous avons obtenu deux résultats essentiels dans la perspective de réduire le courant tunnel des QWIP. A savoir que le courant est dominé par la diffusion liée aux impuretés et deuxièmement que ce taux de diffusion est contrôlable via le profil de dopage.

Nous pourrions être tentés de jouer sur l’amplitude du dopage pour réduire le courant d’obscurité mais cette approche nuirait fortement à l’absorption de la structure. Cette solution est donc à éviter. Nous allons plutôt jouer sur le positionnement et la répartition du dopage. Cette étude à conduit à la réalisation d’une série de nouveaux échantillons optimisés par leur profil de dopage. Ces échantillons nous ont permis de valider nos prédictions.

#### *5.1.2.1. Ingénierie d’élément de matrice*

Je rappelle une fois encore l’expression de l’intégrale de recouvrement liée aux impuretés :

$$F_{II}(Q) = \left| \int dz \xi_f^*(z) e^{-Q|z-zi|} \xi_i(z) \right|^2 \qquad (5\text{-}1)$$

Nous voulons dépasser l'ingénierie de fonction d'onde pour faire de l'ingénierie d'élément de matrice. Notre objectif est donc de dessiner une structure pour laquelle cette intégrale de recouvrement serait réduite et que cela se traduise expérimentalement par une diminution du courant d'obscurité à basse température.

Pour cela il est nécessaire de décortiquer ce facteur de forme. Les fonctions d'onde présentent un maximum au centre des puits, tout au moins en l'absence de champ électrique. Le fait de doper le centre des puits maximise le recouvrement entre les électrons et le dopage ce qui maximise le taux de diffusion inter puits. La situation actuelle qui consiste à doper le centre des puits est donc mauvaise du point de vue du courant d'obscurité. Il serait en effet plus favorable d'éloigner le maximum des fonctions d'onde des impuretés.

#### *5.1.2.2. Nouvelles structures*

Commençons par donner un certain nombre de règles utiles à l'élaboration de nouveaux profils de dopage :

- Mieux vaut éviter de doper les barrières. En effet les impuretés au sein de la barrière peuvent créer des défauts profonds[71] qui vont créer un courant de fuite. De même le réarrangement électrostatique dû au déplacement des impuretés entraîne une modification du profil de bande qui, à son tour, peut créer un niveau dans la barrière, voir graphe FIG. 5-5. Ce qui conduit de la même façon à la création de niveaux quantiques pouvant faciliter la fuite des électrons[130, 166].

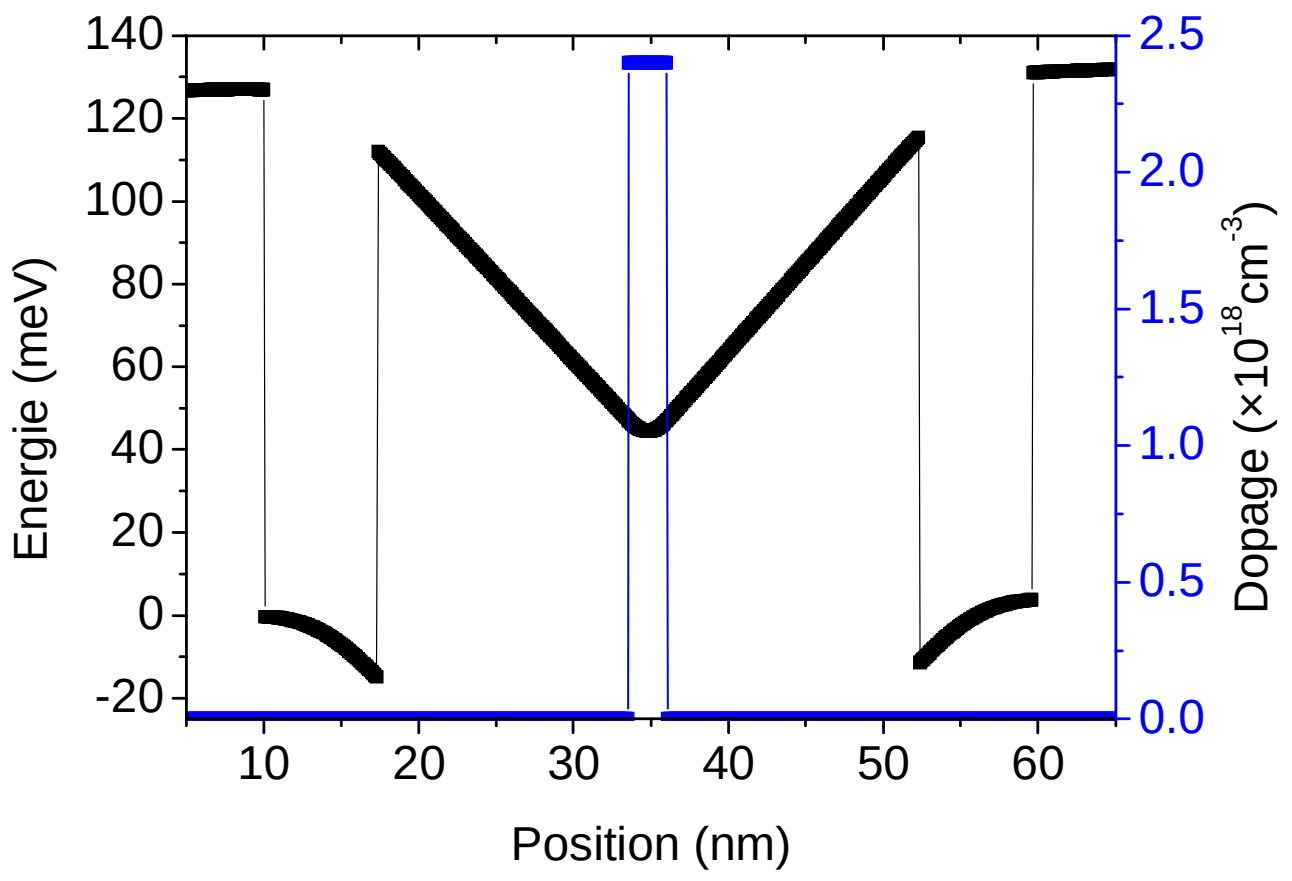

*FIG. 5-5 : Profil de bande et de dopage dans le cas d'un dopage au centre de la barrière. Calcul fait dans une structure à deux puits. Création d'un puits au milieu de la barrière.*

- Il faut tenir compte d'éléments technologiques pour dessiner des profils réalistes. Par ce propos je veux souligner la limite du contrôle que l'on peut avoir sur le profil de dopage lors d'une croissance par MBE. Le dopage représente un atome pour un à $10^5/10^6$, la précision que l'on a sur l'amplitude du dopage n'est pas aussi bonne que celle sur la taille des puits. De la même façon la localisation du dopage reste difficilement contrôlable en raison des phénomènes de ségrégation. Concrètement la réalisation d'un delta dopage est difficile. Les techniques de croissance ont donc leurs limites, mais il n'est pas non plus possible de contrôler *a posteriori* ce profil. Il n'existe en effet aucune technique de caractérisation qui donne accès à ce profil. La SIMS (secondary ion mass spectrometry) est bien capable de mesurer la concentration en impuretés mais sans information de positionnement. Les techniques de microscopie électronique en transmission pourraient au contraire apporter cette résolution mais ne sont pas assez sensibles pour mesurer de si faibles concentrations.

Compte tenu de ces deux règles nous avons dessiné trois structures, qui ne doivent nominalement ne différer que par la position du dopage. L'objectif est de valider expérimentalement nos prédictions théoriques de réduction du courant d'obscurité. La structure nominale est relativement proche de celle du composant 15µm qui a été étudié au chapitre deux à quatre. Ces structures ont été réalisées au III-V Lab sur mes spécifications. C'est une structure à quarante puits pour laquelle la taille des puits et des barrières est respectivement de 6.7nm et 39nm. La composition en aluminium est située autour de 15%. Enfin le dopage surfacique est dans les trois cas de $3\times10^{11}cm^{-2}$. La première structure ($D_1$) sert de référence et est dopée dans son tiers central. La seconde structure est dopée entre le troisième et le cinquième sixième, tandis que le dernier composant est dopé dans son dernier tiers. Nous avons choisi de déplacer le dopage dans la direction du réseau plutôt que dans celle du substrat pour deux raisons. Premièrement la ségrégation tend à faire baver la distribution de dopage dans la direction du réseau. Donc si nous avions éloigné les impuretés dans l'autre sens, ce déplacement aurait été certainement en partie compensé par la ségrégation. Deuxièmement les QWIP sont généralement polarisés avec des tensions négatives. Si l'on souhaite réduire le courant d'obscurité pour cette polarisation, le champ électrique tend à localiser la fonction d'onde du coté du substrat dans cette situation, il est donc favorable de doper de l'autre côté.

Nous voulons à tout prix éviter qu'une réduction du courant puisse être attribuée à un autre phénomène que le déplacement du dopage. Pour cela des mesures de diffraction des rayons X nous ont permis de mesurer précisément les dimensions associés à chaque composant. Les résultats sont rapportés dans le tableau tab. 5-2.

Nous nous sommes intéressés à la précision des mesures Il existe clairement un désaccord sur ce point entre les épitaxieurs et les gens qui font de la caractérisation à partir de technique spectroscopique. Les premiers sont partisans de donner des chiffres avec une précision de l'ordre de la monocouche, donc autour de 3Å dans GaAs. A l'opposée ceux qui utilisent des méthodes de spectroscopie telles que la diffraction des rayons X défendent une résolution égale à celle de leur appareil. Typiquement pour de la diffraction basée sur l'utilisation de la raie K du cuivre la longueur d'onde est de 154pm, donc la résolution théorique est inférieure à la centaine de picomètres. J'ai donc choisi une solution intermédiaire et donné mes données avec une précision de 1Å.

| Echantillon | $D_1$ | $D_2$ | $D_3$ |
|---|---|---|---|
| $L_w$ (nm) | 6.7±0.1 | 6.6±0.1 | 6.7±0.1 |
| $L_b$ (nm) | 39.2±0.1 | 38.7±0.1 | 38.9±0.1 |
| %Al | 15.6±0.1 | 15.6±0.1 | 15.5±0.1 |
| Position nominale du dopage | Tiers central | Du 3$^{éme}$ au 5$^{éme}$ sixième | Dernier tiers |
| Schéma du profil de dopage | 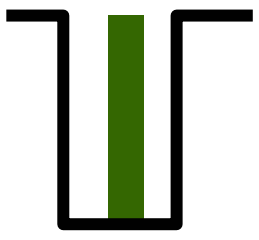 | 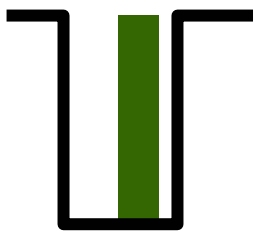 | 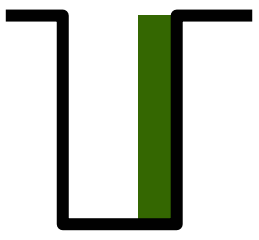 |

*tab. 5-2 : Grandeurs propres aux détecteurs caractérisés. En bas schéma du profil de dopage (en vert) et profil de bande en noir.*

Les barrières du composant de référence ($D_1$) sont plus larges que celles des deux autres composants. Et elles sont plus hautes que celle du composant $D_3$. Il faut donc s'attendre à ce que le confinement des électrons soit plus fort pour le composant de référence. Ces résultats sont confirmés par les réponses spectrales, voir la FIG. 5-6. Cela signifie que si nous ne tenions pas compte du déplacement du dopage il faudrait s'attendre à un courant plus fort dans les composants dans lesquels le dopage a été déplacé. Ce qui n'est pas le cas et donc plaide pour un effet du déplacement du dopage.

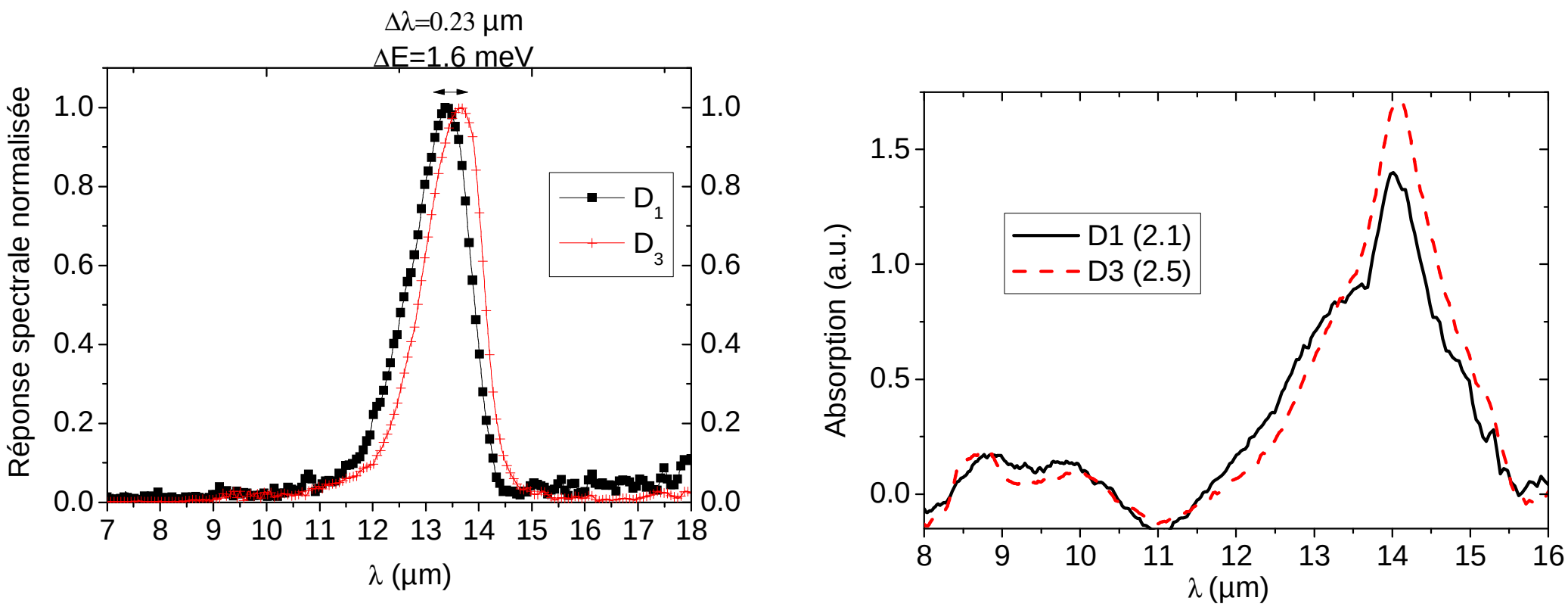

*FIG. 5-6 : A gauche : Réponse spectrale pour les composants $D_1$ et $D_3$. A droite courbe d'absorption à 300K. L'intégrale des spectres est donnes dans la légende. Cette dernière mesure a été faite au III-V Lab.*

Nous avons également vérifié que l'absorption des trois composants était similaire, ce qui signifie que leurs niveaux de dopage sont très proches, voir FIG. 5-6. Nous n'attendons donc pas de différence de courant dû à ce phénomène.

### 5.1.2.3. Résultats

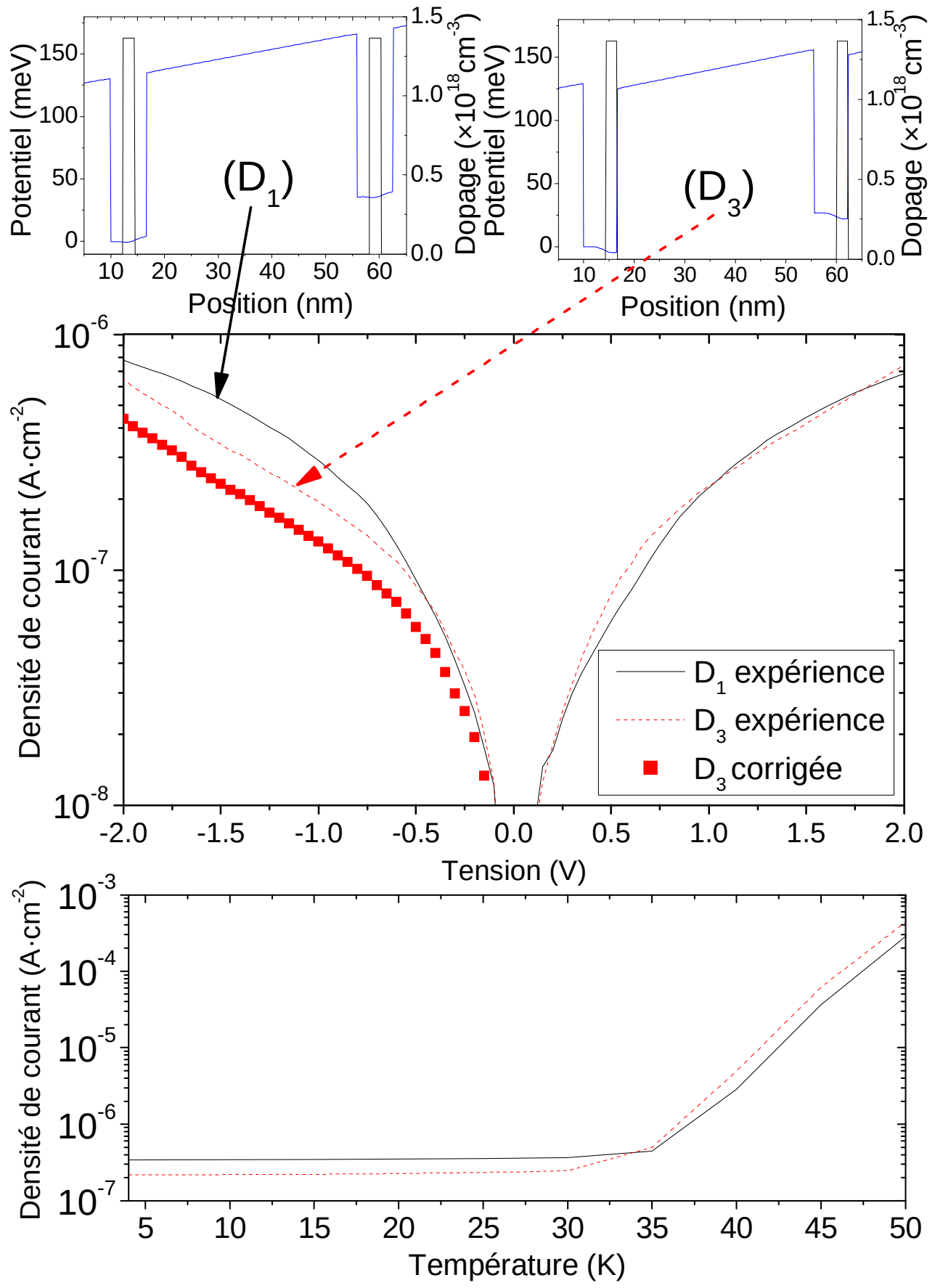

*FIG. 5-7 : En haut : profil de bande obtenu de façon auto consistante pour les composants $D_1$ et $D_3$. Au milieu Courbe I(V) expérimentales à 4K des composants $D_1$ et $D_3$, la courbe en pointillé donne le courant pour $D_3$ si il avait eu parfaitement la même structure que $D_1$. En bas densité de courant en fonction de la température pour les composants $D_1$ et $D_3$ faites à -1V ($F \approx 5.4 kV.cm^{-1}$).*

La FIG. 5-7 montre les courbes *I(V)* pour les composants $D_1$ et $D_3$. Je ne montre pas la courbe associée à $D_2$ car elle est quasi superposée à celle de $D_1$. C'est regrettable mais l'effet de déplacement du dopage est complètement caché par l'effet de la fluctuation des structures. La figure FIG. 5-7 montre une diminution du courant d'obscurité à tension négative ($|V| > 0.6V$), ce qui est bien l'effet attendu. A tension positive, l'effet est plus complexe. En effet d'un part le dopage est éloigné du centre des puits, mais l'effet Starck tend à localiser les fonctions d'onde sur le dopage. La réduction du courant est de l'ordre de 30%. Nous avons voulu quantifier la réduction du courant que nous aurions obtenu si les deux composants avaient été rigoureusement identiques, à l'exception de la position du dopage. Pour cela j'ai tracé sur ce même graphe le courant corrigé, c'est à dire le courant expérimental multiplié par le rapport des transparences tunnel, cf équation (3-7).

$$J_{corrigé} = J_{\text{expérimental}} \times \frac{T(E_1(D_1))}{T(E_1(D_3))} \qquad (5\text{-}2)$$

De cette façon nous pouvons estimer la réduction de courant résultant du déplacement du dopage, qui est ici de 50%. La dépendance du courant en fonction de la température est présentée sur le bas de la figure FIG. 5-7. A haute température, aucun effet du déplacement du dopage n'est attendu, seuls les effets liés à la structure devraient donc être observés. Effectivement nous observons que le courant est plus fort pour le composant $D_3$ que pour $D_1$. A basse température ce comportement s'inverse, en dépit d'un confinement plus faible dans la structure $D_3$. Preuve que notre action sur le taux de diffusion est bien à l'origine de la réduction de courant. La figure FIG. 5-8 compare les taux de diffusion expérimentaux ($\Gamma = \frac{J}{en}$) et théoriques pour les composant $D_1$ et $D_3$. L'accord obtenu sur la réduction du taux de diffusion par le déplacement du dopage est plus que satisfaisant. Par contre les formes des courbes sont assez différentes, ce point est encore en cours d'investigation.

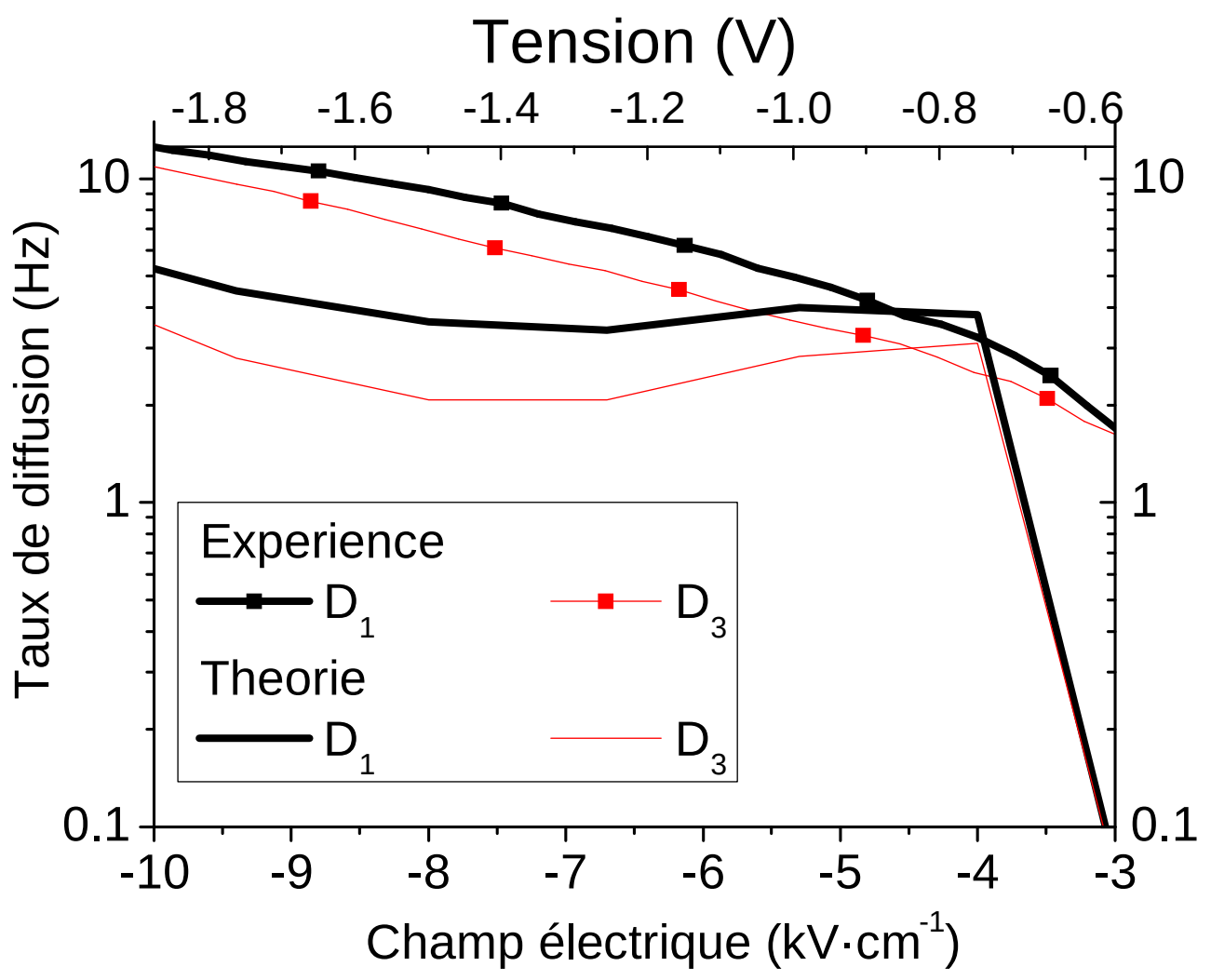

*FIG. 5-8 : Comparaison du taux de diffusion expérimental et théorique pour les composants $D_1$ et $D_3$. Calcul fait à 4K.*

### 5.1.2.4. Commentaire sur l'absence de plateau

Le lecteur attentif aura sans doute noté que les courbes *I(V)* des composants D ne présentent pas de plateau alors que leurs structures ne sont pas très différentes de celle du composant étudié au chapitre deux. La principale différence vient du fait que cette nouvelle série de composants a des barrières plus larges que précédemment. Du coup la largeur de minibande est encore réduite, ce qui réduit l'effet de localisation par le champ et au contraire augmente l'effet d'abaissement de barrière dans la mesure où la chute de potentiel par période est plus grande pour un même champ. Il en résulte un déplacement de l'équilibre entre les deux effets qui ne se compensent plus. L'effet plateau disparaît !

#### 5.1.2.5. Comment aller au delà

Le déplacement du dopage a bien sûr pour effet de réduire le taux de diffusion des électrons en présence d'impuretés, mais ce déplacement du dopage affecte aussi le profil de bande de par le réarrangement électrostatique des charges. En effet dans le composant de référence il y avait une symétrie entre les distributions d'impuretés et d'électrons. Ce n'est plus le cas dans les composants à dopage déplacé. Il faut déterminer si cette modification du profil de bande joue en faveur ou pas de la réduction de courant. Pour cela j'ai développé un code Poisson-Schrödinger auto-cohérent qui donne accès au profil de bande réel de nos structures.

A paramètres de croissance égaux (taille des puits, hauteur et largeur des barrières), le déplacement du dopage augmente l'intégrale de recouvrement $\langle \Psi_n | \Psi_{n+1} \rangle$, où $\Psi_n$ est la fonction d'onde associée au niveau fondamental du n$^{\text{éme}}$ puits. Pour notre structure cette augmentation est importante : un facteur 3. Cela signifie donc que la modification du profil de bande due au réarrangement électrostatique tend à augmenter le couplage entre puits. L'élément de matrice associé à la diffusion par les impuretés $\langle \Psi_n | \frac{e^2}{4\pi\varepsilon_0\varepsilon_r} \frac{1}{r} | \Psi_{n+1} \rangle$ est bien réduit, mais cela signifie que les éléments de matrice associés à d'autres interactions ont augmenté. Autrement dit une partie de ce qui est gagné sur l'interaction électron impuretés est perdue sur les autres interactions.

Afin de réduire plus fortement le courant il faudrait donc éviter que ce qui est gagné sur un élément de matrice soit perdu sur les autres. Pour cela il serait souhaitable que la modification du profil de dopage se fasse en respectant la symétrie entre électrons et impuretés. Deux solutions pourraient être envisagées. La première pourrait consister à doper le puits sur toute sa longueur (voir FIG. 5-9), au lieu de seulement doper le tiers central. Technologiquement cette solution a l'avantage de la simplicité. Mais en terme de réduction du courant d'obscurité les résultats vont rester modestes puisque le recouvrement entre impuretés et électrons reste fort au centre du puits. La seconde solution consisterait à séparer le créneau de dopage en deux créneaux plus petits, chacun étant placé au bord des puits. Cette fois-ci la réduction en terme de courant d'obscurité sera plus importante, mais cette solution risque de se heurter à des difficultés technologiques. La ségrégation du dopage pourrait en effet être particulièrement défavorable à ce type de profil piqué.

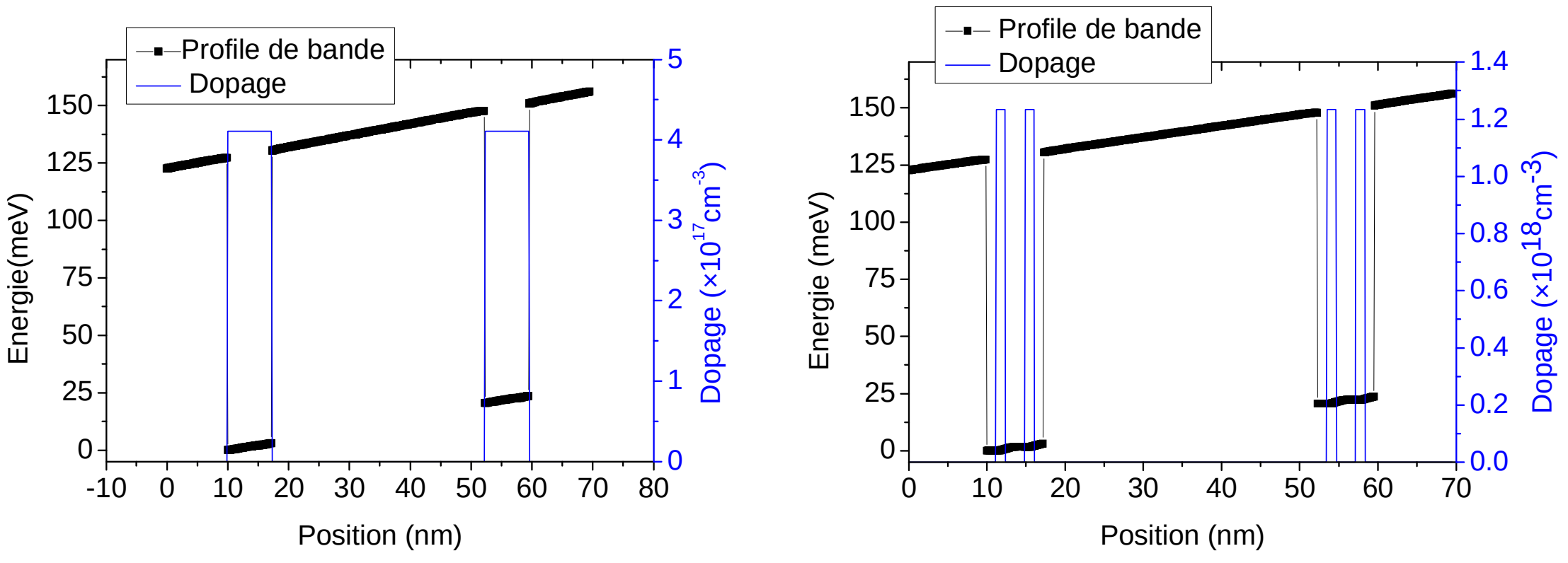

*FIG. 5-9 : A gauche, Profil de bande sous un champ de 5kVcm$^{-1}$, dopage créneau sur toute la largeur du puits. A droite, Profil de bande sous un champ de 5kVcm$^{-1}$,*

*double dopage créneaux entre le 1$^{er}$ et le 2$^{ème}$ sixième du puits, puis entre le 4$^{ème}$ et le 5$^{ème}$ sixième du puits*

### 5.1.3. Conclusion

L'utilisation de notre outil de simulation a donc non seulement permis d'identifier l'interaction à l'origine du transport mais nous a aussi fait comprendre comment en réduire l'intensité. Notre travail ne se limite plus à la « simple » ingénierie de fonction d'onde qui est utilisée d'habitude dans la conception des hétérostructures, mais inclut maintenant de l'ingénierie d'éléments de matrice. Nous avons montré qu'un déplacement du dopage permettait une réduction du courant d'obscurité et ce principe a été vérifié expérimentalement de façon univoque. Une réduction de près de 50% du taux de diffusion associée à l'interaction dominante a ainsi pu être obtenue tout en conservant le rendement quantique quasi inchangé. De nouveaux profils de dopage plus prometteurs ont également été proposés, mais leur réalisation nécessitera certainement une amélioration des techniques de croissance avant de pouvoir être mis en œuvre.

Se pose à présent la question de l'utilisation de cette technique de déplacement du dopage. Il serait en effet possible d'objecter que pour obtenir une réduction similaire du courant un petit allongement de la taille des barrières serait également une solution, cf la section 4.3.3. Il est évident que ces deux techniques sont complémentaires mais il existe des cas ou l'allongement de barrière n'est pas toujours possible. Tout d'abord dans les QCD LWIR, il a été récemment identifié que l'interaction limitante est également celle entre électron et impuretés. Or dans ce cas un changement dans la taille des barrières est difficilement envisageable, un déplacement d dopage semble une solution raisonnable pour augmenter le $R_0A$ de la structure. Par ailleurs dans les QWIP bicolore, la structure la plus basse longueur d'onde opère généralement en régime tunnel. Dans ce cas il peut être difficile d'augmenter la taille des barrières car d'une part l'empilement fait par MBE est déjà important et d'autre part cela pourrait nuire au couplage optique évanescent avec le réseau. Nous avons donc démontré la faisabilité de la méthode, il reste encore à l'implémenter dans des structures complexes.

## 5.2. Non idéalité dans les structures

Nous nous sommes jusque-là intéressés à des structures QWIP idéales, nous avons inclus quelques originalités sur le profil de dopage. Nous voulons dans la suite de chapitre montrer que notre outil de simulation permet d'adresser des problèmes liés à des structures plus complexes. Nous allons donc introduire dans la suite deux sources de complexité. Nous continuerons tout d'abord avec les QWIP mais nous adresserons le problème des structures non idéales du point de vue de leur profil de bande. Puis nous étudierons à la fin de ce chapitre des structures beaucoup plus complexes que sont les QCD.

Dans cette partie je propose d'étudier les effets liés à la non idéalité du profil de bande dans les QWIP. Nous nous sommes focalisés sur deux sources de non idéalité que sont la présence des défauts d'interface et l'introduction de désordre dans la structure. Nous allons voir comment ces deux effets peuvent influencer les propriétés optiques et de transport des QWIP. Une comparaison de l'effet par rapport aux autres écarts à l'idéalité est également proposée. Enfin je finirai cette partie par quelques mots sur la robustesse des structures vis-à-vis des défauts inhérents à leur croissance.

Cette étude résulte de la caractérisation de nos hétérostructures par microscopie électronique en transmission. Cet outil grâce à son grande résolution spatiale et chimique nous a donné accès aux propriétés des interfaces. Ce travail répond surtout au besoin de quantifier l'effet des écarts entre la structure dessinée et celle obtenue en sortie de croissance : Sont-ils négligeables ? Se compensent-ils entre eux ? Peut-on réduire certains effets ? Nos structures sont-elles robustes ? Comment améliorer leur robustesse ? L'idée est donc d'observer ces non idéalités, d'en quantifier l'impact puis d'opérer une rétroaction vers la croissance.

### 5.2.1. Rôles des interfaces

Les performances des détecteurs sont en partie liées à la capacité du moyen de croissance utilisé à reproduire fidèlement le profil de bande nominal. Et cela passe bien sûr par l'obtention d'interfaces d'une grande qualité.

#### *5.2.1.1. Etude des interfaces*

Afin d'évaluer la capacité de notre moyen de croissance à reproduire une hétérostructure dont le profil de bande soit proche de la structure nominale, il faut disposer d'une méthode de caractérisation qui permette l'observation de ces interfaces. Généralement les paramètres d'interface sont déduits de mesures de transport (mobilité[152,153]), ce qui n'est pas possible dans notre cas dans la mesure où c'est ce que nous cherchons à prédire. Il nous faut donc disposer d'une technique présentant une très grande résolution spatiale et qui soit capable de donner des informations sur le cœur du matériau. Et elle doit de plus être capable d'apporter un contraste puits barrière. Il existe bien une technique de caractérisation disposant de niveau de performance, le STEM en mode HAADF. Cette caractérisation a été faite par Gilles Patriarche du LPN.

- *LE STEM*

La STEM pour Scanning Transmission Electron Microscope est un mode particulier de microscopie électronique. Il peut être vu comme le mariage réussi du microscope électronique en transmission haute résolution (HR-TEM) et du microscope électronique à balayage (MEB). La surface de l'échantillon est balayée par un faisceau convergent d'électrons comme dans un MEB, mais l'observation des électrons se fait en transmission comme dans un TEM.

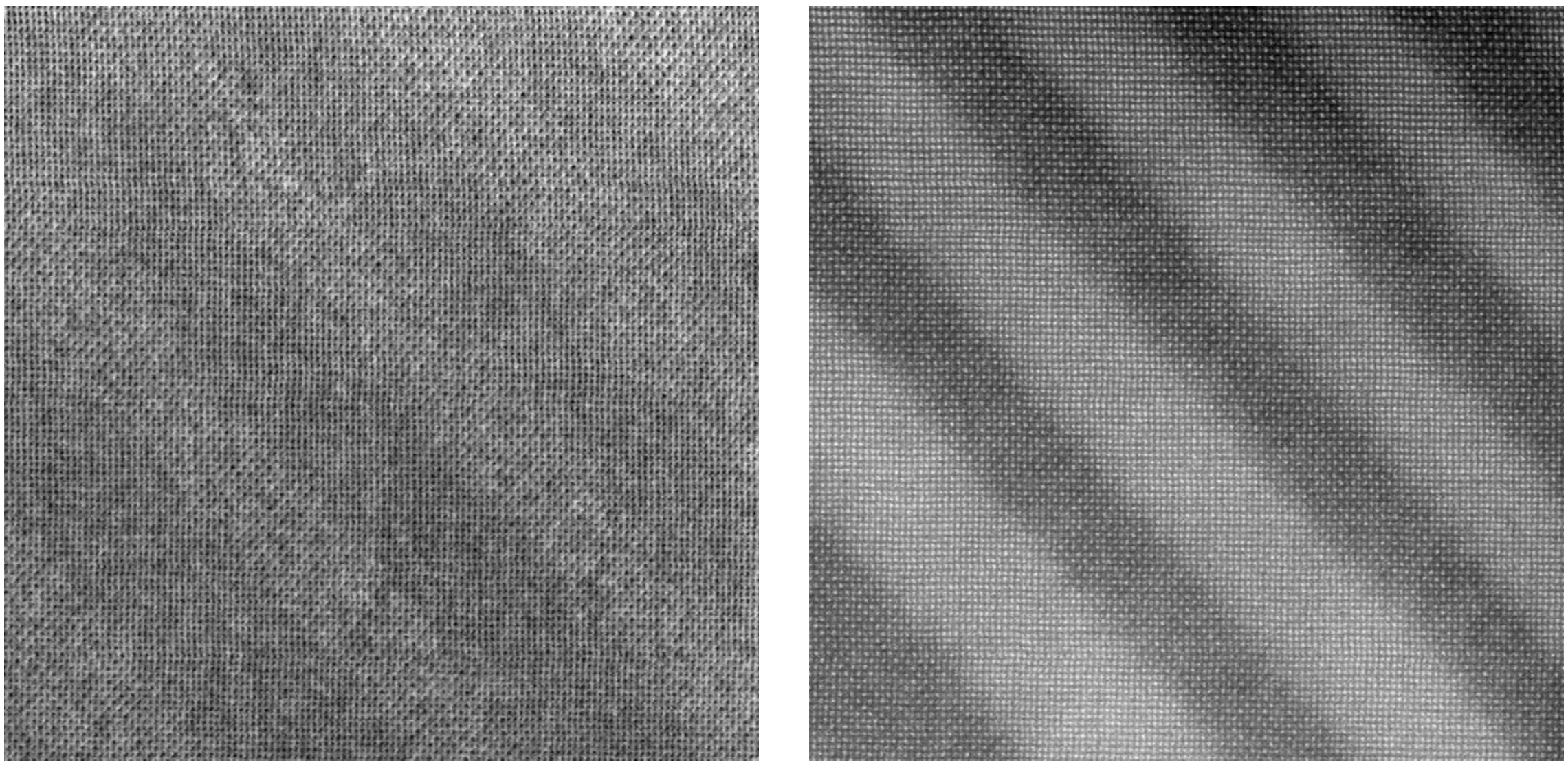

*FIG. 5-10 : A gauche : image STEM classique. A droite : image STEM en mode HAADF. L'image fait typiquement 25nm par 25nm.*

Pour notre étude nous avons utilisé le STEM dans un mode particulier d'imagerie qu'est le HAADF (High Angle Annular Dark Field imaging), cette technique consiste à ne collecter que les électrons qui seront diffusés à travers un anneau. Seuls les grands angles de diffusion sont ainsi captés par le détecteur. L'intérêt d'une telle imagerie réside dans le fait que les électrons diffusés aux grands angles sont ceux diffusés par les noyaux atomiques (plutôt que par les plans de Bragg). Le contraste de ce type d'image est alors très sensible au numéro atomique des atomes diffuseurs (Z contrast imaging). Le mode HAADF donne une image dont le contraste semble varier avec le carré du numéro atomique. L'appareil utilisé est un JEOL JEM 2200FS dans lequel les électrons sont accélérés à 200kV.

Le STEM du LPN est de plus muni d'un correcteur d'aberration sphérique qui lui donne une résolution spatiale autour de l'angström. Valeur qu'il faut comparer aux 1.4Å qui existent entre deux plans atomiques dans la direction 001 de GaAs. Cela signifie donc que l'on peut résoudre avec un tel appareil toutes les colonnes atomiques dans nos hétérostructures.

La figure FIG. 5-10 montre une comparaison de l'imagerie d'une même zone en mode HAADF et en mode classique. Pour conclure le STEM apporte la résolution spatiale nécessaire et le mode HAADF donne le contraste puits barrières.

- *Echantillon utilisé*

L'échantillon étudié lors de cette caractérisation est un QCD 8µm qui a servi dans les références 48 et 167. Il reste pertinent d'utiliser un QCD plutôt que directement notre QWIP dans la mesure où les deux échantillons ont été fabriqués par MBE dans des conditions similaires. Il faut donc s'attendre à ce que les défauts de croissance soient du même type.

- *Visualisation des interfaces*

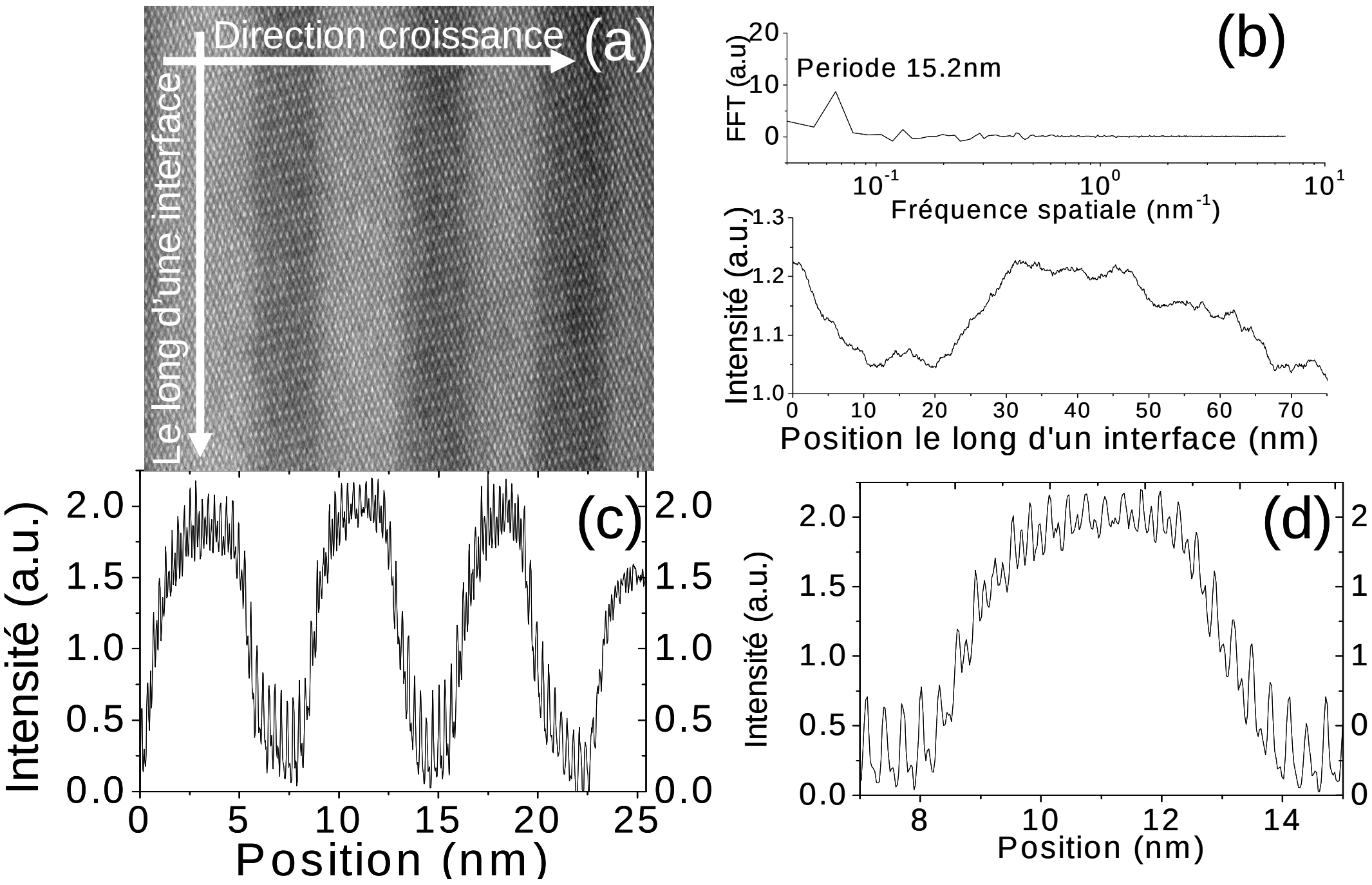

*FIG. 5-11 : A gauche (a) Image STEM HAADF faite dans la direction 100 (b) an bas : Profil d'intensité fait le long du plan des couches. En haut FFT de ce même profil, il est possible d'identifier deux pic de période 15 et 38 nm (c) Profil d'intensité fait le long de l'axe de croissance. (d) zoom sur le profil (c), chaque pic correspond à une colonne atomique.*

Les figures montrent des images de STEM et les profils d'intensité qui leur sont associés. Ces profils ont pu être faits dans la direction de croissance, voir figure FIG. 5-11 (c)et (d) ou le long d'une interface, voir figure FIG. 5-11 (b). Selon la direction de croissance, nous observons que les interfaces sont loin d'être aussi abruptes que nous aurions pu le souhaiter. Le profil semble adouci, il existe donc un gradient de composition dans lequel la composition en aluminium passe continûment de celle de GaAs à celle de l'alliage de AlGaAs souhaitée. Les images dans la direction 100 montrent clairement un pic au niveau de chaque colonne atomique, ce qui nous permet d'évaluer précisément l'extension du gradient de composition aux interfaces, typiquement deux ou trois mono-couches atomiques (6 à 9Å). Ce défaut de composition aux interfaces va se confondre avec le paramètre $\Delta$ de la rugosité d'interface. Les profils d'intensité le long des interfaces vont, quant à eux, nous permettre d'évaluer sur le paramètre $\xi$ de la rugosité d'interface. Nous pouvons observer une fluctuation de la position des interfaces perpendiculairement à la direction de croissance. Cette fluctuation présente une périodicité proche de 10nm (voir FFT sur graphe FIG. 5-11), valeur que nous avons obtenue après moyennage sur de nombreuses interfaces.

Ces valeurs de paramètres de rugosité sont de l'ordre de grandeur de ceux rapportés dans la littérature[152,153], mais plutôt dans la moyenne haute.

Dernier point que je souhaite adresser : les profils des puits présentent une légère asymétrie. Cela laisse donc supposer que le profil d'aluminium subie également la ségrégation.

- *Origine de l'imperfection des interfaces*

L'origine de la rugosité d'interface est clairement à imputer aux fluctuations locales des flux atomiques provenant des sources utilisées dans l'appareil de MBE. L'origine de la non raideur peut paraître un peu plus compliquée à identifier. Les températures de croissance, situées autour de 500-600°C, sont trop faibles pour permettre de la diffusion d'espèce dans le cœur du matériau. Pour obtenir de tels effets il faut se placer généralement au delà de 800°C (dans la gamme 800°C-1100°C)[168]. Il est d'ailleurs possible d'utiliser cette diffusion pour modifier le profil d'aluminium et ainsi ajuster, post-croissance, la longueur d'onde détectée par un QWIP[169,170]. De la diffusion surfacique, également appelée ségrégation, dans laquelle la diffusion se fait seulement entre la dernière couche et l'interface est envisageable à ces températures. Ce processus conduit à des profils asymétriques qui ont tendance à baver dans la direction du réseau. Il doit toutefois rester minoritaire dans notre cas vu que les profils mesurés ne sont que très peu asymétriques.

Nous pensons plutôt que cette non raideur résulte de l'ouverture et de la fermeture des caches de cellules de MBE. Ce processus d'ouverture et de fermeture agit en effet de deux façons sur le flux :

- Bien évidement lors de l'ouverture et de la fermeture du cache, se produit un effet géométrique qui veut que le flux soit partiellement limité par la partie de la cellule qui n'est pas encore ouverte. Ce temps d'ouverture et de fermeture du cache est généralement situé entre 100 et 200 ms[171]. Or la vitesse de croissance est de l'ordre de $2Å\cdot s^{-1}$ ce qui fait que ce processus géométrique ne devrait pas affecter plus qu'une monocouche.
- L'ouverture et la fermeture du cache s'accompagnent également d'une variation de température de la cellule, ce qui à son tour affecte le flux émis. Or le temps typique de ce processus peut atteindre 30 secondes[172], ce qui est donc parfaitement compatible avec l'extension du gradient de composition observé. Il faut néanmoins que cette valeur de 30 secondes est un maximum et qu'elle va fortement dépendre de la taille de la cellule ainsi que du bâti utilisé.

Dans la mesure où le processus qui limite la qualité des interfaces est lié à l'appareil de MBE lui-même, il est donc *a posteriori* justifié d'extrapoler les résultats obtenus sur un QCD vers un QWIP.

Nous avons donc mis en évidence que la qualité des interfaces est directement limitée par la qualité de l'épitaxie. Nous ressentons les limites actuelles de cette méthode de croissance. L'amélioration des performances des détecteurs passera donc soit par une optimisation des procédés de croissance soit par la conception de structures très peu sensibles à ces défauts.

#### 5.2.1.2. Effet du caractère non abrupt des interfaces

Le profil de composition en aluminium étant directement relié à celui en énergie il nous est possible de modéliser le profil réel de nos hétérostructures. Pour cela nous utilisons un profil adouci faisant intervenir la fonction erreur (erf)[165] :

$$V(z) = V_b \left( 1 - 1/2 \sum_{well} erf(\frac{z - Z_r}{L_d}) - erf(\frac{z - Z_l}{L_d}) \right) \qquad (5\text{-}3)$$

Où $Z_r/Z_l$ sont les positions des interfaces des puits et $L_d$ est l'extension de la zone de gradient de composition. Un tel profil de puits est présenté dans l'encart de la FIG. 5-12. Cette même figure montre la dépendance de la longueur d'onde associée à la transition entre le niveau fondamental et le niveau excité avec l'extension du gradient de composition. Ce graphe montre que plus la zone de transition entre puits et barrières est large, plus la longueur d'onde détectée est déviée vers le rouge. Cette déviation de la transition se comprend aisément dans la mesure où un profil plus doux conduit à un confinement moindre de l'électron et donc à des niveaux plus rapprochés en énergie. Comme ce gradient d'interface n'est pas toujours très bien défini (il peut varier d'une interface à l'autre) il s'accompagne d'un élargissement de la transition de 0.15µm, soit 0.9meV (variation de longueur d'onde détecté pour un gradient passant de 2 à 3 monocouches).

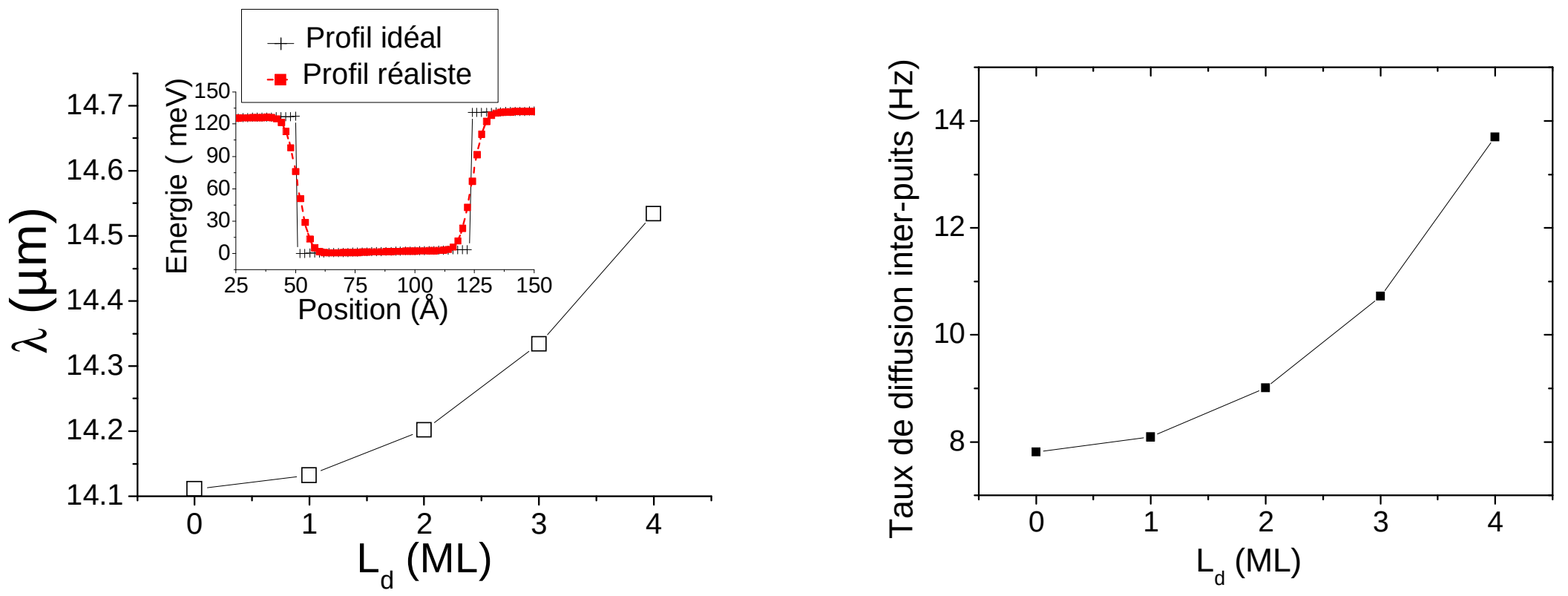

*FIG. 5-12 : A gauche : Longueur d'onde associée à la transition entre le niveau fondamental et le niveau excité en fonction de l'extension du gradient de composition aux interfaces. Encart : Profil de bande d'un puits dans le cas idéal et quand les interfaces ne sont plus rigoureusement abruptes ($L_d$=3ML). A droite : Taux de diffusion inter-puits en fonction de l'extension du gradient de composition aux interfaces*

Intéressons-nous maintenant à la modification des propriétés de transport. Pour cela le graphe FIG. 5-12 montre l'évolution du taux de diffusion inter-puits (somme des différents processus étudiés au chapitre quatre) avec l'extension du gradient de composition. Le taux de diffusion, et donc le courant d'obscurité qui l'accompagne, augmentent quand le profil de bande s'adoucit. L'origine est la même que le déplacement de la longueur d'onde pic à savoir un confinement moindre des électrons.

#### 5.2.1.3. Rôle de la rugosité d'interface

La non raideur est le défaut que peuvent subir les interfaces dans la direction de croissance, mais ces dernières peuvent également être affectées par des inhomogénéités dans le plan. En moyenne cette rugosité d'interface ne conduit pas à une modification de la taille des puits, ni des barrières, elle ne provoque donc pas de déviation de la longueur d'onde détectée. Par contre elle peut provoquer un élargissement de la transition. En estimant le taux de diffusion entre le niveau excité et le fondamental (même procédure qu'au chapitre précédent), nous pouvons avoir une idée de l'élargissement lié à ce processus : environ 2.4meV. Ce qui fait de la rugosité d'interface un des processus d'élargissement de la transition les plus importants.

#### *5.2.1.4. Dépendance des effets liés aux interfaces avec la géométrie de la structure*

Il faut s'attendre à ce que ces effets liés aux interfaces présentent une dépendance avec les propriétés géométriques des structures. Il est important de comprendre cette dépendance pour concevoir des structures plus robustes.

La figure FIG. 5-13 montre l'évolution de l'énergie entre le niveau fondamental et le niveau excité quand l'interface n'est plus abrupte en fonction de la taille du puits, pour une longueur fixée du gradient de composition (trois monocouches). Comme nous pouvions nous y attendre, plus le puits est étroit plus l'effet de la non raideur se fait sentir. La plage d'excursion en terme de taille de puits est modeste, car je voulais conserver seulement deux niveaux par puits. Dans cette gamme de taille le comportement semble être linéaire avec la dimension du puits. Afin de faire varier la taille des puits sur une plus large gamme, j'ai tracé la variation d'énergie due à la non raideur pour trois structures de longueur d'onde pic très différentes. Ces trois structures ($D_4$, $D_5$ et $D_6$) sont présentées dans le tableau tab. 5-3. Leurs longueurs d'onde pics vont du MWIR au THz. Cela passe bien sûr par un élargissement de la taille du puits mais aussi des barrières moins hautes. De façon plus générale cette augmentation de la longueur d'onde va de paire avec un confinement moindre des électrons. L'effet de la non raideur des interfaces est quant à lui d'autant plus fort que ce confinement est fort, voir figure FIG. 5-13.

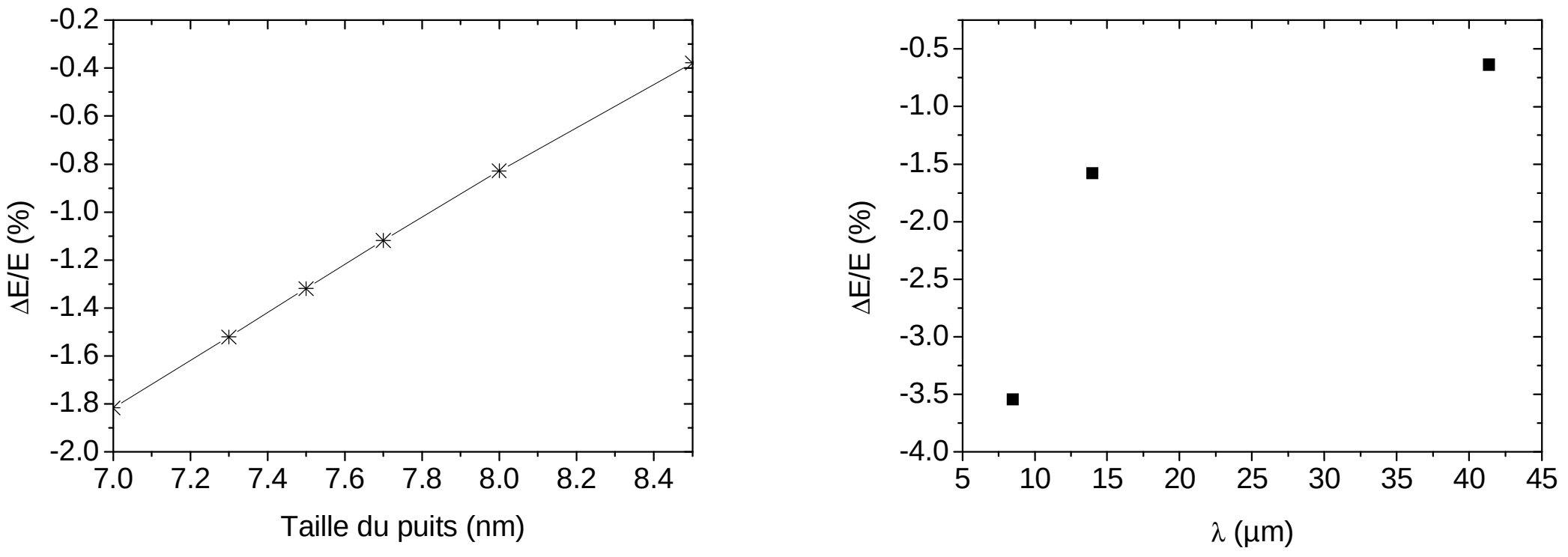

*FIG. 5-13, A gauche : Différence d'énergie de transition entre les cas où les interfaces sont non abruptes et celui ou elles le sont, normalisée par l'énergie dans le cas idéal (interfaces raides) en fonction de la taille du puits. Le gradient de composition aux interfaces s'étend ici sur trois monocouches. A droite : Différence d'énergie de transition entre le cas où les interfaces sont non abruptes et celui ou elles le sont, normalisée par l'énergie dans le cas idéal pour les trois structures décrites dans le tableau tab. 5-3. Le gradient de composition aux interfaces s'étend ici sur trois monocouches.*

| Echantillon | $D_4$ | $D_5$ | $D_6$ |
|---|---|---|---|
| $L_w$ (nm) | 5 | 7.3 | 11.9 |
| $L_b$ (nm) | 35 | 35 | 55.2 |
| %Al | 26 | 15.2 | 5 |
| Longueur d'onde pic (µm) | 8.5 | 14.5 | 41[vii] |
| Gamme de détection | LWIR | VLWIR | THz |
| Référence | 173 | | 35 |

*tab. 5-3 Paramètres de croissance des trois composants étudiés dans la figure FIG. 5-13.*

Pour conclure sur la dépendance des effets d'interface par rapport à la géométrie de la structure, nous pouvons affirmer que plus une structure est confinée plus elle est sensible à une perturbation de ce confinement. Cela va donc induire certaines règles de conception des hétérostructures.

### 5.2.2. Rôle du désordre

Nous avons précédemment discuté des effets liés aux défauts d'interface, et nous avons identifié que ces défauts résultaient d'imperfections liées à la technique de croissance. Il existe toutefois d'autres sources de non idéalité indépendantes de la qualité de réalisation du composant. C'est en particulier le cas du désordre. Nous avons déjà évoqué précédemment le rôle joué par le désordre d'alliage : la distribution aléatoire des atomes d'aluminium dans la matrice de GaAs pouvait être à l'origine de diffusion entre états. Ici nous allons nous intéresser plus particulièrement au désordre lié à la présence de charges. Nous allons en particulier voir que la répartition aléatoire des charges au sein de la zone dopée peut également affecter les propriétés optiques et de transport des détecteurs QWIP.

Pour modéliser le désordre lié à la présence d'impuretés nous devons prendre en compte le profil de dopage, mais aussi ses fluctuations spatiales dans le plan. Nous avons choisi de suivre une procédure proche de celle suggérée par Metzner et al[174] et Willenberg et al[175]. Tout d'abord, nous supposons que les fluctuations latérales de potentiel sont petites par rapport aux variations dans le sens de la croissance. Ainsi, le potentiel peut être divisé en deux parties $V(z,r) = \langle V(z) \rangle + \Delta V(z,r)$ où $\langle V(z) \rangle$ est le potentiel biaisé dans la direction de la croissance en raison de la composition de l'alliage et $\Delta V(z,r)$ la fluctuation latérale due au désordre de charges. Nous faisons l'hypothèse que l'approximation de la fonction enveloppe reste valide, c'est-à-dire que l'on peut encore écrire la fonction d'onde comme le produit d'une fonction d'enveloppe et d'une onde plane latérale. L'électron qui traverse la structure ressent un potentiel qui dépend de sa position dans le plan. Plutôt que d'évaluer la fonction d'onde pour différentes positions dans le plan, nous préférons fixer une position du plan et faire un tirage aléatoire de la distribution de charges vue par notre électron.

---

[vii] Cette valeur ne tient pas compte de l'énergie d'échange ni d'effet de dépolarisation.

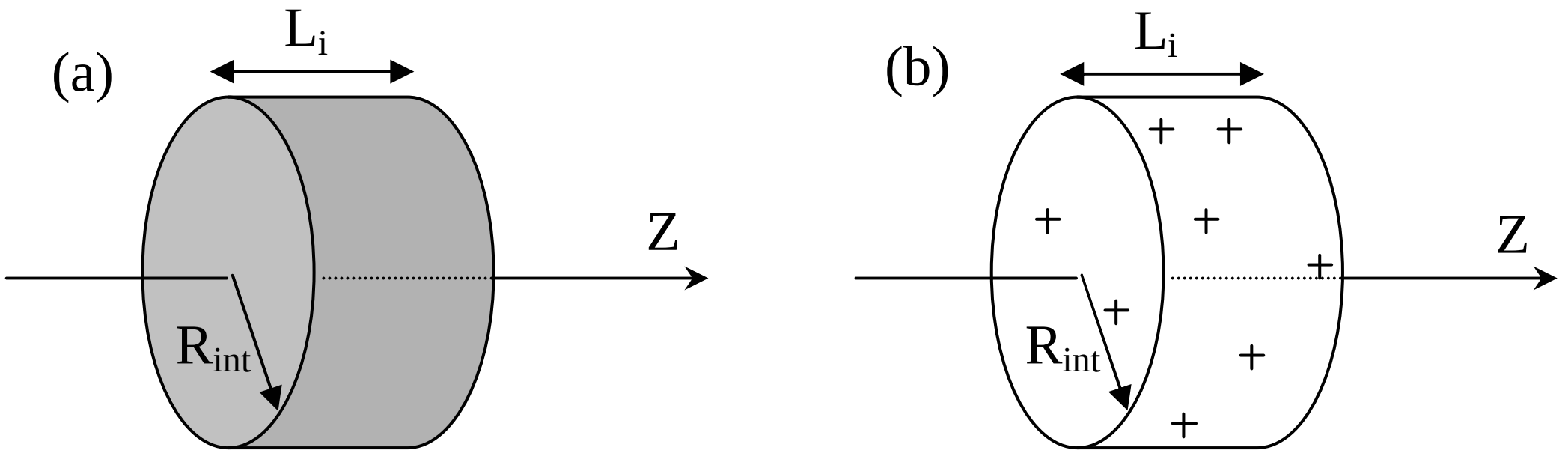

*FIG. 5-14 (a) Volume d'intégration du potentiel écranté dans le cas d'une distribution homogène de charges. (b) Volume d'intégration du potentiel écranté dans le cas d'une distribution discrète de charges.*

Le potentiel écranté est traité de la façon suivante. Nous évaluons le potentiel le long de l'axe z due à un cylindre homogène dopé, voir FIG. 5-14(a). Le potentiel associé est donné par:

$$V_{\text{homogène}}(z) = -\frac{n_{3D}e^2}{4\pi\varepsilon\varepsilon_0}\int_0^{L_i}\int_0^{R_{\text{int}}} 2\pi r'dr'dz'\frac{e^{-q_o r}}{r} \qquad (5\text{-}4)$$

Où $L_i$ est l'extension du dopage dans le sens de la croissance, $r = \sqrt{(z-z')^2 + r'^2}$ et le vecteur d'onde $q_0$ de Thomas-Fermi est donné par $q_0^{\ 2} = \frac{e^2 n_{3D}}{\varepsilon_r \varepsilon_0 k_b T}$, avec $n_{3D}$ la densité du volume, $\varepsilon_r$ la permittivité relative, $\varepsilon_0$ la permittivité du vide, $k_b$ la constante de Boltzmann et T la température. Ici, nous avons considéré que l'interaction de Coulomb est écrantée dans une approche de Thomas-Fermi et est donc donnée par un potentiel de Yukawa. En raison de l'écrantage du potentiel le rayon de la distribution de charges reste limitée : $R_{\text{int}}$ est de l'ordre de quelques dizaines de nanomètres[viii]. Le potentiel lié au caractère discret des charges est donné par, voir la FIG. 5-14 (b).

$$V_{\text{inhomogène}} = -\frac{e^2}{4\pi\varepsilon\varepsilon_0}\sum_{charges}\frac{e^{-q_o r}}{r} \qquad (5\text{-}5)$$

Nous prenons soin de maintenir inchangé le nombre de charges pris en considération pour les deux distributions. En général, le potentiel de désordre devient indépendant du nombre de charges si celui est plus grand que dix par puits. Ainsi, le potentiel lié au désordre est donné par la différence entre ces deux contributions $V_{désordre} = V_{\text{inhomogène}} - V_{\text{homogène}}$. Comme le confinement moyen de l'électron doit être changé, nous annulons la valeur moyenne de cette distribution[174]. Le potentiel du désordre $V_{désordre}$ est alors donné par une fonction en escalier, où la longueur de

[viii] Le rayon de Bohr dans GaAs est de l'ordre de 10nm, même si dans notre cas la longueur pertinente est plutôt $1/q_0$.

chaque marche est égale à une monocouche. Comme nous pouvions nous y attendre, le potentiel de désordre est important à proximité de la région de dopage, mais il s'étend au delà de la zone de dopage, voir figure FIG. 5-15. Une modélisation gaussienne de la distribution liée au désordre conduit à un écart-type de 8 nm. Ce qui est trois fois plus que la taille de la région de dopage nominal. Enfin, pour résoudre l'équation de Schrödinger, ce potentiel de désordre est ajouté au potentiel idéal $\langle V(z) \rangle$.

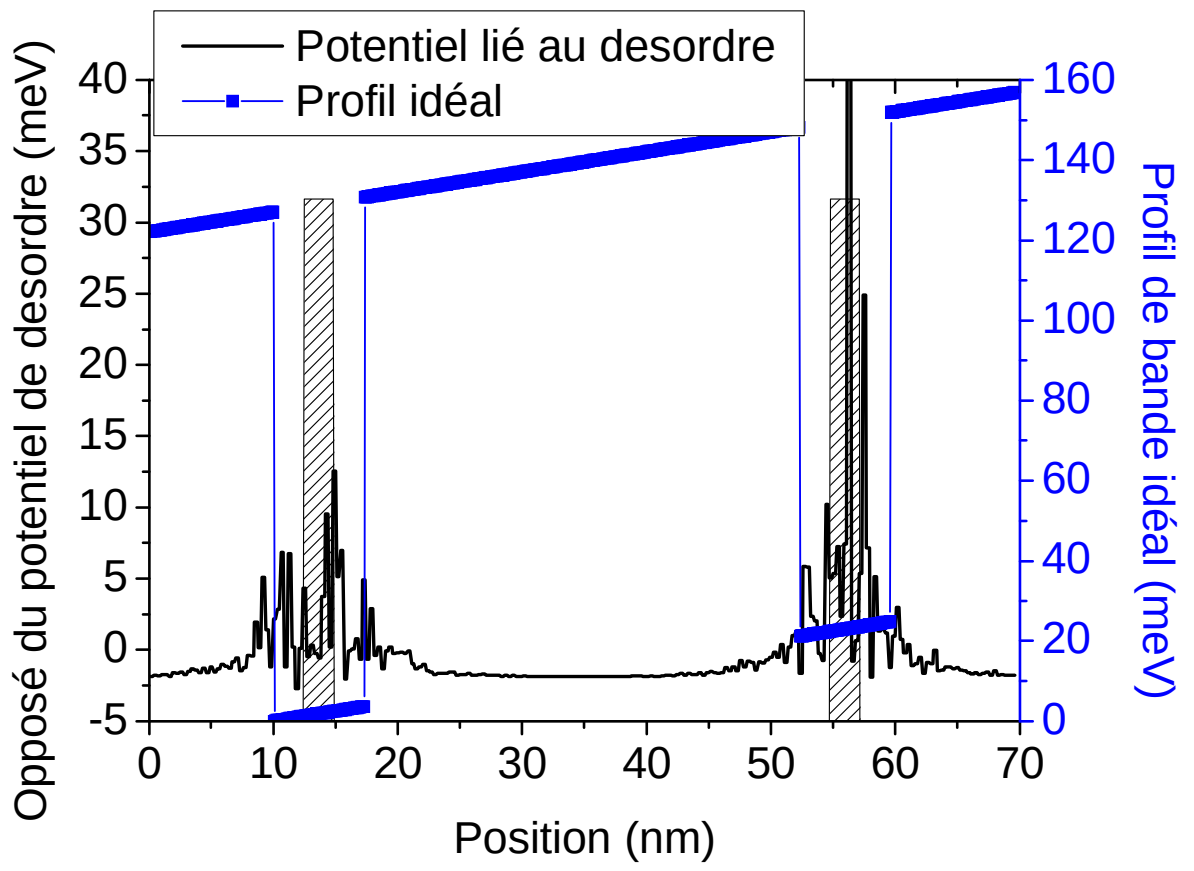

*FIG. 5-15 : Opposé du potentiel lié au désordre pour une distribution de charges tirée au hasard (N=15 charges par puits - R~40nm). Les rectangles striés représentent la position du dopage.*

L'effet de ce potentiel de désordre sur la réponse spectrale (distribution normalisée de la quantité $E_2$-$E_1$) est donné par la figure FIG. 5-16. Nous constatons que la réponse est déviée vers le bleu de 0.5µm et qu'elle subit un élargissement spectral de 0.4µm (3meV). Bien que la valeur moyenne de ce potentiel de désordre soit nulle, il tend à augmenter le confinement des électrons au niveau du dopage et donc du centre du puits, voir figure FIG. 5-15, ce qui augmente l'espacement en énergie des sous-bandes. Il est intéressant de noter que le désordre ne modifie pas de la même façon toutes les sous-bandes, la figure FIG. 5-16 montre que le niveau fondamental est clairement plus affecté. C'est donc sur lui que devrons se porter les efforts de conception des structures. Cela peut par exemple se traduire par le fait qu'il peut être assez favorable d'utiliser des états excités dans un QCD, surtout si il est fortement dopé, plutôt que seulement des états fondamentaux.

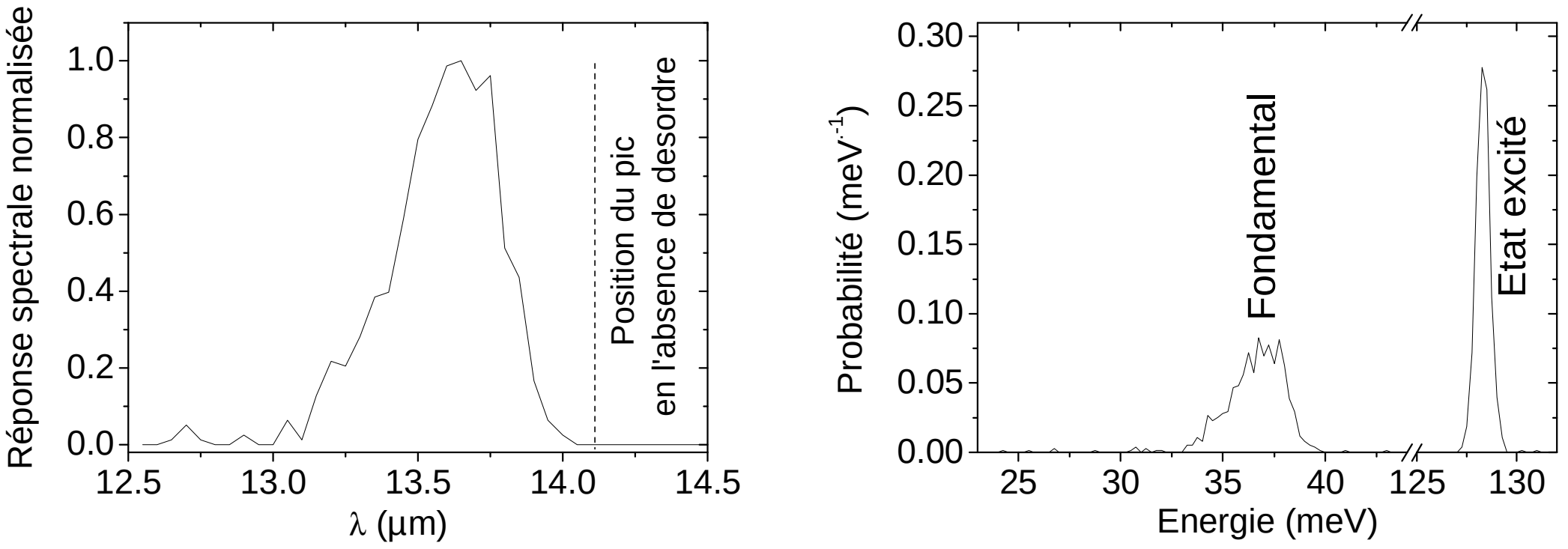

*FIG. 5-16 : A gauche : Réponse spectrale (distribution normalisée de la quantité $E_2$-$E_1$) du QWIP en présence de désordre lié aux charges. 700 configurations de charges ont été utilisées pour ce graphe. A droite : Probabilité de présence des niveaux fondamentaux et excités en fonction de l'énergie. 700 configurations de charges ont été utilisées pour ce graphe.*

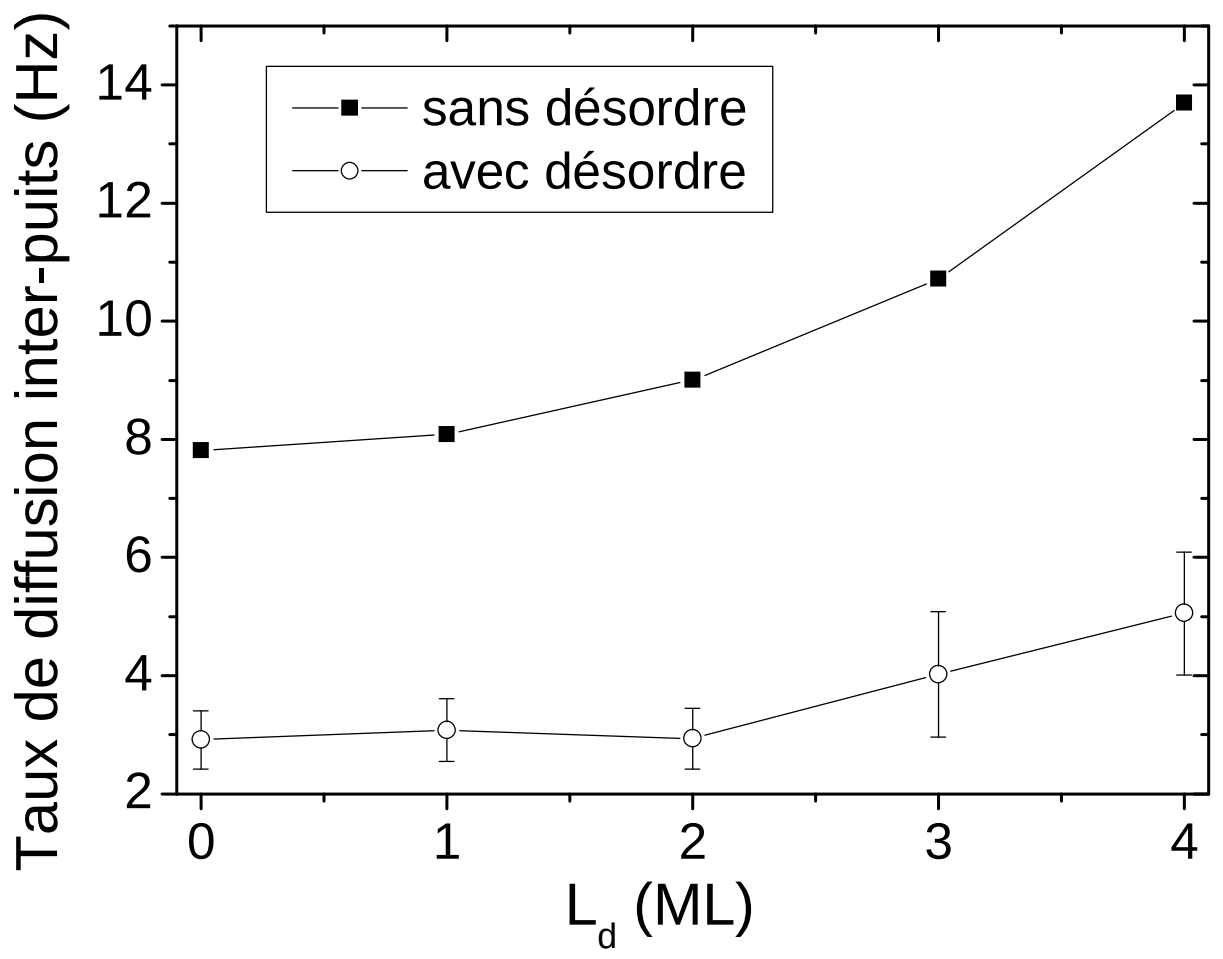

*FIG. 5-17 Taux de diffusion inter-puits avec et sans désordre en fonction de l'extension du gradient de composition aux interfaces*

L'effet du désordre sur les propriétés de transport est donné par la figure FIG. 5-17, il tend à diminuer le courant d'obscurité par rapport au cas idéal et cela résulte simplement du confinement supplémentaire qu'il apporte.

### 5.2.3. Autres sources d'élargissement

Nous avons donc vu que les non idéalités que sont les défauts d'interface et le désordre induisaient un élargissement de la transition optique. Toutefois ces deux sources d'élargissement sont insuffisantes pour expliquer la valeur expérimentale observée (au chapitre deux nous avions estimé que dans ce composant 15µm l'élargissement était de 2.1µm, soit autour de 12meV). Il doit donc exister d'autres sources d'élargissement, qui soient homogènes ou non.

#### 5.2.3.1. Elargissement inhomogène

En plus des fluctuations des dimensions géométriques et du désordre, il peut exister dans les QWIP une autre source d'élargissement inhomogène qui est la fluctuation du champ électrique. Il est généralement admis que le champ n'est pas parfaitement homogène sur l'intégralité de la structure[176]. En particulier il peut être plus élevé à proximité de l'injecteur[177]. Or la réponse spectrale du QWIP dépend du champ électrique appliqué. Cela résulte de la dépendance des fonctions d'onde et donc de l'élément de matrice dipolaire avec le champ. Cette dépendance reste toutefois faible. Nous avons estimé que pour une inhomogénéité de champ de l'ordre de 20%[178], l'élargissement associé était de 0.1meV.

#### 5.2.3.2. Elargissement homogène

Je n'ai pour l'instant adressé que les sources d'élargissement inhomogène, or il existe dans les QWIP un élargissement homogène important qui résulte des différents processus de diffusion, qu'ils soient élastiques ou non.

Notre code de simulation permet simplement d'estimer les contributions liés aux phonons (LA et LO), le désordre d'alliage, la diffusion par une population homogène d'impuretés ionisées. Leurs contributions sont respectivement de 0.01meV, 1.6meV, 0.9meV et de 0.1meV. L'élargissement lié aux effets à grand nombre d'électrons a été estimé en utilisant la formule analytique décrite dans les références 29 et 70. Il semble que cette interaction électron-électron soit celle qui donne la contribution majoritaire à l'élargissement (5meV), mais cette formule a généralement tendance à surestimer cette composante.

Enfin j'ai également estimé la contribution à l'élargissement imputable à la non parabolicité des bandes : celle-ci reste minoritaire (0.2meV)

| Processus | Elargissement (meV) |
|---|---|
| Phonon LA | $1.1\times10^{-2}$ |
| Impuretés ionisées | 0.1 |
| Champ électrique | 0.1 |
| Non parabolicité | 0.2 |
| Désordre d'alliage | 0.9 |
| Interface non abrupte | 0.9 |
| Phonon LO | 1.6 |
| Rugosité d'interface | 2.4 |
| Désordre lié aux charges | 3 |
| Inter-electron | 5 |

*tab. 5-4 Elargissement spectral lié à différents processus possibles.*

Le tableau tab. 5-4 récapitule les différentes sources d'élargissement envisagées. La somme des ces différents processus conduit à un élargissement théorique de 14.2meV, en assez bon accord avec la valeur expérimentale de 12meV. Certains trouveront suspicieux que nous trouvions une valeur théorique supérieure à la valeur expérimentale, mais j'attribue cela à une surestimation de la composante due à l'interaction électron-électron.

### 5.2.4. Conclusion sur la robustesse des structures

Le STEM apparaît comme un excellent outil d'analyse de la qualité des interfaces des hétérostructures. Grâce à sa grande résolution spatiale nous avons pu estimer précisément la raideur des interfaces ainsi que les paramètres liés à la rugosité des interfaces. Nous avons mis en évidence que les non-idéalités d'interface conduisaient à un déplacement de la longueur d'onde détectée vers le rouge ainsi qu'à un courant d'obscurité plus important. Ces deux effets résultent du moindre confinement des électrons quand les interfaces ne sont pas rigoureusement abruptes.

Nous avons ensuite étudié l'impact dû au désordre de répartition des charges qui conduit au contraire à une diminution du courant d'obscurité et à un déplacement vers le bleu de la transition photonique. A nouveau ces effets résultent d'une modification du confinement électronique.

Le désordre et les défauts d'interface conduisent donc à des effets antagonistes qui tendent à se compenser mutuellement. Il ne sera donc pas nécessaire de prendre en compte tous ces effets à moins de vouloir estimer les performances d'un détecteur à mieux que quelques pourcents. Néanmoins le concepteur d'hétérostructures devra apporter un soin particulier au positionnement des niveaux les plus confinés dans la mesure où ce sont eux qui sont les plus sensibles aux défauts.

## 5.3. Transport dans les QCD THz

### 5.3.1. Design du QCD THz

L'utilisation de notre code pour les QCD est un défi dans la mesure où ces structures sont beaucoup plus complexes (grand nombre de puits et barrières) et que l'on perd de nombreuses propriétés de périodicité utilisables dans les super-réseaux. La modélisation du transport dans les QCD était particulièrement bien maîtrisée dans le moyen infrarouge. Ainsi Koeninger et al[49] ont réussi à modéliser fidèlement la résistance de QCD autour de 8µm sur près de huit ordres de grandeurs. Toutefois l'extension de l'utilisation de ces structures à la nouvelle gamme de fréquence qu'est le térahertz remet en cause cette capacité de modélisation dans la mesure où le transport via l'interaction electron-phonon optique n'est plus possible. L'utilisation de mon code devrait donc permettre de déterminer l'interaction à l'origine du transport dans ces détecteurs et cela à des fins de modélisation et d'amélioration des structures.

Ce travail est le fruit d'une étroite collaboration quadripartite. D'une part le laboratoire MPQ (Amandine Buffaz) et le 3-5 lab (Mathieu Carras) ont conçut la structure. L'ONERA intervient pour la phase de modélisation et enfin le LPN (Gilles Patriarche) s'est vu confier la caractérisation par microscopie électronique d'échantillons. La recherche sur les détecteurs QCD en régime THz débute juste. Une seule autre équipe est travaille sur ce sujet, voir les travaux de Graaf et al[179].

La structure que A. Buffaz et M. Carras ont proposée est une structure à quarante périodes composée de GaAs-$Al_{0.27}Ga_{0.73}As$. Chaque période contient cinq puits couplés avec la succession barrières **puits** suivante : 75/**52**/30/**48**/50/**44**/60/**50**/68/**54** (épaisseurs données en Å). Les deux premiers puits de chaque période sont dopés dans leur tiers central par du silicium à hauteur de $3\times10^{17}cm^{-3}$. La structure est représentée sur la figure FIG. 5-18. La détection dans cette structure se fait depuis le niveau $E_1$ vers les niveaux $E_4$ et $E_5$, ce qui conduit à un pic d'absorption autour de 70µm (4.3 THz).

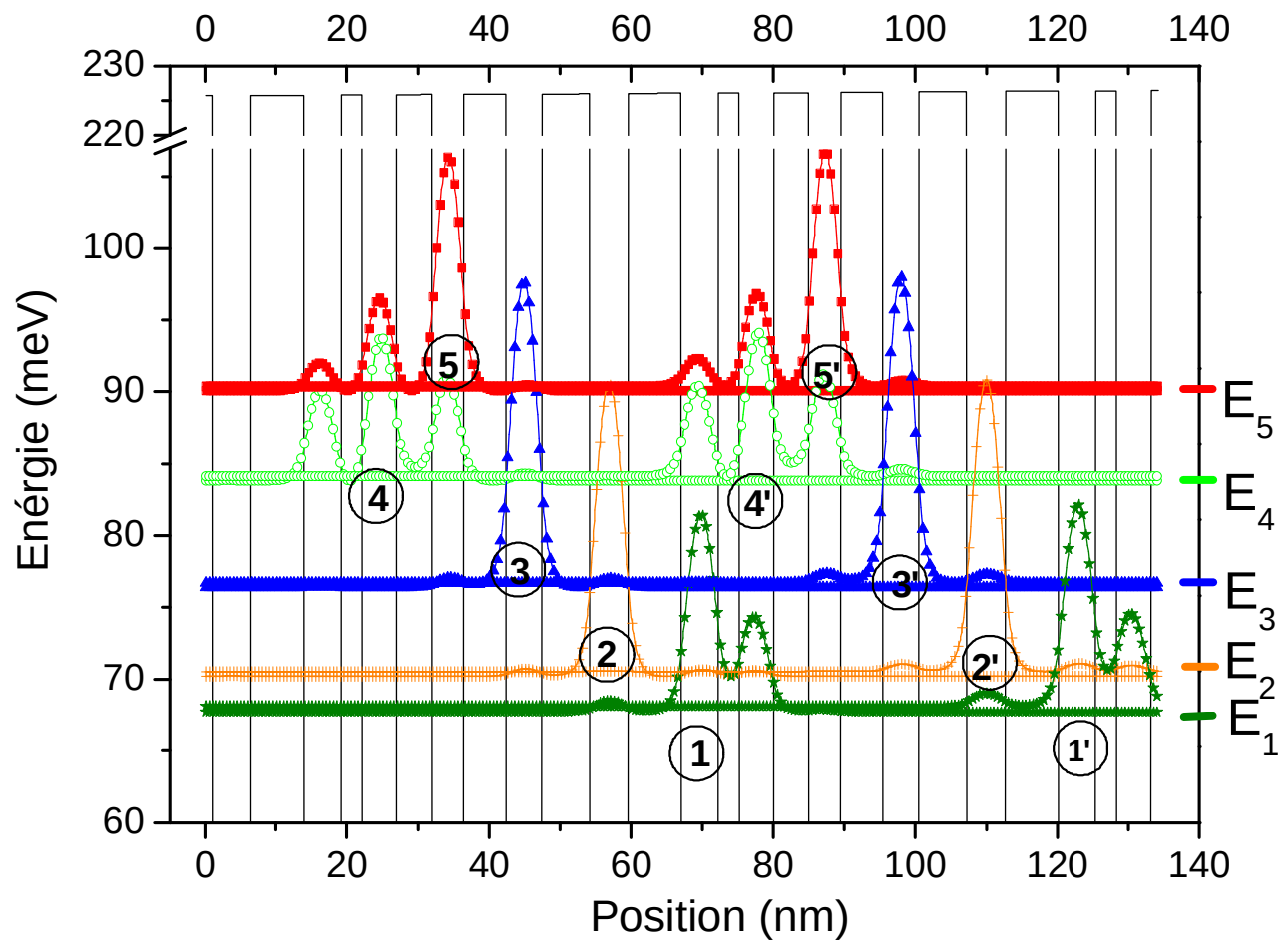

*FIG. 5-18 : Profil de bande et fonctions d'onde de la structure QCD THZ pour deux périodes consécutives.*

Dans le moyen infrarouge les QCD fonctionnent sur un couplage basé sur l'interaction électron-phonon optique. Or l'énergie échangée en régime THz devient plus faible que celle du phonon dans GaAs. Le couplage par phonon optique n'est donc plus possible. Il est alors primordial d'identifier l'interaction à l'origine du transport. Pour cela nous avons calculé le taux de transition associé à différentes transitions pour les processus inclus dans notre code de simulation. Les résultats sont donnés dans tab. 5-5. Le taux de diffusion lié à la rugosité d'interface domine clairement les autres d'au moins une décade. D'un point de vue modélisation une seule interaction aura donc besoin d'être pris en compte. Par contre les performances des échantillons risquent de fortement dépendre de la qualité des interfaces.

| Interac-tion | $1/\tau_{54}$ | $1/\tau_{43}$ | $1/\tau_{32}$ | $1/\tau_{53}$ | $1/\tau_{42}$ | $1/\tau_{31}$ | $1/\tau_{5'1}$ | $1/\tau_{4'1}$ |
|---|---|---|---|---|---|---|---|---|
| Alliage | $9.5\times10^{10}$ | $8.3\times10^{9}$ | $1\times10^{10}$ | $1.2\times10^{10}$ | $2.1\times10^{8}$ | $3.5\times10^{8}$ | $3\times10^{10}$ | $6.3\times10^{10}$ |
| Rugosité | $4.2\times10^{12}$ | $3.4\times10^{11}$ | $4.5\times10^{11}$ | $4.1\times10^{11}$ | $7.2\times10^{9}$ | $1.4\times10^{10}$ | $9.9\times10^{11}$ | $2.4\times10^{12}$ |
| impuretés | $1.5\times10^{11}$ | $6.3\times10^{9}$ | $7.5\times10^{10}$ | $1.7\times10^{9}$ | $2.5\times10^{8}$ | $3.1\times10^{9}$ | $3.1\times10^{7}$ | $2.3\times10^{8}$ |
| Electron | $1.4\times 10^{9}$ | $2.4\times 10^{8}$ | $5.3\times 10^{8}$ | $2.4\times 10^{8}$ | $1.3\times 10^{6}$ | $4.6\times 10^{6}$ | $1.3\times 10^{8}$ | $8.3\times 10^{8}$ |
| Phonon AC | $5.7\times10^{9}$ | $4.7\times10^{8}$ | $7.1\times10^{8}$ | $5.5\times10^{8}$ | $1.2\times10^{7}$ | $2.3\times10^{7}$ | $2.1\times10^{9}$ | $4.8\times10^{9}$ |
| Phonon LO | $3\times10^{-7}$ | $2.6\times10^{-8}$ | $4.3\times10^{-8}$ | $2.8\times10^{-8}$ | $6\times10^{-10}$ | $1.3\times10^{-9}$ | $9.1\times10^{-8}$ | $2.1\times10^{-7}$ |

*tab. 5-5 : Taux de transition associés à différentes transitions intra ou inter périodes pour les différents processus. Le vecteur d'onde initial a été pris nul. La température est prise égale à 10K. L'exposant ' indique que les deux niveaux ne sont pas dans la même cascade.*

### 5.3.2. Estimation de la rugosité

Avant d'aller plus loin dans l'estimation des propriétés de transport il est nécessaire une fois encore d'estimer la qualité des interfaces. L'échantillon QCD THz n'étant pas encore fabriqué au moment de ces simulations il nous fallait disposer d'une structure proche pour faire les mesures de microscopie électronique. Nous disposions par contre de QCD autour de 8µm, ceux qui ont été étudiés dans la seconde partie de ce chapitre. Nous allons donc à nouveau utiliser ces mesures et les extrapoler à notre structure THz. Nous avons fait l'hypothèse que ces deux QCD auraient des propriétés d'interface assez proches. Et deuxièmement les structures sont également proches (pourcentage d'aluminium autour de 30% dans les deux cas). Ce qui légitime d'extrapoler les propriétés d'interfaces obtenues sur un QCD moyen infrarouge à un QCD lointain infrarouge.

Nous prendrons donc les paramètres $\Delta$ et $\xi$ respectivement égaux à 2-3ML et 10nm, voir figure FIG. 5-11.

### 5.3.3. Transport dans les QCD THz

Nous avons identifié le rôle déterminant joué par la rugosité d'interface sur le transport dans les QCD THz. Nous pouvons à présent déduire de nos simulations certaines propriétés de ce QCD et prédire le niveau de performance qu'il pourra atteindre.

Pour cela il est utile de définir le taux de diffusion net entre deux états qui prend en compte les termes de population de l'état initial et de l'état final. Celui-ci est défini de la façon suivante :

$$G_{ij} = \int_0^{\infty} DOS_{2D} \cdot f_{FD}(E_i) \cdot (1 - f_{FD}(E_f)) \cdot \frac{1}{\tau_{ij}(E_i - E_f)} \cdot dE_i \qquad (5\text{-}6)$$

Il donne le nombre d'électrons qui passe par seconde entre deux états. Une étude approfondie des taux de diffusion nets dans la structure a montré que dans ce QCD THz, les sauts entre états se font marche par marche. Cela signifie que l'ensemble des états de la cascade sera visité. C'est une différence majeure avec le QCD moyen infrarouge dans la mesure où le couplage par phonon optique autorisait le fait de sauter un état de la cascade. Avec une interaction élastique comme la rugosité d'interface, il est défavorable de gagner trop d'impulsion, ce qui favorise les petits sauts en énergie.

En ce qui concerne l'évaluation des performances nous avons choisi comme critère de mérite le $R_0A$. La résistance d'un composant photovoltaïque est en effet déterminante quant au niveau de bruit et de courant d'obscurité dans le composant. Il a été montré précédemment que le $R_0A$ est relié aux taux de diffusion nets par la relation suivante[49] :

$$R_0 A = \frac{k_b T}{e^2 \sum_{i \in C} \sum_{j \in C'} G_{ij}} \qquad (5\text{-}7)$$

Où C et C' sont deux cascades consécutives. Il est alors possible de tracer le graphe $R_0A=f(T)$, voir FIG. 5-19. Quatre processus sont montrés sur ce graphe et comme nous pouvions nous y attendre la rugosité d'interface est celle qui présente le $R_0A$ le plus faible, c'est donc bien ce processus qui détermine le niveau de performance du composant. Notons que je n'ai pas présenté la courbe due au phonon optique qui serait beaucoup plus haute sur ce graphe. Enfin la courbe liée à l'interaction électron-électron n'est pas non plus présente mais pour une raison numérique. Le calcul d'une telle courbe aurait pris plusieurs mois de calcul avec les moyens dont je dispose ce qui n'était donc pas envisageable.

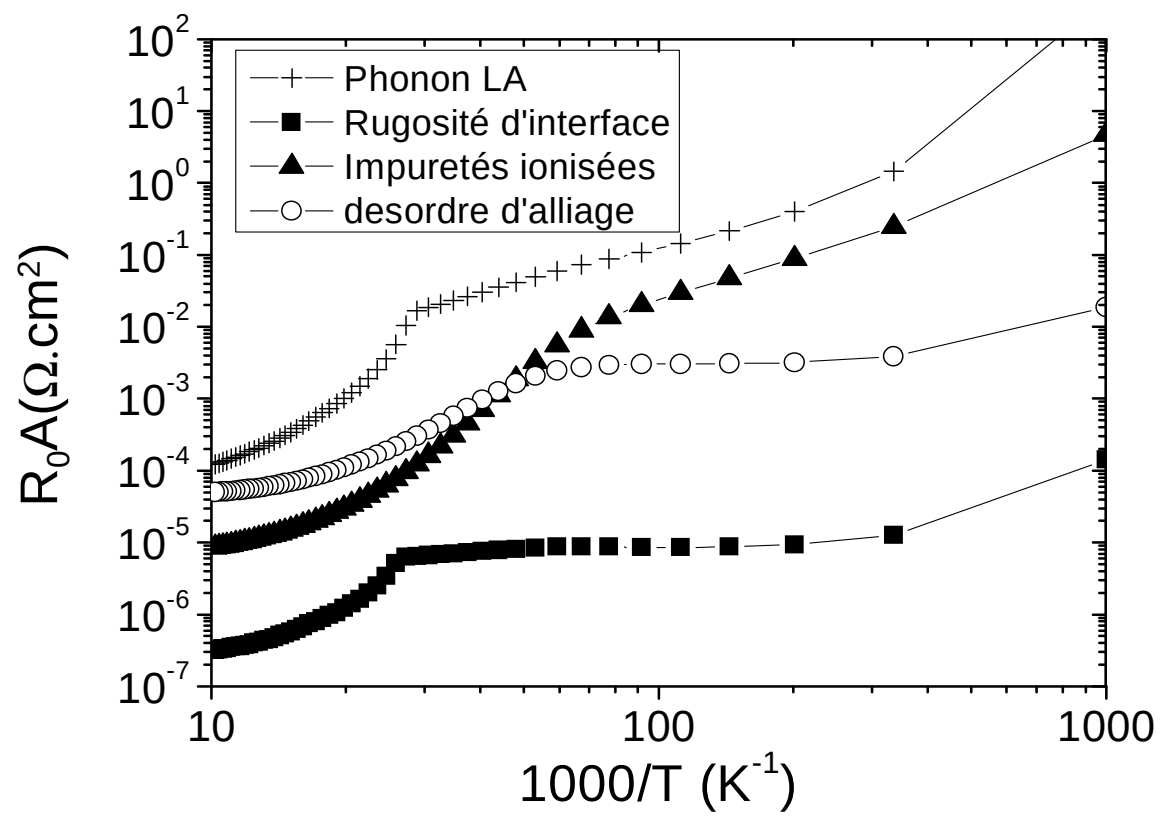

*FIG. 5-19 : $R_0A$ par période en fonction de la température pour quatre processus de diffusion.*

La figure FIG. 5-19 tient compte de l'évaluation qui a été faite par STEM des paramètres de rugosité. Il est toutefois pertinent d'évaluer la sensibilité du $R_0A$ avec ces paramètres. Les figures FIG. 5-20 montrent respectivement l'évolution de la courbe $R_0A(T)$ pour différentes valeurs de Δ et ξ (voir 4.1.3.4). L'évolution avec la hauteur des fluctuations est assez intuitive dans la mesure où plus ceux-ci sont importants, plus la résistance est faible, le tout se faisant avec une dépendance quadratique. L'évolution de $R_0A(T)$ avec ξ est beaucoup moins intuitive. En effet dans notre gamme de valeurs de ξ et pour les énergies échangées dans cette structure, le $R_0A$ augmente en rapprochant les défauts donc en augmentant leur concentration surfacique. Ce comportement résulte du fait que la rugosité d'interface se comporte comme un filtre à impulsion (le terme $\exp(-Q^2\xi^2)$, voir équation du chapitre 4 dont il est difficile de prédire *a priori* le comportement.

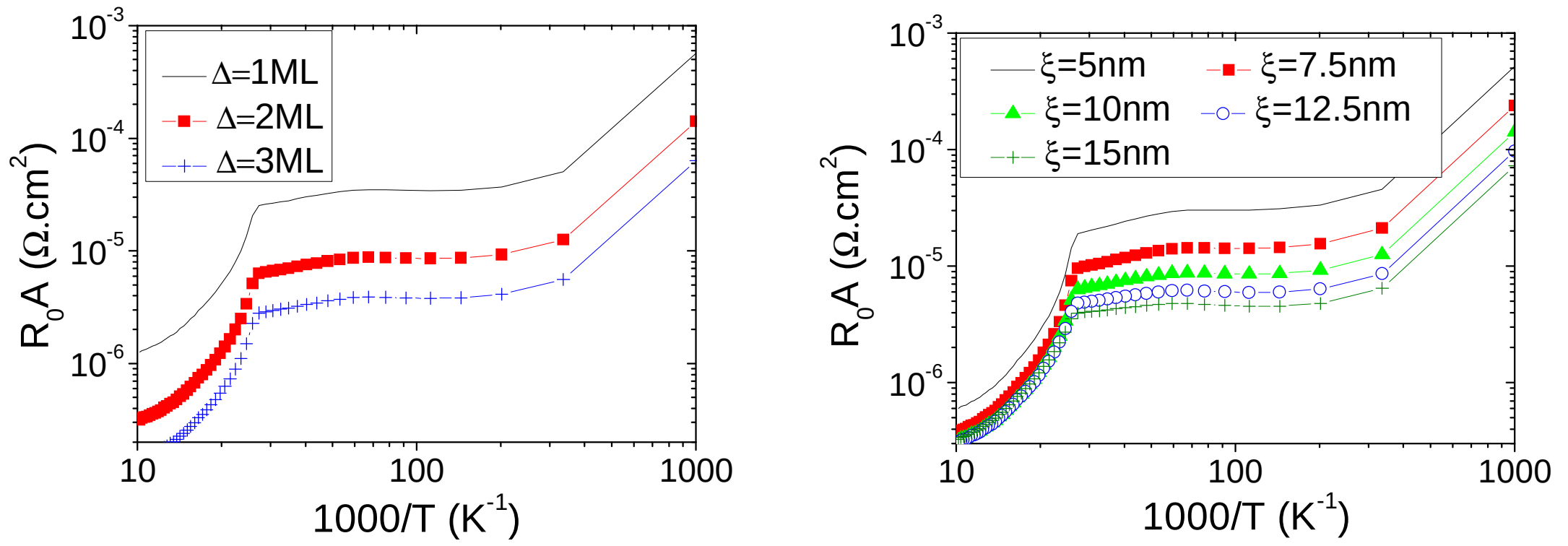

*FIG. 5-20 : $R_0A$ par période en fonction de la température lié au processus de rugosité d'interface. A gauche pour plusieurs valeurs de Δ, à droite pour plusieurs valeurs de ξ.*

### 5.3.4. Niveau de performance du composant

La résistance $R_0A$ se situe autour de $10^{-5}\Omega\cdot cm^2$, ce qui est plutôt faible dans la mesure où des valeurs de 100 $\Omega\cdot cm^2$ [49] ont été obtenues sur des QCD 8µm et que des valeurs de 10 $\Omega\cdot cm^2$ [179] ont été mesurées sur le seul autre QCD THz. Une telle valeur résulte de trois effets. Le premier est clairement lié à la valeur du dopage. Dans notre structure il est plus de trente fois plus élevé que dans la structure décrite par Graaf et al[179]. Cela se traduit par une diminution du $R_0A$ qui se situe entre deux et cinq décades selon la température envisagée. A cela s'ajoute le fait que les interfaces sont moins lisses que nous pouvions l'espérer. Cet effet peut réduire le $R_0A$ d'une demi à une décade. Enfin la structure n'est pas parfaitement optimisée du point de vue de l'injection. Si l'on se réfère au profil de bande, voir FIG. 5-18, le niveau quatre n'est pas localisé entre les niveaux cinq et trois, ce qui peut être favorable pour l'absorption mais nuit à l'injection des électrons dans la cascade. De tels effets devront être corrigés dans les prochaines structures.

### 5.3.5. Comparaison avec l'expérience

La croissance et l'étape de technologie ayant pu être faite, la caractérisation de ce composant devrait suivre rapidement. Ces mesures seront faites au LPA (ENS). Toutefois ces mesures ne sont pas encore disponibles à l'heure actuelle.

Afin de tout de même comparer ces simulations avec des données expérimentales j'ai utilisé la structure publiée dans la référence 179. L'accord avec l'expérience est tout à fait convenable sur une plage d'une trentaine de Kelvin, voir FIG. 5-21. Les simulations conduisent toutefois à une surestimation de la dépendance en température du $R_0A$. Cette différence peut être du au fait que pour ce calcul seule la rugosité d'interface a été considéré, alors que celle-ci ne domine peut être pas aussi fortement quand dans notre structure en raison de différences notables entre les deux échantillons.

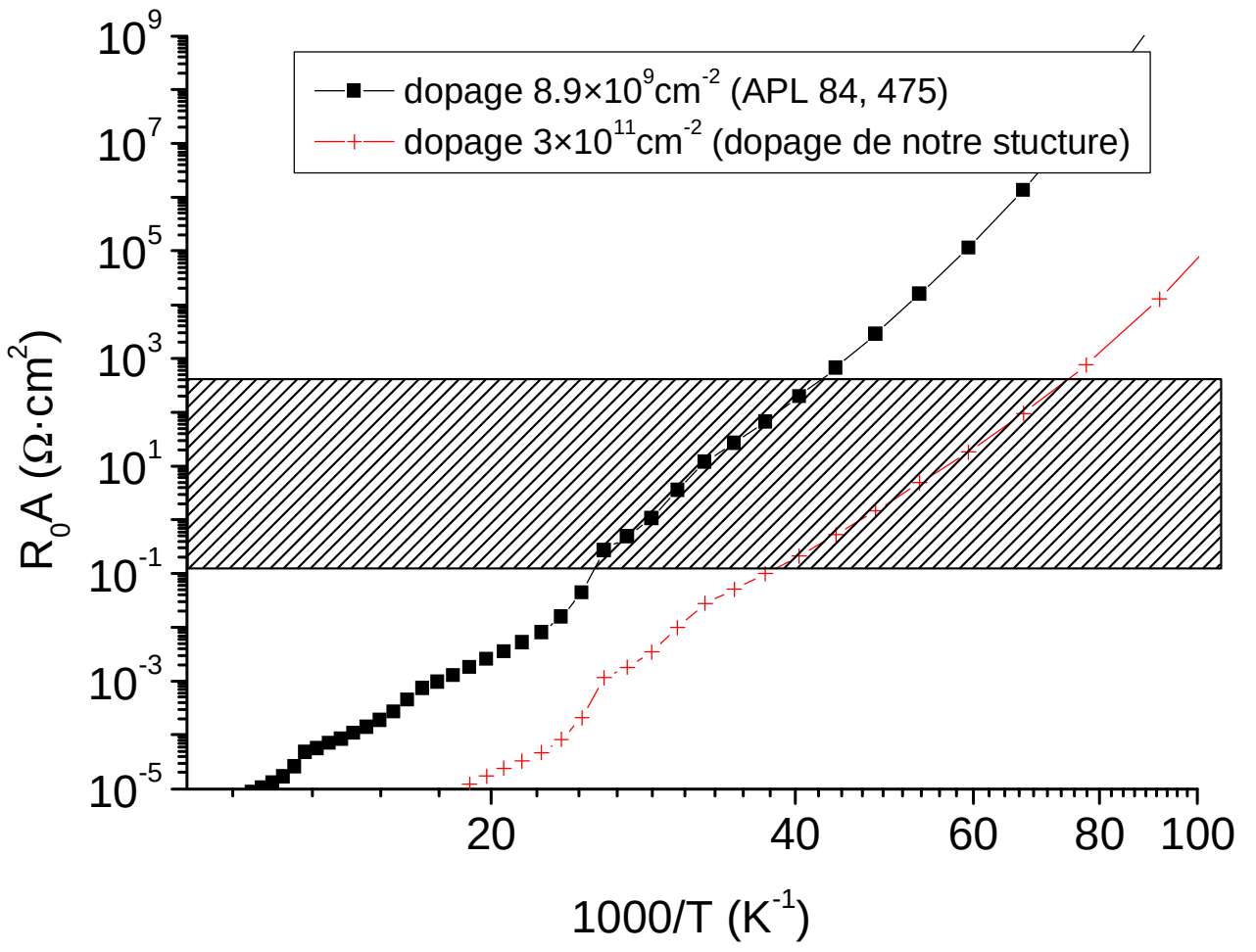

*FIG. 5-21 $R_0A$ en fonction de la température pour deux valeur de dopage et comparaison avec les données expérimentales. Ces dernières n'apparaissent que sous forme d'une plage de valeur car la publication concernes ne donne pas directement le $R_0A$.*

## 5.4. Conclusion

J'ai commencé ce chapitre en montrant le rôle essentiel joué par le profil de dopage sur le niveau de courant d'obscurité. J'ai en particulier montré que l'approche de hopping conduisait à une meilleure dépendance du courant d'obscurité avec l'amplitude du dopage : loi quadratique contre au mieux linéaire dans les modèles précédents. J'ai ensuite montré qu'il était possible de modifier l'élément de matrice de diffusion associé à l'interaction électron dopage en modifiant la position des impuretés. Ce déplacement nous a permis de diminuer de près de 50% le niveau de courant d'obscurité. Cette technique pourra donc être utilisée par la suite lors de la conception des QWIP dédiés aux faibles flux.

Dans la suite du chapitre je me suis intéressé à la modélisation de structures plus complexes. J'ai pour cela commencé à prendre en compte des défauts dans la structure. Deux types d'écart à l'idéalité ont ainsi été traités : les non-idéalités d'interface et le désordre. Le couplage entre une technique de microscopie électronique et notre outil de simulation a permis une estimation de la qualité des interfaces dans nos structures et l'estimation de leur influence sur les propriétés optiques et de transport des QWIP hautes longueurs d'onde. J'ai également montré que les structures QWIP étaient relativement robustes vis-à-vis des non idéalités puisque les modifications apportées par le désordre sont en partie compensées par celles imputables aux interfaces.

Enfin mon code de simulation a été généralisé pour être utilisé dans une structure QCD et a permis d'identifier l'interaction à l'origine du transport et a aussi conduit à la première estimation du $R_0A$ de QCD THz.

Quelles sont les utilisations futures et possibles d'un tel outil de simulation ? Une autre thèse en cours au laboratoire MPQ, qui vise à simuler les effets d'ionisation par impact dans les QWIP THz, se base sur un code dérivé de celui-ci. Mais l'utilisation de cet outil de simulation ne se limite pas nécessairement au transport de charges. On pourrait par exemple utiliser cet outil pour quantifier le transport d'entropie dans la structure. Une telle approche permettrait d'évaluer précisément la température électronique dans les QWIP, comme cela a pu être fait dans les lasers à cascade[180]. Un tel calcul serait particulièrement intéressant sous éclairement.

Mon modèle a donc montré son habileté à rendre compte des effets locaux liés au transport, il faudrait à présent être capable d'inclure à cette capacité de modélisation des effets non locaux tels que des effets de contacts, de profil de champ non uniforme…Ce sera donc l'objet du dernier chapitre de cette thèse.

# 6. TRANSPORT NON-LOCAL DANS LES SUPER-RESEAUX

**Sommaire**

**Publications associées à ce chapitre :**

- Saw tooth patterns on the I(V) curve of quantum well infrared photodetectors: from modeling to the impact on the detector performance, en preparation pour J. Appl. Phys.

# 6.1. Introduction

## 6.1.1. Généralités

Dans les chapitres précédents j'ai traité du transport d'un point de vue purement local. Je suis reparti des lois fondamentales de la mécanique quantique et j'ai cherché à en déduire des propriétés macroscopiques de transport. Cela m'a conduit à la modélisation des courbes de courant d'obscurité en fonction du champ électrique local. Je faisais alors l'hypothèse que le champ électrique sur la structure était homogène pour comparer nos courbes théoriques aux courbes expérimentales. Dans ce chapitre, c'est cette hypothèse forte que je me propose de faire disparaître. L'objectif est d'être capable de rendre compte d'effets non locaux qui peuvent être observés sur certaines courbes *I(V)* : résistance différentielle négative (RDN), dents de scie[74,181,182]…

Pour atteindre cet objectif, il faut rendre compte des effets liés à une distribution de champ électrique inhomogène sur la structure ainsi que des effets de transport résonant. C'est une physique ancienne[183] des super réseaux dans la mesure où ils ont entre autre été créés pour observer des effets de résistances différentielles négatives (RDN)[181,184]. Cette particularité de la courbe *I(V)* des diodes à double barrières tunnel[185] se manifeste dans les structures à multi-puits quantiques par l'apparition de dents de scie sur l'*I(V)* (ou de pic sur la courbe de conductance) et de zone de fort champ dans le profil de bande. Cette physique a également été étudiée par des membres de la communauté des la physique statistique car la bistabilité[186], induite par la présence de la RDN, permet d'étudier des phénomènes de transition de phase entre différentes formes des courbes *I(V)*[186] voire même des phénomènes de chaos[148]. Cet aspect du transport ne sera toutefois pas traité ici. Je vais plutôt chercher à appliquer cette physique au cas des super-réseaux ultra faiblement couplés que sont les QWIP.

L'observation de dents de scie a été rapportée à de nombreuses reprises dans les super réseaux et il est vrai que nous nous attendions à en observer sur le composant du chapitre 2. Ce ne fut pas le cas. Par contre en puisant dans la bibliothèque de composants de Thales, nous avons pu observer ce phénomène sur plusieurs QWIP ($\lambda$ = 9 et 14µm). Il nous revenait alors de comprendre la physique qui en était à l'origine et d'identifier les conditions de leur existence. C'est donc une physique du transport assez différente de celle que nous avons vue jusqu'à présent qui va être étudiée dans ce chapitre. Je vais surtout m'intéresser à la reproduction et à la compréhension qualitative de cette physique. Une fois que j'aurais mis en place les éléments nécessaires à la compréhension du phénomène de dents de scie, j'en étudierai l'impact sur les performances des détecteurs.

Ce travail résulte d'une collaboration avec le III-V Lab d'Alcatel Thales (Vincent Guériaux, Alexandru Nedelcu, Virginie Trinité, Mathieu Carras). Pour les résultats présentés dans ce chapitre les données expérimentales viennent du III-V Lab et j'ai réalisé les simulations.

## 6.1.2. Structures étudiées

Deux séries de structures vont être étudiées dans ce chapitre. Les éléments de la première (série E) diffèrent de par le niveau de dopage des structures, ceux de la seconde (série F) de par la taille des barrières. Les paramètres de ces structures sont donnés dans tab. 6-1.

| Paramétres | Série E | Série F |
|---|---|---|
| $L_w$ (Å) | 72 | 49 |
| $L_b$ (Å) | 340 | 150 ($F_1$)<br>200 ($F_2$)<br>250 ($F_3$) |
| Dopage ($cm^{-2}$) | $1\times10^{11}$ ($E_1$)<br>$1.5\times10^{11}$ ($E_1$)<br>$2\times10^{11}$ ($E_1$) | $2\times10^{11}$ |
| $\lambda_{pic}$ (µm) | 14.2 | 8.4 |

*tab. 6-1 Paramètres de structures des échantillons étudiés dans ce chapitre.*

### 6.1.3. Phénomènes mis en jeu

Afin de comprendre les effets que nous allons modéliser par la suite, commençons par les visualiser. Les courbes *I(V)* présentent un plateau, un phénomène d'hystérésis à son niveau, ainsi que des oscillations de courant, voir FIG. 6-1.

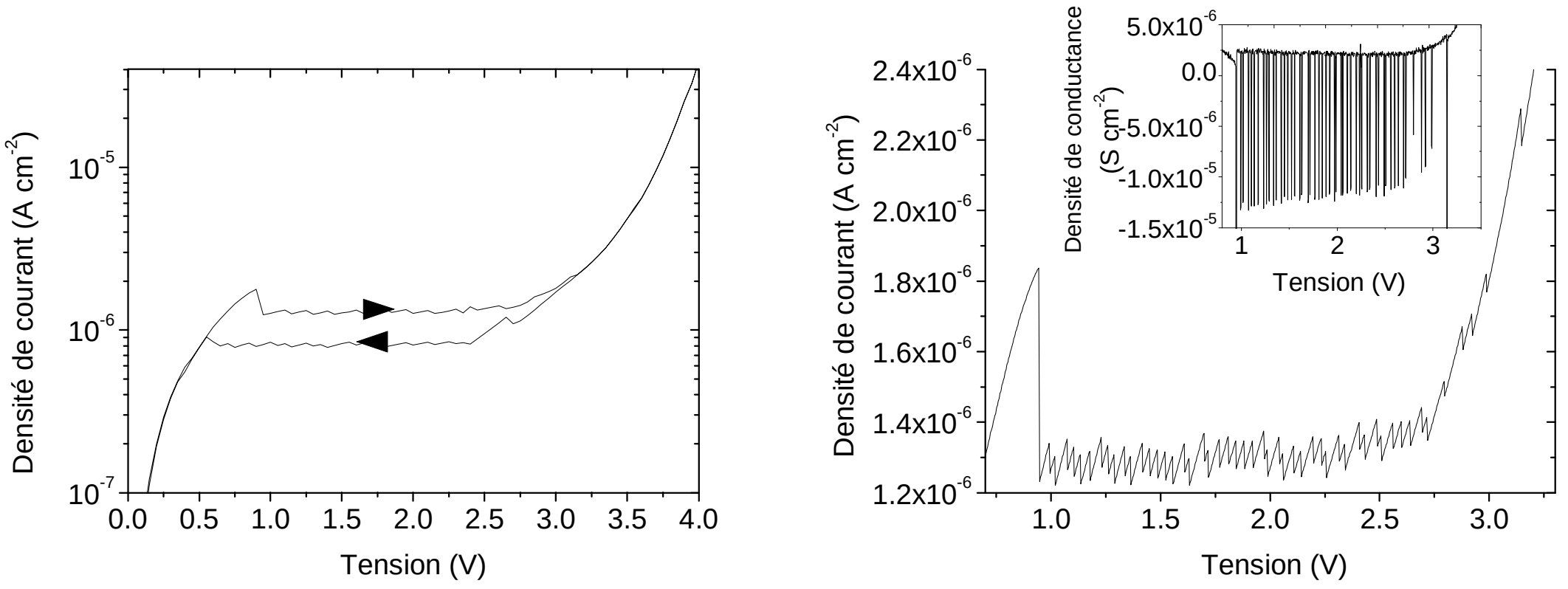

*FIG. 6-1 : A gauche : Courbe I(V) pour l'échantillon $E_3$ avec balayage en tension dans les deux sens. A droite, Cette même courbe I(V) avec une meilleure résolution en tension. Encart : densité de conductance en fonction de la tension appliquée. Mesures faite à 23K.*

## 6.2. Modélisation

Dans ce chapitre nous allons une dernière fois nous intéresser à modéliser le courant d'obscurité. Mais à la différence du travail précédent il nous faut prendre en compte l'ensemble de la structure. De nouveaux effets liés aux contacts et au profil de champ électrique sont donc à attendre. L'un des enjeux est de comprendre comment les propriétés locales (RDN) peuvent affecter l'ensemble de la structure (dents de scie).

### 6.2.1. Physiques du problème

Commençons par définir quelques notations pour des questions de commodité par la suite :

- $n_m$ est la population du m$^{\text{éme}}$ puits.
- $F_m$ est le champ sur la m+1$^{\text{éme}}$ barrière.
- $J_m$ est la notation réduite de $J_{m->m+1}$ pour le courant à travers la m+1$^{\text{éme}}$ barrière.

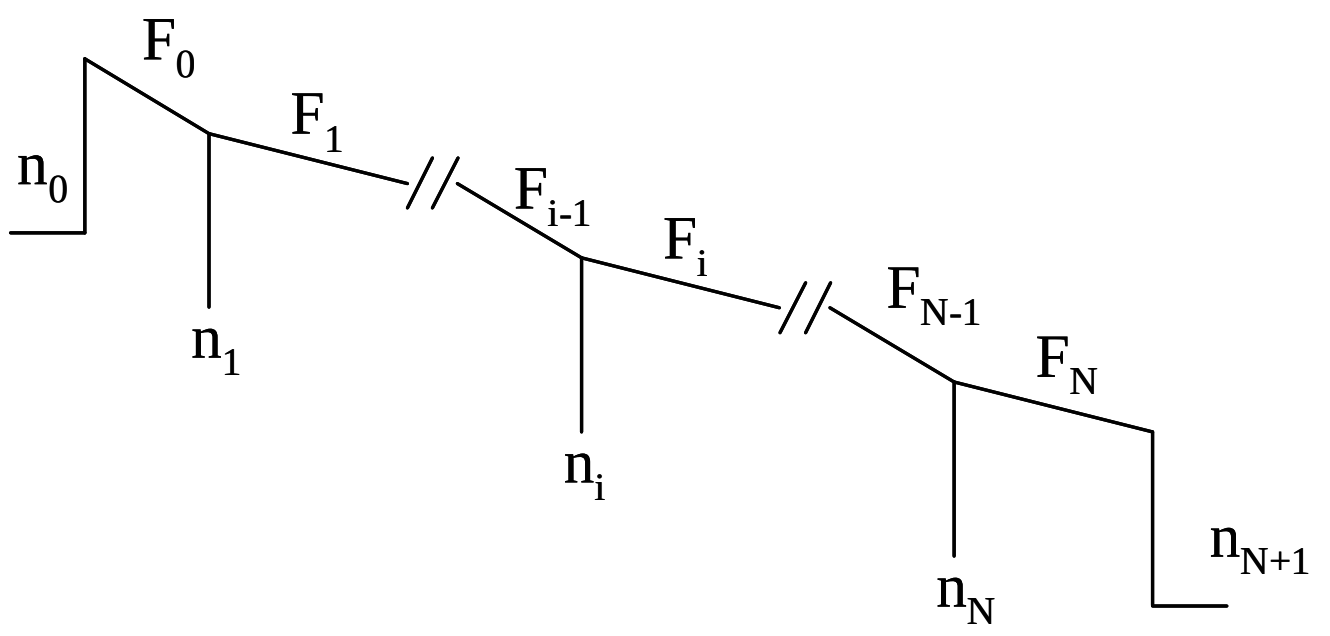

*FIG. 6-2 : Schéma de la structure*

Afin de modéliser la courbe *I(V)* (resp. *V(I)*) globale, il faut se munir d'un certain nombre de règles qui vont régir l'évolution de la structure dans le temps et avec la tension (resp. le courant) appliqué(e). Plusieurs modèles de transport ont pu être proposés[186,187,188] mais ils ont en commun trois éléments :

- Une courbe *I(V)* locale, ou plutôt *J(F)*, qui doit modéliser le courant à travers une barrière. Ce courant dépend de la population des puits qui l'entourent et du champ sur la barrière. J'ai utilisé l'expression de Amann et al[187] qui s'écrit :

$$J_{m->m+1} = \sum_{\upsilon} \frac{e}{\hbar} \left|H^{1,\upsilon}\right|^2 \frac{\Gamma_1 + \Gamma_\upsilon}{(E_\upsilon - E_1 - eF_m d)^2 + (\Gamma_1 + \Gamma_\upsilon)^2} \left\{ n_m - DOS_{2D} * k_b T \ln\left[ \left( \exp\left( \frac{n_{m+1}}{DOS_{2D}.k_b T} \right) - 1 \right) \exp(-eF_m d / k_b T) + 1 \right] \right\} \quad (6\text{-}1)$$

Où ν est l'indice de la sous-bande concernée. C'est une expression de type Kazarinov-Suris. Cette expression rend compte du phénomène de résistance différentielle négative (RDN)[184], voir figure FIG. 6-3. Rappelons que ce phénomène résulte de la localisation des fonctions d'onde par le champ électrique. Le second pic dans la courbe *J(F)* locale provient de la mise en résonance des niveaux fondamentaux et excités de deux puits consécutifs[181]. Il est par ailleurs indispensable de se munir d'un modèle simple pour ce courant local, en pratique une expression analytique, car le temps de calcul va être réparti sur toute la structure et ne doit pas seulement rester au niveau local. Il serait possible d'étendre à l'infini cette expression pour tenir compte de davantage d'effets, mais cela se payerait rapidement sur le temps de calcul et conduirait à l'ajout de paramètres dans un modèle qui n'en manque déjà pas. Si dans un premier temps nous voulons comprendre l'origine des dents de scie, cette expression du courant doit contenir une RDN et des dépendances explicites en $F_m$, $n_m$ et $n_{m+1}$.

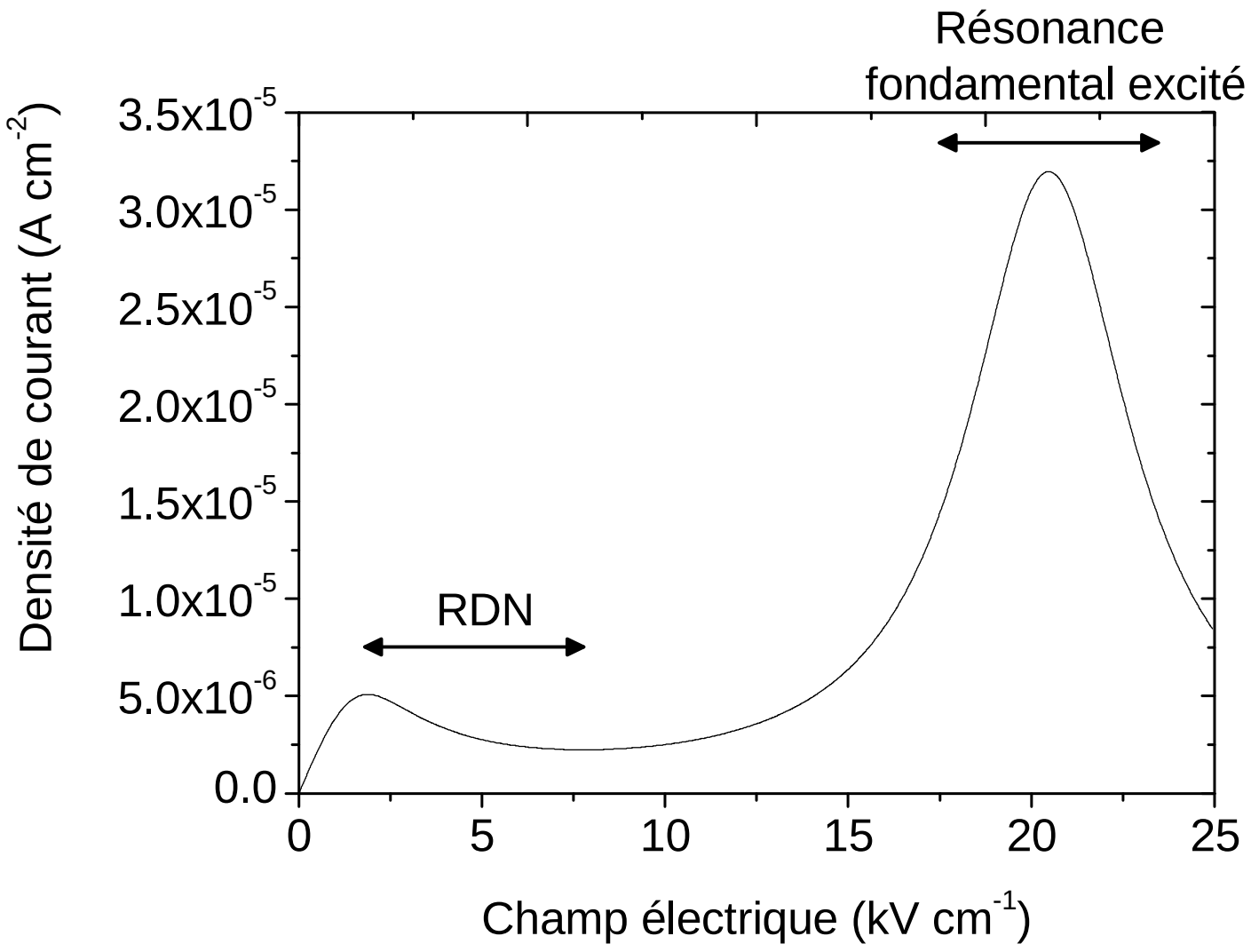

*FIG. 6-3 : Courbe J(F) locale quand les puits amont et aval d'une barrière ont une population égale à leur dopage*

- Une équation de conservation de la charge au niveau de chaque puits $e\frac{dn_m}{dt} = J_{m-1->m} - J_{m->m+1}$ qui sera linéarisée au premier ordre. Le pas de temps qui sert à discrétiser cette équation devra donc être adapté au niveau de courant dans la structure.

- L'équation de Maxwell-Gauss $\varepsilon(F_m - F_{m-1}) = e(n_m - N_D)$ qui régit l'évolution du champ avec la population des puits.

### 6.2.2. Système d'équations

Pour un système à $N$ puits, nous avons :

| | |
|---|---|
| N populations liées aux puits | N équations de Maxwell-Gauss |
| N+1 champs sur les N+1 barrières | N équations de la charge |
| N+1 courants à travers les N+1 barrières | N-1 équations de courant local : J(F) |
| 3N+2 inconnues | 3N-1 équations |

*tab. 6-2 Bilan des équations et des inconnues*

C'est donc un système de *3N+2* inconnues avec *3N-1* équations. Les trois équations manquantes proviennent des conditions aux limites.

### 6.2.3. Conditions aux limites

Les conditions aux limites spatiales et temporelles permettent de boucler notre système d'équations. La première équation vient du fait que l'évolution temporelle se fait à tension ou courant fixé(e). Les deux dernières vont être données par les conditions aux limites sur les contacts. Nous allons maintenant nous pencher plus avant sur la façon de traiter ces équations.

#### 6.2.3.1. Evolution temporelle : I(V) vs V(I)

Le fait d'imposer dans la structure une tension ou un courant régit fortement son évolution et conduit à des courbes assez différentes, comme peut le montrer la figure FIG. 6-4. Dans un cas la contrainte se porte sur la somme des champs électriques, dans l'autre sur le courant injecté à l'émetteur. Le phénomène de dents de scie disparaît quand le courant est imposé.

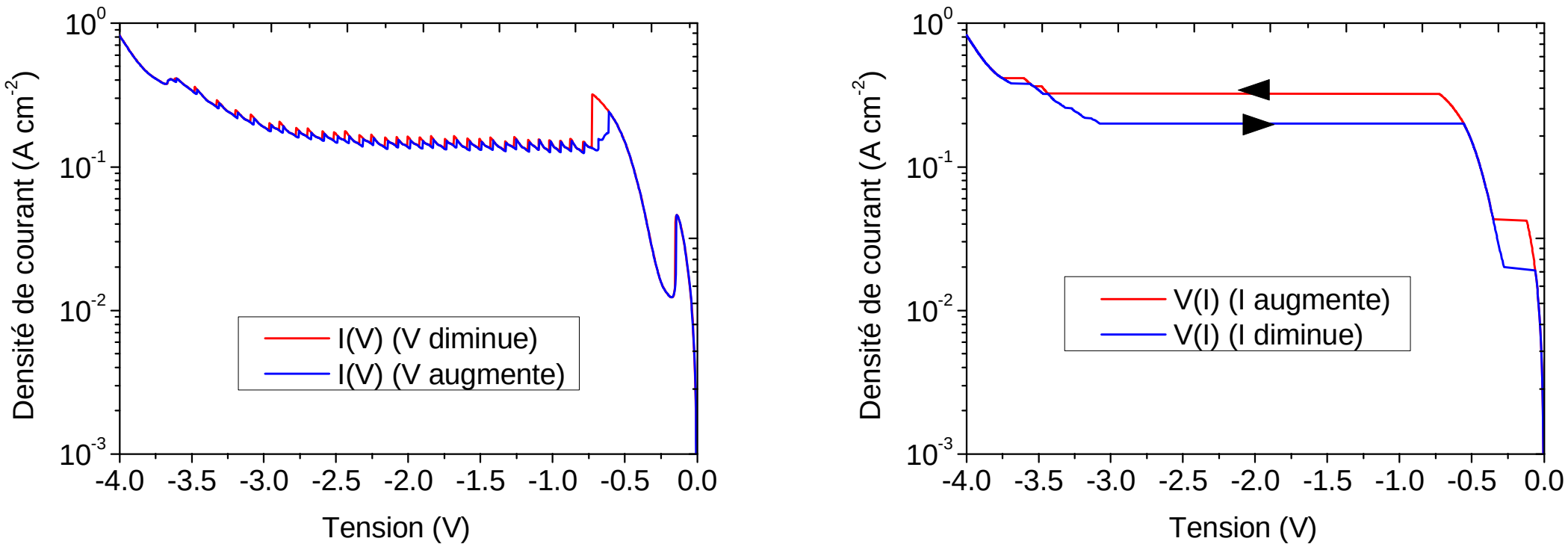

*FIG. 6-4 : A gauche courbes I(V), A droite courbes V(I) pour le composant F1. Données III-V Lab.*

#### 6.2.3.2. Conditions aux limites spatiales : rôle des contacts

Il existe deux façons de traiter les contacts dans la littérature[118]. Il a été montré que les deux méthodes[118] conduisaient à des résultats similaires.

- Condition aux limites constantes

Cette méthode consiste à introduire deux puits fictifs au niveau des contacts de population constante $n_0$ et $n_{N+1}$. Leur population est constante et vérifie $n_0 = (1+c)N_D$, avec *c*>-1. C'est cette voie qui sera utilisée par la suite.

- Condition aux limites ohmique

Dans cette approche le courant sur la première barrière est relié au champ qui règne sur celle-ci par une relation linéaire $J_{0->1} = \sigma F_0 d$ avec $\sigma$ un paramètre effectif. Sur la dernière barrière, une équation similaire est introduite $J_{N->N+1} = \sigma F_0 d \frac{n_N}{n_D}$, le facteur supplémentaire $\frac{n_N}{n_D}$ évite que la population du dernier puits ne devienne négative.

### 6.2.4. Principe de la simulation

#### *6.2.4.1. Généralités*

Cette simulation est très intuitive puisque l'on suit l'évolution temporelle du dispositif. Elle consiste à modifier un paramètre puis à attendre le retour à l'équilibre. Le fonctionnement de la simulation est le suivant dans le cas d'une courbe *I(V)* et voir schéma FIG. 6-5.

1. Initialisation de la structure, il faut se placer dans une zone où le champ est homogène, en pratique autour de zéro ou à tension assez forte. Toutes les valeurs de population de puits sont initialisées telles que leur remplissage soit égal au dopage du puits, le champ sur la structure est pris homogène.
2. Le temps est incrémenté d'une unité.
3. La population des puits est calculée en utilisant l'équation de conservation de la charge.
4. Connaissant la nouvelle population de chaque puits, la variation de champ due à la variation de population $\Delta F_m = \frac{e}{\varepsilon}(n_m - N_D)$ est évaluée. Connaissant les $\Delta F_m$ et la tension totale sur la structure il est possible d'évaluer le champ $F_0$ sur la première barrière. Son expression vérifie $F_0 = \frac{V - L_d \sum_{m=1}^{N-1} \sum_{i=1}^{m} \Delta F_i - L_b \sum_{i=1}^{N} \Delta F_i}{L_{tot}}$ avec $L_{tot}$ la taille totale de la structure, $L_b$ la taille d'une barrière et $L_d$ celle d'une période. Finalement il est alors possible de calculer les $F_m$ de proche en proche.
5. Le courant au niveau du dernier puits est mis en mémoire.
6. Les étapes deux à cinq sont répétées un certain nombre de fois, si l'évolution temporelle du courant à travers la structure se stabilise, on considère qu'il y a convergence. Dans le cas contraire les opérations deux à cinq sont répétées à nouveau jusqu'à la convergence du système.
7. Une fois la convergence obtenue la tension est incrémentée. Au final une courbe *I(V)* est obtenue.

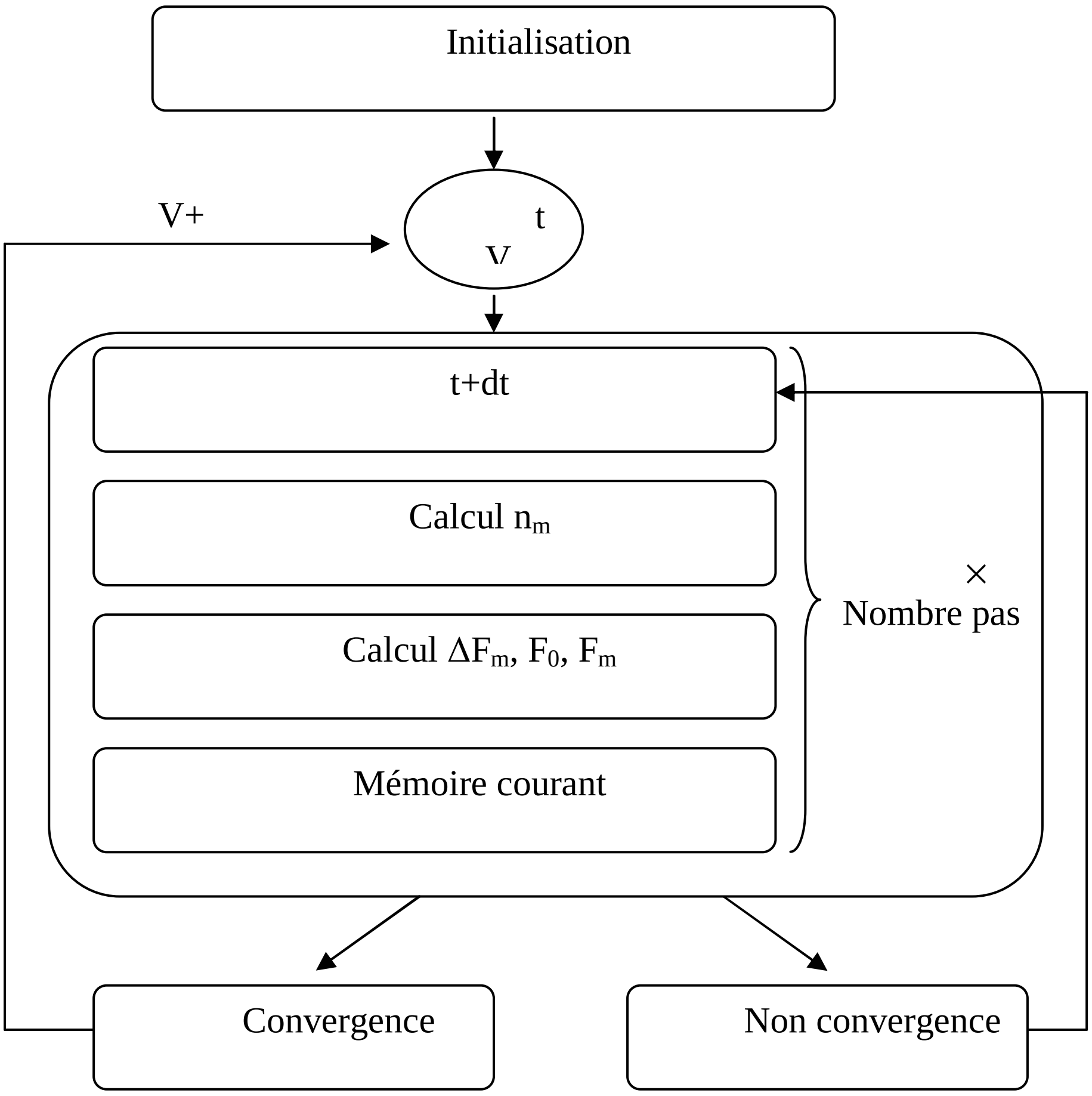

*FIG. 6-5 : Schéma du fonctionnement de la simulation dans le cas d'un I(V)*

Dans le cas d'une *V(I)*, c'est essentiellement l'étape 4 qui est modifiée. En effet les conditions aux limites imposent le courant à l'émetteur. J'utilise alors une fonction qui calcule le champ sur la première barrière connaissant le courant qui la traverse. Le champ sur le reste de la structure est obtenu de la même façon que pour une *I(V)*.

#### 6.2.4.2. Effets des paramètres

Ce modèle nécessite l'évaluation d'un certain nombre de paramètres :

- L'amplitude des couplages entre niveau : $\left|H^{1,1}\right| \left|H^{1,2}\right|$. Ces paramètres peuvent être évalués en résolvant l'équation de Schrödinger. Ces paramètres jouent essentiellement sur le niveau de courant des plateaux des *I(V)*, et un peu sur leur extension en tension pour le paramètre $\left|H^{1,2}\right|$. Voir FIG. 6-6 (a) et (b).

- L'élargissement de chacun des niveaux : $\Gamma_1$ et $\Gamma_2$. Ces paramètres sont pris égaux à la diffusion intrabande de chacune des sousbandes. La valeur de ces paramètres est donc estimable via le code de diffusion décrit au chapitre 4. $\Gamma_1$ et $\Gamma_2$ contrôlent essentiellement la période en tension des dents de scie. Voir FIG. 6-6 (c) et (d).

- Le paramètre c de population des contacts. C'est un paramètre totalement empirique qui sera ajusté afin de faire correspondre la forme de l'*I(V)* avec la courbe expérimentale. C'est un paramètre qui influence la tension de la première dent de scie. Il contrôle également la position de la zone de fort champ. Suivant son signe elle sera placée coté collecteur ou coté émetteur. Voir FIG. 6-6 (e).

- Le temps de convergence et le pas en temps de l'équation de conservation de la charge. Un temps de convergence inadapté peut affecter la forme des dents de scie.

Il est donc en théorie possible d'évaluer avec précision la majorité de ces paramètres, néanmoins un ajustement de leurs valeurs est parfois nécessaire pour obtenir une meilleure adéquation théorie-expérience.

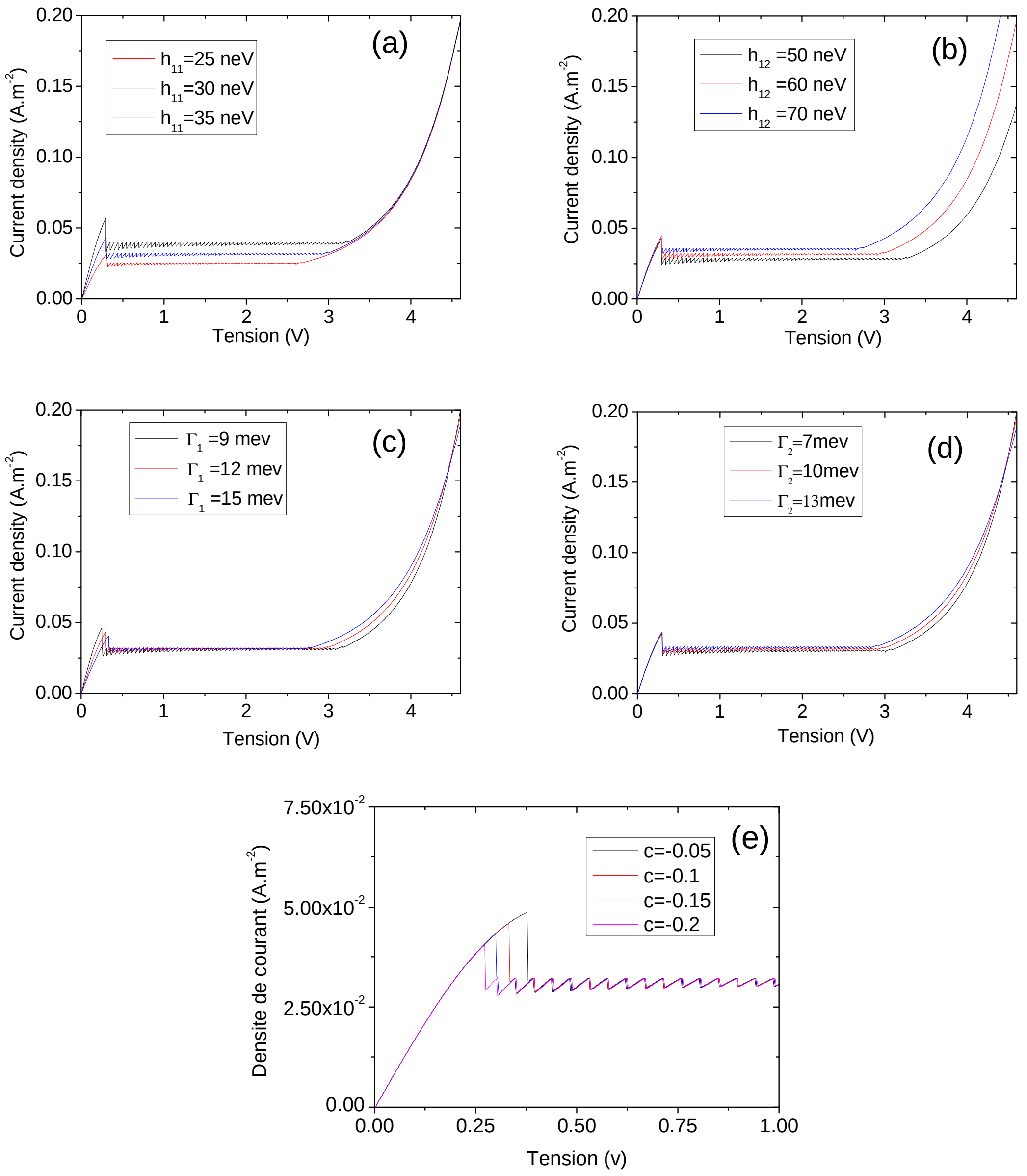

*FIG. 6-6 (a) courbe I(V) pour différentes valeurs de $H_{11}$. (b) courbe I(V) pour différentes valeurs de $H_{12}$. (c) courbe I(V) pour différentes valeurs de $\Gamma_1$. (d) courbe I(V) pour différentes valeurs de $\Gamma_2$. (e) courbe I(V) pour différentes valeurs de c.*

## 6.3. Interprétation des dents de scie

La figure FIG. 6-7 montre une comparaison des courbes *I(V)* expérimentales et théoriques. Notre simulation est donc capable de reproduire les dents de scie et la présence d'hystérésis. Reste donc maintenant à en comprendre l'origine et à exhiber les paramètres qui contrôlent sa forme.

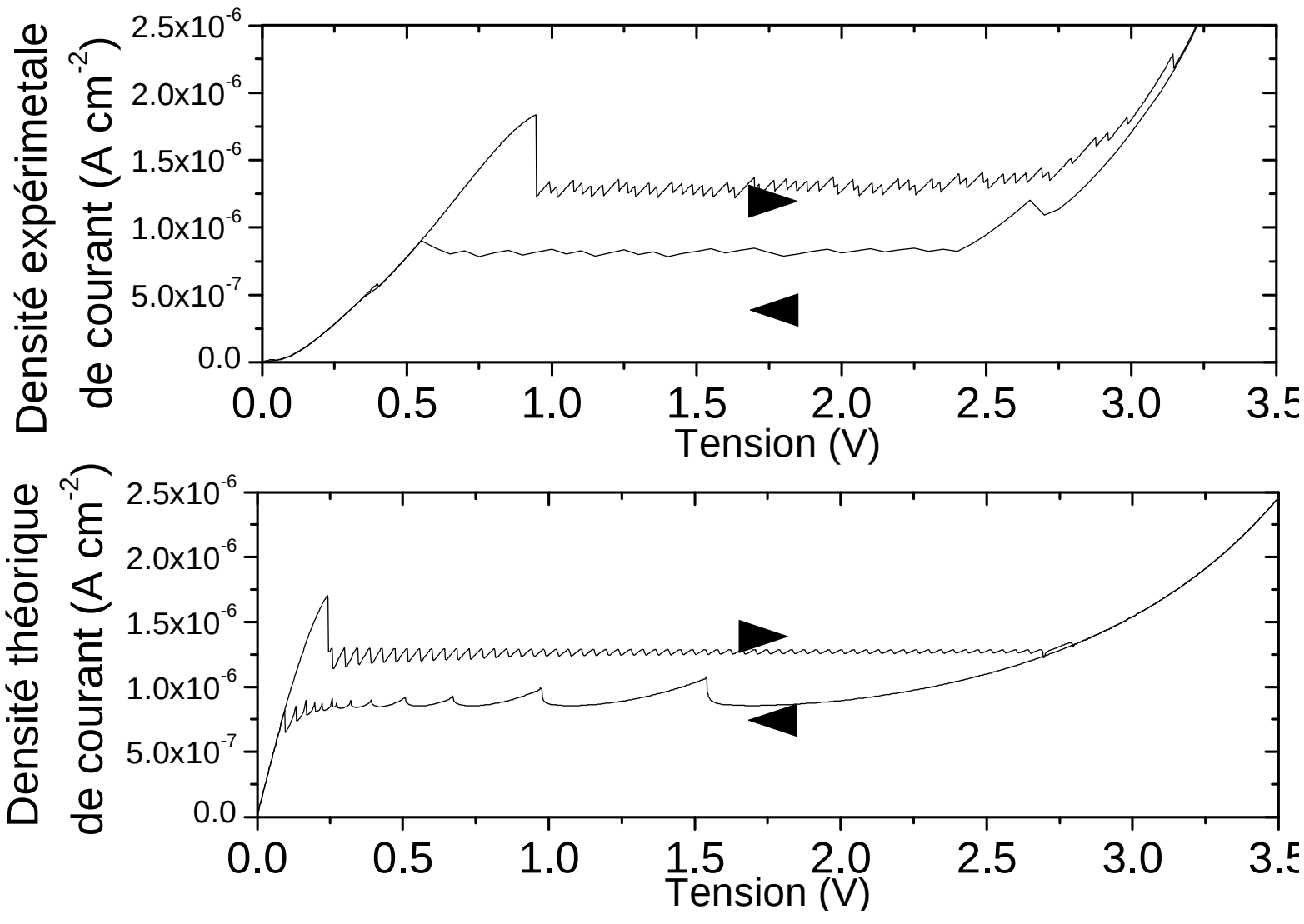

*FIG. 6-7 : Comparaison théorie-expérience des courbes I(V) pour le composant $E_3$. A tension décroissante le pas de tension pris lors de la mesure est insuffisant pour résoudre les dents de scie précisément.*

Nous allons à présent passer en revue plusieurs phénomènes qui vont nous permettre de mieux comprendre l'origine et les paramètres qui contrôlent la forme de la courbe *I(V)*. Nous étudierons également l'effet de l'introduction du désordre lié à la croissance ainsi que les conditions d'existence des dents de scie. Cela va en partie nous permettre d'expliquer les différences que l'on peut observer entre la courbe *I(V)* expérimentale et celle simulée de la figure FIG. 6-7.

### 6.3.1. Origine des dents de scie

Historiquement les super réseaux ont été dessinés pour observer des effets comme la RDN[183], les dents de scie en sont la conséquence directe. Pour en comprendre l'origine revenons à la courbe *J(F)* locale, voir FIG. 6-8. Quand la tension part de zéro, le courant suit cette courbe jusqu'au point A. Mais les points de fonctionnement situés entre les points A et C sont instables. Le système « préfère » donc conserver le courant et faire un saut de tension. Il se crée alors une zone sur laquelle le champ est plus fort que sur le reste de la structure. En continuant à augmenter la tension, cette zone de fort champ se propage à l'ensemble de la structure, voir FIG. 6-9. Quand la tension redescend, le saut de champ électrique se fait au niveau du point C, car les points de fonctionnement entre B et C sont stables. Le courant est alors plus faible que lorsque la tension était croissante, c'est ce qui explique l'hystérésis des courbes. L'amplitude de l'hystérésis est donc contrôlée par le rapport pic/vallée du courant sur la courbe *J(F)* locale.

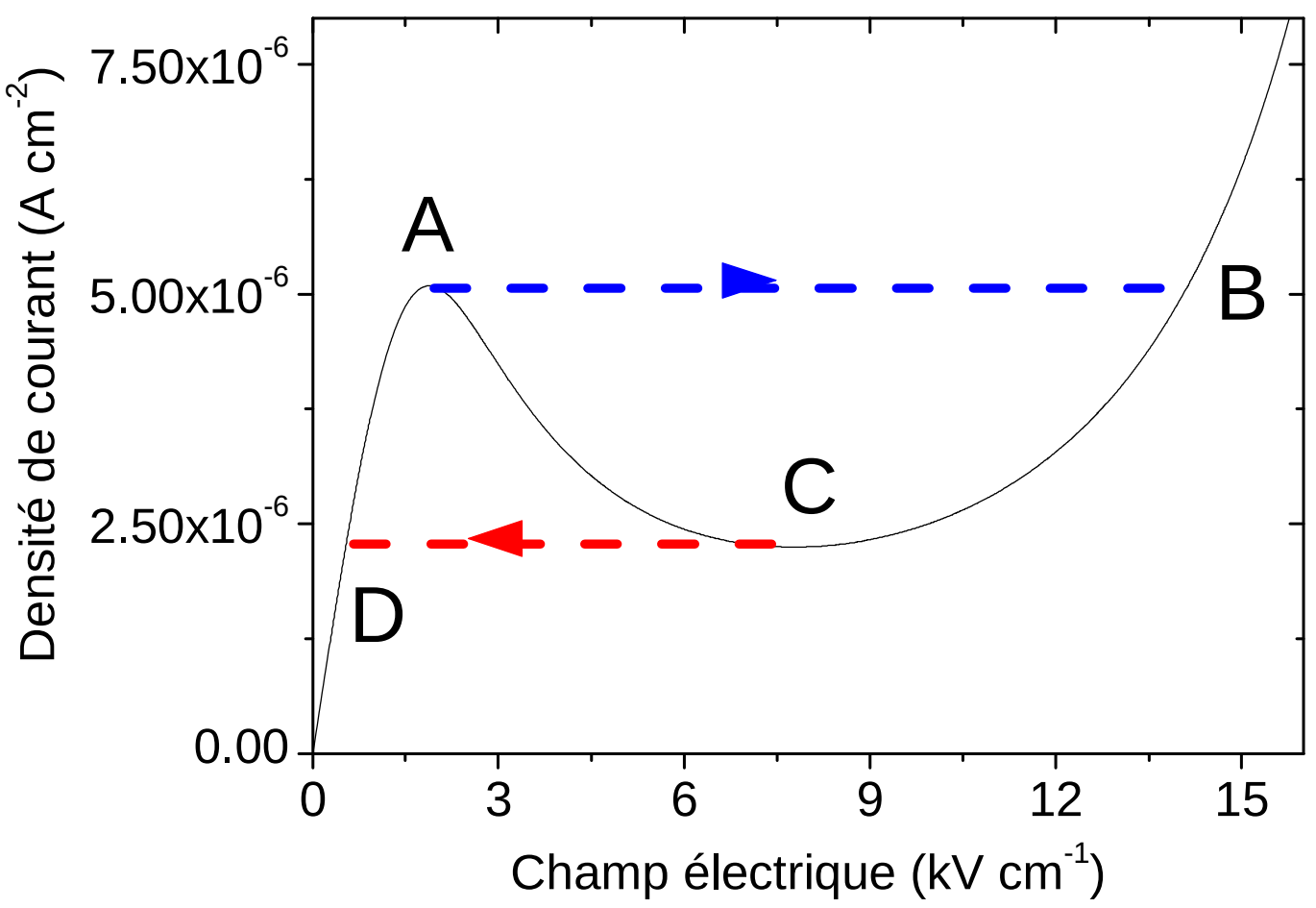

*FIG. 6-8 : Courbe J(F) locale.*

La période en tension des dents de scie est, elle, contrôlée par la différence de champ électrique entre les points A et B à tension croissante et celle entre les points C et D à tension décroissante. Cette différence de tension est en particulier inférieure à $\frac{E_2 - E_1}{eL_d}$ en raison de l'élargissement des niveaux, cette différence est également contrôlée par l'élargissement des niveaux.

La population du puits à l'interface entre la zone de fort et de faible champ n'est pas égale à son dopage, voir FIG. 6-9. Le puits peut être vidé ou rempli, cela dépend du signe du champ électrique et de la position de la zone de fort champ par rapport à l'émetteur. Notons que c'est un phénomène extrêmement localisé qui n'affecte quasiment qu'un seul puits.

Pour conclure les dents de scie résultent de la propagation d'une zone de fort champ à travers la structure qui elle-même résulte de l'existence d'une RDN sur les courbes *J(F)* locales.

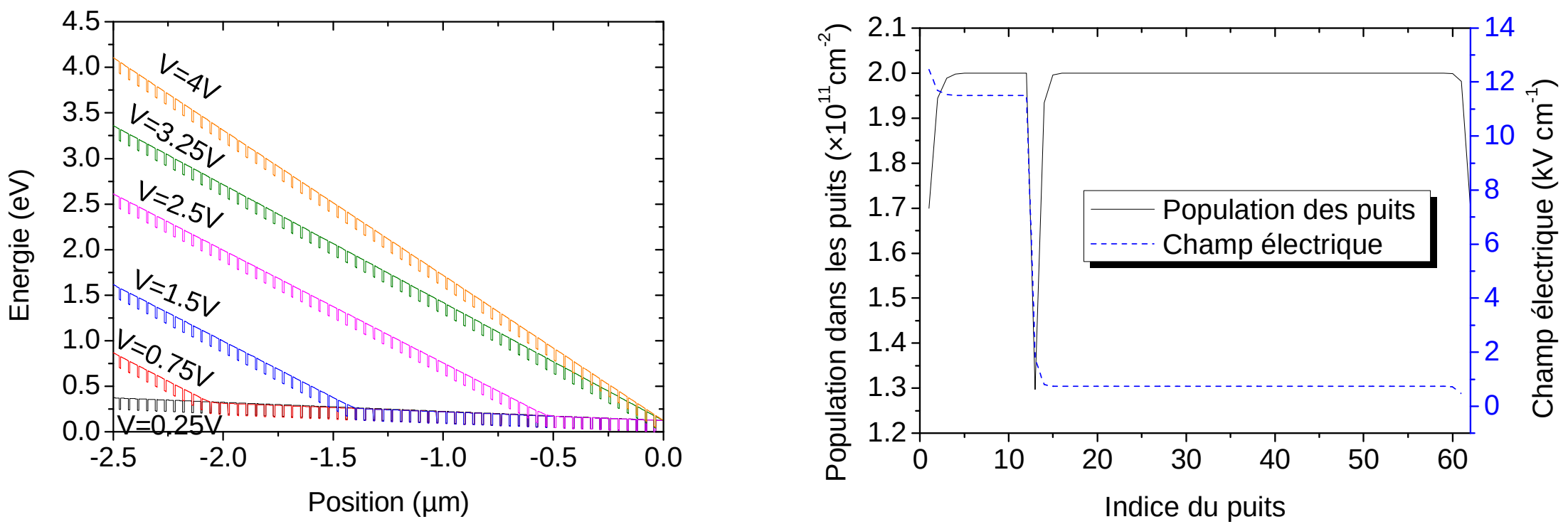

*FIG. 6-9 : A gauche : Structure de bande à différentes tensions. A droite : Population et champ électrique sur la structure (V=0.7V).*

### 6.3.2. Passage d'une dent de scie à l'autre

Nous voulons ici comprendre les paramètres qui contrôlent l'amplitude en courant des dents de scie. Nous avons notamment observé une diminution de l'amplitude des dents de scie sur les *I(V)* simulées quand la valeur absolue de la tension appliquée augmente, voir la FIG. 6-7. Ce n'est pas le cas sur les courbes expérimentales et nous voulons en identifier l'origine. La FIG. 6-9 montre que dans la zone d'existence des dents de scie, la structure se partage en seulement deux zones : une de fort champ pour laquelle le champ est proche de $F_B$ et une de faible champ pour laquelle le champ est proche de $F_A$.

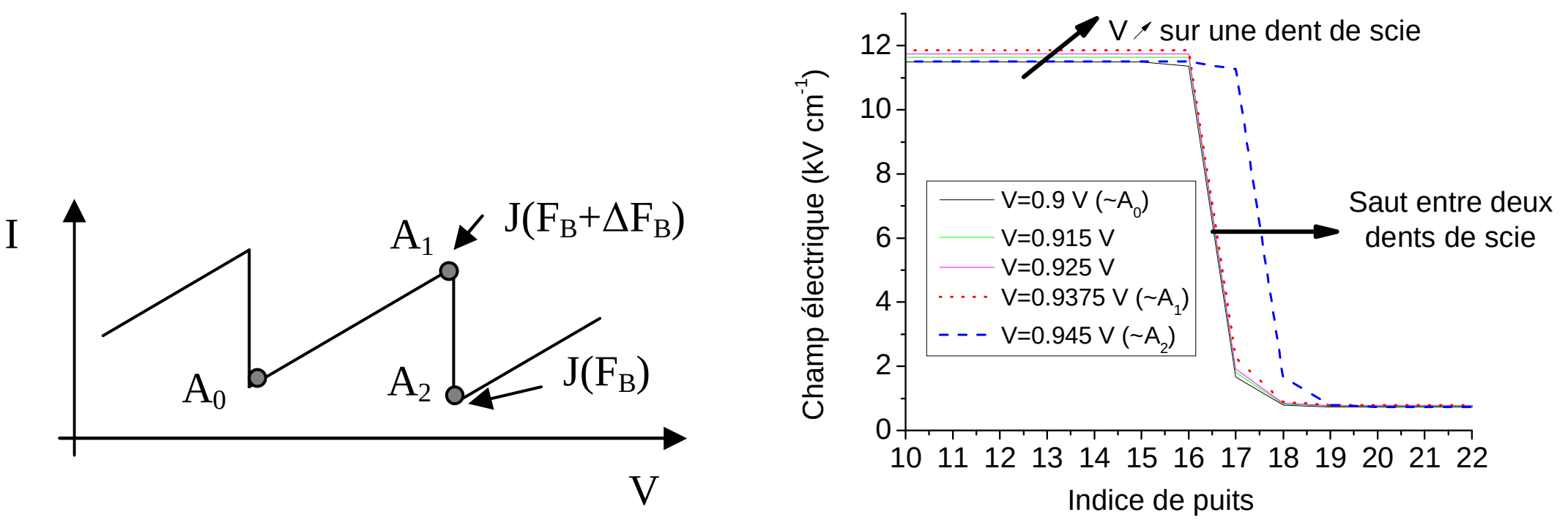

*FIG. 6-10 : A gauche : Schéma de dents de scie. A droite : Profil de champ électrique autour du passage d'une dent de scie à la suivante.*

La FIG. 6-10 montre que lorsque l'on passe du point $A_0$ au point $A_1$ (voir schéma FIG. 6-10) l'excès de tension s'accumule sur la zone de fort champ et pas dans celle de faible champ ou juste à l'interface. Dans ce cas le champ passe de $F_B$ à $F_B + \Delta F_B$. Le courant passe lui, de $J(F_B)$ à $J(F_B + \Delta F_B)$. Au premier ordre, lorsque le système passe sur la dent de scie suivante, c'est-à-dire passe du point $A_1$ au point $A_2$, le champ (courant) repasse à $F_B$ ($J(F_B)$). Lors de cette opération le champ sur la dernière barrière saute de $F_B$-$F_A$. Appelons $n_0$ l'indice du puits où il y a cassure, alors :

- La tension en $A_0$ est : $n_0 \times F_B + (N-n_0) F_A$
- La tension en $A_1$ est : $n_0 \times (F_B + \Delta F_B) + (N-n_0) F_A$
- La tension en $A_2$ est : $(n_0+1) \times F_B + (N-n_0-1) F_A$

L'évolution de $A_1$ vers $A_2$ se fait à tension constante. Il est donc possible d'écrire l'égalité des tensions en ces deux points, ce qui conduit à $n_0 \Delta F_B = \mathrm{F_B} - \mathrm{F_A}$. L'amplitude en courant des dents de scie est alors donnée par :

$$\Delta J = J(F_B + \Delta F_B) - J(F_B) = J(F_B + \frac{F_B - F_A}{n_0}) - J(F_B) \approx \frac{F_B - F_A}{n_0} \frac{\partial J(F_B)}{\partial F} \qquad (6\text{-}2)$$

Il sort alors que l'amplitude des dents de scie dépend du nombre de périodes qui se trouvent dans la zone de fort champ, mais aussi de leur période en tension et de la forme de la courbe *J(F)* locale.

### 6.3.3. Amplitude de la première dents de scie

Nous observons expérimentalement et sur les simulations que la première dent de scie présente un courant plus important que les autres. Afin de comprendre ce phénomène, référons nous à la FIG. 6-9. Dans ce cas les champs électriques sont positifs, l'émetteur est donc à droite et le collecteur à gauche. Au niveau de la cassure, le puits concerné est partiellement vidé. Le courant émis par ce puits vers l'amont s'en trouve réduit, puisque le nombre de porteurs est réduit. Cette zone de cassure du champ électrique est donc plus résistive que le reste de la structure. Le passage d'une structure où le champ est complètement homogène à une structure où le champ n'est plus homogène s'accompagne donc de l'apparition d'une zone plus résistive au niveau de la cassure, le courant est donc réduit au moment de l'apparition des dents de scie.

### 6.3.4. Absence de dents de scie sur les V(I)

Les courbes *V(I)* présentent une allure sensiblement différente de celle des *I(V)* et en particulier les dents de scie ont disparu, voir FIG. 6-11.

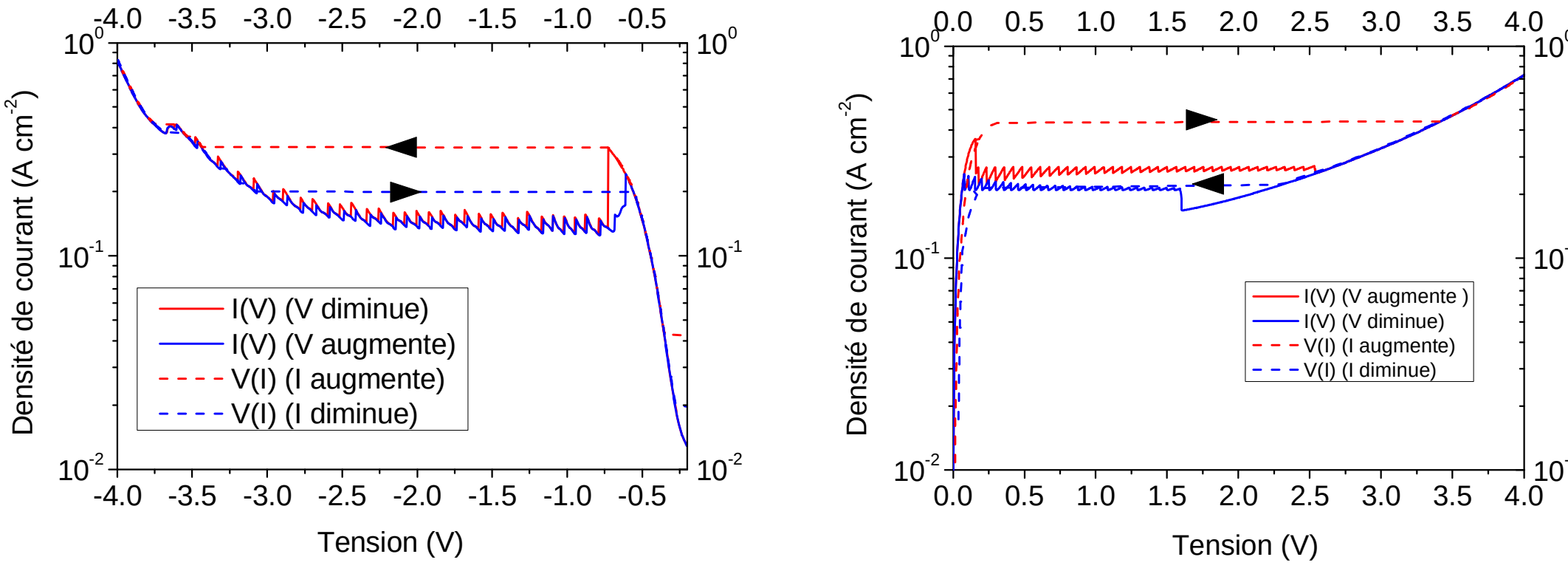

*FIG. 6-11 : Courbe I(V) et V(I) du composant $F_1$, à gauche courbes expérimentales, à droite résultats de simulation*

Dans une courbe *V(I)*, les conditions aux limites imposent la continuité du courant (temporellement et spatialement) dans la structure en régime permanent. Cette constance spatiale ne permet donc plus de remplir/vider un puits localement. La population des puits au coeur de la structure doit rester constante et égale à leur dopage. Il ne peut donc plus se former ni se propager de zone de fort champ, ce qui explique la disparition des dents de scie. Il est toutefois important de noter que ce raisonnement n'est valable qu'en régime permanent ! Il peut en effet exister un transitoire, au moment de l'incrément en courant, pendant lequel la distribution de champ se réarrange sur la structure. Cette évolution doit respecter les conditions aux limites (courant constant en l'occurrence).

### 6.3.5. Effet du désordre

La FIG. 6-12 compare les courbes expérimentales et simulées pour les échantillons de la série E. Comme nous pouvions nous y attendre la dépendance en dopage de ce modèle peut laisser à désirer : il faudrait en effet non seulement ajuster le dopage mais aussi les valeurs du couplage pour des simulations plus précises. Mais la différence majeure entre courbes théoriques et simulées vient de la trop grande régularité dans ces dernières. Nous allons voir que l'introduction de défauts de croissance peut corriger cet écart.

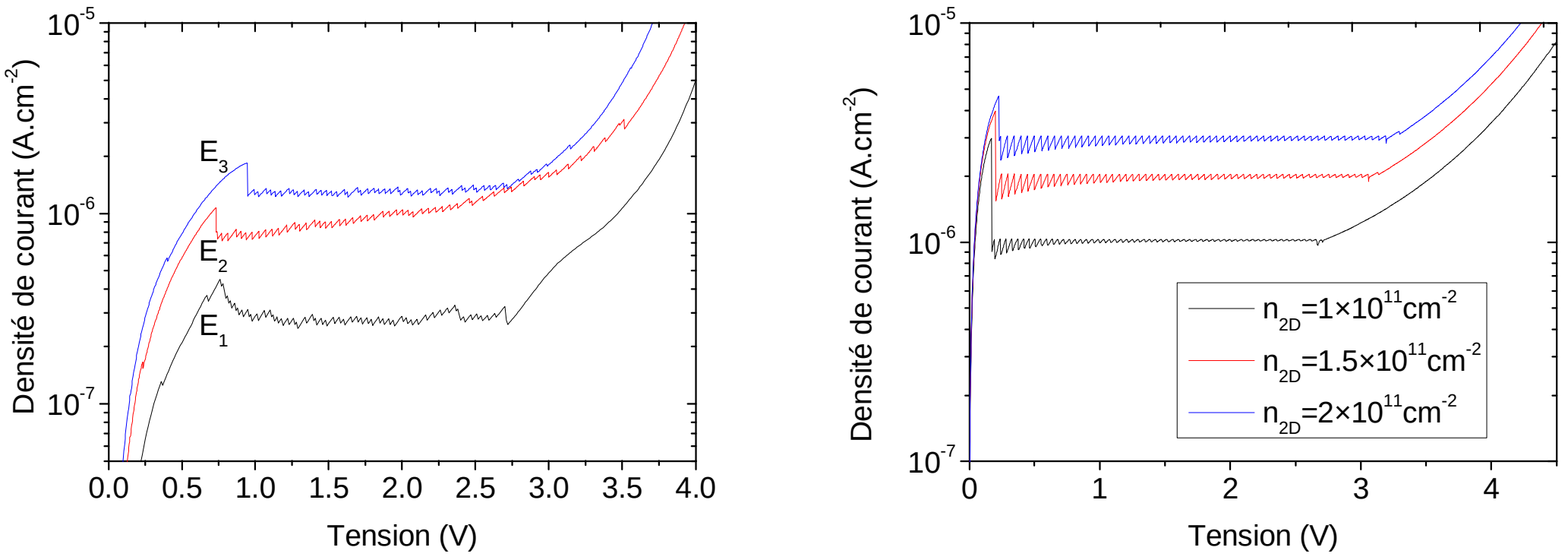

*FIG. 6-12 : Courbe I(V) pour les composants de la série E. A gauche courbes expérimentales et à droite courbes simulées.*

Nous observons en particulier que le nombre de dents de scie est toujours inférieur (entre 55 et 57 sur la série E) au nombre de périodes dans la structure ($N$=60) alors que la simulation prédit l'égalité de ces deux quantités. De plus le composant $E_2$ présente un comportement singulier dans la mesure où il ne présente pas de plateau, voir FIG. 6-12. Nous allons tenter de rendre compte de ces effets en ajoutant à notre modèle quelques défauts de croissance.

Jusqu'ici les super-réseaux étudiés se confondaient avec les structures nominalement proposées à la croissance. Mais nous avons déjà senti, au chapitre précédent, les limites de la MBE. Ici je me propose de rependre les travaux de Wacker et al[189,190] qui ont étudié les effets de fluctuation des paramètres de croissance sur la courbe *I(V)*. Nous nous intéresserons plus particulièrement aux effets liés au dopage et aux écarts de périodicité.

#### ***6.3.5.1. Fluctuation du dopage***

Le contrôle précis du dopage est un art difficile. Des fluctuations entre les populations des puits sont donc envisageables. Commençons par introduire une fluctuation locale de population sur un puits, dans ce cas j'ai choisi le puits du milieu de structure qui a été fortement sur-dopé, voir FIG. 6-13. Les deux courbes *I(V)* se superposent quasiment parfaitement sauf au niveau du 30$^{éme}$ puits. La courbe *I(V)* est donc une sonde à défaut. Connaissant le numéro de la dent de scie défaillante, il est possible de remonter à la période de la structure incriminée.

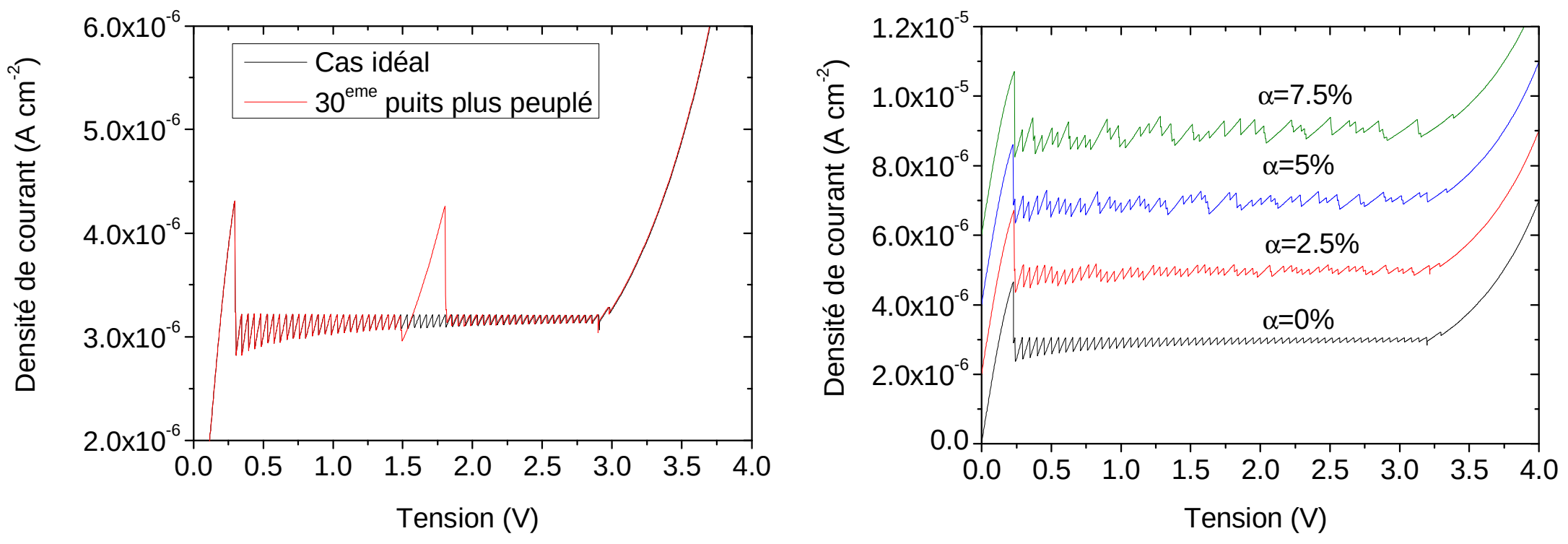

*FIG. 6-13 : A gauche : effet d'un puits plus peuplé (25% de dopage en plus) sur la courbe I(V). A droite : effet de la fluctuation de population des puits sur la courbe I(V), avec décalage de courbes de $2\times10^{-6}$A cm$^{-2}$ entre chaque courbe.*

Afin de généraliser le résultat précèdent nous introduisons une perturbation aléatoire sur la population des puits telle que[190] :

$$N_D^{réel} = N_D^{\text{nominal}} \times (1+\alpha \cdot e_i) \qquad (6\text{-}3)$$

Où $N_D^{nominal}$ est le dopage nominal des puits, $\alpha$ un paramètre qui contrôle l'amplitude du désordre lié au dopage et $e_i$ une variable aléatoire qui varie entre -1 et 1. L'introduction de ce désordre sur les courbes *I(V)* fait en particulier perdre la périodicité parfaite des dents de scie en tension, certaines dents de scie ont une extension en tension plus importante, d'autres sont extrêmement réduites. Le nombre de dents de scie peut ainsi devenir plus faible que celui des périodes.

#### 6.3.5.2. Fluctuation de la périodicité

Il est également possible que le dopage ne soit pas le seul perturbé lors de la croissance mais que la structure présente des irrégularités de périodicité.

J'ai commencé par introduire une fluctuation sur la taille des barrières, sans affecter le couplage entre les niveaux. Pour cela j'ai introduit une variable aléatoire qui peut enlever ou ajouter une monocouche à chaque période. L'effet sur la courbe *I(V)* est présenté sur la figure FIG. 6-14. L'effet obtenu est proche de celui lié à une fluctuation du dopage.

Il faut toutefois également s'attendre à ce qu'une variation de taille de barrière influe sur le couplage. Comme précédemment, nous introduisons à présent une fluctuation aléatoire et continue de la taille de barrière[190] telle que la période de la structure soit

$$L_D^{réel} = L_D^{nominal} \times (1+\Gamma \cdot e_i) \qquad (6\text{-}4)$$

Et j'introduis la fluctuation de couplage qui l'accompagne par une expression empirique.

$$H_{ii}^{réel} = H_{ii}^{nominal} \exp\left(-\frac{L_D^{réel} - L_D^{nominal}}{L_D^{nominal}}\right) \qquad (6\text{-}5)$$

Cette dernière expression traduit simplement la dépendance exponentielle du couplage avec la taille de barrière. Je tiens par ailleurs à préciser que la variation de $L_d$ n'est qu'un paramètre effectif et que la gamme de variation présentée ne correspond pas à des valeurs réalistes de fluctuation lors d'une croissance par MBE. C'est le couplage qui doit, lui, varier dans une gamme réaliste. Les résultats sont rapportés par la figure FIG. 6-14. Comme pour les cas précédents nous voyons apparaître une variation dans l'extension des dents de scie. Nous voyons également que pour de forts taux de désordre, la courbe *I(V)* commence à perdre sa forme de plateau. L'absence de plateau sur le composant $E_2$ serait donc le résultat d'une croissance avec d'importantes fluctuations de taille de périodes.

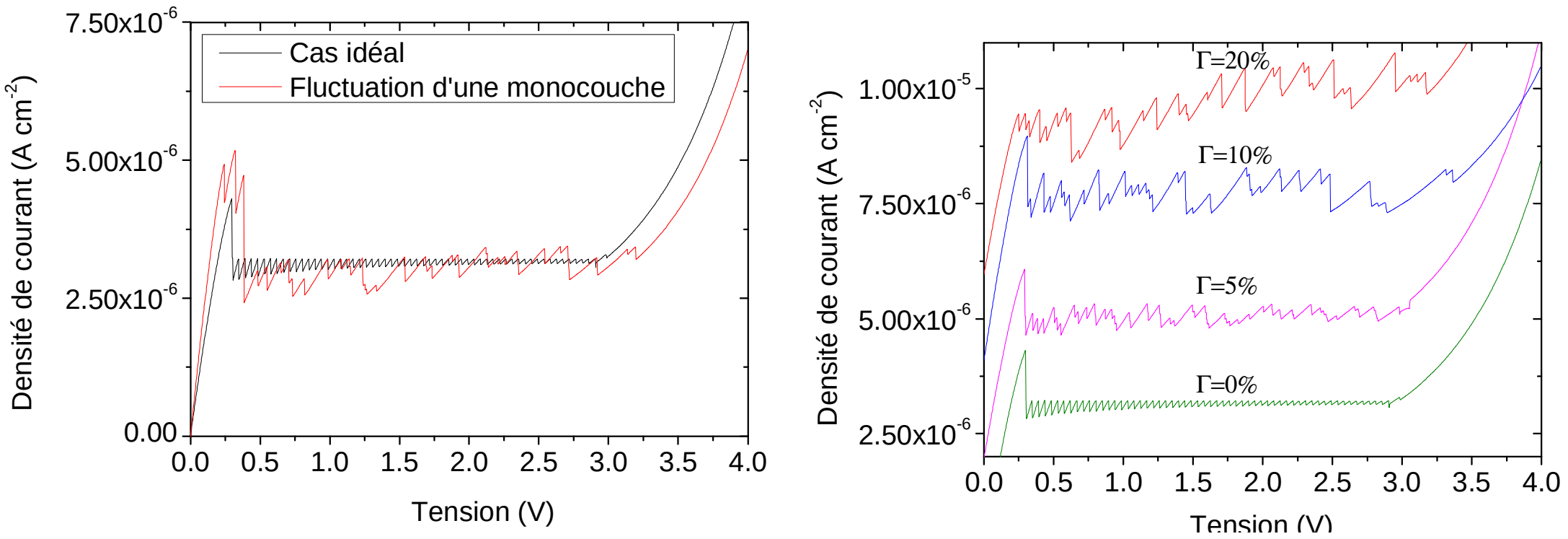

*FIG. 6-14 : A gauche : effet des fluctuations de la taille de la barrière sur la courbe I(V). A droite : Effet de la fluctuation continue de la taille de la barrière sans changement du dopage, avec décalage des courbes de $2\times10^{-6}$A cm$^{-2}$ entre chaque courbe.*

Le désordre dans la structure est donc bien à l'origine des irrégularités des dents de scie. Ce désordre tend à réduire le nombre de dents de scie par rapport au nombre de période. La formes des courbe *I(V)* permet donc d'estimer le niveau de désordre d'une croissance. Quand le désordre reste limité, il est possible de corréler la position du défaut sur l'*I(V)* avec sa localisation spatiale dans la structure. Pour les composants $E_1$ et $E_3$ il est raisonnable d'estimer que les paramètres $\alpha$ et $\Gamma$ sont compris entre 0 et 5%. Par contre, le composant $E_2$ semble avoir subi une croissance moins propre que les deux autres (absence de plateau). On peut certainement imputer cette forme de courbe *I(V)* à des fluctuations dans la périodicité ($\Gamma$>10%).

### 6.3.6. Domaine d'existence des dents de scie.

Les structures de la série E sont relativement proches de celle étudiée au chapitre deux ($L_w$=73Å, $L_b$=350Å, *%Al*=15%, $n_{2D}=3\times10^{11}$cm$^{-2}$) or ces derniers ne présentent pas de dents de scie. Pour expliquer cette différence nous avons fait une simulation des composants de la série E mais en faisant varier la valeur de dopage sur une plus large gamme (presque deux décades). La FIG. 6-15 met en évidence que les dents de scie et l'hystérésis n'existent que sur une très faible gamme de dopage.

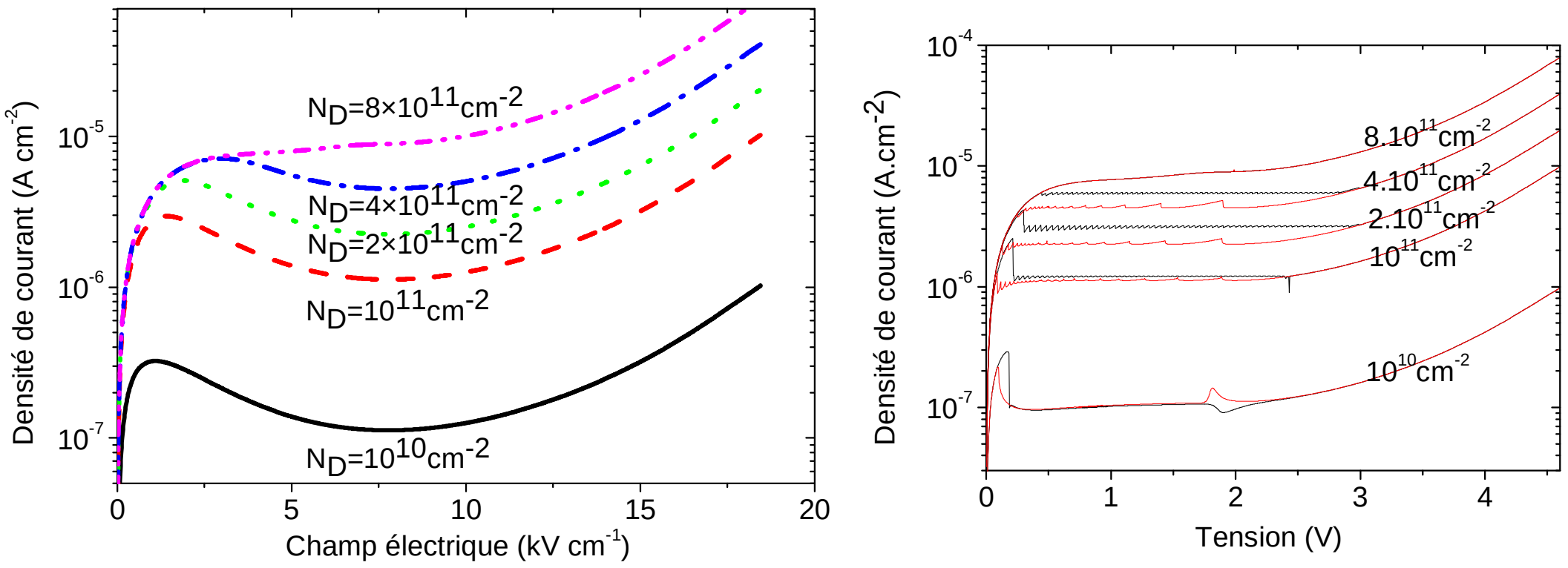

*FIG. 6-15 : A gauche : Courbe J(F) locale, pour différentes valeurs de dopage. A droite : Courbe J(V), pour différentes valeurs de dopage.*

A faible dopage, les dents de scie et l'hystérésis existent en réalité toujours, mais leurs amplitudes ont été considérablement atténuées. En effet dans ce modèle la courbe *J(F)* varie linéairement avec l'amplitude du dopage. Donc en perdant une décade sur le dopage, nous perdons une décade sur l'amplitude de l'hystérésis. De même l'amplitude des dents de scie qui est proportionnelle à la dérivée de la pente est affectée dans les mêmes proportions.

A fort dopage, le phénomène de RDN disparaît de la courbe *J(F)* (absence de vallée), ce qui fait disparaître les dents de scie et l'hystérésis qui leur sont associées. Cette disparition de la RDN traduit le fait qu'à fort dopage la diffusion devient plus forte et qu'alors le ratio courant pic sur courant de vallée tend à devenir plus petit que un. L'absence de dents de scie sur l'*I(V)* de l'échantillon du chapitre deux s'explique par ce dernier phénomène. Le plateau des courbes *I(V)* du chapitre deux a donc bien une origine différente.

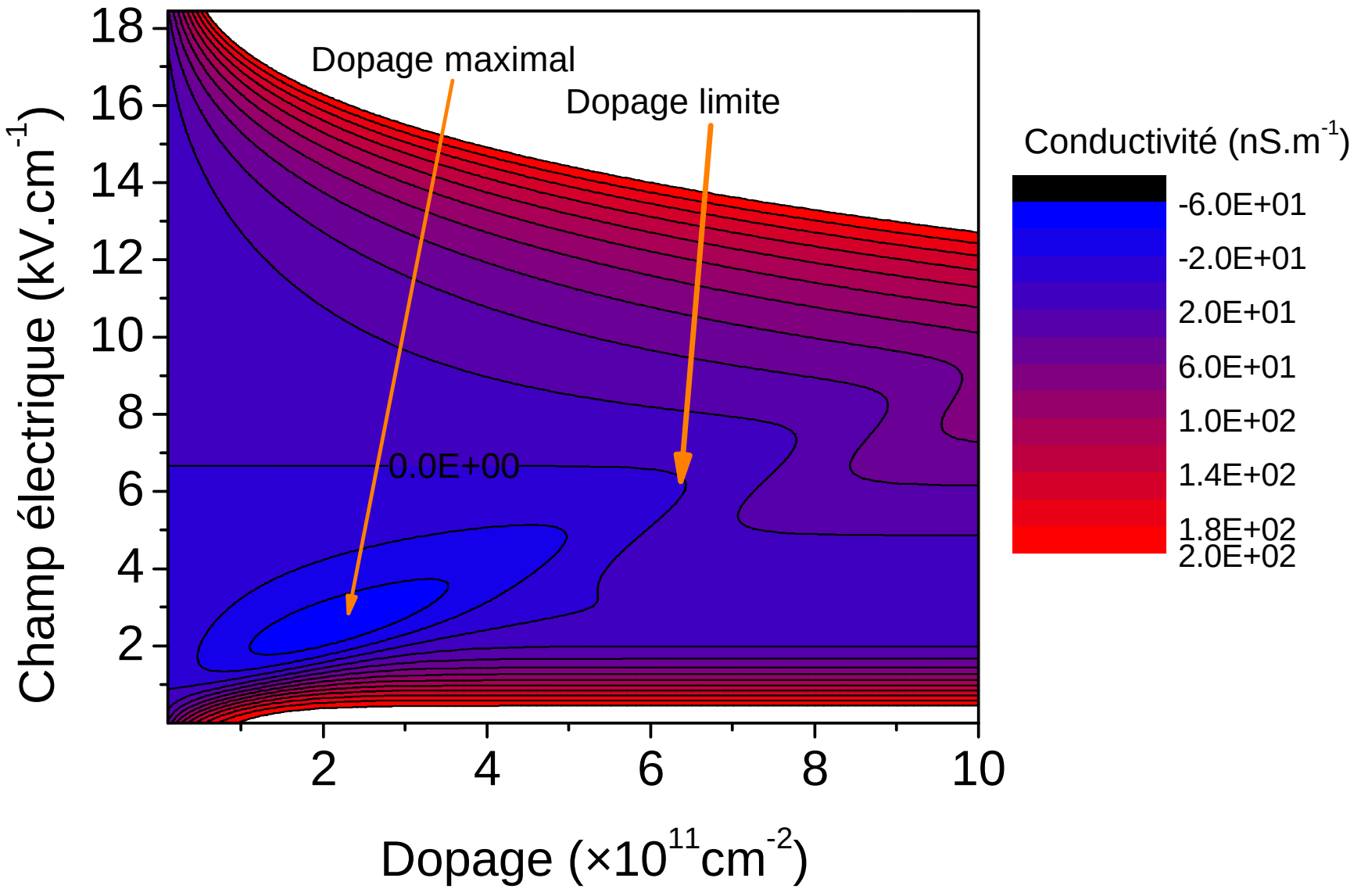

*FIG. 6-16 Conductivité (dérivé de la densité de courant en fonction du champ électrique) pour le composant de la série E en fonction du dopage et du champ électrique. En bleu la zone d'existence de la RDN.*

Plus généralement en traçant la conductivité en fonction du dopage et du champ électrique, voir FIG. 6-16, je suis capable de tracer un diagramme d'existence de la RDN et donc des dents de scie. Ce diagramme fait ressortir l'existence d'un dopage limite pour l'existence de dents de scie au delà duquel les dents de scie disparaisse, ainsi que celle d'un dopage dit maximal pour lequel l'hystérésis sera la plus forte.

A haute température les dents de scie disparaissent également, cela résulte de l'élargissement des niveaux avec la température, qui finit par faire disparaître la RDN[81]. Cela explique pourquoi il n'est jamais fait allusion à de telles dents de scie lors de la caractérisation classique de QWIP.

## 6.4. Influence des dents de scie sur les performances du détecteur.

### 6.4.1. Impact des dents de scie sur la réponse et le bruit

Nous avons été capables de modéliser les dents de scie sur le courant d'obscurité, reste maintenant à identifier leur impact sur les performances du détecteur final. Notons que ce comportement de dents de scie trouve son origine dans le transport en condition d'obscurité mais il n'impacte pas que le courant d'obscurité. Ces dents de scie se retrouvent sur le gain de bruit ainsi que sur la réponse, voir FIG. 6-17. Ceux-ci ont été mesurés avec un pas de 10mV.

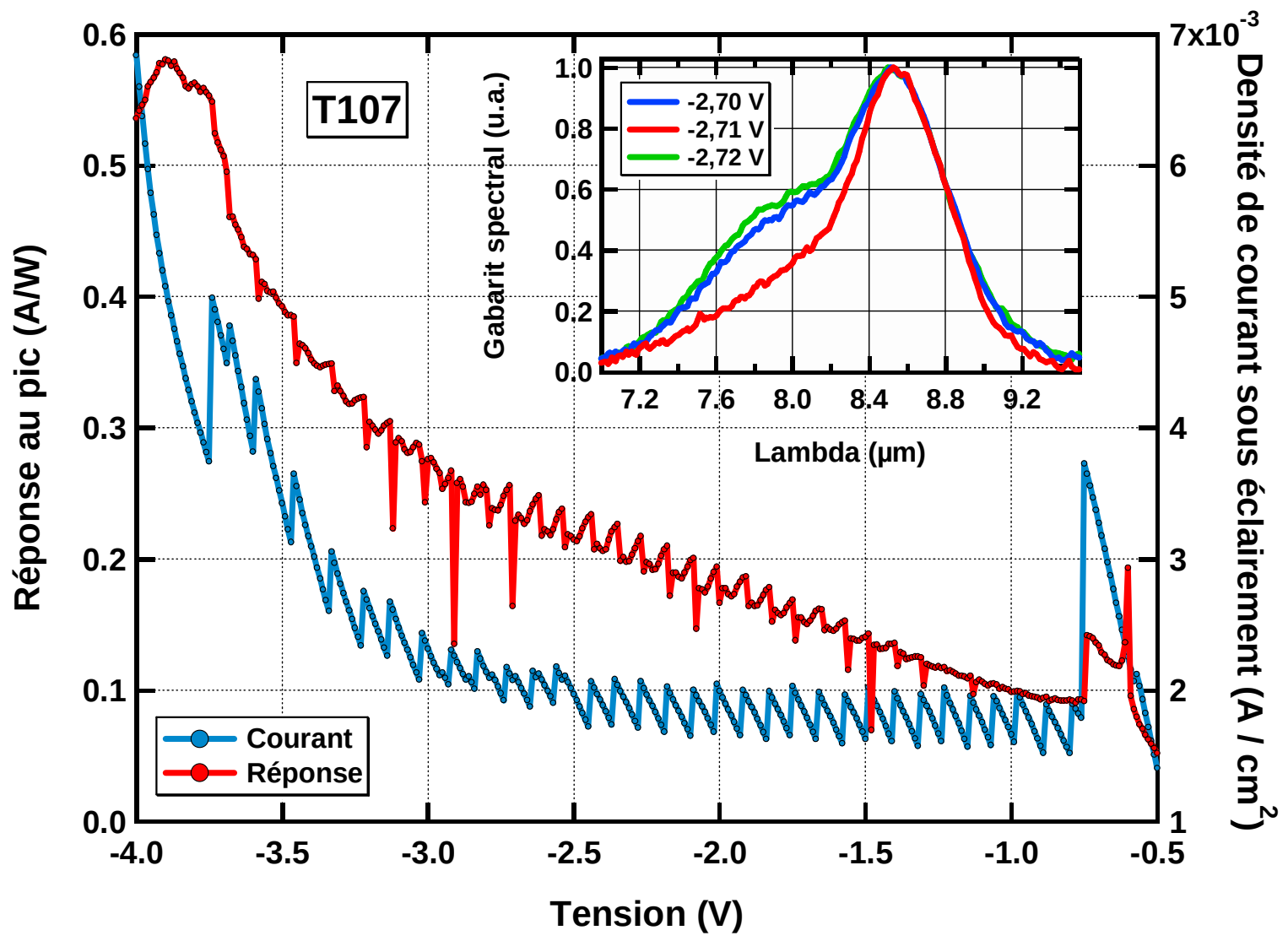

*FIG. 6-17 : Réponse et densité de courant en fonction de la tension pour le composant $F_2$, encart réponse spectrale autour d'une dent de scie – Fourni par V. Gériaux (III-V Lab).*

### 6.4.2. Effet sur la détectivité

Afin de quantifier l'impact de ces oscillations sur les performances du détecteur, nous avons cherché à évaluer la variation de la détectivité autour d'une dent de scie. La présence des dents de scie pose la question de l'uniformité des matrices. Dans le cas d'une uniformité parfaite, les dents de scie ne poseraient pas de problème puisqu'il suffirait de travailler à tension fixée. Les mesures ont mis en évidence que les dents de scie en courant ont une transition verticale qui se fait sur moins de 200µV. Or il peut exister des non uniformités, deux pixels distincts peuvent travailler à des tensions différentes. Cela peut provenir du circuit de lecture (dispersion en tension de polarisation de plusieurs millivolts), ou de légères différences de profondeur de gravure. Sur une matrice il peut en effet exister des variations de profondeur de mesa de l'ordre de 10%, ce qui se traduit par des variations similaires sur la tension ressentie par la couche active. Soit pour des tensions de polarisation située entre -1 et -2V, une dispersion qui peut être de plusieurs dizaines de millivolts. Ces deux sources de non uniformités de la tension peuvent donc créer des variations de tension plus importantes que la tension nécessaire au passage d'une dent de scie à l'autre.

| Grandeurs | Courant d'obscurité | Gain de bruit | Réponse |
|---|---|---|---|
| Valeur moyenne | $1.7\times10^{-2}$ A.cm$^{-2}$ | 0.2 | 0.17 A/W |
| Amplitude d'une dent de scie | $5.4\times10^{-3}$ A.cm$^{-2}$ | 0.03 | 0.03 A/W |

*tab. 6-3 Valeur moyenne et amplitude des dents de scie pour le composant $F_2$ autour de -2V.*

Bien sûr il est particulièrement intéressant d'adresser ce problème quand le détecteur est éclairé par de faibles flux. Dans ce cas il faut s'attendre à ce que les oscillations liées au courant d'obscurité aient un impact plus important. L'expression de la détectivité est, rappelons-le, donnée par :

$$D = \frac{R}{\sqrt{4eg(J_{dark} + R \cdot \frac{hc}{\lambda}\phi)}} \quad (6\text{-}6)$$

Ou *R* est la réponse, *g* le gain de bruit et $J_{dark}$ le courant d'obscurité. Chacun de ces paramètres présentant des dents de scie il faut s'attendre à ce que la détectivité soit également affectée. Le tableau tab. 6-3 donne les valeurs moyenne et l'amplitude d'une dent de scie autour de -2V pour le composant $F_2$. Il faut noter que le courant d'obscurité et le gain de bruit présente la même forme de dents de scie, alors que la réponse présente des dents de scie inversées, c'est-à-dire qu'un maximum du courant correspond à un minium de la réponse, voir FIG. 6-17. Il est donc plus favorable d'un point de vue détectivité de travailler en bas d'une dent de scie en courant, plutôt qu'en haut. J'ai donc tracé la variation de détectivité autour d'une dent de scie en fonction du flux de photons, le résultat est donné par la figure FIG. 6-18. Pour cela la variation de détectivité est calculée selon la formule (6-7).

$$\Delta D = \left| \frac{R}{\sqrt{4e(g+\Delta g)((J_{dark} + \Delta J_{dark}) + R \cdot \frac{hc}{\lambda}\phi)}} - \frac{R + \Delta R}{\sqrt{4eg(J_{dark} + (R+\Delta R) \cdot \frac{hc}{\lambda}\phi)}} \right|$$

*(6-7)*

Quel que soit le flux, la dégradation de détectivité est importante : entre 15 et 35%. La transition autour de $10^{18}$photons·s$^{-1}$·cm$^{-2}$ (ce qui correspond à l'éclairement de ce composant 8.4µm par un corps noir autour de 1000K) vient du fait qu'au-delà de ce flux le courant d'obscurité devient négligeable et qu'alors ses oscillations se font moins ressentir. Néanmoins la variation de détectivité reste de l'ordre de 15%. Notons toutefois que ce calcul peut être pessimiste dans la mesure où il ne prend pas en compte l'éventuelle réorganisation du profil de champ avec l'éclairement.

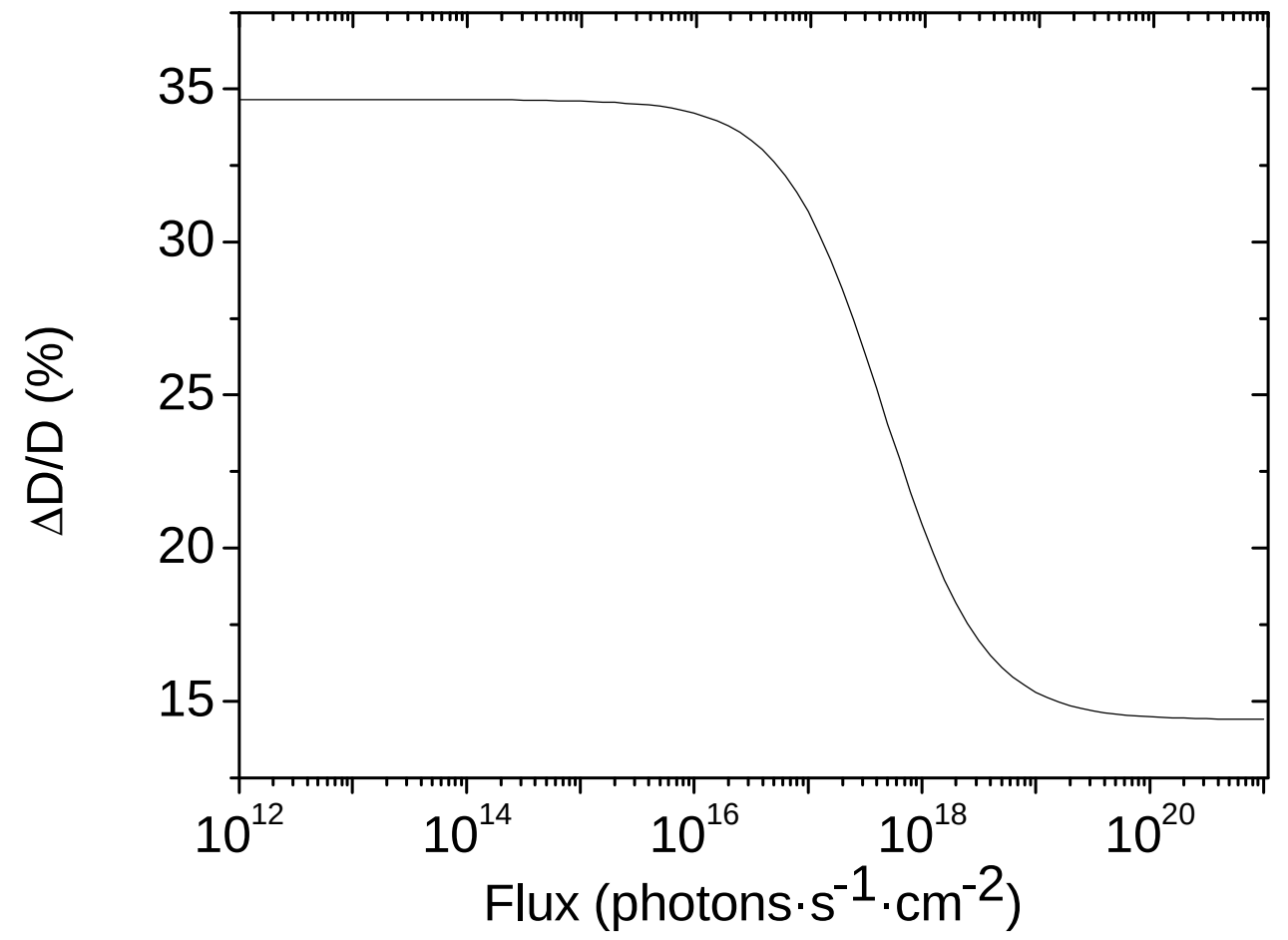

*FIG. 6-18 : Variation de détectivité autour d'une dent de scie en fonction du flux pour le composant $F_2$.*

La conclusion qui s'impose est simple, en présence de dents de scie il est inutile d'espérer faire fonctionner le détecteur en régime tunnel à cause de sa trop forte variation de détectivité. Comme j'ai signalé précédemment qu'à haute température les dents de scie disparaissaient, il ne reste donc plus qu'à réchauffer le composant. On voit donc ici une des limites de l'utilisation basse température des composants QWIP. Une autre possibilité consiste à doper plus fortement la structure, ce qui doit provoquer une envolée du courant tunnel, mais cette augmentation aura le mérite d'être uniforme.

L'impact des dents de scie sur ce composant doit toutefois être modérée car ce composant a des périodes fortement couplés entre elles (faible taille des barrières). Ceci conduit à un courant d'obscurité important (et donc aussi a d'importantes dents de scie) qui même en l'absence de dents de scie aurait déjà été préjudiciable à l'utilisation de ce QWIP. Il faut donc comprendre que ce composant n'est pas optimisé pour des applications bas flux.

## Conclusion

En donnant quelques règles d'évolution simples : courbe *J(F)* locale, conservation de la charge, équation de Gauss, nous avons été capables de reproduire les dents de scie qui peuvent être observée sur certains composants QWIP à basse température. J'ai montré quels étaient les paramètres qui en contrôlaient l'amplitude en tension comme en courant. J'ai mis en évidence que l'introduction du désordre était à l'origine d'une modification de la forme de la courbe *I(V)*. Et enfin j'ai donné des règles quant à l'existence de ces dents de scie. Pour finir nous avons observé que ces dents de scie n'affectaient pas seulement le courant d'obscurité mais aussi la réponse et le gain et nous avons été capable de quantifier l'effet de ces dents de scie sur la détectivité. Ma conclusion étant qu'elles peuvent être extrêmement préjudiciables pour les activités nécessitant une bonne homogénéité de la matrice de détection.

# CONCLUSION

Les applications d'imagerie infrarouge faibles flux requièrent des détecteurs infrarouges performants dans les grandes longueurs d'onde. J'ai montré que les QWIP étaient à présent capables de répondre à ce type de mission. Dans le cas particulier d'une caméra infrarouge destinée à une soufflerie cryogénique, la filière QWIP permettrait de gagner 20K en température de fonctionnement par rapport à la filière Si:Ga. Et de plus, leur large format matriciel évite d'avoir recours à des balayages mécaniques ou à de complexes systèmes optiques.

Toutefois malgré cette réussite, le point faible des détecteurs QWIP dans ce type de mission reste le niveau du courant d'obscurité. C'est donc sur sa réduction qu'ont porté mes efforts en vue d'optimiser les QWIP dédiés aux scénarios faibles flux.

La particularité de ces applications est qu'il est possible de faire opérer les détecteurs à basse température en vue d'obtenir les meilleures performances, quitte à le payer par une cryogénie plus importante. C'est un régime quasi vierge vis-à-vis de l'optimisation des QWIP.

J'ai testé différentes possibilités de modélisation du transport tunnel. Les approches classiques basées sur des mini-bandes ou sur l'approximation WKB ont rapidement montré leurs limites quant à la reproduction des phénomènes observés sur les courbes *I(V)* de nos composants : forme de plateau, dépendance vis-à-vis du dopage.

J'ai donc développé un code de simulation du transport électronique basé sur une approche de hopping dans les super-réseaux faiblement couplés. Ce code est finalement une boîte à outils qui me permet de simuler les propriétés d'une large gamme d'hétérostructures.

Grâce à cet outil de simulation j'ai pu montrer le rôle prédominant de l'interaction entre les électrons et les impuretés ionisées. En conséquence le profil de dopage s'avère déterminant quant au niveau du courant d'obscurité. A l'aide de cet outil j'ai été capable de reproduire la dépendance du courant avec l'amplitude du dopage, ainsi que l'effet de la ségrégation sur le niveau du courant. Mais surtout j'ai montré que le positionnement du dopage était crucial pour la réduction du courant. J'ai donc dessiné et fait réaliser une nouvelle série d'échantillons selon mes spécifications, qui ont conduit à une réduction de 50% du taux de diffusion. Ceci constitue le résultat central de cette thèse. D'autres structures plus prometteuses ont également été proposées, mais leur réalisation peut s'avérer ardue. Maintenant que le principe de cette méthode a été validé il reste à l'appliquer à des structures plus complexes comme les QCD ou les QWIP bicolores dans lesquels une augmentation de la taille de barrière n'est pas toujours envisageable.

Par ailleurs, afin d'expliquer complètement la forme originale de nos courbes *I(V)* (forme de plateau), nous avons dû apporter une modification à l'approche classique de hopping. Cette modification permet la prise en compte de la durée de vie finie des états de Wannier-Stark. J'ai alors été en mesure d'apporter une interprétation quantique au plateau des *I(V)*. Ce dernier résulte de la compétition entre le phénomène de résistance différentielle négative et celui d'abaissement de barrière par le champ électrique. Ceci constitue le second résultat le plus important de cette thèse.

Mon outil de simulation s’est par ailleurs montré capable de simuler des structures plus complexes que le QWIP idéal. Nous l’avons utilisé pour quantifier l’incidence de la non idéalité des structures. Le couplage de cet outil de simulation à la microscopie électronique en mode STEM s’est avéré un duo efficace pour quantifier l’impact des non-idéalitées d’interface. Enfin la prise en compte du désordre coulombien m’a amené à conclure que la structure QWIP classique est relativement robuste vis-à-vis des défauts de croissance.

Par ailleurs j’ai utilisé notre outil de simulation du transport pour évaluer les performances d’un QCD THz. J’ai identifié le rôle déterminant que peut avoir la qualité des interfaces vis à vis de la résistance du système. Une comparaison avec les mesures est prévue dans un futur proche. Il semble que cette première réalisation d’un QCD THz par l’équipe MPQ/Thales ne soit pas complètement optimisée, mais le fait d’à présent pouvoir simuler ces structures devrait permettre rapidement de dessiner des composants présentant une résistance à l’origine très largement supérieure.

Je me suis enfin intéressé aux effets non locaux du transport afin de quantifier les effets liés à une possible distribution inhomogène du champ électrique. Je me suis attaché à reproduire l’hystérésis et les dents de scie qui sont observés sur les courbes *I(V)* de certains QWIP. J’ai identifié les paramètres qui contrôlent la taille et l’amplitude des dents de scie ainsi que de l’hystérésis. Et j’ai donné les conditions de leur existence. J’ai également montré que la présence de ces dents de scie était extrêmement préjudiciable pour la détection bas flux.

Pour conclure j’ai été capable d’identifier le point faible des QWIP dans nos applications et j’ai su modéliser le courant d’obscurité. Par un travail d’ingénierie d’élément de matrice nous avons pu en réduire l’amplitude. Nos prédictions ont été expérimentalement confirmées par la réalisation d’un échantillon fait selon nos spécifications.

Cette thèse ayant été faite en collaboration entre trois laboratoires, il est intéressant de comprendre quelles sont les retombées pour chacun d’eux.

Pour l’ONERA, j’ai développé un banc de caractérisation électro-optique des détecteurs sous faible flux. Ce dernier pourra être utilisé pour la caractérisation d’autres détecteurs, plutôt haute longueur d’onde, qu’ils proviennent de la filière QWIP ou non. J’ai également contribué à la généralisation du programme de simulation des performances d’instruments optiques (voir chapitre 2) afin de le rendre compatible avec les détecteurs de la filière QWIP. Enfin cette thèse s’inscrit parfaitement dans le cadre du travail d’expertise du laboratoire pour la DGA, J’ai en effet caractérisé des composants QWIP à l’état de l’art, tout en contribuant à l’orientation de la filière via la proposition de détecteurs optimisés.

Pour le MPQ, le principal apport de ma thèse réside dans l’outil de simulation du transport que j’ai pu développer. Il a déjà été utilisé au cours de cette thèse pour la prédiction de performances de QCD THz. Et il devrait continuer à être utilisé pour le design d’autres structures.

Pour le III-V Lab, j’ai pu faire des mesures à L’ONERA avec des outils de caractérisation des composants infrarouges souvent complémentaires de ceux du III-V lab. Les mesures de linéarité de la réponse sous de faibles flux sont des mesures qui ne peuvent actuellement pas être faites à Thales. Enfin la conception des composants à dopage déplacé offre un outil de plus pour l'optimisation des structures limitées par l'interaction electron-dopage.

# PERSPECTIVES

Suite à ce travail je peux identifier plusieurs voies dans lesquelles il serait intéressant de poursuivre ces recherches :

- Ce travail a mis en évidence la possibilité de modéliser de façon quantique le transport microscopique dans les QWIP. Par ailleurs nous avons étudié des effets non locaux, en utilisant alors une équation simplifiée pour le transport microscopique. L'étape suivante va donc consister à coupler ces deux échelles. C'est-à-dire à créer un outil de simulation qui modélise de façon quantique le transport sur l'ensemble de la structure.

- Nous souhaitons Réaliser des échantillons à dopage décalé et symétrique. Ce travail devra se faire en étroite collaboration avec la personne en charge de l'épitaxie afin que la structure soit réalisable et le plus proche possible de la structure nominale. C'est un travail dans la continuité directe de celui du chapitre 5 et qui devrait encore permettre de gagner un facteur deux sur le courant d'obscurité.

- Nous nous intéressons à présent à l'application de cette physique à de nouveaux matériaux. Il existe actuellement un engouement important en détection infrarouge pour les super réseaux de type II. Or jusqu'à récemment ces structures étaient limitées par des aspects matériaux. Ce n'est à présent plus le cas. Par contre la modélisation du transport y est encore à ses balbutiements, il devient donc pertinent d'appliquer les modèles physiques développés pour les super réseaux de type I, au type II. Ce travail commence au sein de l'équipe CIO.

- Un travail important est fait sur l'optimisation du transport dans les hétérostructures et parallèlement un travail tout aussi important est fait pour l'optimisation des structures de couplage avec la lumière (plasmonique, super lentille…). L'étape suivante sera donc de disposer d'un outil de simulation qui couple ces deux aspects. Cela demandera un fort investissement numérique, mais ce sera le prix à payer pour disposer d'un outil de simulation complet des détecteurs.

# PUBLICATIONS

## 1. Articles

1. *Ultimate performance of Quantum Well Infrared Photodetectors in the tunneling regime*, E. Lhuillier, I. Ribet-Mohamed, M. Tauvy, A. Nedelcu, V. Berger and E. Rosencher, Infrared phys. tech. 52, 132-137 (2009).
2. *Modeling of dark current in midinfrared quantum well infrared photodetectors*, F. Castellano, F. Rossi, J Faist, E. Lhuillier and V. Berger, Phys. Rev. B 79, 205304 (2009).
3. *Interface roughness transport in THz quantum cascade detectors* E. Lhuillier, I. Ribet-Mohamed, E. Rosencher, G. Patriarche, A. Buffaz, V. Berger, and M. Carras, App. Phys. Lett. 96, 061111 (2010).
4. *Low temperature quantum transport in weakly coupled superlattices*, E. Lhuillier, I. Ribet-Mohamed, A. Nedelcu, V. Berger and E. Rosencher, Phys. Rev. B 81, 155305 (2010).
5. *Quantum well infrared photodetectors hardiness to the non ideality of the energy band profile*, E. Lhuillier, N. Pere-Laperne, I. Ribet-Mohamed, E. Rosencher, G. Patriarche, A. Buffaz, V. Berger, A. Nedelcu, M. Carras, J. Appl. Phys. 107, 123110 (2010) (arXiv:1002.3543v2).
6. *15µm Quantum well infrared photodetector for thermometric imagery in cryogenic windtunnel*, E. Lhuillier, I. Ribet-Mohamed, N. Péré-Laperne, M. Tauvy, J. Deschamps, A. Nedelcu, E. Rosencher, accepted to Infrared phys. tech. (arXiv:1003.2310v1).
7. *Quantum scattering engineering of quantum well infrared photodetectors in the tunnelling regime*, E. Lhuillier, E. Rosencher, I. Ribet-Mohamed, A. Nedelcu, L. Doyennette, V. Berger, accepted in J. App. Phys. (arXiv:1005.1798v1).
8. *Saw tooth patterns on the I(V) curve of quantum well infrared photodetectors: from modeling to the impact on the detector performance*, E. Lhuillier, V. Guériaux, V. Trinité and M. Carras, in preparation for J. Appl. Phys.

## 2. Proceedings

1. *Quantum transport in Quantum Well Infrared Photodetectors in the tunneling regime*, E. Lhuillier, I. Ribet-Mohamed, A. Nedelcu, V. Berger and E. Rosencher, Infrared phys. tech. 52, 247 (2009). (QSIP 2009 conference paper).
2. *Modeling of dark current in mid-infrared quantum-well infrared photodetectors*, F. Castellano, R. C. Iotti, F. Rossi, J. Faist, E. Lhuillier, V. Berger Infrared phys. tech. 52, 220 (2009). (QSIP 2009 conference paper).
3. *Dark current reduction in high wavelength quantum well infrared photodetector operating at low temperature*, E. Lhuillier, E. Rosencher, I. Ribet-Mohamed, A. Nedelcu, L. Doyennette, V. Berger, accepted in Infrared phys. Tech. (QSIP 2010 conference paper).

## 3. Présentations sans proceedings

1. *Optimisation du transport dans les détecteurs infrarouges à puits quantiques*, Emmanuel Lhuillier, Isabelle Ribet-Mohamed, Vincent Berger, Emmanuel Rosencher, Congrès de la SFP, école polytechnique juillet 2009.
2. *Using STEM measurements to evaluate the transport in multi quantum wells structures*, Emmanuel Lhuillier, Isabelle Ribet-Mohamed, Vincent Berger, Emmanuel Rosencher, Ecole ESONN, Grenoble, août 2009.
3. *De la caractérisation à l'optimisation des détecteurs infrarouges à puits quantiques*, Emmanuel Lhuillier, Isabelle Ribet-Mohamed, Vincent Berger, Emmanuel Rosencher, Journée scientifique de la DGA, Paris, mai 2010.
4. *Optimisation du transport dans les détecteurs infrarouges à puits quantiques*, Emmanuel Lhuillier, Isabelle Ribet-Mohamed, Vincent Berger, Emmanuel Rosencher, JNMO 2010, les Issambres, septembre 2010.

# BIBLIOGRAPHIE